\newif\iffigs\figstrue
\documentstyle[12pt,epsf]{report}
\iffigs
   \input{epsf}
\else
\fi
\newcommand{\eqn}[1]{(\ref{#1})}

\newsavebox{\uuunit}
\sbox{\uuunit}
    {\setlength{\unitlength}{0.825em}
     \begin{picture}(0.6,0.7)
        \thinlines
        \put(0,0){\line(1,0){0.5}}
        \put(0.15,0){\line(0,1){0.7}}
        \put(0.35,0){\line(0,1){0.8}}
       \multiput(0.3,0.8)(-0.04,-0.02){12}{\rule{0.5pt}{0.5pt}}
     \end {picture}}

\def\IP{\relax{\rm I\kern-.18em P}}

\font\cmss=cmss10 \font\cmsss=cmss10 at 7pt

\def\inbar{\vrule height1.5ex width.4pt depth0pt}
\def\IC{\relax\,\hbox{$\inbar\kern-.3em{\rm C}$}}
\def\IG{\relax\,\hbox{$\inbar\kern-.3em{\rm G}$}}
\def\IB{\relax{\rm I\kern-.18em B}}
\def\ID{\relax{\rm I\kern-.18em D}}
\def\IL{\relax{\rm I\kern-.18em L}}
\def\IF{\relax{\rm I\kern-.18em F}}
\def\IK{\relax{\rm I\kern-.18em K}}
\def\IH{\relax{\rm I\kern-.18em H}}
\def\II{\relax{\rm I\kern-.17em I}}
\def\IN{\relax{\rm I\kern-.18em N}}
\def\IP{\relax{\rm I\kern-.18em P}}
\def\IQ{\relax\,\hbox{$\inbar\kern-.3em{\rm Q}$}}
\def\bfzero{\relax\,\hbox{$\inbar\kern-.3em{\rm 0}$}}
\def\IR{\relax{\rm I\kern-.18em R}}
\def\ZZ{\relax\ifmmode\mathchoice
{\hbox{\cmss Z\kern-.4em Z}}{\hbox{\cmss Z\kern-.4em Z}}
{\lower.9pt\hbox{\cmsss Z\kern-.4em Z}}
{\lower1.2pt\hbox{\cmsss Z\kern-.4em Z}}\else{\cmss Z\kern-.4em
Z}\fi}
\def\bfone{\relax{\rm 1\kern-.35em 1}}

\def\bfone{\relax{\rm 1\kern-.35em 1}}
\font\cmss=cmss10 \font\cmsss=cmss10 at 7pt

\def\cA{{\cal A}} 
\def\cC{{\cal C}} 
\def\cF{{\cal F}} \def\cG{{\cal G}}

\def\cN{{\cal N}} 
 \def\cQ{{\cal Q}}
 
\newcommand{\be}{\begin{equation}}
\newcommand{\ee}{\end{equation}}
\newcommand{\bea}{\begin{eqnarray}}
\newcommand{\eea}{\end{eqnarray}}
\let\la=\label  
\def\tilde{\widetilde}
\def\bar{\overline}

\def\hat{\widehat}

\def\Coe#1.#2.{{#1\over #2}}

\def\coe#1.#2.{\relax{\textstyle {#1 \over #2}}\displaystyle}

\def\to{\rightarrow}
\def\notin{\hbox{{$\in$}\kern-.51em\hbox{/}}}

\def\IE{\relax{{\rm I\kern-.18em E}}}

\def\IGam{\relax{{\rm I}\kern-.18em \Gamma}}

\begin{document}
\begin{titlepage}
\begin{center}
{\LARGE 
Dualities in Supergravity and Solvable Lie Algebras}\\
\vfill
{\large  Mario Trigiante   } \\
\vfill
{\small
Department of Physics, University of Wales Swansea, Singleton Park,\\
Swansea SA2 8PP, United Kingdom
}
\end{center}
\vfill
\begin{center}
{\bf Abstract}
\end{center}
{\small 
The purpose of the present thesis is to give a self--contained review
of the solvable Lie algebra approach to supergravity problems
related with S, T and U dualities. After recalling the general
features of dualities in both Superstring theory and Supergravity , we introduce the solvable Lie
algebra formalism as an alternative description of the scalar manifold
in a broad class of supergravity theories. 
It is emphasized how
this mathematical technique on one hand allows to achieve a
geometrical intrinsic characterization of Ramond--Ramond,
Neveu--Schwarz and Peccei--Quinn scalars, once the supergravity theory
is interpreted as the low energy limit  of a suitably compactified
superstring theory, on the other hand provides a convenient framework in
which to deal with several non--perturbative problems. Using solvable Lie 
algebras for instance we find a general mechanism for spontaneous 
$N=2$ to $N=1$ local supersymmetry breaking. Moreover solvable Lie algebras are
used to define a general method for studying systematically BPS
saturated Black Hole solutions in supergravity.
}
\vspace{2mm} \vfill \hrule width 3.cm
\end{titlepage}
\tableofcontents
\setcounter{chapter}{-1}
\chapter{Introduction}
Recently considerable results were obtained towards a deeper 
understanding of some non--perturbative aspects of both superstring
and supergravity theories. The basic concept underlying these latest 
developments is {\it duality} \cite{schw}. \par
Superstring theory was originally introduced as a promising 
proposal for a fundamental quantum theory in which gravity is unified to the
other elementary interactions in a consistent and finite framework.
It is formulated in 10 space--time dimensions, but, if compactified 
to a D--dimensional space--time, it's low--energy states are described by
an effective D--dimensional N--extended {\it supergravity} theory. 
For a suitable choice of the initial superstring theory and of its
compactification to 4 dimensions, the low--energy supergravity seems to have
the right phenomenological content of fields to be the candidate 
for a theory describing our physical world. Nevertheless, 
superstring theory soon revealed not to be fundamental, since it is not
unique. Indeed there are 5 known superstring theories (Type IIA, Type IIB,
Type I, Heterotic $E_8\otimes
E_8^{\prime}$, Heterotic $SO(32)$), and a supergravity
 in D--dimensions ($D<10$) usually describes the low--energy content of 
more than one superstring theory compactified on different manifolds.
The discovery of string--string dualities made it possible to
reduce the number of inequivalent superstring theories.\par
By string--string duality one denotes a correspondence between regimes
of two different superstring 
theories which preserves the spectrum and the interactions.
Such a correspondence allows to consider the two related 
theories as different mathematical descriptions of a same one.
By exploiting the duality symmetries it was indeed possible to 
relate all the five known superstring theories 
 and therefore view them as perturbative
relizations on different backgrounds of a larger non--perturbative quantum 
theory in $11$ or $12$ space--time dimensions \cite{witten}, \cite{vasch}
(named M-theory and F-theory respectively). Even if the physical content 
of the latter is not known so far, they are expected to admit all the 
known dualities as exact symmetries.\par
In order to characterize the concept of duality in a more
precise fashion, let us consider a colsed superstring theory compactified on a 
$d$--dimensional compact manifold ${\cal K}_d$. The geometry of ${\cal K}_d$
is completely defined by the vacuum expectation values of certain low energy
excitations, denoted by $t^n$, which define its {\it moduli}.
Moreover, among the scalar $0$--modes of the superstring theory, the dilaton
 $\phi$ plays the important role of defining, through its vacuum expectation 
value, the string coupling constant $g$ : 
$g=exp\left(\langle \phi\rangle\right) $. The main feature of the scalars
$t^n$ and $\phi$ is that they derive from fields which in the original
$10$--dimensional string theory couple directly to the geometry of the 
world--sheet
and are called ({\it Neveu--Schwarz})--({\it Neveu--Schwarz}) (NS). 
The remaining 
scalar $0$--modes are associated with the internal components (i.e. components 
along the directions of ${\cal K}_d$) of states in the original 
$10$--dimensional theory
whose vacuum expectation values do not enter the superstring action. 
These states
correspond to an even boundary condition for both the left and right fermionic
movers 
of the closed string and are called {\it Ramond--Ramond} (RR) states. The 
corresponding scalar $0$--modes of the compactified theory are thus called
{\it Ramond--Ramond} scalars. All the low--energy scalar excitations of the 
string theory compactified on ${\cal K}_d$ define the scalar content
of a suitable {\it effective} N--extended supergravity theory in 
$D=10-d$ space--time dimensions.\par
Duality transformations are mappings between different backgrounds 
on which the superstring theory is realized. They are therefore  
represented by thansformations on the scalar fields of the 
low--energy supegravity, leaving the spectrum and the couplings of the 
string theory invariant. For instance the T--duality acts on the
$t^n$ and RR scalars, leaving the dilaton invariant. Its action amounts 
to a transformation of the inner space geometry
and is a generalization of the mapping $R\rightarrow \alpha^{\prime}/R$ 
($\alpha^{\prime}$ being the string tension) for 
a string theory compactified on a circle. Since it does not affect $\phi$,
 it is {\it perturbative} that is it 
can be verified to be an exact symmetry of the theory order by order 
in the string coupling constant $g$. On the other hand, the effect of
the conjectured
S--duality, roughly speaking, is to map $\phi$ into $-\phi$, which implies
the transformation $g\rightarrow 1/g$. It is therefore a 
{\it non--perturbative} duality which relates  
the strong coupling regime of a theory to the weak coupling regime 
on an other.
In general, verifying the existence of a non--perturbative duality would 
require the knowledge of the complete non--perturbative spectrum of a 
given superstring theory
which is not accessible since string theories are defined perturbatively.
Therefore S--duality could only be conjectured.
A common characteristic of both T and S--duality is that they
do not mix RR and NS scalars. \par
Finally a larger non--perturbative U--duality was recently conjectured
\cite{huto} whose main feature is to transform all the scalars into
 each other
as if they were treated on an equal footing. U--duality therefore 
includes also transformations which do not preserve the RR or NS connotation 
of the scalar fields. The conjecture for the two
non--perturbative S \cite{sen} and U \cite{huto} dualities
 was based on evidences drawn from the comparison between the perturbative 
spectrum  of a string theory and some of its known non--perturbative
 BPS states. \par
Since suitably compactified superstring theories mapped into each other 
by duality transformations have the same low--energy effective field theory, 
dualities between superstring theories should be strictly related to 
global symmetries
of the underlying supergravity. The latter, generally called  
{\it hidden symmetries} or {\it dualities} (in the supergravity framework)
are well known and have been widely studied since the early eighties
\cite{cj},\cite{gz}.
Lets consider this relationship in more detail, for it shows that a 
proper starting point for a discussion on  duality in superstring theory 
is the analysis of the global symmetries of its underlying low--energy
 effective theory.\par
In supergravity the scalar fields ($\Phi^I$) are described by a D--dimensional 
$\sigma$--model (D being the number of space--time dimensions), that is  
they are local coordinates of a {\it non--compact} 
 Riemannian manifold ${\cal M}$ and the scalar action is invariant under 
the {\it isometries} of ${\cal M}$ (i.e. diffeomorphisms leaving 
the metric on ${\cal M}$ invariant). The isometry--group ${\cal G}$
of ${\cal M}$ is promoted to be a global symmetry--group 
of the field equations and 
the Bianchi identities when its action on the scalar fields is associated with
a suitable transformation of the vectors or in general p--forms (duality transformation) entering the same supermultiplets as the scalars (this double action
of ${\cal G}$ is required if supersymmetry is large enough, i.e. $N\ge 2$ in 
$D=4$). The NS scalars $\phi$ and $t^n$ span two submanifolds ${\cal M}_s$
and ${\cal M}_t$ of ${\cal M}$ respectively. In particular ${\cal M}_t$
is the moduli--space of the internal compact space ${\cal K}_d$ and its
isometry group ${\cal G}_t\subset {\cal G}$ is a candidate for the T--duality
group at the superstring level, since it leaves $\phi$ invariant. 
Nevertheless, it turns out 
that, in order to preserve the superstring energy levels, the  
T--duality group must be a suitable restriction to the integers
${\cal G}_t (Z)$ of ${\cal G}_t$. This statement was verified order by order 
in the string coupling expansion. In light of this observation, 
the previously mentioned S and U--duality conjectures may now be restated as
 the hypothesis that suitable descrete
 versions of the isometry group ${\cal G}_s$ of ${\cal M}_s$ and of the whole
${\cal G}$ correspond to the S and U--dualities respectively,
at the superstring level. The restriction to the integers is required already 
in the framework of the effective supergravity once quantization conditions are taken into account i.e. once it is demanded that the hidden symmetries
 preserve the lattice spanned by the integer valued electric and magnetic 
charges carried by the solutions of the theory.\par
In order to retrieve the main ideas expressed so far, let us consider for
 instance type IIA superstring theory compactified on a six--torus 
(${\cal K}_d=T_6$). Its low--energy states are described by an $N=8$
supergravity in $D=4$ dimensions. The scalar manifold \cite{cj} is:
\begin{equation}
{\cal M}\, =\, \frac{E_{7(7)}}{SU(8)}\nonumber
\label{mn8}
\end{equation}
The moduli $t^n$ of $T_6$ are $G_{ij}$
and $B_{ij}$ ($i=1,...,6$),
 internal components of the space--time metric $G_{MN}$
and the antisymmetric tensor field $B_{MN}$ in $10$--dimensions.
They naturally span the moduli--space of $T_6$: 
\begin{equation}
{\cal M}_t\, =\, \frac{O(6,6)}{O(6)\otimes O(6)}\nonumber
\label{mtn8}
\end{equation}
while the dilaton $\phi$ and the {\it axion} $B_{\mu\nu}$ span the manifold:
\begin{equation}
{\cal M}_s\, =\, \frac{SL(2,\IR)}{O(2)}\nonumber
\label{msn8}
\end{equation}
The isometry groups of these three manifolds are respectively:
${\cal G}=E_{7(7)}$, ${\cal G}_t=O(6,6)$ and ${\cal G}_s=SL(2,\IR)$.
The string--string T--duality group has been verified to be a 
restriction to the 
integers of ${\cal G}_t$, that is $O(6,6;Z)$, while the conjectured S and 
U--duality groups are $SL(2,Z)$ and $E_{7(7)}(Z)$ respectively. 
\par 
As previously mentioned, evidences for these conjectures were drawn from the analysis of the 
BPS solitonic spectrum of the effective supergravity theory, under the 
hypothesis that these states already include the known BPS excitations
of the fundamental string. BPS solitons are those solitonic solutions of 
a supergravity theory which saturate the Bogomolnyi bound, i.e. whose masses 
equal one or more eigenvalues of the central charge. If we consider theories 
having a large enough supersymmetry (e.g. $N\ge 4$ in $D=4$), the central charge is not affected by quantum corrections and therefore the value of the
BPS mass computed semiclassically is exact. Moreover a duality transformed 
solution is still a solution of the supergravity theory since 
duality transformations are symmetries of the field equations and 
Bianchi identities. This property, together with the fact that the BPS 
condition is duality invariant, implies that all the BPS solitons of 
a supergravity theory fill a representation of the U--duality group (therefore
 also of the T and S--duality groups) suitably restricted to the integers 
${\cal G}(Z)$ in order to preserve the charge lattice. The conjecture that
${\cal G}_s(Z)$ and the whole ${\cal G}(Z)$ are symmetries of the superstring 
spectrum requires the perturbative electrically charged string excitations
fulfilling the BPS conditions to be in the same duality representation
of a magnetically charged BPS soliton 
state of the low--energy effective supergravity
 (indeed S--duality exchanges electric and
magnetic charges). But this state already 
belongs to a representation of ${\cal G}(Z)$ which is completely filled by BPS 
solitons, therefore the S and U--duality conjecture implies that the 
electrically charged superstring excitations fulfilling the BPS condition
should be identified with equally charged BPS solitons of the low--energy
 effective theory. The fact that two superstring theories which differ
at a perturbative level have, when compactified to a lower dimension, the 
same low--energy effective supergravity theory, and therefore the same
BPS solitonic spectrum, is an evidence that they should correspond 
through a non perturbative duality. An example of two theories which are 
conjectured to be one the S-dual of the other is given by Type IIA superstring 
compactified on $K_3\times T_2$ and heterotic compactified on $T_6$. Their 
common low--energy effective theory is $D=4$, $N=4$ supergravity.
In this case the S and U--duality groups of the two theories are :
${\cal G}_s(Z)=SL(2,Z)$ and ${\cal G}(Z)=SL(2,Z)\times O(6,22;Z)$.\par
\section{Contents of the dissertation.} 
After this sketchy introduction to the concept of duality in superstring and 
supergravity theories, in what follows I will give an outline of the 
main topics discussed in the present thesis, posponing to the next chapters
a more precise and complete analysis of some of the ideas introduced 
so far. All the problems which will be dealt with throughout the thesis
are deeply related to the concept of duality in supergravity theories
and their common denominator is a well extablished mathematical technique
 which has been applied for the first time in supergravity 
by our group: {\it the solvable Lie algebras}. \par
In the first chapter, after an introduction to electric-magnetic duality 
in supergravity and to Gaillard--Zumino model, the formalism of solvable
 Lie algebras will be introduced. The use of this technique, as it will be 
 emphasized in the sequel, on one hand allows a considerable semplification 
of the mathematical structure of the scalar sector, on the other hand
makes it possible to unravel the action of the U--duality on the scalar fields,
revealing therefore to be a suitable mathematical laboratory for studying 
the non perturbative aspects of supergravity and hence of superstring theories.
It will be shown that in many cases of interest, the scalar manifold ${\cal M}$
, which is a non--compact Riemannian manifold, may be described as a slovable 
Lie group manifold, locally generated therefore by a solvable Lie algebra.
Since the supergravity action depends on the scalar manifold only through its local differential geometry, describing ${\cal M}$ as a group manifold allows to
reformulate the scalar dependence of the action in an algebraic way which is 
much simpler since, for instance, differential operators on  ${\cal M}$ 
correspond to linear mappings on the generating solvable algebra. Moreover 
such a new formulation allows to extablish a local correspondence between
scalar fields and generators of the solvable algebra, this in turn yields a 
geometrical characterization of the RR and NS scalars and makes it possible to
exploit the action of the S,T and U--duality groups on each of them.
This analysis will be developed in the second chapter, where explicit examples 
of how the solvable Lie algebra machinery works will be given.\par 
Another relevant potentiality of solvable Lie algebras is that they allow
to determine easily the so called Peccei--Quinn scalars of a given theory, that
 is 
those scalars which appear in the supergravity action 'covered' by a derivative
and therefore are associated with global translational symmetries of the theory
. The Peccei--Qiunn scalars turn out to be the parameters of the {\it maximal
abelian ideal} ${\cal A}$ of the solvable algebra, which indeed 
generates all the possible translations on ${\cal M}$. Once the correspondence 
between scalar fields and  generators of the solvable algebra has been 
extablished, it is strightforward  to determine the RR and NS generators 
of ${\cal A}$. The latter act on RR Peccei-Quinn scalars translating them by a
 constant configuration. 
In order to introduce the topic of the third chapter, let us spend few words 
about RR fields and their charge. As previously mentioned, a characteristic of the RR fields is that they do not couple directly to the string. In $10$ 
dimensions the RR fields are higher order forms which can be thought of as 
the gauge fields corresponding to a certain $U(1)$--charges: the RR charges.
Indeed it was recently conjectured that these fields couple 
solitonic extended objects \cite{polc}
 named D--branes, therefore the RR charges are solitonic (from the string 
point of view). In the 
effective low--energy supergravity, the RR scalars may be interpreted as
D0--branes originated by wrapping the D--branes in higher dimension around the 
internal directions, and therefore they carry a solitonic RR charge.\par
The effect of a RR translation in ${\cal A}$ on a generic bosonic background 
is to generate a RR field. If this global symmetry is suitably gauged, i.e. 
made local by associating a vector field with it in a proper way, one may expect 
that the new theory admit vacua on which the RR scalar field has a non
 vanishing expectation value, i.e. vacua exhibiting a condensation of 
the RR charge \cite{postr}. Since D--branes were shown by Polchinski to 
saturate the Bogomolnyi bound, they share with BPS saturated states the 
property of preserving a certain 
fraction of the initial supersymmetries of the theory. Therefore
 the process of soliton 
condensation may yield to a spontaneous partial
supersymmetry breaking. This mechanism was verified in the case of 
spontaneous $N=2\rightarrow N=1$ supersymmetry breaking in which the effective 
theory had a surviving quite general compact gauge group \cite{fegipo} but
its general check for a generic N--extended supergravity is still work in
 prograss. In chapter 3, besides showing how gauging a RR translational 
isometry yields spontaneous supersymmetry breaking in the particular case
of the $N=2$ theory mentioned above, emphasis will be given to the important 
role played by the solvable Lie algebras in solving the problem,
since they allowed to determine easily not only ${\cal A}$ but also
the flat directions of the scalar potential. \par
Finally the problem of finding the BPS solitonic solutions  of a 
supergravity theory filling a U--duality 
representation is faced in Chapter 4 in the case of an $N=8$
theory \cite{noialtri3}.
 In supergravity BPS solitons are Black Holes saturating the Bogomolnyi
bound and, as previously pointed out, they have the property of preserving 
$1/2$, $1/4$ or $1/8$ of the original supersymmetries. We will be dealing with
 the latter kind of solitons. They are solutions of 
the second order field equations together with  a system of first order 
equations deriving from the condition for the solution to have a
 residual number of supersymmetries (Killing spinor equations). The descriprion
of the scalar manifold in terms of a solvable Lie algebra allows to formulate
the problem  in a geometrical fashion in which the first order equations are much easier to be solved. Moreover, using this formalism, it is possible 
to find the minimal number of scalars characterizing the generating solution
of the U--duality orbit. Explicit calculations are carried out applying this 
approach to a simplified case in which the BPS Black Hole is characterized
by the only dilaton fields.
\chapter{Duality in Supergravity and Solvable Lie Algebras}
The first and most widely known example of duality in field theory
is given by the symmetry of the Maxwell--Einstein equations in the vacuum
\begin{eqnarray}
\partial_{\mu}F^{\mu\nu}\, &=&\,0\nonumber \\
\partial_{\mu}\tilde{F}^{\mu\nu}\, &=&\,0
\label{maxvac}
\end{eqnarray}
with respect to a generic linear transformation mixing the electric 
field strenght $F^{\mu\nu}$ with its dual $\tilde{F}^{\mu\nu}=
(1/2)\epsilon^{\mu\nu\rho\sigma}F_{\rho\sigma}$ (duality rotation).
 It is apparent from eqs.\ref{maxvac} that in presence of an electric current 
on the rhs of the first equation, a magnetic current should be introduced on the rhs of the second in order for the whole system to remain duality invariant.
The requirement of duality symmetry in electromagnetism therefore implies the 
existence of magnetic charges.\par A solution of the Maxwell--Einstein 
equations carrying
magnetic charge $m$ ({\it monopole}) was initially found by P.A.M. Dirac
\cite{dirac}.
He moreover obtained the following quantization condition which should be fulfilled by the electric charge $e$ and the magnetic charge $m$ in order for his solution to be physically consistent:
\begin{equation}
\frac{em}{4\pi \hbar}\,=\, \frac{n}{2}\quad\,\, n\in Z
\label{dirac}
\end{equation}
Later 't Hooft and Polyakov \cite{thoftpoly} found a monopole solution in a gauge 
theory with spontaneous symmetry breaking $SU(2)\rightarrow U(1)$
for which condition \ref{dirac} was automatically fulfilled.
The main ingredient used to find such a solution was the presence in the theory
of scalar fields in the adjoint representation of the gauge group G.
Their result is immediately generalized to a theory with generic compact 
non--abelian gauge group
G describing scalar fields $\phi^{\alpha}\in {\it adj}(G)$. On a vacuum 
in which $\langle \phi^{\alpha}\rangle \neq 0$, $G$ is spontaneously
broken to its Cartan subgroup ${\cal C}=U(1)^r\, ;\, r=rank G$
(indeed, being ${\cal C}$ the little 
group of the adjoint representation it is the symmetry group of the vacuum). The 't Hooft--Polyakov monopole is an 
example of {\it soliton} in field theory.\par 
Since this kind of solution will play an important role in future analyses,
I am going to spend few words about it.
By soliton one denotes  a stationary solution 
of the field equations with finite energy which corresponds to a 
boundary condition defining a topologically non trivial mapping between 
the sphere at 
infinity $S^{2}_{\infty}$ and the manifold of the degenerate 
scalar vacua $G/{\cal C}$.
Such configurations are characterized by a conserved {\it topological charge}
related to the element of $\pi_2(G/{\cal C})$ defined by their asymptotic behaviour.
In the case of a monopole the topological charge coincides with its magnetic
 charge $m$. Denoting by $\phi_o=\vert \langle \phi^{\alpha}_o\rangle\vert $ 
the gauge invariant norm of the scalar vector corresponding to a chosen vacuum
$V_o$, the effective theory on $V_o$ will have a spectrum of monopoles
whose asymptotic behaviour fulfills the condition:
\begin{equation}
\vert \langle \phi^{\alpha}\rangle\vert_{S^{2}_{\infty}}=\phi_o
\label{asymono}
\end{equation}
Bogomolnyi found that the mass of a monopole is bounded from below according 
to the condition:
\begin{equation}
M_{monopole}\,\ge\, 4\pi\phi_o m\approx \frac{n}{e}\quad \,n\in Z 
\label{bogob}
\end{equation} 
therefore in the perturbative regime of the theory ($e\approx 0$) monopoles are
 expected to be very massive.\par
One may think of monopoles as elementary excitations of a {\it dual theory}
with gauge group given by the {\it magnetic} $U(1)^r$. The correspendence 
between a spontaneously broken non--abelian 
gauge theory and its abelian dual was first conjectured by Montonen and Olive
\cite{molive}.
The mapping between the elementary spactra of these two theories is {\it non--perturbative} since it is realized by the following duality transformations :
\begin{eqnarray}
e&\leftrightarrow &  m\approx\frac{n}{e}\nonumber\\
B^{i}_{\mu}&\leftrightarrow &A^{i}_{\mu}
\label{montolive}
\end{eqnarray}
where $ B^{i}_{\mu} $ are the the {\it magnetic gauge fields} defined in the following way:
\begin{equation}
\tilde{F}^{i}=dB^{i}
\end{equation}
The first of eqs.\ref{montolive} tells us that this duality maps the {\it weak
coupling regime} of a theory into the {\it strong coupling regime} of its dual,
while the second equation defines a transformation which is {\it non--local}
and therefore can be a symmetry of, at most, the field equations and Bianchi
 identities, but not of the entire action. Moreover the transformations 
\ref{montolive} 
should belong to a descrete group $\Gamma_D$ in order to preserve the lattices 
described by the magnetic and electric charges fulfilling the Dirac quantization condition \ref{dirac}.\par In the case of
electromagnetism, it is strightforward to check that the electric--magnetic
duality transformations $\Gamma_D$ are a symmetry of the Maxwell equations
in presence of electric and magnetic currents.
\section{Duality in Supergravity: \\ the Gaillard--Zumino model.}
Let us now turn on duality in  a supersymmetric gauge theory.
The following introduction to this topic will mainly refer to the presentation 
given in the
review \cite{pietrolectures}. \par
In order for a theory to exhibit electric-magnetic duality, it must have 
a non--perturbative spectrum of magnetic monopole solutions besides the 
electric fundamental excitations. As previously pointed out, the basic
 ingredient for the existence of 't Hooft--Polyakov monopole solutions 
in a gauge theory is the presence of scalar fields transforming in the adjoint 
representation of the gauge group G. This requirement is automatically
fulfilled in a four dimensional $N\ge 2$ supersymmetric gauge theory in which
the gauge vectors $A^{\Lambda}_{\mu}\, ;\, \Lambda=1,...,dim(G)$ are related 
through supersymmetry transformations to scalar fields $\phi^{\Lambda}$ in the 
same representation of G (i.e. the adjoint one).
Moreover the fact that vector fields and scalars sit in the same supermultiplet
requires the duality group $\Gamma_D$ to be a suitable discrete form of the 
isometry group of the scalar manifold, and gives these theories a particularly 
interesting mathematical structure, as I am going to show next.\par
Let us consider an N--extended supersymmetric model in $D=2p$ dimensions
describing among the other fields a set of $n$ $(p-1)$--forms $A^{\Lambda}$ and $m$ sclalar fields $\phi^{I}$ belonging to the same supermultiplet.
 I shall restrict my treatment  to  the part of the action describing only the 
 $A^{\Lambda}$ and $\phi^{I}$, which generalizes the Gaillard--Zumino action
 \cite{gz} to space--time dimensions greater than four.
Moreover in this theory the $(p-1)$--forms do not gauge any non--abelian
 gauge group, therefore we can think of it as an abelian gauge 
theory with gauge group $U(1)^n$. As a starting hypothesis we will assume that 
all the fields are neutral with respect to this gauge group and therefore 
no electric or magnetic current will couple to the $(p-1)$--forms.\par
The scalar sector of a supergravity theory is 
described by a D--dimensional $\sigma$--model, which means that
 the scalar fields
$\phi^{I}$ are local coordinates of a non--compact Riemannian manifold 
${\cal M}$ and the scalar action $S_{scal}$ is required to be 
invariant with respect 
to the {\it isometries} ($t\in {\cal I}som({\cal M}$)) of ${\cal M}$, i.e. 
diffeomorphisms ($t\in {\it Diff}({\cal M}$)) on ${\cal M}$ leaving 
the metric $g$ invariant ($t^{*}g=g$). The action describing the scalars and 
the $(p-1)$--forms has the following structure:
\begin{eqnarray}
S\, &=&\, S_{scal}\, +\, S_{(p-1)}\, =\, -\frac{1}{2}\int d^D\sqrt{-det(g)
}g_{IJ}\partial_{\mu}\phi^I\partial^{\mu}\phi^J\,+\nonumber\\
&& \frac{1}{2}\int d^D\left[ (-)^p\gamma_{\Lambda\Sigma}(\phi)F^{\Lambda}
\wedge\tilde{F}^{\Sigma}\,+\,\theta_{\Lambda\Sigma}(\phi)F^{\Lambda}
\wedge F^{\Sigma}\right]\nonumber\\
&&\mbox{where:}\nonumber\\
F^{\Sigma}\, &=&\, dA^{\Sigma}\nonumber\\
\tilde{F}^{\Sigma}\, &=&\,\frac{1}{p!}\epsilon_{\mu_{1}...\mu_{p}
\sigma_{1}...\sigma_{p}}F^{\Sigma\vert\sigma_{1}...\sigma_{p}}dx^{\mu_{1}}
\wedge....\wedge dx^{\mu_{p}}\nonumber\\
\gamma_{\Lambda\Sigma}\,&=&\, \gamma_{\Sigma\Lambda}\nonumber\\
\theta_{\Lambda\Sigma}\,&=&\, \theta_{\Sigma\Lambda}\,\,(p \, even)\nonumber\\
\theta_{\Lambda\Sigma}\,&=&\, -\theta_{\Sigma\Lambda}\,\,(p \, odd)
\label{gaizum}
\end{eqnarray}
The fact that 
the field--strenghts of the $(p-1)$--forms couple to the scalars through
a generalized coefficient of the kinetic term $\gamma_{\Lambda\Sigma}(\phi)$
and a generalized theta--angle $\theta_{\Lambda\Sigma}(\phi)$ is a requirement 
of supersymmetry. 
Let us specialize the expression of the lagrangian describing the $(p-1)$--forms
for $p$ even and $p$ odd and define in each case the {\it magnetic} field--strenght $G^{\Lambda}$ corresponding to the {\it electric} $F^{\Lambda}$:
\begin{eqnarray}
\mbox{{\bf p even}}\quad\quad\quad\,\,\,\,\,&&\nonumber\\
{\cal L}\, &=&\,\frac{1}{2}\left[ -\gamma_{\Lambda\Sigma}(\phi)F^{\Lambda}
 F^{\Sigma} \, -\,{\rm i}\theta_{\Lambda\Sigma}(\phi)F^{\Lambda}
F^{*\Sigma}\right]\nonumber\\
G^{*}_{\Lambda}\, &=&\, {\rm i}\frac{\delta {\cal L}}{\delta  
F^{\Lambda}}\, =\, {\rm i}\left( -\gamma_{\Lambda\Sigma}(\phi)
 F^{\Sigma} \, -\,{\rm i}\theta_{\Lambda\Sigma}(\phi)
F^{*\Sigma}\right)\nonumber\\
\label{peven}
\end{eqnarray}
\begin{eqnarray}
\mbox{{\bf p odd}}\quad\quad\quad\,\,\,\,\,&&\nonumber\\
{\cal L}\, &=&\,\frac{1}{2}\left[ -\gamma_{\Lambda\Sigma}(\phi)F^{\Lambda}
 F^{\Sigma} \, +\,\theta_{\Lambda\Sigma}(\phi)F^{\Lambda}
F^{*\Sigma}\right]\nonumber\\
G^{*}_{\Lambda}\, &=&\, \frac{\delta {\cal L}}{\delta  
F^{\Lambda}}\, =\, \left( -\gamma_{\Lambda\Sigma}(\phi)
 F^{\Sigma} \, +\,\theta_{\Lambda\Sigma}(\phi)
F^{*\Sigma}\right)\nonumber\\
\label{peven2}
\end{eqnarray}
In both cases the field equations and the Bianchi identities have the following
 form:
\begin{eqnarray}
\partial^{\mu}G^{*}_{\Lambda\vert \mu \mu_{1}...\mu_{p-1}}\,&=&\, 0
\nonumber\\
\partial^{\mu}F^{*\Lambda}_{\mu \mu_{1}...\mu_{p-1}}\,&=&\, 0
\label{feqbid}
\end{eqnarray}
where we have introduced an {\it involutive} duality operation (*) on the 
field strenght $p$--forms defined as follows: $F^{*\Lambda}={\rm i}\tilde{F}^{\Lambda}\,\,(p\, even)$ and $F^{*\Lambda}=\tilde{F}^{\Lambda}\,\,
(p\, odd)$.
Eqs.\ref{feqbid} are a generalization of eqs.\ref{maxvac}. 
By definition $S_{scal}$ is invariant with respect to the isometries of ${\cal M}$, but since the $(p-1)$--forms are related to the scalars by supersymmetry
($\Leftrightarrow$ they are coupled in $S_{p-1}$),
a transformation on $\phi^{I}$ should have a couterpart also on the 
$(p-1)$--forms in order for at least the field equation and the Bianchi identities to remain invariant. Therefore, in order for 
${\cal I}som({\cal M})$ to be promoted to a global symmetry group of the 
field equation and the Bianchi identities, it should have a suitable action 
 also on the 
field strenghts $F^{\Lambda}$ and $G_{\Sigma}$ 
(duality transformation). The aim of the following calculations is to 
define such transformations.\par
It is useful to rewrite the quantities introduced so far in terms of the 
selfdual and anti--selfdual components $F^{(\pm)\Lambda}$ and 
$G_{\Sigma}^{(\pm)}$ of the electric and magnetic field strenghts:
\begin{eqnarray}
F^{(\pm)\Lambda}\, &=&\,\frac{1}{2}\left(F^{\Lambda}\pm F^{*\Lambda}\right)\nonumber\\
\mbox{{\bf p even:}}\quad\quad\quad\,\,\,\,\,&&\nonumber\\
{\cal L}\, &=&\,\frac{{\rm i}}{2}\left[{\cal N}_{\Lambda\Sigma}F^{+\Lambda}
F^{+\Sigma}\,-\, {\bar {\cal N}}_{\Lambda\Sigma}F^{-\Lambda}
F^{-\Sigma}\right]\nonumber\\
{\cal N}_{\Lambda\Sigma}\, &=&\,\theta_{\Lambda\Sigma}-{\rm i}\gamma_{\Lambda\Sigma}\nonumber\\
G_{\Lambda}^{+}\, &=&\, {\cal N}_{\Lambda\Sigma}F^{+\Sigma}\,\,\,\,
G_{\Lambda}^{-}\, =\, {\bar {\cal N}}_{\Lambda\Sigma}F^{-\Sigma}\nonumber\\
\mbox{{\bf p odd:}}\quad\quad\quad\,\,\,\,\,&&\nonumber\\
{\cal L}\, &=&\,\frac{1}{2}\left[{\cal N}_{\Lambda\Sigma}F^{+\Lambda}
F^{+\Sigma}\,+\, {\cal N}^{T}_{\Lambda\Sigma}F^{-\Lambda}
F^{-\Sigma}\right]\nonumber\\
{\cal N}_{\Lambda\Sigma}\, &=&\,\theta_{\Lambda\Sigma}-\gamma_{\Lambda\Sigma}\nonumber\\
G_{\Lambda}^{+}\, &=&\, {\cal N}_{\Lambda\Sigma}F^{+\Sigma}\,\,\,\,
G_{\Lambda}^{-}\, =\, -{\cal N}^{T}_{\Lambda\Sigma}F^{-\Sigma}
\label{selfd}
\end{eqnarray}
where we have introduced a complex $n\times n$ field--dependent matrix ${\cal N}_{\Lambda\Sigma}(\phi)$. 
Our problem is to define, for a generic isometry $t$ on $\phi^{I}$, the corresponding transformation rule for ${\cal N}_{\Lambda\Sigma}$ and the linear transformation mixing $F^{(\pm)\Lambda}$ with $G_{\Sigma}^{(\pm)}$ of the form:
\begin{equation}
\left(
\matrix{ F^{(\pm)\Lambda} \cr G_{\Sigma}^{(\pm)}}\right)
\,\rightarrow \,
\left(\matrix{ F^{(\pm)\prime\Lambda} \cr G^{(\pm)\prime}_{\Sigma}}
 \right)
\, =\, \left(\matrix{ A_t & B_t \cr C_t & D_t } \right)
\left(\matrix{ F^{(\pm)\Lambda}\cr G_{\Sigma}^{(\pm)}} \right)
\label{dualtra1}
\end{equation}
\begin{equation}
{\cal N}_{\Lambda\Sigma}(\phi)\, \rightarrow \,{\cal N}^{\prime}_{\Lambda\Sigma}(t(\phi))
\label{dualtra2}
\end{equation} 
which are demanded to be consistent with the definition of $G^{\Lambda}$ and,
moreover, 
to leave eqs.\ref{feqbid} invariant. The latter requirement is automatically fulfilled by the linear transformation defined in eq.\ref{dualtra1}, 
while the former implies that the period matrix should transform by means of 
the following {\it fractional linear transformation}:
\begin{eqnarray}
{\cal N}^{\prime}(t(\phi))\,&=&\,\left(C_t +D_t {\cal N}(\phi)\right)\left(A_t +B_t {\cal N}(\phi)\right)^{-1}\nonumber\\
\mbox{{\bf p even:}}\quad\quad\quad\,\,\,\,\,&&\nonumber\\
{\bar {\cal N}}^{\prime}(t(\phi))\,&=&\,\left(C_t +D_t {\bar {\cal N}}(\phi)\right)\left(A_t +B_t {\bar {\cal N}}(\phi)\right)^{-1}\nonumber\\
\mbox{{\bf p even:}}\quad\quad\quad\,\,\,\,\,&&\nonumber\\
-{\cal N}^{T\prime}(t(\phi))\,&=&\,\left(C_t -D_t {\cal N}^{T}(\phi)\right)\left(A_t -B_t {\cal N}^{T}(\phi)\right)^{-1}
\label{fraclin}
\end{eqnarray} 
Concistency of the second and the first of eqs.\ref{fraclin} in the case
$p$ even or of the third and the first in the case $p$ odd implies, for the
 linear transformation \ref{dualtra1} the following condition:
\begin{equation}
 \left(\matrix{ A_t & B_t\cr C_t & D_t } \right)\in
\matrix{ Sp(2n, \IR) & \mbox{p even}\cr SO(n,n) & \mbox{p odd} }
\label{dualgroup}
\end{equation} 
Summarizing the results described so far, the isometry group of the scalar manifold
is promoted to be a global symmetry group of the field equations and the Bianchi 
identities once a double action on both scalar fields and field strenghts $p$--forms is defined by means of a suitable {\it embedding} $i_{\delta}$, which,
in the two relevant cases, is:
\begin{equation}
i_{\delta}\, :\, \cases{ {\cal I}som({\cal M})\longrightarrow Sp(2n, \IR)& 
\mbox{p even}\cr {\cal I}som({\cal M}) \longrightarrow SO(n,n) & \mbox{p odd}} 
\label{embedding}
\end{equation}
Under the action \ref{dualtra1} of the duality group ${\cal G}=
i_\delta({\cal I}som({\cal M}))$, the field strenghts $F^{\Lambda}$ and $G_{\Sigma}$ therefore transform as components
of a {\it symplectic}
or {\it pseudo--orthogonal} vector.\par
The model which we have considered so far may be thought of as describing the
massless bosonic field content of a theory whose gauge group G, 
of rank m, was spontaneously broken to its Cartan subgroup $U(1)^m$ by a non vanishing expectation value of its scalar fields transforming in the $adj$ representation of G
through a Higgs mechanism. Such mechanism would leave massless only those 
fields $A^{\Lambda}$ and $\phi^I$ which are neutral with respect to the 
residual $U(1)^m$ (the index $I$ labeling the {\it 
Cartan directions} of the adjoint representation on G), and generate
 a spectrum of massive  electrically and magnetically charged  states (the 
latter corresponding to 't Hooft--Polyakov monopoles) coupled to the 
vector bosons $A^{\Lambda}$. Adopting this interpretation of the model, 
 eqs.\ref{feqbid} should be modified by the presence of electric and magnetic currents. We now define the following electric and magnetic charges:
\begin{equation}
{\cal Q}=\left(\matrix{ p^{\Lambda} \cr q_{\Sigma}}\right)=\left(\matrix{ 
\int_{S^{2}_{\infty}}F^{\Lambda} \cr \int_{S^{2}_{\infty}}G_{\Sigma}}\right)
\in\cases{Sp(2n, \IR)& \mbox{p even}\cr SO(n,n) & \mbox{p odd}}
\label{noechar}
\end{equation} 
Besides the fundamental electrically charged excitations and the magnetic 
monopoles there are also {\it dyonic} colutions carrying both electric and magnetic charges $(p,q)$. Moreover, quantization implies that the charge vector
 ${\cal Q}$ must fulfill a symplectic (or pseudo--orthogonal)
invariant generalization of condition \ref{dirac} called Dirac--Schwinger--Zwanziger (DSZ) condition:
\begin{equation}
\left({\cal Q},{\cal Q}^\prime\right)=p^{\Lambda}q^{\prime}_{\Lambda}-
p^{\prime\Lambda}q_{\Lambda}\in Z
\label{DSZ}
\end{equation}
where $\left(\, ,\, \right)$ is the symplectic (or pseudo--orthogonal) invariant scalar product. Let us assume as hypothesis that all types of charges exist
 in the spectrum
and let us fix a charge vector ${\cal Q}$, then the charge vector ${\cal Q}^\prime$ fulfilling eq.\ref{DSZ} will describe a lattice, i.e.
DSZ--condition implies that $p$ and $q$ are quantized.
Requiring the action of the duality group ${\cal G}$ on
the vector ${\cal Q}$ to leave the charge lattice invariant, the group
${\cal G}$ is spontaneously broken to a suitable restriction to the integers:
\begin{equation}
{\cal G}\longrightarrow {\cal G}(Z)={\cal G}\cap \cases{Sp(2n, Z)& \mbox{p even}\cr SO(n,n;Z) & \mbox{p odd}}
\label{uz}
\end{equation}
The infinite discrete group ${\cal G}(Z)$ is the conjectured U--duality group in superstring theory, 
while the S and T duality, as we will see, will be suitable restrictions to the integers of well defined 
continous subgroups ${\cal G}_s$, ${\cal G}_t$ of  ${\cal G}$.\par
For the sake of simplicity, let us restrict ourselves to the case of $D=4$
space--time dimensions.
Our initial abelian $U(1)^n$ gauge theory may be also regarded as the fundamental theory prior to gauging. Performing the {\it gauging} procedure means
to associate a certain number $k$ of vector fields $A^{\alpha}_{\mu}\, ;
\,\alpha=1,...,k$ with the generators of a non--abelian gauge 
group $G$ (dim$G=k$), subgroup of ${\cal I}som({\cal M})$ and to modify the lagrangian of the theory, by introducing $G$--covariant derivatives, coupling constants
(one for each simple factor of $G$) and a suitable scalar potential in order to
 make the action locally invariant with respect to $G$. Gauging therefore, 
roughly speaking, means to promote a certain subgroup of isometries of ${\cal M}
$ from {\it global} symmetries of the action to {\it local} symmetries.
After gauging electric and magnetic currents will be present in the theory and the U--duality group will break to its discrete subgroup ${\cal G}(Z)$.
As I mentioned earlier,  not all the duality transformations in ${\cal G}$ 
are a symmetry of the action. Some of them indeed (those with $B_t\neq 0$ )
act on the charge vector
${\cal Q}$ by transforming the electric charges $q^\Lambda$ into a linear 
combination of electric and magnetic charges. Such transformations are 
obviously {\it non--perturbative} and they amount to a non--local transformation of the vector fields. Therefore they are not a global symmetry of the action
and hence cannot be gauged. For instance the S--duality group
${\cal G}_s\subset {\cal G}$ is non--perturbative while the T--duality group
${\cal G}_t\subset {\cal G}$ is {\it perturbative} on the NS charges for its symplectic action on the corresponding field strenghts \ref{dualtra1} is characterized by $ B_t = 0$. T--duality indeed leaves the perturbative expansions in the electric (NS) co

uplings invariant and in general it can be shown that upon restriction to the integers (as in \ref{uz}) ${\cal G}_t$ is a symmetry of the action.
By {\it electric} transformations one denotes those perturbative duality transformations having a {\it diagonal} action on the electric and magnetic charges
and they are global symmetries of the lagrangian, with no need of any restriction to the integers.
Finding the electric subgroup of ${\cal G}$ is therefore crucial in 
order to perform a gauging procedure and in next chapter we are going to deal 
with this problem.\par
Before electric and magnetic charged were {\it switched on}, the operation of conjugating the embedding $i_\delta$ by a symplectic (or pseudo--orthogonal)
transformation (i.e. choice of a symplectic (or pseudo--orthogonal) gauge) is 
 {\it immaterial} for it would lead to 
physically equivalent theories. In presence of charged states however,
different choices of this gauge would lead to inequivalent theories
(the intersection in eq.\ref{uz} is indeed dependent on the choice of
 embeddings related to each other by means of a symplectic (or pseudo--orthogonal)
transformation). The physical relevance, after {\it gauging}, of which symplectic (or pseudo--orthogonal) gauge is chosen for a given theory will be apparent  in capter 3 where it will be shown how, in a four dimensional theory, the choice of a particular

 symplectic 
gauge will be crucial in order for a mechanism of spontaneous supersymmetry breaking to occur.
Our analysis so far has shown that  studying the global symmetries of
 supergravity at the {\it classical} level,
i.e. the isometries of the scalar manifold, is the first step towards a deeper understanding of dualities in superstring theories. 
 For this reason, in what follows, I shall focus my analysis on
the continuous versions of the S, T and U--dualities, which will be denoted by the same symbols as their 
discrete counterparts, seen as isometries of ${\cal M}$ and introduce a new description of the scalar manifold by means of 
{\it solvable Lie algebras} which reveals to have several advantages, as it will be apparent in the sequel.\par 
An important feature of supergravity theories with sufficently large supersymmetry (e.g. in $D=4$ $N\ge 3$) is that
the scalar manifold is an homogeneous manifold of the form: 
\begin{equation}
{\cal M}={\cal G}/{\cal H}\, ;\quad\quad m=dim{\cal M}=dim{\cal G}-dim{\cal H}
\label{homoman}
\end{equation}
where ${\cal G}$ has already been defined and ${\cal H}$ is its maximal compact subgroup. Since ${\cal M}$
is a non--compact Riemannian manifold, ${\cal G}$ is a {\it non--compact 
real form} of a semisimple Lie group. Writing the scalar manifold in the form \ref{homoman} amounts to associate with each point $P$ of ${\cal M}$,
parametrized by the scalar fields $(\phi^I)$, a {\it coset--representative} 
$\IL(\phi)\in {\cal G}$ generated by the non-compact generators of ${\cal G}$
on which  the action of a generic isometry $g\in {\cal G}$ on $\phi^I$ 
is defined as follows:
\begin{equation}
g:\phi\longrightarrow \phi^\prime\Leftrightarrow
g\cdot \IL(\phi)\, =\, \IL(\phi^\prime)\cdot h(g,\phi)
\label{coset}
\end{equation}
where $h(g,\phi)\in {\cal H}$ is called the {\it compensator} of the transformation $g$ on $\IL(\phi)$ (Figure \ref{homo}).\par
Since, by definition, ${\cal G}$ is the isometry group of ${\cal M}$ embedded in $Sp(2n,\IR)$ ($ SO(n,n)$), eq.\ref{homoman} means that we are choosing
a description of the scalar manifold in which the coset--representative 
is a symplectic (pseudo--orthogonal) $2n\times 2n$ matrix: $\IL(\phi)\equiv \IL(\phi)^A_B\in Sp(2n,\IR)$ ($\in SO(n,n)$).
So far we have considered the linear action of U--duality only on 
$(p-1)$--form potentials in an even D--dimensional space--time for which $D=2p$. As far as $n$ generic $(p^\prime -1)$--form potentials ($D\neq 2p^\prime$) are concerned, U--duality will act only on the corresponding
 $p^\prime$--forms field strenghts $F^\Lambda$
by means of a linear transformation belonging to the $n$--dimensional 
irreducible representation of ${\cal I}som({\cal M})$ which the $F^\Lambda$
belong to.\par
In next section I will show that a generic non--compact Riemannian manifold is isomorphic to a {\it Lie group manifold} ${\cal G}^{(s)}
$ and 
therefore can be described {\it locally} in terms of a new coset--representative which is now a group element and whose parameters
are the new local coordinates. The latter can be thought of as {\it geodesic} coordinates in the neighborhood of each point on 
${\cal M}$. Indeed the transitive action of ${\cal G}^{(s)}$ on ${\cal M}$ needs no {\it compensator}, being the action of a
 group on its own manifold. In next chapter I will show how describing the scalar manifold in terms of a group manifold leads to 
a deeper understanding of different supergravity theories in diverse dimensions from a geometrical point of view and allow 
to exploit useful relations between them.
\section{Homogeneous manifolds and\\solvable Lie algebras.}
\iffigs
\begin{figure}
\caption{}
\label{homo}
\epsfxsize = 10cm
\epsffile{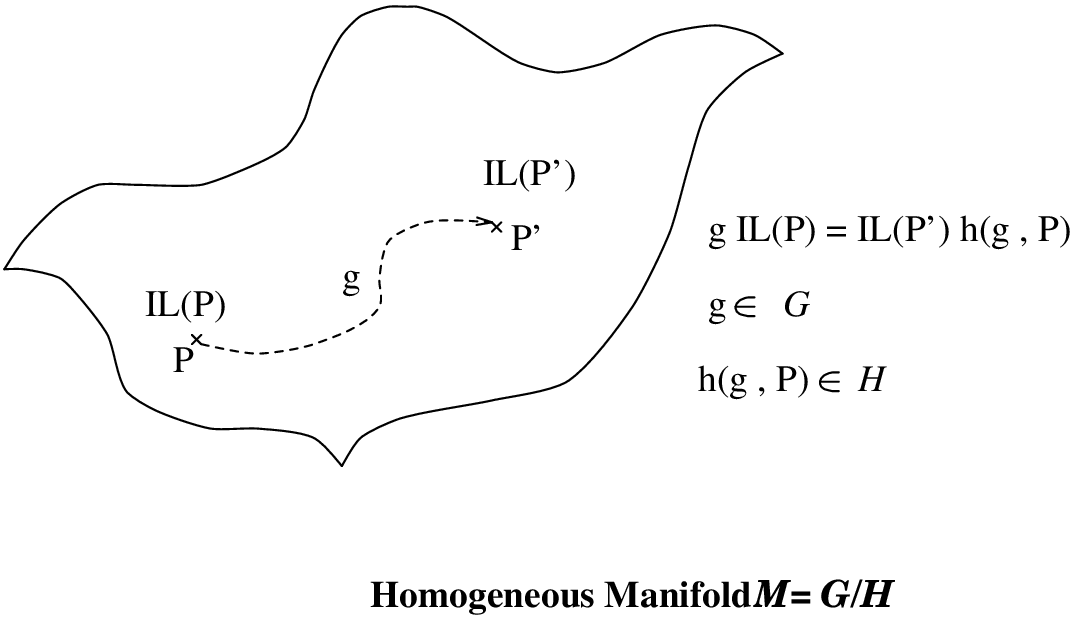}
\vskip -0.1cm
\unitlength=1mm
\end{figure}
\fi
In this section we will deal with a general property according to which
any homogeneous non-compact coset manifold may be expressed as
a group manifold
 generated by a suitable solvable Lie algebra. \cite{alex}
\par
Let us start by giving few preliminar definitions.
A {\it solvable } Lie algebra $Solv$ is a Lie algebra whose $n^{th}$ order
(for some $n\geq 1$) derivative algebra vanishes:
\begin{eqnarray}
{\cal D}^{(n)}Solv&=&0 \nonumber \\
{\cal D}Solv=[Solv,Solv]&;&\quad {\cal D}^{(k+1)}Solv=[{\cal D}^{(k)}Solv,
{\cal D}^{(k)}Solv]\nonumber
\end{eqnarray}
A {\it metric} Lie algebra $(\IG,\langle,\rangle)$ is a Lie algebra endowed with an
euclidean metric $\langle,\rangle$. An important theorem states that if a Riemannian
manifold
$({\cal M},g)$ admits a transitive group of isometries ${\cal G}_s$ generated by
a solvable Lie algebra $Solv$ of the same dimension as ${\cal M}$, then:
\begin{eqnarray}
{\cal M}\sim {\cal G}_s&=&exp(Solv)\nonumber\\
 g_{|e\in {\cal M}}&=&\langle,\rangle \nonumber
\end{eqnarray}
where $\langle,\rangle$ is an euclidean metric defined on $Solv$.
 Therefore there is a one to one correspondence between
 Riemannian manifolds fulfilling the hypothesis
stated above and solvable metric Lie algebras $(Solv,\langle,\rangle)$.\\
Consider now an homogeneous coset manifold ${\cal M}={\cal G} /{\cal H}$,
${\cal G}$ being a non compact real
 form of a semisimple Lie group and ${\cal H}$ its maximal compact
subgroup. If $\IG$ is the Lie algebra generating ${\cal G}$, the so
called {\it Iwasawa
decomposition} ensures the existence of a solvable Lie subalgebra
$Solv\subset
\IG$, acting transitively on ${\cal M}$, such that \cite{helgason}:
\begin{equation}
\IG=\IH\oplus Solv \qquad \mbox{dim }Solv=\mbox{dim }{\cal M} \nonumber
\end{equation}
$\IH$ being the maximal compact subalgebra of $\IG$ generating ${\cal H}$. \\
In virtue of the previously stated theorem, ${\cal M}$ may be expressed
 as a solvable
 group manifold generated by $Solv$. The algebra $Solv$ is constructed
as follows \cite{helgason}. Consider the Cartan decomposition
\begin{equation}
\IG = \IH \oplus \IK
\end{equation}
where $\IK$ is the subspace consisting of all the non compact 
generators of $\IG$.
Let us denote by ${\cal C}_K$ the  maximal abelian subspace
of $\IK$ and by ${\cal C}$
the Cartan subalgebra of $\IG$.
It can be proven \cite{helgason} that ${\cal C}_K = {\cal C} \cap \IK$,
that is it consists of all non compact elements of ${\cal C}$.
Furthermore let $h_{\alpha_i}$ denote the elements
of ${\cal C}_K$, $\{\alpha_i\}$ being
 a subset of the positive roots of $\IG$ and $\Delta^+$
 the set of positive roots $\beta$ not orthogonal to
all the $\alpha_i$ (i.e. the corresponding ``shift'' operators
$E_\beta$ do not commute with ${\cal C}_K$).
It can be demonstrated that the solvable algebra $Solv$
defined by the Iwasawa decomposition
may be expressed in the following way:
\begin{equation}
  \label{iwa}
  Solv = {\cal C}_K \oplus \{\sum_{\alpha \in \Delta^+}E_\alpha \cap \IG \}
\end{equation}
where  the intersection with $\IG $ means that $Solv$ is generated
by those suitable complex combinations of the ``shift'' operators
which belong to the real form of the isometry algebra $\IG$.
\par
The {\it rank} of an homogeneous  coset manifold is defined as
the maximum number of commuting semisimple
elements of the non compact subspace $\IK$. Therefore it
coincides with the dimension of ${\cal C}_K$,
i.e. the number of non compact Cartan generators of $\IG$.
A  coset manifold is {\it maximally non compact} if
${\cal C} ={\cal C}_K \subset Solv$. As we will see this kind of 
manifolds corresponds to the scalar manifold of the so called {maximally extended supergravities}, that is D--dimensional supergravity theories,
whose  supersymmetry is the maximal allowed by the space--time dimension
 D. \par
It is useful to spend few words about the relation between the two 
descriptions of 
${\cal M}$ considered so far, i.e. coset manifold and the group 
manifold ${\cal G}^{(s)}$. In the former case each point $P$ of ${\cal M}$
is associated with a coset--representative $\IL(P)\in Exp(\IK)$ which is
{\it not} a group element, being $\IK$ not an algebra, in the latter case, on the other hand, the same point is described by a group element
 $\IL_s(P)\in {\cal G}^{(s)}\approx Exp(Solv)$.
The two transformations $\IL(P)$ and $\IL_s(P)$ belong to the same 
{\it equivalence class} with respect to the right multiplication by an element
of ${\cal H}$:
\begin{equation}
\IL(P)\,=\,\IL_s(P)\cdot h(P)\, ;\quad  h(P)\in {\cal H}
\label{solcos}
\end{equation}
Differently from $\IL$, whatever point $P$ on ${\cal M}$ may be reached from a 
fixed origin $O\in {\cal M}$ by the left action of a suitable group 
element of ${\cal G}^{(s)}$ on $\IL_s(O)$ ,
 with no need of any compensator, being ${\cal G}^{(s)}$ a group of transitive 
isometries on ${\cal M}$:
\begin{equation}
\IL_s(P)\,=\, g(P, O)\cdot \IL_s(O)\, ;\quad   g(P, O)
\in {\cal G}^{(s)}
\end{equation}
As far as the action of the whole isometry group ${\cal G}$ on ${\cal M}$
is concerned, the best description of the latter is the 
coset manifold one, since $Solv$ is not stable with respect 
to the adjoint action of the compact subalgebra $\IH$ while $\IK$ on the other 
hand is. Therefore, starting from the parametrization of a point $P\in{\cal M}$
by means of {\it solvable coordinates} $\phi^I(P)$ defined by:  
\begin{equation}
\IL_s(\phi(P))\,=\,\exp{\left( \phi^I T_I\right)}\, ;\quad \phi^I T_I\in Solv
\label{solvcoor}
\end{equation}  
$T_I$ being the generators of $Solv$ in eq.\ref{iwa}, the transformation rules of $\phi^I$ under the action of a generic isometry $g\in {\cal G}$ in are defined in the following way:
\begin{eqnarray}
g:\, P\, &\longrightarrow &\, P^\prime\nonumber\\
\IL_s(\phi(P))& \rightarrow &\IL(P)\,{\stackrel{g}{\longrightarrow}}\,
\IL(P^\prime)\, \rightarrow\,\IL_s(\phi^{\prime}(P^\prime))
\label{solvtra}
\end{eqnarray}
where the first and the last arrows refer to a transformation from the 
group description to the coset manifold description and vice versa, while 
the middle arrow refers to the isometry transformation in the coset manifold formalism as described in eq. \ref{coset}.\par
One of the main reasons for choosing to describe a scalar manifold ${\cal M}$ 
in terms of a solvable Lie algebra is that in this formalism the scalar fields 
$\phi^I$ are interpreted as the local solvable coordinates defined in  eq.
\ref{solvcoor} which means to extablish a local one to one correspondence between the scalars and the generators $T_I$:
\begin{equation}
\phi^I\longleftrightarrow T_I\, ;\quad I=1,...,m=dim{\cal M}
\label{onetoone}
\end{equation}
As we will se in next chapter, such a correspondence comes out in a quite 
natural way and allows on one hand to define, through the procedure 
represented in eq. \ref{solvtra}, the trasformation rule of each scalar field 
under the U--duality group ${\cal G}$, on the other hand to achieve a 
geometrical intrinsic characterization of the scalars and in particular of the
RR and NS sectors.
An other considerable advantage of using metric solvable Lie algebras 
$(Solv, \langle,\rangle)$ is that the 
local differential geometry of the scalar manifold is described in terms of 
algebraic structures. In particular differential operators on ${\cal M}$
such as the covariant derivative and the Riemann tensor correspond to 
linear operators on the generating solvable Lie algebra $Solv\sim T_p{\cal M}$
in the neighborhood of a generic point $p\in {\cal M}$:
\begin{eqnarray}
\bigtriangledown_X && \longrightarrow L_X\,\,(\mbox{Nomizu operator})\,\,\,\forall x\in Solv\nonumber\\
R(X,Y) && \longrightarrow Riem(X,Y)\,\,(\mbox{Riemann operator})\,\,\,\forall X
, Y\in Solv
\label{nomiri}
\end{eqnarray}
where the {\it Nomizu} operator $ L_X\,:Solv\rightarrow Solv$ is defined in the following way:
\begin{eqnarray}
  \forall X,Y,Z \in Solv & : &2 <Z,L_X Y> \nonumber\\
&=& <Z,[X,Y]> - <X,[Y,Z]> - <Y,[X,Z]>
\label{nomizu}
\end{eqnarray}
and the Riemann curvature 2--form $ Riem(X,Y)\,:Solv\rightarrow Solv$ 
is given by the commutator of two Nomizu operators:
\begin{equation}
 <W,\{[L_X,\IL_Y]-L_{[X,Y]}\}Z> = Riem^W_{\ Z}(X,Y)
\label{nomizu2}
\end{equation}
Since the action of a supergravity theory depends on the scalar manifold only through its local differential properties, such a new formalism allows
to describe the scalar dependence of the theory through an algebraic structure
which is much simpler to master. \par
To end this section I will give an explicit example of Iwasawa decomposition Vs
 Cartan decomposition  for the semisimple group ${\cal G}=SL(3,\IR)$.
The positive roots of ${\cal G}$ are:
\begin{equation}
\alpha_1=\epsilon_1-\epsilon_2\,;\,\,\alpha_2=\epsilon_2-\epsilon_3\,;\,\,
\alpha_3=\epsilon_1-\epsilon_3
\label{posrot}
\end{equation}
Let $\lambda_i\, ;\, i=1,...,8$ denote a basis for the Lie algebra $\IG$ of 
${\cal G}$ consistent with the Cartan decomposition.
\begin{eqnarray}
\mbox{{\bf Cartan decomposition:   }} \IG\,=\,\IH\oplus \IK\nonumber\\
\nonumber \\
\IH\,=\, \left[\matrix{ \lambda_6=\left(\matrix{0 & 1 & 0\cr
-1 & 0 & 0\cr 0 & 0 & 0}\right) &\lambda_7=\left(\matrix{0 & 0 & 1\cr
0 & 0 & 0\cr -1 & 0 & 0}\right)\cr \lambda_8=\left(\matrix{0 & 0 & 0\cr
0 & 0 & 1\cr 0 & -1 & 0}\right) &\phantom{\matrix{0 & 0 & 1\cr
0 & 0 & 0\cr 1 & 0 & 0}}}\right]\nonumber\\
\IK\,=\, \left[\matrix{\lambda_1 =\left(\matrix{1 & 0 & 0\cr
0 & 0 & 0\cr 0 & 0 & -1}\right) &\lambda_2=\left(\matrix{0 & 0 & 0\cr
0 & 1 & 0\cr 0 & 0 & -1}\right)\cr \lambda_3=\left(\matrix{0 & 1 & 0\cr
1 & 0 & 0\cr 0 & 0 & 0}\right) &\lambda_4=\left(\matrix{0 & 0 & 1\cr
0 & 0 & 0\cr 1 & 0 & 0}\right)\cr \lambda_5=\left(\matrix{0 & 0 & 0\cr
0 & 0 & 1\cr 0 & 1 & 0}\right) &\phantom{\matrix{0 & 0 & 1\cr
0 & 0 & 0\cr 1 & 0 & 0}}}\right]\nonumber\\
\nonumber \\
\mbox{{\bf Iwasawa decomposition:   }}\IG\,=\,\IH\oplus Solv \nonumber\\
\nonumber \\
Solv\,=\, \left[\matrix{\lambda_1 =\left(\matrix{1 & 0 & 0\cr
0 & 0 & 0\cr 0 & 0 & -1}\right) &\lambda_2=\left(\matrix{0 & 0 & 0\cr
0 & 1 & 0\cr 0 & 0 & -1}\right)\cr E_{\alpha_1}=\left(\matrix{0 & 1 & 0\cr
0 & 0 & 0\cr 0 & 0 & 0}\right) &E_{\alpha_2}=\left(\matrix{0 & 0 & 0\cr
0 & 0 & 1\cr 0 & 0 & 0}\right)\cr E_{\alpha_3}=\left(\matrix{0 & 0 & 1\cr
0 & 0 & 0\cr 0 & 0 & 0}\right) &\phantom{\matrix{0 & 0 & 1\cr
0 & 0 & 0\cr 0 & 0 & 0}}}\right]
\label{carvsiva}
\end{eqnarray}
Siunce the Cartan subalgebra ${\cal C}$ is generated by $\{ \lambda_1, \lambda_2\}$
which are non--compact, it follows that ${\cal C}={\cal C}_K$ and the algebra 
is maximally non compact. The solvable algebra $Solv$ is the direct sum of ${\cal C}$
and the {\it nilpotent} generators corresponding to {\it all} the positive roots, according to eq.\ref{iwa}.
\section{K\"ahlerian, Quaternionic algebras and c--map.}
In this last section, I will specialize the solvable algebra machinery
to the case of K\"ahlerian and Quaternionic manifolds which are of great interest in some
supergravity theories. The main reference which I will follow in this treatment
is \cite{alex}, even if I will give just the main results which will be useful 
for a later analysis, skipping the mathematical details.\par
Let us breifly recall the definitions of {\it Special K\"ahler } (of {\it local
} type) 
and of {\it Quaternionic} manifolds whose geometries characterize the scalar 
sector of $N=2$ supergravity theories. For a complete analysis of their 
properties see \cite{pietrolectures}.
\subsection{Special K\"ahler manifolds.}
Consider a complex n--dimensional manifold ${\cal M}$ endowed with a {\it complex structure}
$J\in End(T{\cal M})$, $J^2=-Id$ and a metric $g$, {\it hermitian}
with respect to $J$, i.e.
\begin{equation}
g(J\cdot w\, ,\, J\cdot u)\, =\, g(w\, ,\, u)\, ; \quad\forall w,u\in T{\cal M}
\label{hermite}
\end{equation}
Referring to a {\it well adapted } basis of $T{\cal M}$ $(\partial
/\partial z^i\, ;i=1,...,n\,\,: J\cdot \partial /\partial z^i={\rm
i}\partial /\partial z^i)$, hermiticity of $g$ implies that 
$g_{ij}=g_{i^\star j^\star}=0$ and its reality that $g_{i\star j}=g_{ij^\star}$. Let us now introduce the 
differential $2$--form $K$:
\begin{equation}
K(w\, ,\,u)\, =\, \frac{1}{2\pi}g(J\cdot w\, ,\, u)\, ; \quad\forall w,u\in T{\cal M}
\label{Kblob}
\end{equation}
whose expression in a local coordinate patch is:
\begin{equation}
K\, =\, \frac{{\rm i}}{2\pi}g_{i j^\star}dz^i\wedge d\bar{z}^{j^\star}
\label{Klocal}
\end{equation}
The hermitian manifold ${\cal M}$ is called a {\it K\"ahler} manifold
iff:
\begin{equation}
dK\, =\,0
\label{kahler}
\end{equation}
Equation \ref{kahler} is a differential equation for $g_{i j^\star}$ whose general solution in a generic local chart is given by the expression:
\begin{equation}
g_{i j^\star}\, =\,\partial_i\partial_{j^\star}{\cal K}
\label{kahlpot}
\end{equation}
where the real function ${\cal K}={\cal K}^\star={\cal K}(z,\bar{z})$ 
is named {\it K\"ahler potential}. and is defined up to the real part of an holomorohic function.
Let us now introduce a bundle ${\cal Z}\rightarrow {\cal M}$ of the form
${\cal Z}={\cal L}\otimes{\cal SV}$ where ${\cal L}\rightarrow {\cal M}$
is a line bundle whose first chern class equals the cohomology class of $K$
($\Rightarrow {\cal M}$ is {\it Hodge K\"ahler} manifold), and ${\cal SV}\rightarrow {\cal M}$ is a symplectic bundle with structure group $Sp(2n+2,\IR)$.\par
{\bf Definition:  }{ \it ${\cal M}$ is a {\bf Special K\"ahler} manifold of 
{\bf local} type iff there exist a holomorphic  section $\Omega\in \Gamma({\cal Z},{\cal M})$
\begin{equation}
\Omega(z)\, =\,\left(\matrix{X^\Lambda (z) \cr F_{\Sigma} (z)}\right)
\label{Zsection}
\end{equation}
such that:
\begin{eqnarray}
K\, &=&\, \frac{{\rm i}}{2\pi}\partial\bar{\partial}\left({\rm i}
\langle \Omega , \bar{\Omega}\rangle\right)\nonumber\\
\langle \partial_i \Omega , \partial_j\Omega\rangle\, &=&\,0\nonumber \\
{\rm i}\langle \Omega , \bar{\Omega}\rangle & {\stackrel{def}{=}} &
{\rm i}\left(\bar{X}^\Lambda F_\Lambda\, -\,\bar{F}_\sigma X^\sigma\right)
\label{speckahl}
\end{eqnarray}
}
The $n$ complex scalar fields belonging to $n$ vector multiplets of a $N=2$
supergravity theory span a Special K\"ahler manifold.
\subsection{Quaternionic manifolds.}
The {\it hyperscalars} of an  $N=2$ supergravity theory on the other hand turn out to parametrize a {\it Quaternionic} manifold ${\cal QM}$ 
which I am about to define.\par
 Consider a $4m$ dimensional real manifold ${\cal QM}$ endowed with three complex structures $J^\alpha:T{\cal QM}\rightarrow T{\cal QM}
\,;\,\alpha=1,2,3$ such that:
\begin{equation}
J^\alpha J^{\beta}\,=\,-\delta^{\alpha\beta}\bfone\, +\, \epsilon^{\alpha\beta
\gamma}J^{\gamma}
\label{quatstru}
\end{equation}
and a metric $h$ which is hermitian with respect to them:
\begin{equation}
h(J^\alpha\cdot w\, ,\, J^\alpha\cdot u)\, =\, h(w\, ,\, u)\, ; \quad\forall w,u\in T{\cal QM}\, ;\, \alpha=1,2,3
\label{qhermite}
\end{equation}
Let us introduce a triplet of $SU(2)$ Lie--algebra valued $2$--forms 
$K^\alpha$ ({\it Hyper K\"ahler} forms):
\begin{equation}
K^{\alpha}\, =\, K^{\alpha}_{u v}dq^u\wedge dq^v\, =\, h_{u w}{J^{\alpha}}^{w}_v
dq^u\wedge dq^v\, ;\, u,v=1,...,4m
\label{kalpha}
\end{equation}
Moreover let a principal $SU(2)$--bundle ${\cal SU}\rightarrow {\cal QM}$
be defined, whose connection is denoted by $\omega^\alpha$ and such that:
\begin{equation}
\bigtriangledown K^\alpha \, {\stackrel{def}{=}} \, dK^\alpha \, +\, \epsilon^{
\alpha\beta\gamma}\omega^\beta \wedge K^\gamma\, =\,0
\label{deltaqk}
\end{equation}
{\bf Definition:  } {\it ${\cal QM}$ is a {\bf Quaternionic} manifold iff
the curvature of the ${\cal SU}$ bundle $\Omega^\alpha$
\begin{equation}
\Omega^\alpha \, {\stackrel{def}{=}} \,d\omega^\alpha\, +\,\frac{1}{2}\epsilon^{\alpha\beta\gamma}\omega^\beta\wedge\omega^\gamma
\label{cursu}
\end{equation}
is proportional to the Hyper K\"ahler 2--forms:
\begin{equation}
\Omega^\alpha\, =\,\frac{1}{\lambda}K^\alpha
\label{quatcon}
\end{equation}
$\lambda$ being a non vanishing real number.}
The {\it holonomy} group of ${\cal QM}$ turns out to be the direct product of 
$SU(2)$ and of $Sp(2m;\IR)$.
\subsection{Alekseevskii's formalism.}
There is a wide class of Quaternionic and Special K\"ahler manifolds 
(${\cal QM}$, ${\cal SK}$) (not necessarely homogeneous) which admit a solvable group of 
transitive isometries (with no fixed point). By definition, these manifold are called {\it normal}.
In virtue of the main theorem stated in last section, they are in one to one correspondence with suitable solvable Lie algebras which are hence named
{\it quaternionic} and {\it K\"ahlerian} algebras respectively.
In the sequel I will denote a quaternionic metric algebra by $(V,\langle ,\rangle)$ and a K\"ahlerian metric algebra by $(W,\langle ,\rangle)$ :
\begin{eqnarray}
{\cal QM}\, &\approx &\,\exp{V}\nonumber\\
{\cal SK}\, &\approx &\,\exp{W} 
\label{qmsk}
\end{eqnarray}
In the seventies Alekseevskii faced the problem of classifying the normal quaternionic manifolds through their generating quaternionic algebras and found 
an interesting correspondence between the latter and certain K\"ahlerian
algebras, later called the {\it c--map}.
All the quantities defined for ${\cal QM}$ and ${\cal SK}$ have their algebraic couterpart on $(V,\langle ,\rangle)$ and $(W,\langle ,\rangle)$, such as 
the quaternionic structure 
$J^\alpha:\, V\rightarrow V$ for the former and the complex structure 
$J:\, W\rightarrow W$ for the latter, fulfilling eq. \ref{quatstru} and 
$J^2=-Id$ respectively.\par
The quaternionic algebras $V$ which we will be interested in, have the following structure:
\begin{eqnarray}
V\, &=&\, U\oplus \tilde{U}\nonumber\\
U\, &=&\, F_0 \oplus W\nonumber\\
\tilde{U}\, &=&\, \tilde{F}_0 \oplus \tilde{W}\nonumber\\
J^1\cdot U\, &=&\,U\,\,\,J^2\cdot U\, =\, \tilde{U}
\label{Vstru}
\end{eqnarray} 
where $F_0$ and $\tilde{F}_0$ are defined in the following way:
\begin{eqnarray}
 F_0\, &=&\,\{e_0\, ,\, e_1\} \nonumber\\
\tilde{F}_0\, &=&\,\{e_2\, ,\, e_3\} \nonumber\\
\tilde{F}_0\, &=&\,J^2\cdot F_0 \nonumber\\
\left[e_0\, ,\, e_1\right]\, &=&\, e_1\,\,;\,\,\left[e_1\, ,\, e_2\right]\, =\,0\nonumber\\
\left[e_0\, ,\, e_2\right]\, &=&\, e_2\,\,;\,\,\left[e_1\, ,\, e_3\right]\, =\,0\nonumber\\
\left[e_0\, ,\, e_3\right]\, &=&\, e_3\,\,;\,\,\left[e_2\, ,\, e_3\right]\, =\,0
\label{f0tf0}
\end{eqnarray} 
where $e_i\, \, i=0,1,2,3$ are orthonormal with respect to $\langle ,\rangle$.
Moreover the following relations hold:
\begin{eqnarray}
\left[e_1\, ,\, W\oplus \tilde{U} \right]\, &=&\,0\nonumber\\
\left[e_0\, ,\,\tilde{U}\right]\, &=&\,\frac{1}{2}\,\tilde{U}\nonumber\\
\left[e_0\, ,\,W\right]\, &=&\,0\nonumber\\
\left[U\, ,\, U\right]\, \subset\, U\, \quad \left[U\, ,\, \tilde{U}\right]\, &\subset&\, \tilde{U}
\, \quad \left[\tilde{U}\, ,\, \tilde{U}\right]\, \subset\,\{e_1\}
\label{Vstru2}
\end{eqnarray} 
$(F_0, J^1_{\vert F_0})$ and  $(W, J^1_{\vert W})$ are K\"ahlerian algebras
whose complex structure is the restriction of $J^1$ on them. $F_0$, defined by \ref{f0tf0} is called 
{\it key algebra} with weight one
(it is strightforward to check that a key algebra is the solvable algebra generating the homogeneous manifold
$SU(1,1)/U(1)$). The mapping $Adj_{U}\, :\,\tilde{U}
\rightarrow \tilde{U}$ defines a representation of the K\"ahlerian algebra 
$U$ on the subspace $\tilde{U}$. With respect to a new complex structure 
introduced on $\tilde{U}$ as follows:
\begin{equation}
\tilde{J}_{\tilde{F}_0}{\stackrel{def}{=}}J^1\, ;\,\,\quad 
\tilde{J}_{\tilde{W}}{\stackrel{def}{=}}-J^1
\label{newcofo}
\end{equation}
one finds that $Adj$ fulfills the axioms defining a {\it Q--representation}.
The most relevant of them tells that $Adj$ is a symplectic representation with 
respect to $\tilde{J}$. Thus $U$ and in particular $W$ is a K\"ahlerian algebra
having a Q--representation and it was shown that there is a one to one 
correspondence ({\it c--map}) between quaternionic algebras of the kind 
\ref{Vstru} and K\"ahlerian algebras $W$ admitting a Q--representation.
\begin{equation}
V(\mbox{quat. algebra}){\stackrel{\mbox{c--map}}{\leftrightarrow}}
W(\mbox{k\"ah. algebra + Q--repr.})
\label{cmap}
\end{equation}
Alekseevskii indeed reduced the problem of classifying quaternionic algebras
to the simpler problem of classifying  K\"ahlerian algebras $W$ admitting a Q--representation. Among the latter, those which we will be dealing with
throughout this thesis are the {\it type 1}, {\it rank 3} ones. They have the following structure:
\begin{eqnarray}
W\,&=&\, F_1\oplus F_2\oplus F_3 \oplus X\oplus Y\oplus Z\nonumber\\
X\,&=&\, X^+\oplus  X^-\nonumber\\
Y\,&=&\, Y^+\oplus  Y^-\nonumber\\
Z\,&=&\, Z^+\oplus  Z^-\nonumber\\
\label{Wstru}
\end{eqnarray}
the $2$--dimensional real subalgebras $F_i\, =\, \{h_i\, ,\,g_i\}\,;\,
\left[h_i\, ,\,g_i\right]\, =\,g_i\,;\, i=1,2,3$ are all key algebras of weight one
(that's why $W$ is said to be a type one algebra). Moreover  the three $h_i$
are the only semisimple generators in $W$, which means that 
 $W$ has rank 3. It was shown that $W$ admits a Q--representation only in the following two cases:
\iffigs
\begin{figure}
\caption{}
\label{quat}
\epsfxsize = 10cm
\epsffile{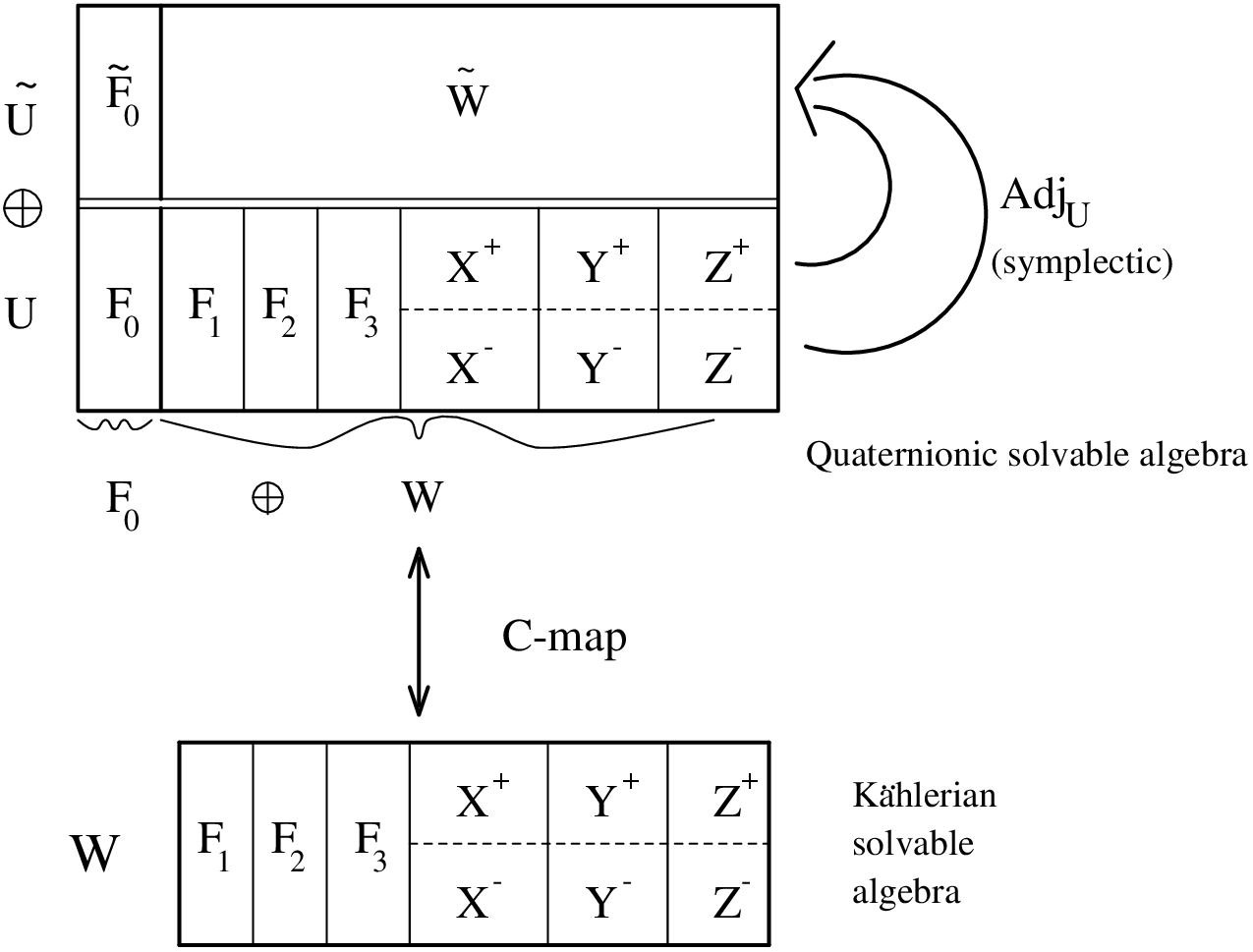}
\vskip -0.1cm
\unitlength=1mm
\end{figure}
\fi
\begin{itemize}
\item $X=0$, dim $Y=p$, dim $Z=q$. In this case the K\"ahlerian algebra
$W$ is denoted by $K(p,q)$;
\item $X,Y,Z\neq 0$ , dim $Y^{\pm}=$dim $Z^{\pm}$
\end{itemize}
We will be dealing with the former kind of algebra in the third chapter, where 
an $N=2$ supergravity theory will be considered in which the vector scalars
describe a manifold of the type:
\begin{equation}
{\cal K}(0,q)\,=\, \exp{K(0,q)}\, =\, \frac{SU(1,1)}{U(1)}\otimes
\frac{SO(q+2,2)}{SO(q+2)\otimes SO(2)}
\label{n2vec}
\end{equation}
and the hyperscalars the following quaternionic manifold:
\begin{equation}
{\cal W}(0,r)\,=\, \exp{W(0,r)}\, =\,\frac{SO(r+4,4)}{SO(r+4)\otimes SO(4)}
\label{n2quat}
\end{equation}
the two algebras $K(0,q)$ and $W(0,q)$ correspond to each other through the c--map.
K\"ahlerian algebras of the second type on the other hand will be considered
in the last chapter for describing the following Special K\"ahler manifolds:
\begin{eqnarray}
H(1,4)\,&=&\,\frac{SO^{\star}(12)}{U(6)}\nonumber\\
H(1,2)\,&=&\,\frac{U(3,3)}{U(3)\otimes U(3)}
\end{eqnarray}
A visual representation of the structure of these quaternionic and
 K\"ahlerian algebras is given in Figure.\ref{quat}
Let us complete this section by writing the main algebraic relations among 
the different parts in \ref{Wstru}, which will be used mainly in
 the last chapter:
\begin{eqnarray}
\left[h_i\, ,\,g_i\right]\, &=&\, g_i\quad i=1,2,3\nonumber\\
\left[F_i\, ,\,F_j\right]\, &=&\, 0 \quad i\neq j\nonumber\\
\left[h_3\, ,\,Y^{\pm}\right]\, &=&\,\pm\frac{1}{2}Y^{\pm}
\nonumber\\
\left[h_3\, ,\,X^{\pm}\right]\, &=&\,\pm\frac{1}{2}X^{\pm}
\nonumber\\
\left[h_2\, ,\,Z^{\pm}\right]\, &=&\,\pm\frac{1}{2}Z^{\pm}
\nonumber\\
\left[g_3\, ,\,Y^{+}\right]\, &=&\,\left[g_2\, ,\,Z^{+}
\right]\, =\,\left[g_3\, ,\,X^{+}\right]\, =\, 0\nonumber\\
 \left[g_3\, ,\,Y^{-}\right]\, &=&\,Y^+\, ;\,\,
\left[g_2\, ,\,Z^{-}\right]\, =\,Z^+\, ;\,\,
\left[g_3\, ,\,X^{-}\right]\, =\,X^+\nonumber\\
\left[F_1\, ,\,X\right]\, &=&\,\left[F_2\, ,\,Y\right]\, =\,
\left[F_3\, ,\,Z\right]\, =\, 0\nonumber\\
\left[X^-\, ,\,Z^{-}\right]\, &=&\, Y^-
\label{Alekah}
\end{eqnarray}
\chapter{Maximally Extended Supergravities}
In the present chapter I will deal with the main applications of the
 solvable Lie algebra machinery to the analisys of the scalar structure 
of N--extended supergravities  in various dimensions and interesting relations 
among them will be found. My treatment will mainly refer to the two papers by our group
\cite{noialtri}, \cite{noialtri2}.\par
As anticipated in the previous chapter, the first goal of the solvable
 Lie algebra technique is to characterise in a geometrically intrinsic
 way the R--R and  N--S scalar sectors. Classifying the scalar fields as R--R or N--S is a meaningful
operation only in the limit in which the supergravity we are considering
is interpreted as a description of the low--energy limit of a suitably compactified superstring theory. In the geometrical procedure which will allow to distinguish
between R--R and N--S generators of a solvable Lie algebra the information
about the stringy origin of the scalar fields is encoded in the S, T--duality groups. In particular the T--duality group ${\cal G}_t$ is the isometry group 
of the moduli space ${\cal M}_t$ of the internal manifold ${\cal K}_d$ which 
the superstring theory has been compactified on, while the S--duality group 
${\cal G}_s$ on the other hand, acts on the dilaton field $\phi$ 
(whose vacuum expectation value defines the superstring coupling constant) and,
 in $D=4$, also on the {\it axion} $\chi$, which span another submanifold 
${\cal M}_s$ of the scalar manifold ${\cal M}$. As we are going to see,  
${\cal M}_s$ and ${\cal M}_t$ are indeed the only {\it input} 
needed in order to 
tell, within a certain supergravity framework, which scalar field is of N--S and which is of RR type, since all the N--S scalars $\phi$, ($\chi$) and $t^i$ (related to the moduli of ${\cal K}_d$) parametrize the solvable algebras $Solv_s\oplus Solv_t$
generating the submanifold ${\cal M}_s\otimes {\cal M}_t$ of ${\cal M}$. All the other scalar fields will transform in an irreducible representation
of the T--duality group and will define the R--R sector. The problem of defining the N--S and the R--R generators therefore reduces to that of decomposing the solvable algebra $Solv$ generating ${\cal M}$ with respect to  $Solv_s\oplus Solv_t$. \par
For the sake of simplicity I will focus my attention mainly on the so called maximally extended supergravities \cite{pope}, whose mathematical
structure is considerably constrained by their high degree of supersymmetry (the maximal allowed for  a certain space--time dimension
). Nevertheless it has to be stressed that all the results obtained for this kind of theories by means of solvable Lie algebra machinery, may be succesfully extended to {\it non}--maximal supergravities, even if mastering solvable algebras 
with non--maximal rank is a bit more complicate. \par One of the analyses we shall be particularly interested in when dealing with a particular supergravity theory
is the determination of the {\it maximal abelian ideal} ${\cal A}$
 within the solvable algebra generating the scalar manifold. Indeed, as already mentioned in the introduction, the knowledge of ${\cal A}$ and of its content in terms of N--S and R--R generators is relevant to the study of a partial supersymmetry breaking

 mechanism based on the gauging of certain translational symmetries
of the theory and which has been  so far succesfully tested in the case of a generic $N=2$ theory 
broken to $N=1$ \cite{fegipo}. We are going to deal with this problem in
 next chapter.\par
 In the first section of the present chapter I am going to define in a more rigorous fashion
the procedure for characterizing in an algebraic language the R--R and N--S sectors of maximally extended supergravities, quoting case by case the dimension and the R--R and N--S content of 
${\cal A}$. In the second section the probelm of finding the {\it electric} subalgebra $Solv_{el}$
of $Solv$ will be considered in the particular case of the $N=8,\, D=4$ theory. Such an analysis will allow to define the {\it gaugeable} isometries $Solv_{el}$ within $Solv$ and furthermore, decomposing $Solv_{el}$ with respect to the S and T--duality groups, to tell which of these generators are of R--R type and which of NS.  In the third section an example of how to extend the previously 
described results to non--maximally extended theories will be given. 
A further example will be considered in chapter four, where
the problem of characterizing geometrically the scalar content of a non--maximal $N=2$ 
supergravity theory in $D=4$ dimensions (obtained through a consistent truncation of the $N=8$
theory) will be dealt with using the procedure defined in the present chapter.\par
In the last three sections 
I shall try to recover the previously obtained results in a more detailed and 
systematic treatment in which the procedure of dimensional reduction on tori, which relates 
all the maximally extended supergravities, will be described in a geometrical way, by defining a chain of sequential embedding of the U--duality solvable algebra $Solv_{r}$ in $D+1=11-r$ dimensions into
the corresponding solvable algebra $Solv_{r+1}$ in $D$--dimensions.
Such chains of {\it regular embeddings} will be defined in the three relevant cases in which a 
given maximally extended supergravity in D--dimensions is thought of as obtained by dimensional reduction of
type IIA, IIB and M--theory. They  allow to keep trace of the scalar fields
inherited from the maximally extended theories in higher dimensions through 
sequential compactifications on tori. Exploiting in such a systematic 
way the geometrical  structure of the maximally extended supergravities and the
 relationship among them, will create the proper mathematical ground on which to formulate 
the problem of gauging compact and non--compact isometries. 
An other interesting result is the definition of 
the maximal abelian ideal ${\cal A}_D$ in each dimension D through the 
decomposition of $ Solv_{r+1}$ with respect to
 $Solv_{r}$. It will be shown how the R--R and N--S content of ${\cal A}_D$ 
may be immediately computed by decomposing it with respect to the S,T--duality group in $D+1$ dimensions. Finally the explicit pairing 
between the generators of $Solv_{r+1}$ and the scalar $0$--modes of type 
IIA theory compactified to D--dimensions will be achieved and the result
is displayed in Appendix A.
\section{Maximally extended supergravities and their solvable algebras:
NS and R--R generators}
\subsection{Counting the N--S and R--R degrees of freedom in a maximally extended supergravity}
A supergravity $(N,D)$ is called {\it maximally extended} when the order N of supersymmetry is the maximum allowed for the space--time dimension D, or, in a dimensionally independent way, when it has $32$ conserved supercharges. In any dimension $D=3,...,9$
such theories are uniquely defined and the scalar manifold ${\cal M}_D$ is constrained
by the large supersymmetry to have the characteristic form:
\begin{equation}
{\cal M}\, =\, \frac{E_{r+1(r+1)}}{{\cal H}_{r+1}}\, ;\,\,\, r=10-D
\label{scalmme}
\end{equation} 
where $E_{r(r)}$ is the maximally non--compact form of the {\it exceptional} series
and ${\cal H}_{r}\subset E_{r(r)}$ the maximal compact subgroup. 
The $E_{r(n)}$ 
algebras have a peculiar structure only for $r=6,7,8$, while for all $r< 6$
they coincide with the known algebras $A_r$ or $D_r$. The index $n$ defines the particular non--compact real form of the algebra and is the difference between the number of non--compact generators of $E_{r(n)}$ and the number of compact ones. In particular 
when $n=r$ this difference is given by the Cartan generators which are 
therefore all non--compact. Hence the scalar manifold of maximally extended supergravities is maximally non compact. In the sequel we will always denote by $r$ the number of compactified dimensions from $D=10$. In $D=10$ there are two kinds of maximally extended theories,
according to the relative chirality of the two supersymmetry generators: the type
IIA and type IIB theory. The former has $O(1,1)$ as U--duality group 
(acting on the only scalar which is the dilaton $\phi$) while in the latter case the 
U--duality group is $SL(2,\IR)$ (acting on two scalar fields: the dilaton $\phi$ and the 
RR scalar $\rho$). All the maximal supergravities in a lower dimension $D=10-r$
are obtained eighter from type IIA or from type IIB by means of compactification on 
an r--torus $T_r$. On the other hand type IIA theory is obtained from the $11$--dimensional supergravity (low--energy limit of the M--thery) through compactification on a circle. Therefore 
it is also possible to think of a maximal extended supergravity
in $D<10$ dimensions as the low--energy effective theory of the M--theory compactified on 
$T_{r+1}$. \par
The scalar fields of a maximally extended supergravity in $D<9$--dimensions will then be identified with the $0$--modes of suitably compactified type IIA, type IIB or M--theory 
according to which of the three possible interpretations of the theory is adopted. When the 
latter is interpreted as the low--energy limit of type IIA or type IIB compactified on tori,
its scalar content is divided into N--S and R--R sector. The number of R--R and N--S scalars 
is independent on the choice of type IIA or type IIB as the superstring 
origin of the theory.
Indeed those two superstring theories compactified to the same dimension on two tori
$T_r$ and $T_r^\prime$ correspond to each other through a T--duality 
transformation which maps the moduli
of $T_r$ into those of  $T_r^\prime$. A feature of T--duality is to leave
 the N--S and  RR
connotation of the scalar fields invariant.\par 
If we consider the bosonic massless spectrum \cite{gsw} of  type IIA theory in $D=10$
in the N--S sector we have the metric, the axion and the dilaton,
while in the R--R sector we have a 1--form and a 3--form:
\begin{equation}
 D=10 \quad : \quad  \cases{
 NS: \quad g_{\mu\nu}, B_{\mu\nu} , \Phi \cr
 RR: \quad  A_{\mu} , A_{\mu\nu\rho} \cr}
 \label{d10spec}
\end{equation}
corresponding to the following counting of degrees of freedom:
$\# $ d.o.f. $g_{\mu\nu} = 35$, $\# $ d.o.f. $B_{\mu\nu} = 28$, $\# $ d.o.f. $A_{\mu} = 8$, $\# $ d.o.f. $A_{\mu\nu\rho} = 56$
so that the total number of degrees of freedom is $64$  both in the Neveu--Schwarz
and in the Ramond:
 \begin{eqnarray}
  \mbox{Total $\#$ of N--S degrees of freedom}&=&{ 64}={ 35}+{ 28}+ { 1} \nonumber\\
  \mbox{Total $\#$ of R--R degrees of freedom}&=&{ 64}={ 8}+{ 56}
  \label{64NSRR}
 \end{eqnarray}
 \par
Let us now organize the degrees of freedom as they appear after toroidal compactification
on a $r$--torus \cite{pope}:
\begin{equation}
{\cal M}_{10} = {\cal M}_{D-r} \, \otimes T_r
\end{equation}
Naming with Greek letters the world indices on the $D$--dimensional
space--time and with Latin letters the internal indices referring to
the torus dimensions we obtain the results displayed in Table \ref{tabu1} and number--wise we
obtain the counting of Table \ref{tabu2}:
\par
\vskip 0.3cm
\begin{table}[ht]
\begin{center}
\caption{Dimensional reduction of type IIA fields}
\label{tabu1}
 \begin{tabular}{|l|l|c|c|c|c|r|}\hline
\vline & \null & \vline & Neveu Schwarz & \vline & Ramond Ramond & \vline \\
 \hline
 \hline
\vline & Metric & \vline &  $g_{\mu\nu}$& \vline  & \null & \vline   \\ \hline
\vline & 3--forms & \vline &  \null & \vline  & $A_{\mu\nu \rho}$ & \vline \\ \hline
\vline & 2--forms & \vline   & $B_{\mu\nu}$ & \vline   & $A_{\mu\nu i}$ & \vline  \\ \hline
\vline & 1--forms & \vline   &  $g_{\mu i}, \quad B_{\mu i}$ & \vline  & $A_{\mu},
 \quad A_{\mu ij}$ & \vline \\ \hline
\vline & scalars  & \vline  & $\Phi, \quad g_{ij}, \quad B_{ij}$ &
 \vline  & $A_{i}, \quad A_{ijk}$ & \vline \\ \hline
 \end{tabular}
 \end{center}
\end{table}
 \par
 \vskip 0.3cm
 \par
\vskip 0.3cm
\begin{table}[ht]
\begin{center}
\caption{Counting of type IIA fields}

\label{tabu2}
 \begin{tabular}{|l|l|c|c|c|c|r|}\hline
\vline & \null & \vline & Neveu Schwarz & \vline & Ramond Ramond & \vline \\
 \hline
 \hline
\vline & Metric & \vline &  ${ 1}$& \vline  & \null & \vline   \\ \hline
\vline & $\#$ of 3--forms & \vline &  \null & \vline  & ${  1}$ & \vline \\ \hline
\vline &$\#$ of  2--forms & \vline   & ${  1}$ & \vline   & ${  r}$ & \vline  \\ \hline
\vline &$\#$ of 1--forms & \vline   &  $ {  2 r}$ & \vline  & $ {  1} +
\frac 1 2 \, r \, (r-1)$ & \vline \\ \hline
\vline & scalars  & \vline  & $1 \, +  \, \frac 1 2 \, r \, (r+1)$&
\vline  & $ r \, +\, \frac 1 6 \, r \, (r-1) \, (r-2)  $ & \vline \\
\vline &  \null   & \vline & $  +  \, \frac 1 2 \, r \, (r-1)  $ & \vline & \null   & \vline \\
\hline
\end{tabular}
\end{center}
\end{table}
\vskip 0.2cm
We can easily check that the total number
of degrees of freedom in both sectors is indeed $64$ after
dimensional reduction as it was before.
\subsection{Geometrical characterization of the N--S and R--R sectors} 
Our next step will be to construct the solvable algebra generating a manifold
of the form \eqn{scalmme}.
As previously pointed out, in the U--duality algebra of maximally extended supergravities
($E_{r+1(r+1)}$ in $D=10-r$) all Cartan generators are non compact , namely $\cC_K= \cC$, and from section 2
 of last chapter we know that the corresponding solvable Lie algebra has the 
universal simple form:
\begin{equation}
Solv \, = \, {\cal C} \, \oplus \,
\sum_{\alpha \, \in \, \Phi^+}  \, E^{\alpha}
\label{simstru}
\end{equation}
where ${\cal C}$ is the Cartan subalgebra, $E^{\alpha}$ is
the root--space
corresponding to the root $\alpha$ and $\Phi^+$
denotes the set of
positive roots of $E_{r+1(r+1)}$.
 \par
From the M--theory interpretation of the supergravity, 
the Cartan semisimple piece $\cC=O(1,1)^{r+1}$ of the
solvable Lie algebra
has the physical meaning of
\footnote{Similar reasonings appear in refs.\cite{pope}}
  diagonal moduli for the $T_{r+1}$
compactification torus
(roughly speaking the radii of the $r+1$ circles)
\cite{witten}.
\par
From  a stringy (type IIA) perspective one of them is the dilaton
and the others are the Cartan piece of the maximal rank solvable
Lie algebra  generating the moduli space ${O(r,r)\over O(r)
\otimes O(r)}$  of the $T_{r}$ torus.
\par
This trivially implies that the Cartan piece is always
in the N--S sector.
\par
We are interested in splitting the maximal solvable subalgebra
\eqn{simstru} into its N--S and R--R parts.
To obtain this splitting, as already mentioned in the introduction,
we just have to decompose
the U--duality algebra $U$ with respect to its ST--duality
subalgebra $ST\subset U$ \cite{feko}, \cite{witten}.\footnote{
Note that at $D=3$, ST--duality merge in a simple Lie algebra
\cite{sm}\cite{sen}.}
We have:
\begin{eqnarray}
  5 \leq D \leq 9 \quad :\quad & ST = &O(1,1)
  \otimes O(r,r) \nonumber\\
 D = 4 \quad :\quad & ST = &Sl(2,\IR) \otimes O(6,6) \nonumber\\
 D =3 \quad :\quad & ST = & O(8,8)
\label{stdual}
\end{eqnarray}
Correspondingly we obtain the decomposition:
\begin{eqnarray}
 5 \leq D \leq 9 \quad :\quad  \mbox{adj} \,
 E_{r+1(r+1)}&= & \mbox{adj}O(1,1) \oplus \mbox{adj}O(r,r)
 \nonumber\\
&&\oplus \left(2,\mbox{spin}_{(r,r)} \right)\nonumber\\
 D=4 \quad :\quad  \mbox{adj} \, E_{7(7)}&= & \mbox{adj}Sl(2,\IR)
 \oplus \mbox{adj}O(6,6)
\oplus \left(2,\mbox{spin}_{(6,6)} \right)\nonumber\\
 D=3 \quad :\quad  \mbox{adj} \, E_{8(8)}& = &  \mbox{adj}O(8,8)
\oplus \mbox{spin}_{(8,8)}
 \label{stdec}
 \end{eqnarray}
From (\eqn{stdec}) it follows that:
\begin{eqnarray}
  \label{stdim}
  5 \leq D \le 9 \quad &:& \quad  \mbox{dim} E_{r+1(r+1)}= 1 +
  r(2r-1) + 2^{r}  \nonumber\\
D= 4  \quad &:& \quad  \mbox{dim}E_{7(7)}=
\mbox{dim}[ \left ( {\bf 66},{\bf 1} \right )
\oplus \left ( {\bf 1},{\bf 3} \right ) \oplus \left ({\bf 2},{\bf 32}
\right )]\nonumber\\
D=3 \quad &:& \quad \mbox{dim}E_{8(8)}=  \mbox{dim}[ {\bf 120}
\oplus {\bf 128}]
\end{eqnarray}
The dimensions of the maximal rank solvable algebras are instead:
\begin{eqnarray}
  \label{solvdim}
  5 \le D\le 9 \quad &:& \quad  \mbox{dim} Solv_{r+1}=  r^2 +  1 +
  2^{(r-1)} = \mbox{dim} {U\over H} \nonumber\\
D= 4 \quad &:& \quad   \mbox{dim}Solv_7 =    32 +  37 + 1 = \mbox{dim}
{U\over H} \nonumber\\
D = 3 \quad &:& \quad \mbox{dim}Solv_8=  64 +  64  = \mbox{dim} {U\over H}
\end{eqnarray}
The above parametrizations of the dimensions of the cosets
listed in Table 1
can be traced back to the fact that the N--S and R--R
generators are given respectively by:
\begin{equation}
  \mbox{ N--S} = \mbox{Cartan generators} \oplus
  \mbox{positive roots of }\mbox{adj}\,ST
\end{equation}
and
\begin{equation}
   \mbox{R--R}= \mbox{positive weights of }\mbox{spin}_{ST}
\end{equation}
In this way we have:
\begin{eqnarray}
 \mbox{dim}(\mbox{ N--S})&=&\cases{ r^2 + 1 \quad\quad (5 \le D\le 9)
 \cr
  38= 7+1+30  \quad\quad (D=4)\cr
64=8+56 \quad\quad (D=3)\cr}\nonumber\\
\mbox{dim}(\mbox{ R--R})&=&\cases{ 2^{(r-1)} \quad\quad
(5 \le D \le 9) \cr
 32  \quad\quad(D=4)\cr
64 \quad\quad (D=3)\cr}
\end{eqnarray}
\par

For $D=3$ we notice that the ST--duality group $O(8,8)$
is a non compact form of the 
maximal compact subgroup $O(16)$ of the U--duality group $E_{8(8)}$.
\par
This explains why R--R = N--S = 64 in this case.
Indeed, 64 are the positive weigths of
$\mbox{dim}(\mbox{spin}_{16}) = 128$.
This coincides with the counting of the bosons
in the Clifford algebra of $N=16$
supersymmetry at $D=3$.

\par
\begin{table}[h]
\begin{center}
\begin{tabular}{|c|c|c|c|}
\hline
$D=9$ &      $E_{2(2)}  \equiv SL(2,\IR)\otimes O(1,1)$ & $H =
O(2) $ &
$\mbox{dim}_{\bf R}\,(U/H) \, =\, 3$ \\
 \hline
$D=8$ &      $E_{3(3)}  \equiv SL(3,\IR)\otimes Sl(2,\IR)$ & $H =
O(2)\otimes O(3) $ &
$\mbox{dim}_{\bf R}\,(U/H) \, =\, 7$ \\
\hline
$D=7$ &      $E_{4(4)}  \equiv SL(5,\IR) $ & $H = O(5) $ &
$\mbox{dim}_{\bf R}\,(U/H) \, =\, 14$ \\
\hline
$D=6$ &      $E_{5(5)}  \equiv O(5,5) $ & $H = O(5)\otimes O(5) $ &
$\mbox{dim}_{\bf R}\,(U/H) \, =\, 25$ \\
\hline
$D=5$ &      $E_{6(6)}$   & $H = Usp(8) $ &
$\mbox{dim}_{\bf R}\,(U/H) \, =\, 42$ \\
\hline
$D=4$ &      $E_{7(7)}$   & $H = SU(8) $ &
$\mbox{dim}_{\bf R}\,(U/H) \, =\, 70$ \\
\hline
$D=3$ &      $E_{8(8)}$   & $H = O(16) $ &
$\mbox{dim}_{\bf R}\,(U/H) \, =\, 128$ \\
\hline
\end{tabular}
\caption{U--duality groups and maximal compact subgroups of maximally
extended supergravities.}
\label{costab}
\end{center}
\end{table}
In  Table 2 we give, for each of the previously listed cases,
 the dimension of the maximal abelian ideal $\cA$ of the solvable
 algebra and its
N--S, R--R content
\cite{cre}, \cite{huto}.
\begin{table}[h]
  \begin{center}
    \begin{tabular}{|c|c|c|c|}
\hline
D  & dim $\cA$ & N--S & R--R \\
\hline
3 & 36 & 14 & 22 \\
\hline
4 & 27 & 11 & 16 \\
\hline
5 & 16 & 8 & 8 \\
\hline
6 & 10 & 6 & 4 \\
\hline
7 & 6 & 4 & 2 \\
\hline
8 & 3 & 2 & 1 \\
\hline
9 & 1 & 0 & 1 \\
\hline
    \end{tabular}
    \caption{Maximal abelian ideals.}
    \label{tab:2}
  \end{center}
\end{table}


\section{Electric subgroups}
In view of possible applications to the gauging
of isometries of the four dimensional U--duality group, which may
give rise to spontaneous partial supersymmetry breaking
with zero--vacuum energy \cite{fegipo}, it is relevant
to answer the following question: what is
the electric subgroup\footnote{It is worth reminding that by ``electric''
we mean the group which has a lower triangular
symplectic embedding, i.e. is a symmetry of the
lagrangian \cite{dwvp1}, \cite{cdfvp}.}
of the solvable group? Furthermore, how many of
 its generators are
of N--S type and how many are of R--R type?
Here as an example we focus on the maximal
$N=8$ supergravity in $D=4$.
To solve the problem we have posed
we need to consider the splitting of the
U--duality symplectic representation pertaining
to vector fields, namely the  ${\bf 56}$ of $E_{7(7)}$,
under reduction with respect to the ST--duality
subgroup.
 The fundamental ${\bf 56}$ representation
defines the symplectic embedding:
\begin{equation}
E_{7(7)} \, \longrightarrow \, Sp(56,\IR)
\label{embed}
\end{equation}
We have:
\begin{equation}
{\bf 56} \, {\stackrel{Sl(2,R)\otimes SO(6,6)}
{\longrightarrow}} \, \left ({\bf 2} ,
{\bf 12}\right ) \, \oplus \, ({\bf 1},{\bf 32})
\end{equation}
This decomposition is understood from the physical
point of view by noticing that
the $28$ vector fields split into $12$ N--S fields
which, together with their
magnetic counterparts, constitute the $\left ({\bf 2} ,
{\bf 12}\right )$ representation plus $16$
R--R fields whose electric and magnetic
field strenghts build up the {\it irreducible} ${\bf 32}$ spinor
 representation of
$O(6,6)$. From this it follows that the T--duality
 group is purely electric only in
the N--S sector \cite{witten}. On the other hand the group
 which has an electric action both on
the N--S and R--R sector is $Sl(8,R)$.
 This follows from the alternative decomposition of the
${\bf 56}$ \cite{cj}, \cite{hull}:
\begin{equation}
{\bf 56} \, {\stackrel{Sl(8,R) }{\longrightarrow}} \,
  {\bf 28}
  \oplus \, {\bf 28}
\label{562828b}
\end{equation}
We can look at the intersection of the ST--duality
 group with the maximal electric group:
\begin{equation}
SL(2,\IR)\otimes O(6,6) \, \cap \, Sl(8,\IR) \,
= \, Sl(2,R)\, \otimes\, Sl(6,\IR) \,
\otimes \, O(1,1).
\end{equation}
Consideration of this subgroup allows to split
 into N--S and R--R parts the maximal
electric solvable algebra. Let us define it.
 I shall denote, in what follows,
by $Solv_{r+1}$ the solvable algebra generating 
the manifold $E_{r+1(r+1)}/H_{r+1}$. Hence  $Solv_7$ will stand for
${Solv}\left (E_{7(7)}/SU(8)\right )$. Its electric part is defined by:
\begin{equation}
{Solv}_{el} \, \equiv \, {Solv}\left (E_{7(7)}/SU(8)\right )
 \, \cap \, Sl(8,\IR) \, =
\,{Solv}\left (Sl(8,\IR)/O(8)\right )
\label{sl8so8}
\end{equation}
Hence we have that:
\begin{equation}
\mbox{dim}_{\bf R} \, {Solv}_{el} \, = 35
\end{equation}
One immediately verifies that the non--compact coset
manifold $Sl(8,\IR)/O(8)$ has
maximal rank, namely $r=7$, and therefore the electric
solvable algebra has once more
the standard form as in eq.\eqn{simstru} where $\cC$
is the Cartan subalgebra
of $Sl(8,\IR)$, which is the same as the original Cartan
subalgebra of $E_{7(7)}$
and the sum on positive roots is now restricted to those
that belong to $Sl(8,R)$.
These are $28$. On the other hand the adjoint representation
 of $Sl(8,\IR)$ decomposes
under the $Sl(2,\IR)\otimes Sl(6,\IR) \otimes O(1,1)$
  as follows
\begin{equation}
{\bf 63} ~~~ {\stackrel{Sl(2,\IR)\otimes Sl(6,\IR)
\otimes O(1,1)}{\longrightarrow}}
\, \left ({\bf 3},{\bf 1} ,{\bf 1} \right ) \, \oplus \,
\left ({\bf 1},{\bf 35} ,{\bf 1} \right ) \, \oplus \,
\left ({\bf 1},{\bf 1} ,{\bf 1} \right )
\, \oplus \, \left ({\bf 2},{\bf 6} ,{\bf 2} \right )
\label{qui}
\end{equation}
Therefore the N--S generators of the electric solvable
 algebra are the $7$ Cartan generators
plus the $16=1 \oplus 15$ {\it positive roots} of
$Sl(2,\IR)\otimes Sl(6,\IR)$.
 The R--R generators
are instead the {\it positive weights} of the
$\left ({\bf 2},{\bf 6} ,{\bf 2} \right )$
representation. We can therefore conclude that:
\begin{equation}
\mbox{dim}_{\bf R} \, {Solv}_{el} \, = 35 \, = 12 \mbox{R--R} \,
\oplus  [(15+1)+7]\, \mbox{N--S}
\end{equation}
\par
Finally it is interesting to look for the maximal abelian
subalgebra
 of the electric solvable
algebra.
It can be verified that the dimension of this algebra is 16,
corresponding to 8 R--R and 8 N--S.

\section{Considerations on non--maximally extended supergravities}
Considerations similar to the above can be made for all
the  non maximally extended or matter coupled
supergravities for which the solvable Lie algebra is not of maximal rank.
 Indeed, in the present case, the set of positive roots entering in formula
\eqn{iwa} is a proper subset of the positive roots
of $U$, namely those which are not orthogonal to the
whole set of roots defining the non--compact Cartan generators.
As an example, let us analyze the coset ${O(6,22) \over O(6)
\otimes O(22) }\otimes {Sl(2,\IR)\over U(1)}$
corresponding  to a $D=4$, $N=4$ supergravity theory obtained
compactifying type IIA string theory
 on $K_3 \times T_2$ \cite{dn}, \cite{sei}, \cite{huto}.
The product $Sl(2,\IR) \otimes O(6,22) $ is the U--duality group
of this theory, while the
  ST--duality group is
 $Sl(2,\IR) \otimes O(4,20 ) \otimes O(2,2)$. The latter acts on the
 moduli space of $K_3 \times T_2$ and on the dilaton--axion system.
Decomposing the U--duality group with respect to the ST duality group
 $Sl(2,\IR) \otimes O(4,20 ) \otimes O(2,2)$
we get:
\begin{eqnarray}
  \label{o622}
  \mbox{adj}(Sl(2,\IR) \otimes O(6,22)) &=& \mbox{adj}Sl(2,\IR) \nonumber\\
&+&  \mbox{adj}O(4,20)+\mbox{adj}O(2,2)+
({\bf 1},{\bf 24},{\bf 4})
\end{eqnarray}
The R--R fields belong to the subset of positive roots
of U contributing to $Solv$ which are also positive
 weights of the ST--duality group,
namely in this case those defining the $({\bf 1},{\bf 24},{\bf 4})$
representation. This gives us 48 R--R fields.
The N--S fields, on the other hand, are selected by taking those
positive roots of U entering
 the definition of $Solv$, which are also positive roots of ST, plus those
 corresponding to the non--compact generators ($\cC_k$) of the
 $U$--Cartan subalgebra.
\par
In our case we have:
\begin{eqnarray}
  &&\mbox{dim} U =\mbox{dim}O(6,22) + \mbox{dim}Sl(2,\IR) = 381 \nonumber\\
&&  \mbox{\# of positive roots of U }= 183 \nonumber\\
&&  \mbox{\# of positive roots of U not contributing to $Solv$ }
\nonumber\\
&& \qquad \qquad\quad = 183 -(\mbox{dim}U/H - \mbox{rank}U/H) =56
\nonumber\\
&& \mbox{dim} ST =  \mbox{dim}O(4,20) +  \mbox{dim}O(2,2) +
\mbox{dim}Sl(2,\IR) = 285
\nonumber\\
&&  \mbox{\# of positive roots of ST } = {1\over 2} (285 -15) = 135
\nonumber\\
&&   \mbox{\# of positive roots of ST  contributing to $Solv$ }= 135-56=79
\nonumber\\
&&   \mbox{\# of N--S }= 79 + \mbox{rank}U/H = 79+7 = 86 \nonumber\\
&& \mbox{dim} (U/H) =\mbox{dim}Solv = 48+86 = 134.
\end{eqnarray}
The maximal abelian ideal $\cA$ of $Solv$ has dimension
 64 of which 24 correspond to R--R fields while 40 to N--S fields.
\par
In an analogous way one can compute the number of N--S and R--R
fields for other
non maximally extended supergravity theories.
%
%
\section{ $E_{r+1}$ subalgebra chains and their string interpretation}
As anicipated in the introduction, we shall now turn on a more systematic 
treatment of the solvable Lie algebra structure of maximally extended 
supergravities and exploit useful relationships among them.
To this aim the first step is to inspect the algebraic properties of the solvable Lie algebras
$Solv_{r+1}$ defined by eq. \eqn{simstru} and illustrate the match between
these properties and the physical properties of the sequential compactification.
\par
Due to the specific structure \eqn{simstru} of a maximal rank solvable Lie algebra
every chain of {\it regular embeddings}:
\begin{equation}
E_{r+1} \, \supset \,K^{0}_{r+1} \, \supset \, K^{1}_{r+1}\, \supset \, \dots \, \supset \,
 K^{i}_{r+1}\, \supset \, \dots
\label{aletto1}
\end{equation}
where $K^{i}_{r+1}$ are subalgebras of the same rank and with
the same Cartan subalgebra $\cC_{r+1}$ as
$E_{r+1}$ reflects into  a corresponding sequence of embeddings
of solvable Lie algebras and,
 henceforth, of  homogenous non--compact scalar manifolds:
\begin{equation}
E_{r+1}/H_{r+1} \, \supset \,K^{0}_{r+1}/Q^{0}_{r+1} \,\supset \,  \dots  \,\supset \,
K^{i}_{r+1}/Q^{i}_{r+1}
\label{caten1}
\end{equation}
which must be endowed with a physical interpretation.
In particular we can consider embedding chains such that \cite{witten}:
\begin{equation}
K^{i}_{r+1}= K^{i}_{r} \oplus X^{i}_{1}
\label{spacco}
\end{equation}
where $K^{i}_{r}$ is a regular subalgebra of $rank= r$ and $X^{i}_{1}$
is a regular subalgebra of rank one.
Because of the relation between the rank and the number
of compactified dimensions such chains clearly
correspond to the sequential dimensional reduction of either typeIIA (or B) or of M--theory.
Indeed the first of such regular embedding chains we can consider is:
\begin{equation}
K^{i}_{r+1}=E_{r+1-i}\, \oplus_{j=1}^{i} \, O(1,1)_j
\label{caten2}
\end{equation}
This chain simply tells us that the scalar manifold of
supergravity in dimension $D=10-r$ contains the
direct product of the supergravity scalar manifold  in dimension $D=10-r+1$
with the 1--dimensional moduli
space of a $1$--torus (i.e. the additional compactification radius one gets by making a further
step down in compactification).
\par
There are however additional embedding chains that originate from the different choices
of maximal
ordinary subalgebras admitted by the exceptional Lie algebra of the $E_{r+1}$ series.
\par
All the $E_{r+1}$ Lie algebras contain a subalgebra $D_{r}\oplus O(1,1)$ so
that we can write the chain \cite{noialtri}:
\begin{equation}
K^{i}_{r+1}=D_{r-i}\, \oplus_{j=1}^{i+1} \, O(1,1)_j
\label{dueachain}
\end{equation}
From the discussion presented earlier in the present chapter
it is clear that the embedding chain \eqn{dueachain}
corresponds to the decomposition of the scalar manifolds into submanifolds spanned by either
 N-S or  R-R fields, keeping moreover track of the way they originate at each level of the
sequential dimensional reduction. Indeed the N--S fields correspond to generators of the
solvable Lie algebra that behave as integer (bosonic) representations of the
\begin{equation}
D_{r-i} \, \equiv \, SO(r-i,r-i)
\label{subalD}
\end{equation}
while R--R fields correspond to generators of the solvable Lie algebra assigned to the spinorial
representation of the subalgebras \eqn{subalD}.
A third chain of subalgebras is the following one:
\begin{equation}
K^{i}_{r+1}=A_{r-1-i}\,\oplus  \, A_1 \, \oplus_{j=1}^{i+1} \, O(1,1)_j
\label{duebchain}
\end{equation}
and a fourth one is
\begin{equation}
K^{i}_{r+1}=A_{r-i}\,  \oplus_{j=1}^{i+1} \, O(1,1)_j
\label{elechain}
\end{equation}
The physical interpretation of the \eqn{duebchain}, illustrated in the next subsection, has its
origin in type IIB string theory. Indeed, as I previously pointed out, 
the same supergravity effective lagrangian can be viewed as
the result of compactifying either version of type II string theory. If we take the IIB
interpretation
the distinctive fact is that there is, already at the $10$--dimensional level a complex scalar
field $\Sigma$ spanning the non--compact coset manifold $SL(2,\IR)_U/O(2)$.
The $10$--dimensional U--duality
group  $SL(2,\IR)_U$ must therefore be present in all lower dimensions and it
corresponds to the addend
$A_1$ of the chain \eqn{duebchain}.
\par
The fourth chain \eqn{elechain} has its origin  in an M--theory interpretation or in a
 physical problem posed by the
$D=4$ theory.
\par
If we compactify the $D=11$ M--theory to $D=10-r$ dimensions using an $(r+1)$--torus $T_{r+1}$,
the flat metric on this is parametrized by the coset manifold $GL(r+1) / O(r+1)$.
The isometry group of the $(r+1)$--torus moduli space is therefore $GL(r+1)$ and its
Lie Algebra is $A_r + O(1,1)$, explaining the chain \eqn{elechain}.
Alternatively, we may consider the origin of the same chain from a $D=4$ viewpoint.
As it has been previously mentioned, in four dimensions
 the electric vector field strengths do not span an irreducible representation
of the U--duality group $E_7$ but sit together with their magnetic counterparts in the irreducible
fundamental ${\bf 56}$ representation.  With respect to the electric subgroup
$ SL(8, \IR )$ of $E_{7(7)}$ the ${\bf 56}$ decomposes as in \eqn{562828b}. 
 The Lie algebra of the electric subgroup is
$A_7 \, \subset \, E_7$ and it contains an obvious subalgebra $A_6 \oplus O(1,1)$.
The intersection
of this latter with the subalgebra chain \eqn{caten2} produces the electric chain \eqn{elechain}.
In other words, by means of equation \eqn{elechain} we can trace back in each upper dimension
which
symmetries will maintain an electric action also at the end point of the dimensional reduction
sequence,
namely also in $D=4$.
\par
We have  spelled out the embedding chains of subalgebras that are physically significant from
a string theory viewpoint. The natural question to pose now  is  how to understand their
algebraic
origin and how to encode them in an efficient description holding true sequentially in all
dimensions,
namely for all choices of the rank $r+1=7,6,5,4,3,2$. The answer is provided by reviewing the
explicit construction of the $E_{r+1}$ root spaces in terms of $r+1$--dimensional
euclidean vectors
\cite{gilmore}.
\subsection{Structure of the  $E_{r+1(r+1)}$ root spaces and of the
associated solvable algebras}
The root system  of type $E_{r+1(r+1)}$  can be described
for all values of $1\le r \le 6$ in the following way. As any other
root system it is a finite subset of vectors $\Phi_{r+1}\, \subset\, \IR^{r+1}$
such that $\forall \alpha ,\beta \, \in \Phi_{r+1}$ one has
$ \langle \alpha , \beta \rangle \, \equiv  2 (\alpha , \beta )/ (\alpha , \alpha) \,
\in \, \ZZ $ and such that $\Phi_{r+1}$ is invariant with respect to
the reflections generated by any of its elements.
\par
\vskip 0.2cm
The root system is
given by the following set of length 2 vectors:
\par
\leftline{ \underline {\sl For $2 \le r \le 5$}}
\begin{equation}
 \Phi_{r+1} \, = \,  \left \{
\matrix { \mbox{roots} & \mbox{number} \cr
\null & \null \cr
{\underbrace {\quad \pm \, \epsilon_k \,\quad \pm \, \epsilon_\ell \quad}} & 4 \times
\left (\matrix { r \cr 2} \right )\cr
1 \,\le \,  k \, < \, \ell \, \le r  & \null \cr
\null & \null \cr
 {
\frac 1 2 \, \left ( \pm\epsilon_1 \pm \epsilon_2 \pm \dots \epsilon_r
\right ) \, \pm \, \sqrt{2 - \frac r 4 } \, \epsilon_{r+1} }  &   2^r \cr
 \null & \null \cr
}    \right \}
\label{erodd}
\end{equation}
\vskip 0.2cm
\par
\leftline{ \underline {\sl For $r=6$  }}
\begin{equation}
 \Phi_{7} \, = \,  \left \{
\matrix { \mbox{roots} & \null &\mbox{number} \cr
\null & \null & \null \cr
{\underbrace {\quad \pm \, \epsilon_k \,\quad \pm \, \epsilon_\ell \quad}} & \null & 60
 \cr
1 \,\le \,  k \, < \, \ell \, \le 6  & \null & \null \cr
\null & \null \cr
 \pm \, \sqrt {2} \epsilon_7 & \null & 2 \cr
 \null & \null & \null\cr
 {\underbrace {
\frac 1 2 \, \left (\pm \epsilon_1 \pm \epsilon_2 \pm \dots \epsilon_6
\right ) }}& \pm \, \sqrt{2 - \frac 3 2 } \, \epsilon_{7}   &   64 \cr
\mbox{even number of + signs} & \null & \null \cr
 \null & \null & \null \cr
}    \right \}
\label{ersix}
\end{equation}
where  $\epsilon_i$ ($i=1,\dots , r+1)$ denote a complete set
of orthonormal vectors.
 As far as the roots of the form
$(1/2)( \pm\epsilon_1 \pm \epsilon_2 \pm \dots \epsilon_r ) \,
\pm \, \sqrt{2 - (r/4) } \, \epsilon_{r+1}$ in \eqn{erodd} are concerned,
the following conditions on the number of plus signs in their expression are
 understood: in the case {\it r=even} the number of plus signs within the
round brackets must be even, while in the case {\it r=odd} there must be an
 overall even number of plus signs. These conditions are implicit also in
\eqn{eroddsol}.
The {\it r$=1$} case is degenerate for $\Phi_2$ consists of the only
roots $\pm [(1/2)\epsilon_1+\sqrt{7}/2 \epsilon_2]$.
\par
For all values of $r$ one can find a set of simple roots
$\alpha_1 , \alpha_2 . \dots \, \alpha_{r+1}$ such that the corresponding
Dynkin diagram is the standard one given in figure \eqn{standar}
\iffigs
\begin{figure}
\caption{}
\label{standar}
\epsfxsize = 10cm
\epsffile{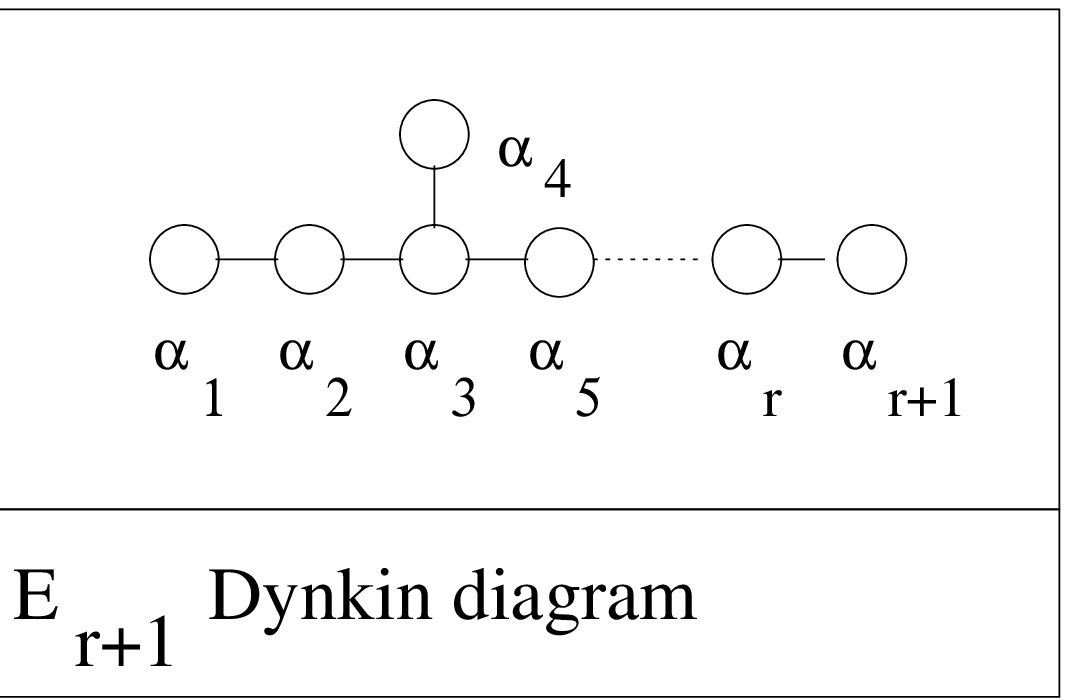}
\vskip -0.1cm
\unitlength=1mm
\end{figure}
\fi
\par
Consequently we can explicitly list the generators of all the
relevant solvable algebras for $r=1,\dots,6$ as follows:
\par
\leftline{ \underline {\sl For $2 \le r \le 5$ }}
\begin{eqnarray}
  & Solv_{r+1} \, = \, & \nonumber \\
  & \left \{
\matrix { \null & \mbox{number} & \mbox{type}\cr
\null & \null & \null \cr
\mbox{Cartan gener.} & r+1 & \mbox{NS} \cr
\null & \null & \null \cr
\mbox{roots} & \null  & \null \cr
\null & \null & \null \cr
{\underbrace {\quad  \, \epsilon_k \,\quad \pm \, \epsilon_\ell \quad}} & 2 \times
\left (\matrix { r \cr 2} \right ) & \mbox{NS} \cr
1 \,\le \,  k \, < \, \ell \, \le r  & \null & \null \cr
\null & \null & \null \cr
 {
\frac 1 2 \, \left (\pm \epsilon_1 \pm \epsilon_2 \pm \dots \epsilon_r
\right ) \, + \, \sqrt{2 - \frac r 4 } \, \epsilon_{r+1} }  &   2^{r-1} & \mbox{RR} \cr
\null & \null & \null \cr
 \null  & 2^{r-1}+r^2+1 &  \mbox{= Total } \cr
}    \right \} & \nonumber\\
\label{eroddsol}
\end{eqnarray}
\vskip 0.2cm
\par
\leftline{ \underline {\sl For $r=6$}}
\begin{eqnarray}
 & Solv_{7} \, = \, & \nonumber\\
 & \left \{
\matrix { \null & \null &\mbox{number} & \mbox{type}\cr
\null & \null & \null & \null\cr
\mbox{Cartan gener.} & \null & 7 & \mbox{NS} \cr
\null & \null & \null & \null \cr
\mbox{roots} & \null & \null & \null \cr
\null & \null & \null & \null \cr
{\underbrace {\quad  \, \epsilon_k \,\quad \pm \, \epsilon_\ell \quad}}& \null & 30
& \mbox{NS} \cr
1 \,\le \,  k \, < \, \ell \, \le 6  & \null  & \null & \null \cr
\null & \null & \null & \null\cr
\sqrt{2} \, \epsilon_7 & \null & 1 &  \mbox{NS} \cr
\null & \null & \null & \null\cr
 {\underbrace {
\frac 1 2 \, \left (\pm \epsilon_1 \pm \epsilon_2 \pm \dots \epsilon_6
\right )}} &   +   \sqrt{2 - \frac 3 2 } \, \epsilon_{7}    &   32 & \mbox{RR}\cr
\mbox{even number of + signs} & \null & \null & \null \cr
 \null & \null & \null & \null\cr
 \null & \null  & 70 &  \mbox{= Total } \cr
}    \right \}& \nonumber \\
\label{er7sol}
\end{eqnarray}
Comparing eq.s \eqn{eroddsol} and \eqn{tabu2} we realize that the match between
the physical and algebraic counting of scalar fields relies on the following numerical
identities, applying to the R--R and N--S sectors respectively:
\begin{eqnarray}
  RR &: & \left\{\matrix{2^{r-1} &=& r + {1\over 6} r(r-1) (r-2) & (r=2,3,4)\cr
2^{r-1} &=& 1 + r + {1\over 6} r(r-1) (r-2) & (r=5)\cr
2^{r-1} &=& r + r + {1\over 6} r(r-1) (r-2) & (r=6)\cr}\right. \\
NS &:& \left\{ \matrix{2 \pmatrix{r\cr 2 \cr}+ r + 1 & = & 1 + r^2 & (r=2,3,4,5)\cr
2 \pmatrix{r\cr 2 \cr}+ r + 1 + 1  & = &  2 + r^2  & (r=6)\cr} \right.
\end{eqnarray}
The physical interpretation of these identities from the string viewpoint is further discussed
in the next section.
\vskip 0.2cm
\subsection{Simple roots and Dynkin diagrams}
The most efficient way to deal simultaneously with all the above root systems and
see the emergence of the above mentioned embedding chains is to embed them in the
largest, namely in the $E_7$ root space. Hence the various root systems $E_{r+1}$
will be represented by appropriate subsets of the full set of $E_7$ roots. In this
fashion for all choices of $r$ the $E_{r+1}$ are anyhow represented by 7--components
Euclidean vectors of length 2.
\par
To see the $E_7$ structure we just need to choose, among the positive roots of
\eqn{er7sol}, a set of seven simple roots $\alpha_1 , \dots \, \alpha_7$ whose
scalar products are those predicted by the $E_7$ Dynkin diagram. The appropriate choice
is the following:
\begin{eqnarray}
\alpha_1 =\left \{-\frac {1}{2},-\frac {1}{2},-\frac {1}{2}, -\frac {1}{2}, -\frac {1}{2},
-\frac {1}{2}, \frac{1}{\sqrt{2}}\right \}\nonumber\\
\alpha_2 = \left \{ 0,0,0,0,1,1,0 \right \}\nonumber\\
\alpha_3 = \left \{ 0,0,0,1,-1,0,0 \right \}\nonumber\\
\alpha_4 = \left \{ 0,0,0,0,1,-1,0 \right \} \nonumber\\
\alpha_5 = \left \{ 0,0,1,-1,0,0,0 \right \} \nonumber\\
\alpha_6 = \left \{ 0,1,-1,0,0,0,0 \right \} \nonumber\\
\alpha_7 = \left \{ 1,-1,0,0,0,0,0 \right \} \nonumber\\
\label{e7simple}
\end{eqnarray}
The embedding of chain \eqn{caten2} is now easily described: by considering the subset of
$r$ simple roots
$\alpha_1 , \alpha_2 \, \dots \, \alpha_r$ we realize the Dynkin diagrams of type $E_{r+1}$.
Correspondingly,
the subset of all roots pertaining to the root system $\Phi(E_{r+1}) \, \subset \,
\Phi(E_7)$ is given by:
\begin{eqnarray}
x&=&6-r+1\nonumber\\
\Phi(E_{r+1}) & \equiv & \cases{ \pm \epsilon_i \pm \epsilon_j \quad \quad x \le i < j \le 7 \cr
 \pm \left [ \frac{1}{2}\, \left ( -\epsilon_1,-\epsilon_2,\dots \, \pm\epsilon_x \,
 \pm \epsilon_{x+1},\dots
, \pm \epsilon_6 \right )+{\sqrt{2}\over 2} \, \epsilon_7 \right ] \cr }
\nonumber\\
\label{erp1let}
\end{eqnarray}
At each step of the sequential embedding one  generator of the $r+1$--dimensional
Cartan subalgebra
$\cC_{r+1}$ becomes orthogonal to the roots of the subsystem
$\Phi(E_{r})\subset\Phi(E_{r+1})$,
while the remaining $r$ span the Cartan subalgebra of $E_{r}$
(Figure \ref{fig22}). 
\iffigs
\begin{figure}
\caption{}
\label{fig22}
\epsfxsize = 10cm
\epsffile{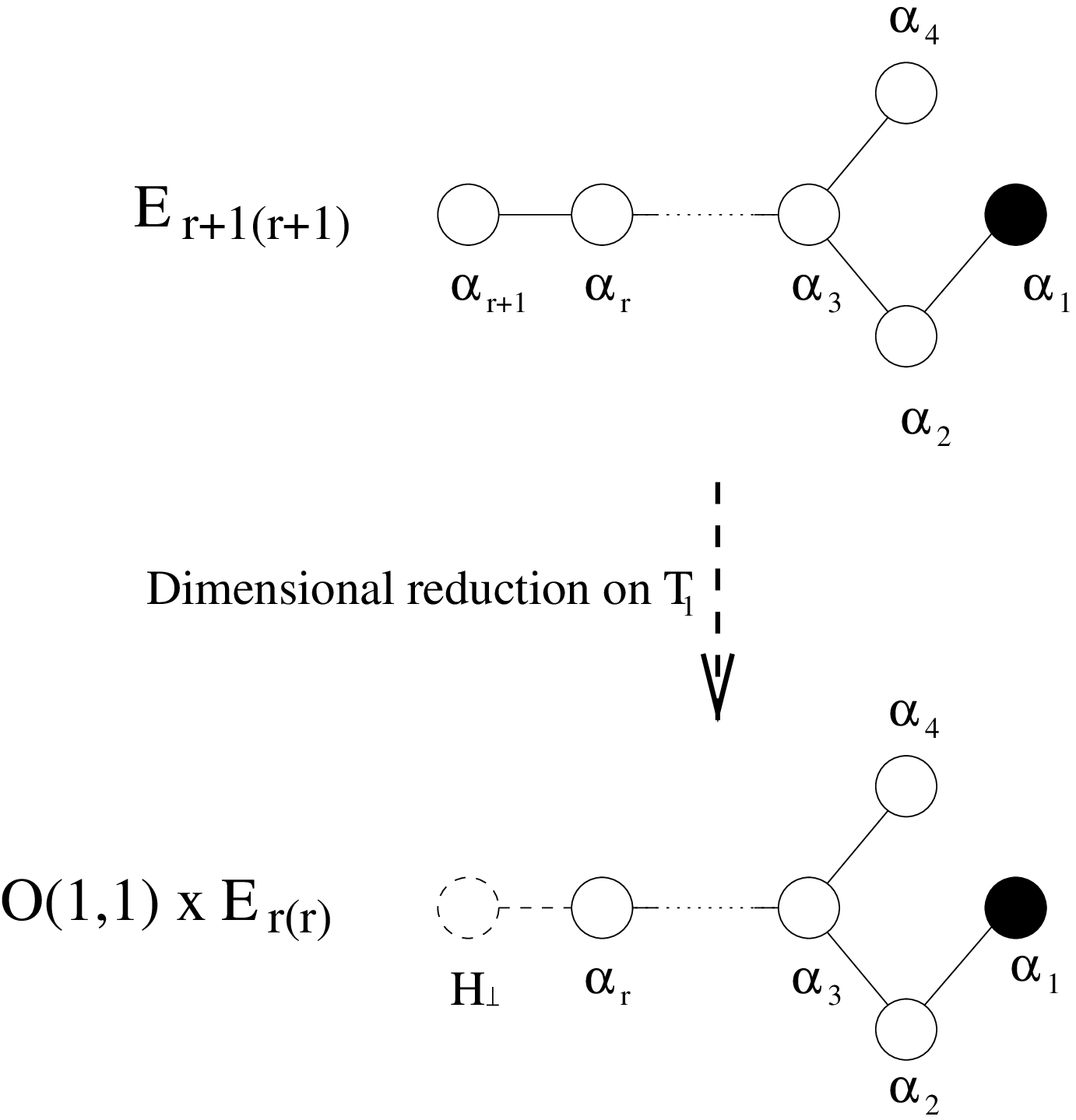}
\vskip -0.1cm
\unitlength=1mm
\end{figure}
\fi
If we name $H_{i}$ ($i=1,\dots ,7$)
the original
orthonormal basis of Cartan generators for the $E_7$ algebra, the Cartan generators
that are orthogonal
to all the roots of the $\Phi(E_{r+1})$ root system
at level $r$ of the embedding chain are the following $6-r$:
\begin{equation}
X_{k} = \left (\frac{1}{\sqrt{2}}\, H_7 + \frac {1}{k} \sum_{i=1}^{k} \, H_i \right )
\quad \quad k=1,
\dots \, 6-r
\label{extracart}
\end{equation}
On the other hand a basis for the Cartan subalgebra of the $E_{r+1}$ algebra embedded in $E_7$ is
given by :
\begin{eqnarray}
Y_i  & = & H_{6-r+i} \quad i=1,\dots \, r-1 \nonumber\\
Y_r  & = & (-)^{6-r} \, H_6 \nonumber\\
Y_{r+1} &=& \frac {1}{8-r} \, \left ( \sqrt{2} H_7 \, - \, \sum_{i=1}^{6-r}\, H_i \right )
\label{incart}
\end{eqnarray}
In order to visualize the other chains of subalgebras it is convenient to make two observations.
The first is to note that the simple roots selected in eq. \eqn{e7simple} are of two types: six
of them have integer components and span the Dynkin diagram of a $D_6 \equiv SO(6,6)$ subalgebra,
while the seventh simple root has half integer components and it is actually a spinor weight
with respect to this subalgebra. This observation leads to the embedding chain \eqn{dueachain}.
Indeed it suffices to discard one by one the last simple root to see the embedding of the
$D_{r-1}$ Lie algebra into $D_{r}\subset E_{r+1}$. As discussed in the next section $D_{r}$
is the Lie algebra of the T--duality group in type IIA toroidally compactified string theory.
\par
The next  observation is that the $E_7$ root system contains an exceptional pair of
roots $\beta =\pm \sqrt{2} \epsilon_7$, which does not belong to any of the other $\Phi (E_r)$
root systems. Physically the origin of this exceptional pair is very clear. It is associated
with the axion field $B_{\mu\nu}$ which in $D=4$ and only in $D=4$ can be dualized to an
additional scalar field. This root has not been chosen to be a simple root in eq.\eqn{e7simple}
since it can be regarded as a composite root in the $\alpha_i$ basis. However we have the
possibility
of discarding either $\alpha_2$ or $\alpha_1$ or  $\alpha_4$ in favour of $\beta$ obtaining a new
basis for the $7$-dimensional euclidean space $\IR^7$. The three choices in this operation
lead to the three different Dynkin diagrams given in fig.s
\eqn{stdualf} 
and \eqn{elecal}, corresponding to
the Lie Algebras:
\begin{equation}
 A_5 \oplus A_2\, , \quad   D_6\oplus A_1  \, , \quad
  A_7
\label{splatto}
\end{equation}

\iffigs
\begin{figure}
\caption{}
\label{stdualf}
\epsfxsize = 10cm
\epsffile{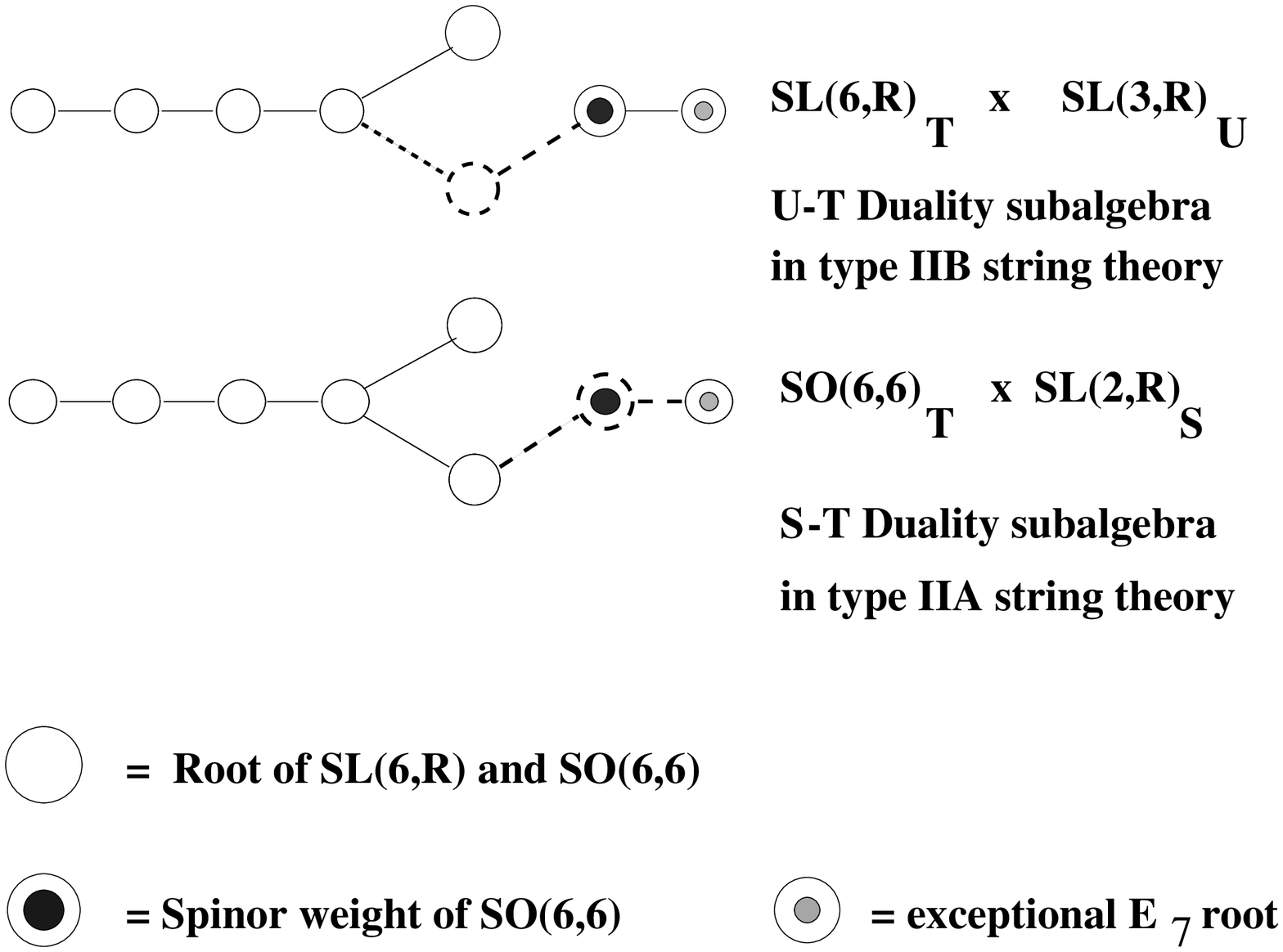}
\vskip -0.1cm
\unitlength=1mm
\end{figure}
\fi

\iffigs
\begin{figure}
\caption{}
\label{elecal}
\epsfxsize = 10cm
\epsffile{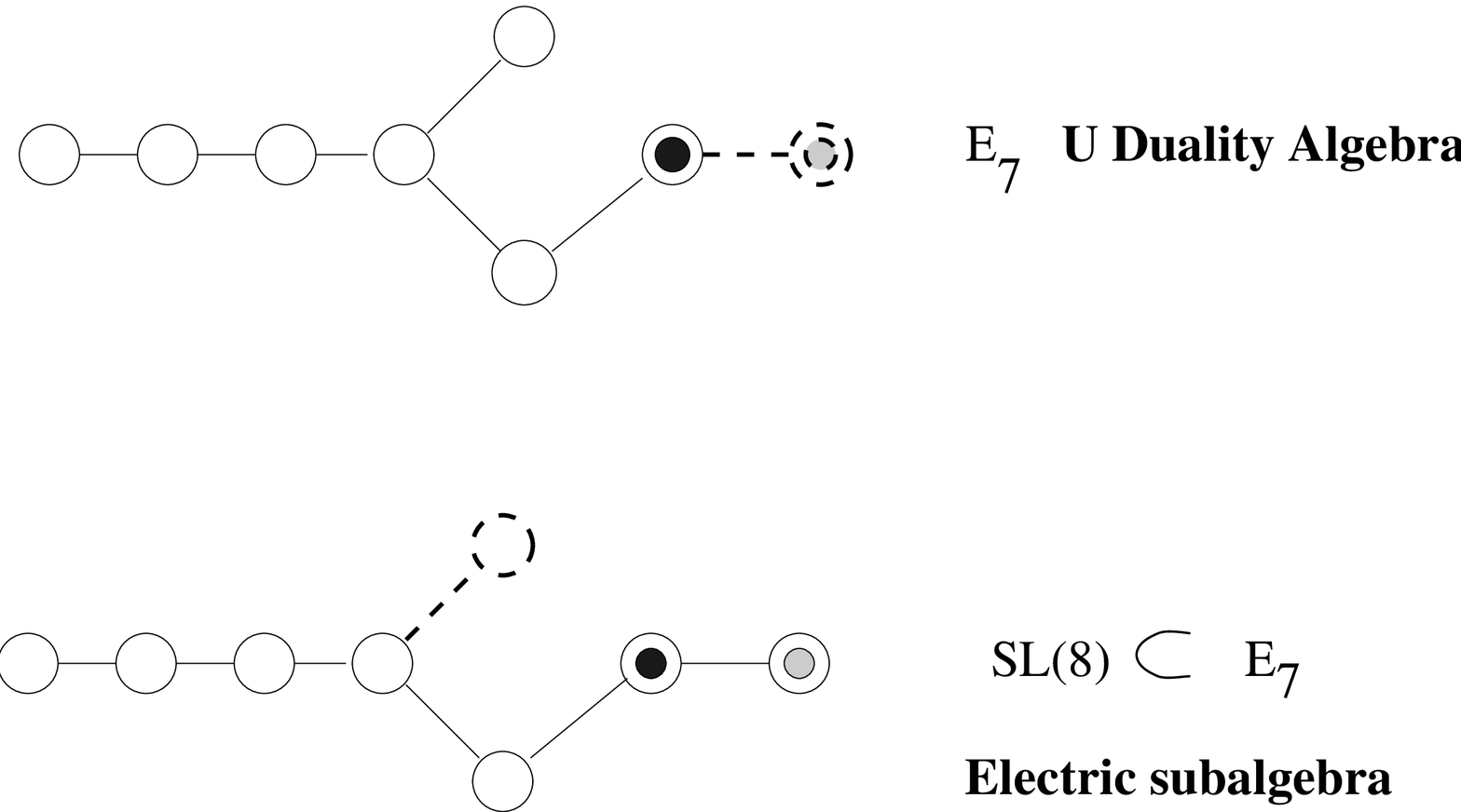}
\vskip -0.1cm
\unitlength=1mm
\end{figure}
\fi
From these embeddings occurring at the $E_7$ level, namely in $D=4$,
one deduces the three embedding chains
\eqn{dueachain},\eqn{duebchain},\eqn{elechain}: it just suffices to peal
off the last $\alpha_{r+1}$ roots
one by one and also the $\beta$ root that occurs only in $D=4$.
One observes that the appearance of the
$\beta$ root is always responsible for an enhancement of the S--duality group.
In the type IIA case
this group is enhanced from $O(1,1)$ to $SL(2,\IR)$ while in the type IIB case
it is enhanced from
the $SL(2,\IR)_U$ already existing in $10$--dimensions to $SL(3,\IR)$.
Physically this occurs by
combining the original dilaton field with the compactification radius of
the latest compactified
dimension.
\subsection{String theory interpretation of the sequential embeddings:
Type $IIA$, type $IIB$ and $M$ theory chains}
We now turn to a closer analysis of the physical meaning of
the embedding chains we have been illustrating.
\par
Let us begin with the chain of eq.\eqn{duebchain} that, as anticipated, is
related with the type IIB interpretation of supergravity theory.
The distinctive feature of this chain of embeddings is the presence
of an addend $A_1$ that is already present in 10 dimensions. Indeed
this $A_1$ is the Lie algebra of the $SL(2,R)_\Sigma $ symmetry of type $IIB$
D=10 superstring. We can name this group the U--duality symmetry $U_{10}$ in
$D=10$. We can use the chain \eqn{duebchain} to trace it in lower dimensions.
Thus let us  consider the decomposition
\begin{eqnarray}
E_{r+1(r+1)} & \rightarrow & N_r \otimes SL(2,\IR) \nonumber\\
N_r & = &   A_{r-1} \otimes O(1,1)
\la{sl2r}
\end{eqnarray}
Obviously $N_r$ is not contained in the  $T$-duality group $O(r,r)$ since the
$NS$ tensor field $B_{\mu \nu}$ (which  mixes with the metric under
$T$-duality) and the $RR$--field $B^c_{\mu \nu}$ form a doublet with
respect $SL(2,\IR)_U$. In fact, $SL(2,\IR)_U$ and  $O(r,r)$ generate
the whole U--duality group $E_{r+1(r+1)}$. The appropriate interpretation
of the normaliser of $SL(2,R)_\Sigma$ in $E_{r+1(r+1)}$  is
\begin{equation}
 N_r  = O(1,1) \otimes SL(r,\IR) \equiv GL(r,\IR)
 \label{agnosco}
\end{equation}
where $GL(r,\IR)$ is the isometry group of the  classical moduli
space for the $T_r$ torus:
\begin{equation}
 \frac{GL(r,\IR)}{Or}.
\end{equation}
The decomposition of the U--duality group appropriate for the type $IIB$ theory is
\be
E_{r+1} \rightarrow U_{10} \otimes GL(r,\IR) = SL(2,\IR)_U \otimes O(1,1) \otimes SL(r,\IR).
\la{sl2rii}
\ee
Note that since $GL(r,\IR) \supset O(1,1)^r$, this translates into $E_{r+1} \supset
SL(2,\IR)_U \otimes O(1,1)^r$.  (In Type $IIA$, the corresponding chain would
be $E_{r+1} \supset O(1,1) \otimes O(r,r) \supset  O(1,1)^{r+1}$.)  Note that while
$SL(2,\IR)$ mixes $RR$ and $NS$ states, $GL(r,\IR)$ does not. Hence we can write the following
decomposition for the solvable Lie algebra:
\bea
Solv_{r+1} &=& Solv \left(\frac{GL(r,\IR)}{Or} \otimes
\frac{SL(2,\IR)}{O(2)} \right) + \left(\frac{\bf r(r-1)}{\bf 2}, {\bf 2} \right) \oplus {\bf X}
\oplus {\bf Y}  \nonumber \\
\mbox{dim }Solv_{r+1}&=& \frac{d(3d-1)}{2} + 2 + x + y.
\la{solvii}
\eea
where $x=\mbox{dim }{\bf X} $ counts the scalars coming from the internal part of the $4$--form
$A^+_{{ \mu}{ \nu}{ \rho}{\sigma}}$ of type IIB string theory.
We have:
\begin{equation}
x =  \left \{
\matrix { 0 & r<4 \cr
\frac{r!}{4!(r-4)!} & r\geq 4 \cr}\right.
\la{xscal}
\end{equation}
and
\begin{equation}
y =\mbox{dim }{\bf Y} = \cases{
\matrix { 0 &  r  < 6   \cr 2 &  r = 6  \cr}\cr}.
\la{yscal}
\end{equation}
counts the scalars arising from dualising the two-index tensor
fields in $r=6$.
\par
For example, consider the $ D=6$ case. Here the type $IIB$  decomposition is:
\begin{equation}
E_{5(5)}=\frac{O(5,5)}{O(5) \otimes O(5)} \rightarrow \frac{GL(4,\IR)}{O(4)}
\otimes \frac{SL(2,\IR)}{O(2)}
\la{exii}
\end{equation}
whose compact counterpart is given by $O(10) \rightarrow SU(4) \otimes SU(2) \otimes U(1)$,
corresponding to the decomposition: ${\bf 45} =
{\bf (15,1,1)}+ {\bf (1,3,1)} + {\bf (1,1,1)} +{\bf (6,2,2)} + {\bf (1,1,2)}$. It follows:
\be
Solv_5 = Solv (\frac{GL(4,\IR)}{O(4)}
\otimes \frac{SL(2,\IR)}{O(2)}) +({\bf 6},{\bf 2})^+ + ({\bf 1},{\bf 1})^+.
\la{solvexii}
\ee
where the factors on the right hand side parametrize the internal part of the metric $g_{ij}$,
the dilaton and the $RR$ scalar ($\phi$, $\phi^c$), ($B_{ij}$, $B^c_{ij}$) and $A^+_{ijkl}$
respectively.
\par
There is a connection between the decomposition \eqn{sl2r} and the corresponding chains
in M--theory. The type IIB chain is given by eq.\eqn{duebchain},
namely by
\begin{equation}
E_{r+1(r+1)} \rightarrow SL(2,\IR) \otimes GL(r,\IR)
\end{equation}
 while the $M$ theory is given by eq.\eqn{elechain}, namely by
 \begin{equation}
E_{r+1} \rightarrow O(1,1) \otimes SL(r+1,\IR)
\end{equation}
coming from the moduli space of $T^{11-D} = T^{r+1}$.
We see that these decompositions involve the classical moduli spaces of $T^r$
 and of $T^{r+1}$ respectively.
Type $IIB$ and $M$ theory decompositions
become identical if we decompose further $SL(r, \IR) \rightarrow O(1,1)
\times SL(r-1,\IR)$ on the type $IIB$ side and  $SL(r+1, \IR) \rightarrow O(1,1)
\otimes SL(2,\IR) \otimes SL(r-1,\IR)$ on the $M$-theory side. Then we obtain for both theories
\be
E_{r+1} \rightarrow SL(2,\IR) \times O(1,1) \otimes O(1,1) \otimes SL(r-1,\IR),
\la{sl2rall}
\ee
and we see that the group $SL(2,\IR)_U$ of type $IIB$ is identified with the
complex structure of the $2$-torus factor of the total
compactification torus $T^{11-D} \rightarrow T^2 \otimes T^{9-D}$.
\par
Note that according to  \eqn{splatto} in 8 and 4 dimensions, ($r=2$ and $6$)
in the decomposition \eqn{sl2rall} there is the following enhancement
(Figure \ref{fig24}):
\begin{eqnarray}
&  SL(2,\IR) \times O(1,1)  \rightarrow SL(3,\IR) \quad (\mbox{for} \, r=2,6) & \\
 &\left\{\matrix{O(1,1) & \rightarrow & SL(2,\IR) \quad (\mbox{for} \, r=2) \cr
SL(5,\IR) \times O(1,1) & \rightarrow & SL(6,\IR) \quad (\mbox{for} \, r=6) \cr}\right. &
\end{eqnarray}
\iffigs
\begin{figure}
\caption{Enhancement for type IIB in D=4 (r=6).}
\label{fig24}
\epsfxsize = 10cm
\epsffile{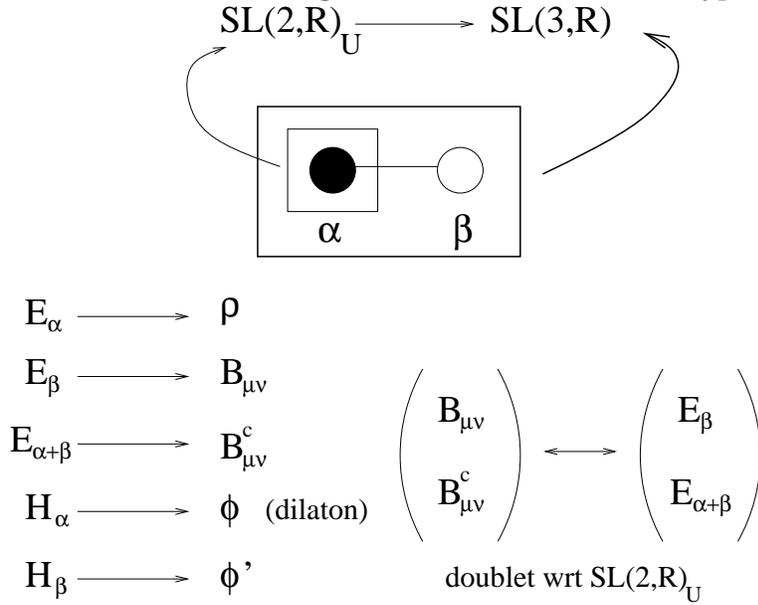}
\vskip -0.1cm
\unitlength=1mm
\end{figure}
\fi
Finally, by looking at fig.(\eqn{wite5}) let us observe that
$E_{7(7)}$ admits also a subgroup $SL(2,\IR)_T$ $\otimes
(SO(5,5)_S$ $\equiv E_{5(5)})$ where the $SL(2,\IR)$ factor is a
T--duality group, while the factor $(SO(5,5)_S$ $\equiv E_{5(5)})$
is an S--duality group which mixes R--R and N--S states.
\iffigs
\begin{figure}
\caption{}
\label{wite5}
\epsfxsize = 10cm
\epsffile{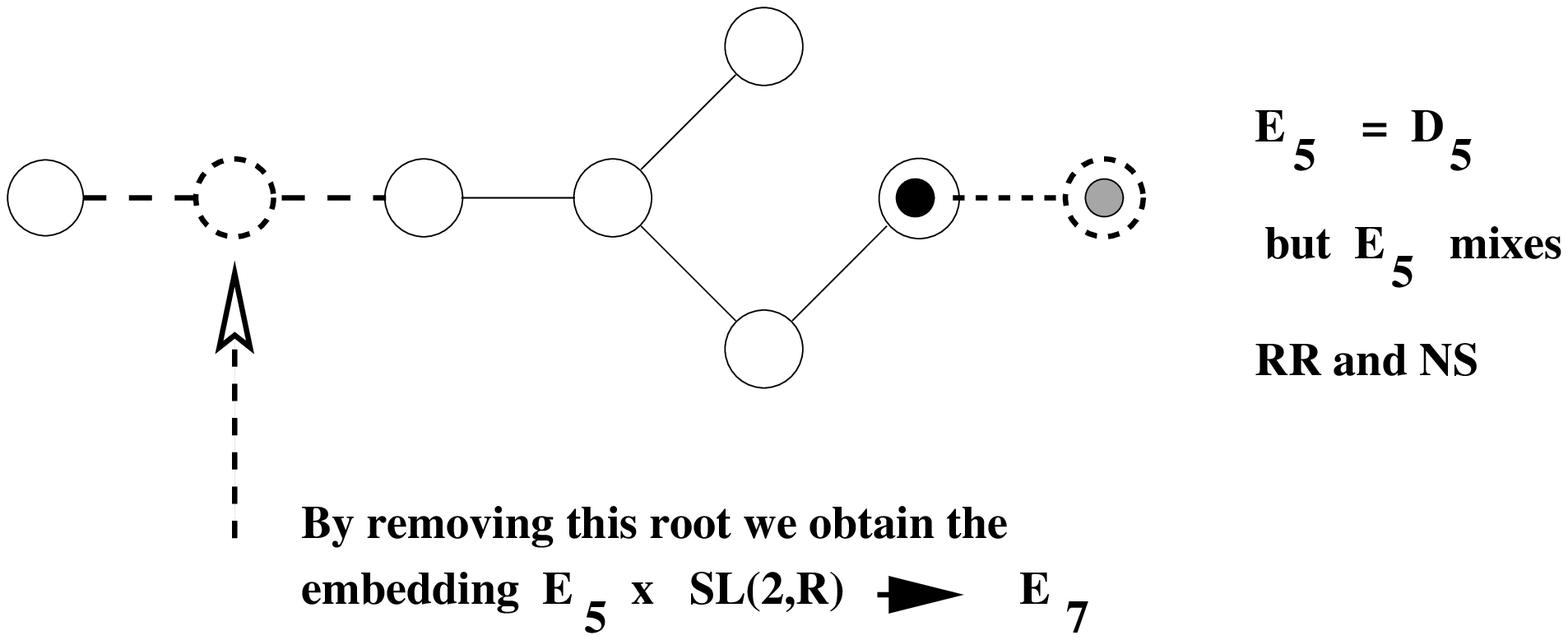}
\vskip -0.1cm
\unitlength=1mm
\end{figure}
\fi
\section{The maximal abelian ideals ${\cal A}_{r+1} \subset Solv_{r+1}$
of the solvable Lie algebra}

It is interesting to work out the maximal abelian ideals
${\cal A}_{r+1} \subset Solv_{r+1}$  of the
solvable Lie algebras generating the scalar manifolds
of maximal supergravity in dimension $D=10-r$.
The maximal abelian ideal of a solvable Lie
algebra is defined as the maximal subset of nilpotent generators  commuting among themselves.
From a physical point of view this is the largest abelian Lie algebra
that one might expect to be able to gauge in the supergravity theory. Indeed, as it turns
out, the number of vector fields in the theory is always larger
or equal than $\mbox{dim}{\cal A}_{r+1}$. Actually, as we are going to
see, the {\it gaugeable} maximal  abelian algebra
is always a proper subalgebra ${\cal A}^{gauge}_{r+1} \subset {\cal A}_{r+1}$
of this ideal.
\par
The criteria to determine ${\cal A}^{gauge}_{r+1}$ will be discussed
in the next section. In the present section we
derive ${\cal A}_{r+1}$ and we explore its relation with the
space of vector fields in one dimension above the dimension
we each time consider.  From such analysis we obtain a
filtration of the solvable Lie algebra which provides us
with a canonical polynomial parametrization of the supergravity
scalar coset manifold $U_{r+1}/H_{r+1}$
\par
\subsection{The maximal abelian ideal from an algebraic viewpoint}
Algebraically the maximal abelian ideal can be characterized by
looking at the decomposition of the U--duality algebra $E_{r+1(r+1)}$ with
respect to the U--duality algebra  in one dimension above.
In other words we have to consider the decomposition of $E_{r+1(r+1)}$ with
respect to the subalgebra $E_{r (r )} \, \otimes \, O(1,1)$. This
decomposition follows a general pattern which is given by the
next formula:
\begin{equation}
 \mbox{adj }  E_{r+1(r+1)} \, = \, \mbox{adj }  E_{r(r)}  \, \oplus
 \, \mbox{adj }  O(1,1) \, \oplus ( \ID^+_{r} \oplus \ID^-_{r} )
 \label{genpat}
\end{equation}
where $\ID^+_{r}$ is at the same time an irreducible representation of
the  U--duality algebra $E_{r (r )}$ in $D+1$ dimensions and coincides with the
maximal abelian ideal
\begin{equation}
\ID^+_{r}    \, \equiv \, {\cal A}_{r+1}   \, \subset \, Solv_{(r+1)}
\label{genmaxab}
\end{equation}
of the solvable  Lie algebra we are looking for. In eq. \eqn{genpat}
the subspace $\ID^-_{r}$ is just a second identical copy of the representation
 $\ID^{+}_{r}$ and it is made of negative rather than of positive weights
 of $E_{r (r )}$. Furthermore   $\ID^{+}_{r}$  and   $\ID^-_{r}$
 correspond to the eigenspaces belonging respectively to the eigenvalues
 $\pm 1$ with respect to the adjoint action of the S--duality group
 $O(1,1)$.
\subsection{The maximal abelian ideal from a physical perspective: the vector fields in one
dimension above and translational symmetries}
Here, we would like to show that the dimension of the abelian
ideal in $D$ dimensions is equal to the number of vectors in dimensions $D+1$.
Denoting the number of compactified dimensions by $r$ (in string theory,
$r=10-D$), we will label the $U$-duality group in $D$ dimensions by $U_D
= E_{11-D} = E_{r+1}$. The $T$-duality group is $O(r,r)$, while the
$S$-duality group is $O(1,1)$ in  dimensions higher than four, $SL(2,R)$
in $D=4$ (and it is inside $O(8,8)$ in $D=3$).
\par
It follows from \eqn{genpat}
that the total dimension of the abelian ideal is given by
\be
{\rm dim} \, \cA_{D} \,\equiv \,  {\rm dim} \, \cA_{r+1}
 \,\equiv \,  {\rm dim} \,
\ID_{r}
\la{abeliii}
\ee
where $\ID_{r} $ is a representation of $U_{D+1}$ pertaining to the vector fields.
 According to \eqn{genpat} we have (for $D \ge 4$):
\be
\mbox{adj } U_D = \mbox{adj } U_{D+1} \oplus {\bf 1} \oplus ({\bf 2}, \ID_r).
\la{irrepu}
\ee
This is just an immediate consequence of the embedding chain \eqn{caten2}
which at the first level of iteration yields
$E_{r+1} \rightarrow E_r \times O(1,1)$. For example, under
$E_7 \rightarrow E_6 \times O(1,1)$ we have the branching rule:
${\rm adj} \, E_7 = {\rm adj} \, E_6 + {\bf 1} + ({\bf 2},{\bf 27})$ and
the abelian ideal is given by the ${\bf 27^+}$ representation of the $E_{6(6)}$ group.
The $70$ scalars of the $D=4, N=8$ theory are
naturally decomposed as ${\bf 70} = {\bf 42} +{\bf 1} +{\bf 27^+}$.
To see the splitting of
the abelian ideal scalars into $NS$  and $RR$ sectors, one has to consider
the decomposition of $U_{D+1}$ under the T--duality group $T_{D+1} = O(r-1, r-1)$,
namely the second iteration of the embedding chain \eqn{caten2}: $E_{r+1}
\rightarrow O(1,1) \times O(r-1,r-1)$. Then the vector  representation of $O(r-1, r-1)$
gives  the $NS$ sector, while the  spinor representation yields  the $RR$
sector. The  example of $E_7$ considered above is somewhat exceptional, since
we have ${\bf 27} \rightarrow ({\bf 10} + {\bf 1} +{\bf 16})$.
Here in addition to the expected ${\bf 10}$ and
${\bf 16}$ of $O(5,5)$ we find an extra $NS$ scalar:  physically this is due
to the fact that in
four dimensions the two-index antisymmetric tensor field  $B_{\mu \nu}$
is dual to a scalar, algebraically this  generator is associated with the
exceptional root $ \sqrt {2} \epsilon_7$.
To summarize, the $NS$ and $RR$ sectors are separately  invariant under
$O(r,r)$ in $D=10-r$ dimensions, while the abelian $NS$  and $RR$ sectors
are invariant under $O(r-1, r-1)$. The standard parametrization of
the
$U_D/H_D$ and $U_{D+1}/H_{D+1}$ cosets gives a clear illustration of this
fact:
\be
\frac{U_D}{H_D} \sim ( \frac{U_{D+1}}{H_{D+1}}, r_{D+1}, {\bf V}_r^{D+1}).
\la{cosetideal}
\ee
Here $r_{D+1}$ stands for the compactification radius, and ${\bf V}_r^{D+1} $
are the compactified vectors yielding the abelian ideal in $D$ dimensions.
\par
Note that:
\begin{equation}
  \mbox{adj}\, H_D = \mbox{adj} \, H_{D+1} \, +\,
\mbox{adj} \,\mbox{Irrep} \, U_{D+1}
\end{equation}
so it appears  that the abelian ideal forms a representation not only of
$U_{D+1}$ but also of the compact isotropy subgroup $H_{D+1}$ of the scalar coset
manifold.
\par
In the  above $r=6$ example we find
${\rm adj} \, SU(8) = {\rm adj} \, USp(8) \oplus {\bf 27^-}$,
$ \Longrightarrow $ ${\bf 63}= {\bf 36}+ {\bf 27^-}$.
\subsection{Maximal abelian ideal and brane wrapping}
Now we would like to turn to a uniform counting  of the ideal dimension in
diverse space--time dimensions. The fact that the ($D+1$)-dimensional
vectors have $0$-branes as electric sources  (or equivalently,
 ($D-3$)-branes as magnetic ones) reduces the analysis of the $RR$ sector
to a simple exercise in counting the ways of wrapping  higher dimensional
$d$-branes around the cycles of the compact manifold.  This procedure spares
one from doing a case-by-case counting and worrying about the scalars
arising from the dualization of the tensor fields. It also easily
generalizes for manifolds other then $T^r$. The latter choice corresponds
to the case of maximal
preserved supersymmetry, for which the counting is presented here.
\par
Starting from Type $IIA$ theory with 0, 2, 4, 6 -Dbranes \cite{polc}, the total number of
($D+1$)-dimensional  0-Dbranes (i.e. the maximal abelian ideal in $D$--dimensions)
is  obtained
by wrapping the Dbranes around the even cycles of the ($9-D$)-dimensional torus.
One gets:
\be
n_{\cA}^{RR} = \sum_{k} b_{2k} (T^{9-D}) = 2^{8-D}
\la{rrideal}
\ee
where $b_{2k}$ are the Betti numbers. The same result is obtained by
counting the magnetic sources: in this case the sum is taken over
alternating series of even (odd) cohomology for $9-D$ even (odd), since
the $6$-Dbrane is wrapped on the  $(9-D)$-dimensional cycle of $T^{9-D}$,
the $4$-Dbrane on $(7-D)$-dimensional cycles and so on.  Note that  wrapping
 Dbranes around the cycles of the same dimensions as above but on a
$T^{10-D}$ yields the total number of the $RR$ scalars  in $D$ dimensions.
\par
The type $IIB$ story is exactly the same with the even cycles replaced by
the odd ones. The only little subtlety is in going from ten dimensions
to nine - there is no $0$-Dbrane in type $IIB$, but instead there is a
$RR$ scalar in the ideal
already in ten dimensions, since the $U$-duality group is non-trivial. Of course the
results agree on $T^r$ as they should on any manifold with a vanishing Euler number.
\par
The $NS$ parts of  the ideal ($(D-3)$-branes in ($D+1$)-dimensions) are
obtained either by  wrapping the ten-dimensional fivebrane or as magnetic
sources for Kaluza-Klein vectors (note that for the $NS$ part, the
reasonings for Type $IIA$ and $IIB$ are identical).  The former are the
fivebrane wrapped on $(8-D)$ cycles of $T^{9-D}$ (there are $(9-D)$ of
them), while the latter are given by the same number since it is the
number of the Kaluza--Klein vectors (number of $1$-cycles). Thus
\be
n_{\cA}^{NS} = 2 b_1 (T^{9-D}) = 2(9-D).
\la{nsideal}
\ee
The only exception to this formula is the $D=4$ case where, as discussed
above, we have to add an extra scalar due to the $B_{\mu \nu}$ field.

\section{Gauging}
In this last section we will consider the problem of gauging some isometries
of the coset $G/H$ in the framework of solvable Lie algebras.
\par
In particular we will consider in more detail the gauging of maximal compact groups
and the gauging of nilpotent abelian (translational) isometries.
\par
This procedure is a way of obtaining
partial supersymmetry breaking in extended supergravities \cite{hull},\cite{warn},\cite{huwa}
and it may find applications in the context of non perturbative phenomena in string
and M-theories.
\par
Let us consider the left--invariant 1--form $\Omega = L^{-1} dL$ of the coset
manifold $U_D/H_D$, where $L$ is the coset representative.
\par
The gauging procedure \cite{castdauriafre} amounts to the replacement of $dL$ with the gauge covariant
differential $\nabla L$ in the definition of the left--invariant 1--form $\Omega = L^{-1} d L$:
\begin{equation}
  \label{gaugedconn}
  \Omega  \rightarrow  \hat\Omega = L^{-1} \nabla  L =  L^{-1}( d + A )  L
= \Omega + L^{-1} A L
\end{equation}
As a consequence $\hat\Omega $ is no more a flat connection, but its curvature is given by:
\begin{equation}
  \label{gaugedchiusa}
 R(\hat\Omega)= d \hat\Omega +  \hat\Omega \wedge  \hat\Omega =  L^{-1}\cF  L \equiv  L^{-1}( dA + A\wedge A )  L
= L^{-1}(F^I T_I +L^I_{AB}T_I \bar\psi^A \psi^B )  L
\end{equation}
where $F^I$ is the gauged  supercovariant 2--form and $T_I$ are the generators of the gauge group
 embedded in the U--duality representation  of the vector fields.

 Indeed, by very definition, under the
full group $E_{r+1(r+1)}$ the gauge vectors are contained
in the representation $\ID_{r+1}$.
Yet, with respect to the gauge subgroup they must
transform in the adjoint representation, so that $\cG_D $ has to be chosen in such a way that:
\begin{equation}
\ID_{r+1} \, \stackrel{\cG_D}{\longrightarrow} \,\mbox{adj}\,\cG_D \, \oplus \, \mbox{rep}\,\cG_D
\label{brancione}
\end{equation}
where $\mbox{rep}\,\cG_D $ is some other representation of $\cG_D$ contained in the
above decomposition.

It is important to remark that vectors which are in $\mbox{rep} \cG_D$ (i.e. vectors which do not gauge $\cG_D$)
 may be required, by consistence
of the theory \cite{topiva}, to appear through their duals $(D-3)$--forms, as for instance happens
for $D=5$ \cite{gurowa}.
In an analogous way $p$--form potentials ($p\neq 1$) which are in non trivial representations
of $\cG_D$ may also be required to appear through their duals $(D-p-2)$--potentials,
nas is the case in $D=7$ for $p=2$ \cite{pepiva}.

The charges and the boosted structure constants discussed in the next subsection can be retrieved from the two terms
appearing in the last expression of eq. \eqn{gaugedchiusa}

\subsection{Filtration of the $E_{r+1}$ root space, canonical
parametrization of the coset representatives and boosted structure constants}
As it has already been emphasized, the complete
structure of $N > 2$ supergravity in diverse dimensions is fully encoded in
the local differential geometry of the scalar coset manifold
$U_D/H_D$. All the couplings in the Lagrangian are described in terms
of the metric, the connection and the coset representative \eqn{solvcoor} of
$U_D/H_D$. A particularly significant consequence of extended
supersymmetry is that the fermion masses and the scalar potential the
theory can develop occur only as a consequence of the gauging and can
be extracted from a decomposition in terms of irreducible $H_D$ representations
of the {\sl boosted structure constants}\cite{fundpaper} \cite{castdauriafre}. Let us define these latter.
Let $\ID_{r+1}$ be the irreducible representation of the $U_D$
U--duality group pertaining to the vector fields and denote by ${\vec
{\bf w}}_\Lambda$ a basis for $\ID_{r+1}$:
\begin{equation}
\forall {\vec {\bf v}} \, \in \, \ID_{r+1}  \qquad : \qquad  {\vec v}\, = \,
v^\Lambda \, {\vec {\bf w}}_\Lambda
\label{baset}
\end{equation}
In the case we consider of maximal supergravity theories, where the
U--duality groups are given by  $E_{r+1(r+1)}$
 the basis vectors ${\vec {bf w}}_\Lambda$ can be identified
with the $56$ weights of the fundamental $E_{7(7)}$ representation or with
the subsets of this latter corresponding to the irreducible
representations of its  $E_{r+1(r+1)}$ subgroups, according to the
branching rules:
\begin{equation}
{\bf 56}{\stackrel{E_6}{\longrightarrow}}\cases{{\bf 27 + 1}{\stackrel{E_5}{\longrightarrow}}
 \cases{{\bf 16}{\stackrel{E_4}{\longrightarrow}}{\bf \dots}\cr
 {\bf 10}{\stackrel{E_4}{\longrightarrow}}{\bf \dots}\cr
 {\bf 1+1}{\stackrel{E_4}{\longrightarrow}}{\bf \dots}\cr} \cr
 {\bf 27 + 1}{\stackrel{E_5}{\longrightarrow}}
 \cases{{\bf 16}{\stackrel{E_4}{\longrightarrow}}{\bf \dots}\cr
 {\bf 10}{\stackrel{E_4}{\longrightarrow}}{\bf \dots}\cr
 {\bf 1+1}{\stackrel{E_4}{\longrightarrow}}{\bf \dots}\cr} \cr
 }
\label{brancio}
\end{equation}
Let:
\begin{equation}
< ~, ~ > \quad : \quad \ID_{r+1} \, \times \ID_{r+1} \, \longrightarrow \,
\IR
\label{norma}
\end{equation}
denote the invariant scalar product in $\ID_{r+1}$ and let
$ {\vec {\bf w}}^\Sigma$ be a dual basis such that
\begin{equation}
 < {\vec {\bf w}}^\Sigma , {\vec {\bf w}}_\Lambda > \, = \,
 \delta^\Sigma_\Lambda
 \label{dualba}
\end{equation}
Consider then the $\ID_{r+1}$ representation of the coset representative
\eqn{solvcoor}:
\begin{equation}
L(\phi) \quad : \quad  \vert {\vec {\bf w}}_\Lambda >  \, \longrightarrow
L(\phi)_\Lambda^\Sigma \, \vert {\vec {\bf w}}_\Sigma >,
\label{matricia}
\end{equation}
and let $T^I$ be the generators of the gauge algebra $\cG_D \,
\subset E_{r+1(r+1)}$.

  The only admitted generators are those with
index $\Lambda = I \, \in \, {adj }\,\cG_D $,
and there are no gauge group generators with index $\Lambda \, \in \,\mbox{rep}\,\cG_D  $.
Given these definitions
the {\it boosted structure constants} are the following three--linear
$3$--tensors in the coset representatives:
\begin{equation}
\IC_{\Sigma\Gamma}^{\Lambda}\left(\phi\right)\, \equiv \,
\sum _{I=1}^{{\rm dim }\cG_D} \,
< {\vec {\bf w}}^\Lambda \, , \, L^{-1}\left(\phi\right) \, T_I \,
  L\left(\phi\right)  \, {\vec {\bf w}}_\Sigma > \, < {\vec {\bf w}}^I \,
 , \, L\left(\phi\right) {\vec {\bf w}}_\Gamma >
 \label{busto}
\end{equation}
and by decomposing them into irreducible $H_{r+1}$ representations we
obtain the building blocks utilized by supergravity in the fermion
shifts, in the fermion mass--matrices and in the scalar potential.
\par
In an analogous way, the charges appearing in the gauged
covariant derivatives are given by the following general form:
\begin{equation}
Q_{I\Sigma}^{\Lambda} \, \equiv \,
< {\vec {\bf w}}^\Lambda \, , \, L^{-1}\left(\phi\right) \, T_I \,
  L\left(\phi\right)  \, {\vec {\bf w}}_\Sigma >
 \label{buscar}
\end{equation}
\par
The coset representative $L\left(\phi\right)$ can be written in a canonical polynomial
parametrization which should give a
simplifying  tool in
mastering the scalar field dependence of all  physical relevant
quantities.
This includes, besides mass matrices, fermion shifts and scalar potential,
also the central charges
 \cite{amicimiei}.
\par
The alluded parametrization is precisely what the solvable Lie
algebra analysis produces.
\par
To this effect let us decompose
the solvable Lie algebra of $E_{7(7)}/SU(8)$ in a sequential way utilizing
eq. \eqn{genpat}. Indeed we can write the equation:
\begin{equation}
Solv_7=\cC_7 \oplus \Phi^{+}(E_{7})
\label{decompo1}
\end{equation}
where $\Phi^{+}(E_{7})$ is the $63$ dimensional positive part of the $E_7$ root space.
By repeatedly using eq.  \eqn{genpat} we obtain:
\begin{equation}
\Phi^{+}(E_7)=\Phi^+(E_2) \oplus \ID^{+}_{2} \oplus \ID^{+}_{3}
\oplus \ID^{+}_{4} \oplus \ID^{+}_{5}  \oplus \ID^{+}_{6}
\label{decompo}
\end{equation}
where $ \Phi^+(E_2) $ is the one--dimensional root space of the
U--duality group in $D=9$ and $\ID^{+}_{r+1}$ are the weight-spaces
of the $E_{r+1}$ irreducible
representations to which the vector field in $D=10-r$
are assigned. Alternatively, as we have already explained, ${\cal A}_{r+2}
\equiv \ID^{+}_{r+1}$ are
the maximal abelian ideals of the U--duality group in $E_{r+2}$ in
$D=10-r-1$ dimensions.
\par
We can easily check that the dimensions sum appropriately as follows from:
 \begin{eqnarray}
\mbox{dim}\, \Phi^{+}(E_7)\, &=& { 63} \nonumber\\
\mbox{dim}\, \Phi^+(E_2)\, &=& { 1}\quad\quad~\mbox{dim}\,
\ID^{+}_{2}\, = { 3}\nonumber\\
\mbox{dim}\, \ID^{+}_{3}\, &=& { 6}\quad\quad~\mbox{dim}\,
\ID^{+}_{4}\, = { 10}\nonumber\\
\mbox{dim}\, \ID^{+}_{5}\, &=& { 16}\quad\quad\mbox{dim}\,
\ID^{+}_{4}\, = { 27}\nonumber\\
\label{ideadim}
\end{eqnarray}
Relying on eq. \eqn{decompo1},
\eqn{decompo} we can introduce a canonical set of scalar field
variables:
\begin{eqnarray}
\phi^i & \longrightarrow &  Y_i \,\in \cC \quad \quad i=1,\dots \,
r \nonumber\\
\tau^i_{k} & \longrightarrow &  \,D^{(k)}_{i} \, \in  \ID_k \quad i=1,\dots \,
\mbox{dim }\,\ID_k  \quad (k=2,\dots,6) \nonumber\\
\tau_1   & \longrightarrow & \ID_1 \, \equiv  \, E_2
\label{canoni}
\end{eqnarray}
and adopting the short hand notation:
\begin{eqnarray}
 \phi \cdot  \cC & \equiv & \phi^i  \,   Y_i  \nonumber\\
 \tau_k \cdot \ID_k & \equiv & \tau^i_{k}  \,D^{(k)}_{i} \nonumber\\
 \label{fucili}
\end{eqnarray}
we can write the coset representative for maximal supergravity in dimension
$D=10-r$ as:
\begin{eqnarray}
L & = & \exp \left [ \phi \cdot  \cC \right ] \,  \prod_{k=1}^{r}
\, \exp \left[   \tau_k \cdot \ID_k \right ] \nonumber\\
& = & \prod_{j=1}^{r+1} S^i \,  \prod_{k=1}^{r}   \,
\left( { 1} + \tau_r \cdot \ID_r \right)
\end{eqnarray}
The last line
follows from the abelian nature of the ideals $\ID_k$ and from the position:
\begin{equation}
S^i  \, \equiv \, \exp[\phi^i Y_i]
\end{equation}
All entries of the matrix $L$ are therefore polynomials of order at most  $2 \, r\, +\, 1$ in the
$S^i, \tau^i_k, \tau_1$ ``canonical'' variables.   Furthermore when the gauge group is chosen within the
maximal abelian ideal it is evident from the definition of the boosted
structure constants \eqn{busto} that they do not depend on the scalar
fields associated with the  generators of the same ideal. In such gauging one
has therefore {\it a flat direction} of the scalar potential for each
generator of the maximal abelian ideal.
\par
In the next section we turn to considering the possible gaugings more
closely.
\subsection{Gauging of compact and translational isometries}
A necessary condition for the gauging of a subgroup $\cG_D\subset U_D$ is that the representation of the vectors
$\ID_{r+1}$ must contain $\mbox{adj} \cG_D$.
Following this prescription, the list of maximal compact gaugings $\cG_D$ in any dimensions
 is obtained in  the third column of Table \ref{tabcomp}.
In the other columns we list the $U_D$-duality groups, their maximal compact subgroups
 and the left-over representations for vector fields.

\begin{table}[ht]
\caption{Maximal gauged compact groups}
\label{tabcomp}
\begin{center}
\begin{tabular}{|c|c|c|c|c|}
\hline
$D$ & $U_D$ & $H_D$ & $\cG_D$ & $\mbox{rep}\cG_D$ \\
\hline
\hline
9 & $SL(2,\IR) \times O(1,1)$ & $O(2)$ &  $O(2)$ & 2 \\
\hline
8 & $SL(3,\IR) \times SL(2,\IR)$ & $O(3) \times O(2)$   & $ O(3)$ & $3$  \\
\hline
7 & $SL(5,\IR)$ & $USp(4) $ &  $ O(5) \sim USp(4) $ & 0 \\
\hline
6 & $O(5,5)$ & $ USp(4) \times USp(4) $  &  $O(5)$ & $5+1$ \\
\hline
5 & $E_{6,(6)}$ & $USp(8)$ &  $ O(6) \sim SU(4) $ & $2\times 6$ \\
\hline
4 & $ E_{7(7)} $ & $SU(8)$ & $O(8)$ & 0 \\
\hline
\end{tabular}
\end{center}
\end{table}

\begin{table}[ht]
\caption{Transformation properties under $\cG_D$ of 2- and 3-forms}
 \label{tabrepba}
  \begin{center}
    \begin{tabular}{|c|c|c|}
\hline
$D$ &  $\mbox{rep} \, B_{\mu\nu}$  & $\mbox{rep} \, A_{\mu\nu\rho}$ \\
\hline
\hline
9 & 2 & 0 \\
\hline
8 & 3 & 0 \\
\hline
7 & 0 & 5 \\
\hline
6 & 5 & 0 \\
\hline
5 & $2\times 6$ & 0 \\
\hline
 \end{tabular}
  \end{center}
\end{table}

We notice that, for any $D$, there are $p$--forms ($p$ =1,2,3) which are charged under the gauge group $\cG_D$.
Consistency of these theories requires that such forms become massive.
It is worthwhile to mention how this can occur in two variants of the Higgs mechanism.
Let us define the (generalized) Higgs mechanism for a $p$--form  mass generation through the absorption
of a massless ($p-1$)--form (for $p=1$ this is the usual Higgs mechanism).
The first variant is the anti-Higgs mechanism for a $p$--form \cite{antihiggs},
which is its absorption by a massless ($p+1$)--form.
It is operating, for $p=1$, in $D=5,6,8,9$
for a sextet of $SU(4)$, a quintet of $SO(5)$, a triplet
of $SO(3)$ and a doublet of $SO(2)$, respectively.
The second variant is the self--Higgs mechanism \cite{topiva},
which only exists for $p = (D-1)/2$, $D= 4k-1$.
This is a massless $p$--form which acquires a mass through a
topological mass term and therefore it
becomes a massive ``chiral'' $p$--form.
The latter phenomena was shown to occur in $D=3$ and $7$.
It is amazing to notice that the representation assignments dictated by
$U$--duality for the various $p$--forms is
precisely that needed for consistency of the gauging procedure (see Table \ref{tabrepba}).
\par
The other compact gaugings listed in Table \ref{tabcomp} are the $D=4$
\cite{dwni} and $D=8$ cases \cite{sase2}.
\par
It is possible to extend the analysis of gauging semisimple groups
also to the case of solvable Lie groups \cite{fegipo}.
For the maximal abelian ideals of $Solv(U_D/H_D)$ this amounts
to gauge an $n$--dimensional subgroup of the translational
symmetries under which at least $n$ vectors are inert. Indeed the
vectors the set of vectors that can gauge an abelian algebra
(being in its adjoint representation) must be neutral under the
action of such an algebra.
We find that in  any dimension $D$ the dimension of this abelian
group $\mbox{dim} \cG_{abel}$ is given precisely by ${\rm dim} (\mbox{rep} \cG_D)$
which appear in the decomposition of $\ID_{r+1}$ under $O(r+1)$.
We must stress that this criterium gives a necessary but not sufficient
condition for the existence of
the gauging of an abelian isometry group,  consistent with supersymmetry.
\begin{table}[ht]
    \caption{Decomposition of fields in representations of the compact  group $\cG_D=O(11-D)$}
    \label{tabcompab}
    \begin{center}\begin{tabular}{|c|c|c|c|c|}
    \hline
    & vect. irrep & adj($O(11-D)$) & $\cA$  & $\mbox{dim} \cG_{abel}$  \\
    \hline
    $D = 9$ & $1+2$ & 1 & 1 & 1 \\
   \hline
     $D= 8$& $ 3+3$ & 3 & 3 & 3  \\ \hline
     $D=7$ & $6 + 4 $ & 6 & 6 & 4  \\
    \hline
     $D=6$ & $ 10+ 5 + 1 $ & 10 & 10 &  $5 + 1 $ \\
    \hline
     $D=5$ & $15+6+6$ & 15 & $ 15+1$ &   $6+6$  \\
     \hline
     $D=4$ & $(21+7)\times 2$ & 21 & $21 + 1 \times 6$ & 7 \\
     \hline
    \end{tabular}\end{center}\end{table}

\chapter{Partial $N=2\rightarrow N=1$ SUSY Breaking}
As it was previously anticipated, in the present chapter I will deal with the 
problem of {\it partial } $N=2\rightarrow N=1$ {\it supersymmetry breaking}
within the general framework of $N=2$ supergravity coupled to an arbitrary
number of vector multiplets and hypermultiplets and with arbitrary gauging 
of the scalar manifold isometries. In the following treatment I will 
mainly refer to 
a work by P. Fr\'e, L. Girardello, I. Pesando and myself \cite{fegipo}.
The relevance of this analysis to the logic of the present dissertation
is to show how solvable Lie algebra machinery can be succesfully applied
to a concrete supergravity problem, giving it a geometrical formulation
and therefore simplifying considerably its solution. Indeed, describing 
the scalar sector of the theory in terms of algebraic objects allows to 
``visualize'' geometrically the mechanism of partial supersymmetry breaking
making it more intuitive.
Moreover, applying the results discussed in last chapter, a physical meaning 
can be given to these geometrical objects, leading to a deeper understanding 
of the physics underlying the mechanism itself.\par 
I shall start giving a brief introduction to the topic of partial supersymmetry
breaking in the context of $N=2$ supersymmetric theories.
$N=2$ supergravity and $N=2$ rigid gauge theory have recently played a major
role in the discussion of string--string dualities
\cite{stdua_1,fhsv,stdua_3,stdua_4,stdua_5,stdua_6,sesch} and in the analysis
of the non--perturbative
phases of Yang--Mills theories \cite{SW_1,SW_2,SWmore_1,SWmore_2}. Furthermore in its
ten years long history, $N=2$ supergravity has attracted the interest of
theorists because of the rich geometrical structure of its
scalar sector, based on the manifold:
\begin{equation}
{\cal M}_{scalar} = {\cal SK}_n \, \otimes \, {\cal QM}_m
\label{scalma}
\end{equation}
where ${\cal SK}_n$ denotes a complex $n$--dimensional special K\"ahler
manifold \cite{dwvp1,speckal_2,speckal_3,skgsugra_1,speckal_4} (for a review of
Special K\"ahler geometry see either \cite{pietrolectures} or
\cite{toinelectures}) and ${\cal QM}_m$
a quaternionic $m$--dimensional quaternionic
manifold, $n$ being the number of vector multiplets and $m$ the  number of
hypermultiplets \cite{quatgeom_1,quatgeom_2,quatgeom_3,skgsugra_1}.
\par
Unfortunately applications of
$N=2$ supergravity to the description of the real world have  been
hampered by the presence of mirror fermions
and by a tight structure which limits severely
the mechanisms of spontaneous breaking of local $N=2$ SUSY.
Any attempt to investigate the fermion problem requires a thorough
understanding of the spontaneous breaking with zero vacuum energy.
In particular, an interesting feature is the sequential breaking to $N=1$
and then to $N=0$ at two different scales.\par
As far as rigid supersymmetry is concerned, a well known theorem \cite{wit81}
forbids the possibility of partial supersymmetry breaking. A heuristic proof 
of this ``no--go'' theorem may be given in the following way. Consider 
the anti--commutation relation between N--extended supersymmetry generators
$Q^i_\alpha\, ;\,\, i=1,...,N$:
\begin{equation}
\{ \bar{Q}^i_{\dot \alpha}\, ,\,Q^j_\alpha\}\, =\, 2\sigma_{{\dot \alpha}\alpha
}^{\mu}\delta^{ij}P_{\mu}
\label{susycom}
\end{equation}
where $P_{\mu}$ is the $4$--momentum generator. Computing the vacuum expectation value (vev) of both sides of the above equation, one gets the following expression for the vacuum energy:
\begin{equation}
E_{vac}\, =\, \frac{1}{2}\sum_{\alpha}\langle 0\vert Q^{\dagger i}_\alpha Q^i_\alpha \vert O \rangle\, =\, \frac{1}{2}\sum_{\alpha}\vert\vert Q^i_\alpha \vert O \rangle \vert\vert \,\,
\,\,(\mbox{no summation over i})
\label{evac}
\end{equation}
If one supersymmetry, say $i=1$, is broken, one has that $Q^1_\alpha\,\vert O
\rangle\neq 0$. This in turn implies, because of \eqn{evac}, that $E_{vac}\neq 0$
and therefore that all the supersymmetries are broken. This argument can be made 
rigorous \cite{por96} by considering the anti--commutation relation between 
local supercurrents in a theory formulated in finite volume and eventually 
performing the infinite volume limit. A way out from the ``no--go'' theorem
was found in the rigid case by I. Antoniadis, H. Partouche and T.R. Taylor
\cite{n2break_2}.
They considered an $N=2$ rigid theory with one vector multiplet and included
in the lagrangian an {\it electric} and {\it magnetic} Fayet--Iliapulos term
(APT model). This amounted to modifying the supercurrent version of \eqn{susycom}
by means of an additive constant matrix on the right hand side, making the latter 
proportional to a non trivial $SU(2)$ matrix which could therefore have, 
on suitable vacua, one vanishing and one non vanishing eigenvalue.\par 
On the local supersymmetry front, sometimes ago a negative result  on partial breaking was established within
the $N=2$ supergravity formulation based on conformal tensor calculus \cite{lucia_nogo}.
A particular way out was indicated in an ad hoc model \cite{newlucia} which prompted
some generalizations based on Noether couplings \cite{berianew}.
\par
 With the developments in special K\"ahler geometry
\cite{cdfvp,stdua_4,stdua_5} stimulated by the studies on
S--duality, the situation can now be cleared in general terms. Indeed
it has appeared from
\cite{n2break_1,n2break_2,n2break_3,n2break_4}  that the negative results
on $N=2$ partial supersymmetry breaking were the consequence of unnecessary
restrictions imposed on the formulation of special K\"ahler geometry and could
be removed. Moreover the minimal $N=2$ supergravity model ( coupled to one vector
multiplet and one hypermultiplet) exhibiting partial supersymmetry breaking
and introduced for the first time by S. Ferrara, L. Girardello and M. Porrati
 in \cite{n2break_1}, was shown to admit the APT model as a suitable flat limit.
\par
As I anticipated at the beginning,  in what follows the generic structure of 
partial supersymmetry breaking within the general form of $N=2$ supergravity coupled with $n+1$ vector multiplets and $m$ hypermultiplets will be discussed in detail.
 In particular an explicit example is worked out in section 3, based on
the choice for the vector multiplets of the special K\"ahler manifold
$SU(1,1)/U(1) \, \otimes \, SO(2,n)/SO(2) \times SO(n)$ and of the
quaternionic manifold $SO(4,m)/SO(4) \times SO(m)$ for the
hypermultiplets.\footnote{This choice of manifolds is inspired by
string theory since it corresponds to $N=2$ truncations of string
compactifications on $T^6$.} In this section the mechanism of partial supersymmetry breaking is formulated mathematically by using the coset manifold description for 
${\cal M}_{scalar}$. In section four the same result will be achieved by adopting
 the solvable Lie algebra description for both the Special K\"ahler and 
Quaternionic manifolds in \eqn{scalma} (Alekseevskii's formalism) and it will be shown how this formulation of the problem, besides other advantages, allows to find, in a strightforward way,
the {\it flat directions} of the scalar potential. 
\subsection{The bearing of the coordinate free definition of
Special K\"ahler geometry}
A main point in our subsequent discussion is the use of the
symplectically covariant, coordinate free definition of special K\"ahler
geometry \eqn{speckahl} given in Chapter 1. 
\par
Consider an $n$ dimensinal Special K\"ahler manifold ${\cal SK}$ 
spanned by the scalar fields belonging to $n$ vector multiplets.
Definition \eqn{speckahl} of ${\cal SK}$ implies the existence of 
a correspondence 
$i_\delta$ 
between isometries on the base manifold ${\cal M}$ and $Sp(2n+2,\IR)$
 transformations on sections of the symplectic bundle 
${\cal Z}\rightarrow {\cal M}$. These transformations turn out to be
 the same duality transformations mixing the electric field--strenghts $F^\Lambda$ 
with the magnetic ones $G_\Sigma$ which have been discussed in Capter 1
and $i_\delta$ is the embedding defined by \eqn{embedding}. 
A change of basis for the symplectic sections of ${\cal Z}$ amounts to 
conjugating $i_\delta$ through a symplectic transformation $S$. It has been 
pointed out in Chapter 1 that such an operation is physically immaterial
only if the theory is {\it ungauged}.
 As soon as an electric current is introduced
the distinction is established and, at least at the classical (or semiclassical)
level, the only
symplectic transformations $S$ that yield equivalent theories
(or can be symmetries) are the perturbative ones generated by lower triangular
symplectic matrices:
\begin{equation}
\left ( \matrix{{\bf A} & {\bf 0} \cr {\bf C} & {\bf D}\cr }\right )\,
\label{perturb}
\end{equation}
\par
It follows that different bases of symplectic sections for the bundle
${\cal Z}$ yield, after gauging, inequivalent physical theories. The
possibility of realizing or not realizing partial supersymmetry breaking
$N=2 \, \to \, N=1$ are   related to this choice of symplectic bases.
In the tensor calculus formulation the lower
part $F_{\Lambda}(z)$ of the symplectic section $\Omega(z)$ defined in 
\eqn{Zsection} should be, necessarily,
derivable from a holomorphic prepotential $F(X)$ that is a degree two
homogeneous function of the upper half of the section:
\begin{equation}
F_{\Sigma}(z) = \frac{\partial }{\partial X^{\Sigma}(z)} \, F(X(z))
\quad ; \quad
F\left ( \lambda \, X^{\Sigma}(z)\, \right ) = \,\lambda^2 \,
F\left (   X^{\Sigma}(z)\, \right )
\label{prepotent}
\end{equation}
This additional request is optional in the more general geometric
formulation of $N=2$ supergravity \cite{skgsugra_1,fundpaper} where only the intrinsic
definition of special geometry is utilized for the construction of the
lagrangian. It can be shown that the condition for the existence of the
holomorphic prepotential $F(X)$ is the non degeneracy of the
jacobian matrix $e^{I}_i(z) \equiv \partial_i\left (X^{I}/X^0\right );
I=1,\dots , n$. There are symplectic bases where this jacobian has vanishing
determinant and there no $F(X)$ can be found. If one insists on the
existence of the prepotential such bases are discarded {\it a priori}.
\par
There is however another criterion to select symplectic bases which has
a much more intrinsic meaning and should guide our choice. Given the
isometry group $G$ of the special K\"ahler manifold ${\cal SV}$, which is
an intrinsic information, which subgroup of $G_{class} \subset G$ is
realized by classical symplectic matrices
$\left ( \matrix{{\bf A} & {\bf 0} \cr {\bf 0} & {\bf D}\cr }\right )$
and which part of $G$ is realized by non perturbative symplectic matrices
$\left ( \matrix{{\bf A} & {\bf B} \cr {\bf C} & {\bf D}\cr }\right )$ with
${\bf B} \ne 0$ is a symplectic base dependent fact. It appears that in
certain cases, very relevant for the analysis of string inspired supergravity,
the maximization of $G_{class}$, required by a priori symmetry
considerations, is incompatible with the condition ${\rm det}
\left ( e^{I}_i(z) \right ) \, \ne \, 0$ and hence with the existence of
a prepotential $F(X)$.  If the unneccessary condition on $F(X)$ existence
is removed and the maximally symmetric symplectic bases are accepted
the no-go results on partial supersymmetry breaking can also be removed.
Indeed in \cite{n2break_1} it was shown that by gauging a group
\begin{equation}
G_{gauge}= \IR^2
\end{equation}
in a N=2 supergravity model with just one vector multiplet and one hypermultiplet
based on the scalar manifold:
\begin{equation}
{\cal SK}=\frac{SU(1,1)}{U(1)} \quad \quad ; \quad \quad {\cal Q}= \frac{SO(4,1)}{SO(4)}
\label{giraferpor}
\end{equation}
supersymmetry can be spontaneously broken from $N=2$ down to $N=1$,
provided one uses the symplectic basis  where the embedding of
$SU(1,1)\equiv SL(2,\IR)$ in $Sp(4,\IR)$ is the following \footnote{ Here
$\eta$ is the standard constant metric with $(2,n)$ signature}:
\begin{equation}
 \forall \,   \left ( \matrix{a & b \cr  c &d \cr }\right )
\, \in \, SL(2,\IR) \quad {\stackrel{\i_\delta}
{\hookrightarrow}}
\left ( \matrix{ a \, \bfone & b \, \eta \cr c \, \eta  &
d \, \bfone \cr } \right )
\, \in \, Sp(4,\IR)
\label{ortolettodue}
\end{equation}
\par
Let us now turn on the general case of an $N=2$ supergravity with $n+1$ 
vector multiplets and $m$ hypermultiplets
based on the scalar manifold:
\begin{equation}
{\cal SK}_{n+1} = \frac{SU(1,1)}{U(1)} \otimes
\frac{SO(2,n)}{SO(2)\otimes SO(n)} \quad \quad ; \quad\quad
{\cal QM}_m = \frac{SO(4,m)}{SO(4)\otimes SO(m)}
\label{ourchoice}
\end{equation}
We show that we can obtain partial breaking $N=2 \, \to \, N=1$ by
\begin{itemize}
\item {choosing the Calabi--Vesentini symplectic basis where the embedding of
$SL(2,\IR)\otimes SO(2,n)$ into $Sp(4+2n,\IR)$ is the following:
\begin{eqnarray}
\forall \,   \left ( \matrix{a & b \cr  c &d \cr }\right )
\, \in \, SL(2,\IR) & {\stackrel{\i_\delta}
{\hookrightarrow}} &
\left ( \matrix{ a \, \bfone & b \, \eta \cr c \, \eta  &
 d \, \bfone \cr } \right )
\, \in \, Sp(2n+4,\IR) \nonumber\\
\forall \,  L \, \in \, SO(2,n) & {\stackrel{\i_\delta}
{\hookrightarrow}} &
\left ( \matrix{ L & {\bf 0}\cr {\bf 0} & (L^T)^{-1} \cr } \right )
\, \in \, Sp(2n+4,\IR)
\label{ortoletto}
\end{eqnarray}
namely where all transformations of the group $SO(2,n)$ are linearly
realized on electric fields}
\item {gauging a group:
\begin{equation}
G_{gauge}= \IR^2 \, \otimes \, G_{compact}
\end{equation}
where
\begin{eqnarray}
\IR^2 \, & \cap & SL(2,R)\otimes SO(2,n) = 0  \nonumber\\
\IR^2 & \subset & \mbox{ an abelian ideal of the solvable Lie
subalgebra } V \,  \subset  so(4,m) \nonumber\\
G_{compact} & \subset & SO(n) \subset   SO(2,n) \nonumber\\
G_{compact} & \subset & SO(m-1) \subset  SO(4,m)
\label{condizie}
\end{eqnarray}
namely where $\IR^2$ is a group of two abelian translations acting on the hypermultiplet
manifold but with respect to which the vector multiplets have {\it zero charge}, while
the compact gauge group $G_{compact}$ commutes with such translations
and has a linear action on both
the hypermultiplet and the vector multiplets.}
\end{itemize}
In the following two sections I shall derive the result summarized above
starting from the recently obtained complete form of N=2 supergravity
with general scalar manifold interactions \cite{fundpaper}.
Let us conclude the
present introduction with some physical arguments why the result
should be
obtained precisely in the way described above.
\subsection{General features of the partial $N=2$ supersymmetry
breaking}
\par
The mechanism of partial supersymmetry breaking that I am about to describe
is referred to as {\it super--Higgs mechanism} and is ``visually'' represented 
in Figure \ref{super}.
\iffigs
\begin{figure}
\caption{}
\label{super}
\epsfxsize = 10cm
\epsffile{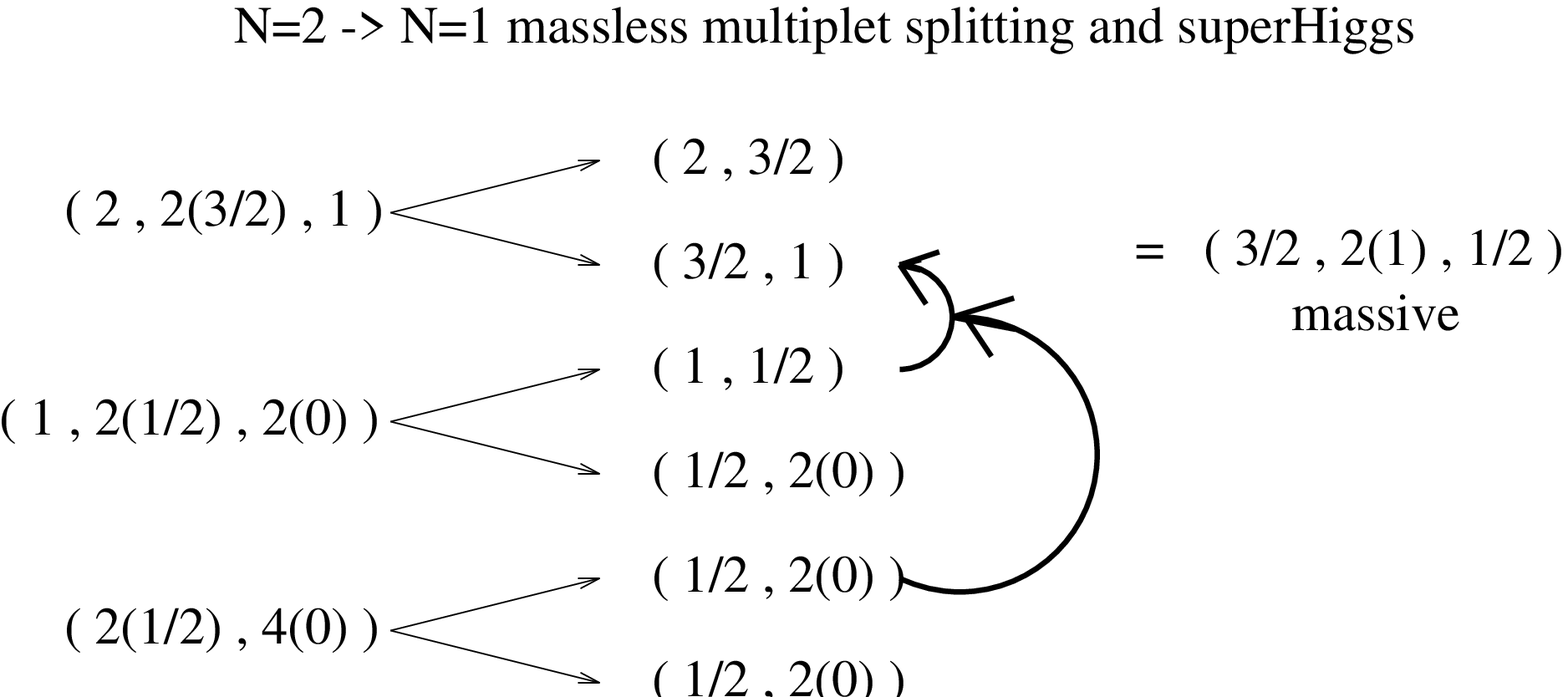}
\vskip -0.1cm
\unitlength=1mm
\end{figure}
\fi
To break supersymmetry from $N=2$ down  to $N=1$ we must break the
$O(2)$ symmetry that rotates one gravitino into the other. This
symmetry is
gauged by the graviphoton $A^0_\mu$. Hence the graviphoton must become massive.
At the same time, since we demand that $N=1$ supersymmetry should be
preserved,
the second gravitino must become the top state of an $N=1$ massive spin
$3/2$ multiplet which has the form $\left \{ \left ( {\o{3}{2}} \right ), {2 }
\left ( 1 \right ) , \left ( {\o{1}{2}} \right )  \right \}   $.
Consequently not only the graviphoton but also a second gauge boson
$A^1_{\mu}$
must become massive through ordinary Higgs mechanism. This explains
while the partial supersymmetry breaking involves the gauging of a
two--parameter group. That it should be a non compact $\IR^2$ acting as
a translation group on the quaternionic manifold is more difficult to
explain a priori, yet we can see why it is very natural.
  In order to obtain a Higgs mechanism for
the graviphoton $A_\mu^0$ and the second photon $A_\mu^1$ these vectors must
couple to the hypermultiplets.  Hence these two fields should
 gauge isometries of the quaternionic manifold ${\cal QM}$. 
That such isometries should be translations is understood by observing that in this way
one introduces a flat direction in the scalar potential,
corresponding to the vacuum expectation value of the hypermultiplet scalar, coupled
to the vectors in such a way. Finally the need to use the correct
symplectic basis is explained by the following remark. Inspection of
the gravitino mass-matrix shows that it depends on both the momentum
map ${\cal P}^0_\Lambda (q)$ for the quaternionic action of the gauge group on
the hypermultiplet manifold and on the upper (electric) part of the
symplectic section $X^\Lambda(z)$. In order to obtain a
mass matrix with a zero eigenvalue we need a contribution from both $X^0$ and $X^1$ at the
breaking point, which can always be chosen at $z^i=0$, since the vector multiplet scalars
are neutral (this is a consequence of $\IR^2$ being abelian). Hence
in the correct symplectic basis we should have both $X^0(0) \ne 0$
and $X^1(0) \ne 0$. This is precisely what happens in the
Calabi--Vesentini basis for the special K\"ahler manifold $SU(1,1)/U(1) \otimes
SO(2,n) / SO(2) \otimes SO(n)$. Naming $y^{i}$ $(i=1,\dots , n)$ a standard set of complex
coordinates for the $SO(2,n)/ (SO(2)\otimes SO(n))$ coset manifold, characterized by linear
transformation properties under the $SO(2)\otimes SO(n)$ subgroup and naming $S$
the dilaton field, i.e. the complex coordinate spanning the coset manifold
$SU(1,1)/U(1)$, the explicit form of the symplectic section (\ref{Zsection})
corresponding to the symplectic embedding (\ref{ortoletto}) is
(see \cite{cdfvp} and \cite{pietrolectures}):
\begin{equation}
\Omega \, = \, \left ( \matrix{ X^0 \cr X^1 \cr X^{i} \cr F_0 \cr F_1 \cr F_i\cr } \right ) \, = \,
\left ( \matrix{ \frac{1}{2}(1+y^2)\cr  \cr {\rm i}\frac{1}{2}(1-y^2)\cr \cr y^{i} \cr S \,
\frac{1}{2}(1+y^2)\cr S \, {\rm i}\frac{1}{2}(1-y^2)\cr
- \, S \, y^{i} \cr } \right )  \, {\stackrel{y\, \to\,  0}{\longrightarrow}}
\, \left ( \matrix{ \frac{1}{2} \cr {\rm i}\frac{1}{2} \cr 0 \cr \frac{1}{2}\, S \cr
{\rm i}\frac{1}{2} \, S \cr 0\cr } \right )
\label{calvesesec}
\end{equation}
\section{Formulation of the  $N=2 \to N=1$ SUSY breaking problem}
As it is well understood in very general terms (see \cite{kilspinold}), a
classical vacuum of an $N=r$ supergravity theory preserving $p \le r$ supersymmetries
 in Minkowski  space--time  is just a constant scalar field
configuration $ \phi^I (x) = \phi^I_0 \, (I = 1,\dots ,\mbox{dim} \, {\cal M}_{scalar}) $
corresponding to an extremum of the scalar potential and such that it
admits $p$ {\it Killing spinors}. In this context, Killing spinors are covariantly constant
spinor parameters of the supersymmetry transformation
$\eta^A_{(a)}$ $(A=1,\dots,n)$, $(a=1,\dots,p)$ such that the
SUSY variation of the fermion fields in the bosonic background
$g_{\mu\nu}=\eta_{\mu\nu} \, , \, A^\Lambda_\mu=0\, , \,  \phi^I =
\phi^I_0$ is zero for each $\eta^A_{(a)}$:
\begin{eqnarray}
\delta_a \, \psi_{\mu A} & \equiv & {\rm i}\gamma_\mu \, S_{AB}(\phi_0)
\, \eta^B_{(a)} \, = \, 0\nonumber\\
\delta_a \xi^i & \equiv &\Sigma^i_A (\phi_0 )\, \eta^A_{(a)} \, = \, 0   \quad \quad (a=1,\dots,p)
\label{kilspigen}
\end{eqnarray}
In  eq.(~\ref{kilspigen})
  the spin $3/2$ fermion shift  $S_{AB}(\phi)$ and
the spin $1/2$ fermion shifts  $\Sigma^i_A (\phi)$ are the
non-derivative contributions to the supersymmetry transformation
rules of the gravitino $\psi_{\mu A}$ and of the spin one half fields $\xi^i$, respectively.
The integrability conditions of supersymmetry transformation rules
are just the field equations. So it actually happens that the
existence of Killing spinors, as defined by eq.(~\ref{kilspigen}), forces the
constant configuration $\phi^I = \phi^I_0$ to be an extremum of the scalar potential.
\par
Hence we can just concentrate on the problem of solving eq.(~\ref{kilspigen}) in the
case $N=2$ with $p=1$.
\par
In $N=2$ supergravity there are two kinds of spin one half fields:  the gauginos
$\lambda^{j^\star}_A$ carrying an $SU(2)$ index $A=1,2$ and a world--index
$j^\star=1^\star,\dots , n^\star$
of the tangent bundle $T^{(0,1)}{\cal SK}$ to the special K\"ahler manifold
($n=\mbox{dim}_{\bf C}\, {\cal SK}=
\# \mbox{vector multiplets}$ ) and the
hyperinos $\zeta^\alpha$ carrying an index $\alpha$=$(1,\dots,n)$ running in the
fundamental representation of $Sp(2m,\IR)$  ($m=\mbox{dim}_{\bf Q}\, {\cal Q}=
\# \mbox{hypermultiplets}$ ). Hence there are three kind of fermion
shifts:
\begin{eqnarray}
\delta\psi_{\mu A}&=& i\gamma_{\mu} S_{A B} \eta^{B}
\nonumber\\
\delta(g_{i^* j} \Lambda^{i^*}_{A}) &=& W_{j |A B} \eta^{B}\nonumber \\
\delta \zeta^{\alpha} &=& N_A^{\alpha} \eta^{A}
\label{ciliegina}
\end{eqnarray}
According to the analysis and the conventions of \cite{fundpaper,skgsugra_1},
the shifts are
expressed in terms of the fundamental geometric structures defined over the special
K\"ahler and quaternionic manifolds as follows:
\begin{eqnarray}
S_{AB}&=&-\frac{1}{2} {\rm i} (\sigma_{x}\epsilon )_{A B} {\cal P}^{x}_{\Lambda} L^{\Lambda}\nonumber \\
W_{j|A B}&=&- ({\epsilon_{A B}} \partial_{j}{\cal P}_{\Lambda} L^{\Lambda} +{\rm i}
(\sigma_{x}\epsilon)_{A B}
 {\cal P}^{x}_{\Lambda} f^{\Lambda}_{j})\nonumber\\
 N_{A}^{\alpha}& =& -2 {\bf i}_{\vec k_{\Lambda}}{\cal U}^{\alpha}_{A}  L^{\Lambda} \, =
 \, -2 \, {\cal U}^\alpha_{A \vert u} \, k^u_{\Lambda} \, L^\Lambda
\label{shifmatrici}
\end{eqnarray}
In eq.(\ref{shifmatrici}) the index $u$ runs on $4m$ values
corresponding to any set of of $4m$ real coordinates
for the quaternionic manifold. Further ${\cal P}_{\Lambda}$ is the holomorphic momentum map
for the action of the gauge group ${\cal G}_{gauge}$ on the special
K\"ahler manifold ${\cal SK}_n$, ${\cal P}^{x}_{\Lambda} \, (x=1,2,3)$ is the
triholomorphic momentum map for the action of the same group on the
quaternionic manifold, $L^{\Lambda}= e^{{\cal K}/2} X^\Lambda(z)$ is
the upper part of the symplectic section (\ref{Zsection}), rescaled
with the exponential of one half the K\"ahler potential ${\cal K}(z,{\bar z})$
(for more
details on special geometry see \cite{fundpaper},
\cite{pietrolectures}, \cite{toinelectures}), ${\vec k}_{\Lambda}$ is the
Killing vector generating the action of the gauge group on both
scalar manifolds and finally ${\cal U}^{\alpha}_{A}$ is the vielbein 1--form on
the quaternionic manifold carrying an $SU(2)$ doublet index and an
index $\alpha$ running in the fundamental representation of
$Sp(2m,\IR)$.
\par In the case of effective supergravity theories
that already take into account the perturbative and non--perturbative
quantum corrections of string theory the manifolds ${\cal SK}$ and
${\cal Q}$ can be complicated non--homogeneous spaces
without continuous isometries. It is not in
such theories, however, that one performs the gauging of non abelian
groups and that looks for a classical breaking of supersymmetry.
Indeed, in order to gauge a non abelian group, the scalar manifold
must admit that group as a group of isometries. Hence ${\cal
M}_{scalar}$ is rather given by the homogeneous coset manifolds that
emerge in the {\it field theoretical limit} to {\it the tree level
approximation} of superstring theory. In a large variety of models the
tree level approximation yields the choice (~\ref{ourchoice}) and we
concentrate on such a case to show that a constant configuration with
a single killing spinor can be found. Yet, as it will appear
from our subsequent discussion, the key point of our construction
resides in the existence of an $\IR^2$ translation isometry group on
${\cal Q}$ that can be gauged by two vectors associated with section
components $X^0 \, X^1$ that become  constants in the vacuum
configuration of ${\cal SK}$. That these requirements can be met
is a consequence of the algebraic structure \'a la Alekseevski of both
the special K\"ahlerian and the quaternionic manifold. Since such
algebraic structures exist for all homogeneous special and
quaternionic manifolds, we are lead to conjecture that the partial
supersymmetry breaking described below can be extended to most N=2
supergravity theories on homogenous scalar manifolds.  \par For the
rest of the chapter, however, I shall concentrate on the study of case
(\ref{ourchoice}).
\par In the Calabi-Vesentini basis
(\ref{calvesesec}) the origin $y=0$ of the vector multiplet manifold
$SO(2,n)/SO(2) \otimes SO(n)$ is a convenient point where to look for
a configuration breaking $N=2 \, \to \, N=1$.  We shall argue that for
$y=0$ and for an arbitrary point in the quaternionic manifold $\forall
q \, \in \, SO(4,m)/SO(4) \times SO(m)$ there is always a suitable
group $\IR^2_q$ whose gauging achieves the partial supersymmetry
breaking. Actually the group $\IR^2_q$ is just the conjugate, via an
element of the isometry group $SO(4,m)$, of the group $\IR^2_0$ the
achieves the breaking in the origin $q=0$. Hence we can reduce the
whole analysis to a study of the neighborhood of the origin in both
scalar manifolds.  To show these facts we need to cast a closer look
at the structure of the hypermultiplet manifold.
\section{Explicit solution}
\subsection{The quaternionic manifold $SO(4,m) / SO(4) \otimes SO(m)$}
We start with the usual parametrization of the coset
$SO(4,m)\over SO(4)\otimes SO(m)$ \cite{cosetto}:
\begin{equation}
\IL(q)=
\left(
\matrix{
\sqrt{1+qq^t} & q \cr
q^t &\sqrt{1+q^tq}}
\right)
=\left(\matrix{r_1 & q \cr q^t & r_2}\right)
\label{parametrizzazione}
\end{equation}
where $q=||q_{a t}||$ is a $4\times m$ matrix
\footnote{In the following letters from the beginning of the alphabet
will range over $1\dots 4$ while letters  from the end of the alphabet
will range over $1\dots m$}.
This coset manifold has a riemannian structure defined by the
vielbein, connection and  metric given below \footnote{For more details
on the following formulae see Appendix C.1 of ref.~ \cite{fundpaper}}:
\begin{eqnarray}
\IL^{-1} d\IL &=&
\left(
\matrix{
\theta & E \cr
E^t  & \Delta}
\right) \in so(4,m) \nonumber \\
ds^2 &=& E^t \otimes E
\end{eqnarray}
The explicit form of the vielbein and connections which we utilise in the sequel
is:
\begin{equation}
E \, = \, r_1 dq -q dr_2 \quad ; \quad
\theta \, = \, r_1 dr_1 -q dq^t \quad ; \quad
\Delta \, = \, r_2 dr_2 -q^t dq
\end{equation}
where $E$ is the coset vielbein, $\theta$ is the $so(4)$-connection
and $\delta$ is the $so(n)$-connection.
The quaternionic structure of the manifold is given by
\begin{equation}
K^x \, = \,  \frac{1}{2} tr(E^t \wedge J^x E) \quad ; \quad
\omega^x \, = \,-\frac{1}{2} tr (\theta J^x) \quad ; \quad
{\cal U}^{A \alpha} \, = \, \frac{1}{\sqrt{2}} E^{a t} (e_a)^A_B
\label{3.4}
\end{equation}
$K^x$ being the triplet of hyperK\"ahler 2--forms, $\omega^x$ the
triplet of $su(2)$--connection 1--forms and ${\cal U}^{A\alpha}$ the
vielbein 1--form in the symplectic notation. Furthermore
$J^x$ is the triplet of $4\times 4$  self--dual 't Hooft matrices
$J^{+|x}$ normalized as in \cite{fundpaper}, $e_a$ are the quaternionic units
as given in \cite{fundpaper} and   the symplectic index $\alpha=1,\dots , 2m$
is identified with a pair of an $SU(2)$ doublet index $B=1,2$ times an $SO(4)$ vector
index $t=1,\dots, m$: $\alpha\equiv B t$.
Notice also the factor $\frac{1}{2}$ in the definition of $K^x$ with respect to the
conventions used in \cite{fundpaper},  which is necessary in order to have
$ \nabla \omega^x= -K^x$
\subsection{Explicit action of the isometries on the coordinates and the
killing vectors.}
Next we compute the killing vectors; to this purpose we need to
know the action of an element $g\in so(4,m)$ on the coordinates $q$.
To this effect we make use of the standard formula
\begin{equation}
\delta \IL \equiv k^{at}_{g} \, \frac{\partial}{\partial q^{at}}= g \IL-\IL w_g
\label{cosettario}
\end{equation}
where
\begin{equation}
w_g=
\left(
\matrix{
w_1 &  \cr
 & w_2}
\right) \in so(4) \oplus so(m)
\end{equation}
is the right compensator and the element $g\in so(4,m)$ is given by
\begin{equation}
g=
\left(
\matrix{
a & b \cr
b^t & c}
\right)\in {\bf so(4,m)}
\label{donpeppone}
\end{equation}
with $a^t=-a$, $c^t=-c$.
The solution to (\ref{cosettario}) is:
\begin{eqnarray}
\delta q =  a q + b r_2 - q c -q \hat w_2 & = & a q +r_1 b -q c + \hat w_1 q\nonumber\\
\delta r_1  = [ a, r_1 ] + b q^t - r_1  \hat w_1 & ; &
\delta r_2  = [ c, r_2 ] + b^t q - r_2  \hat w_2\nonumber\\
w_1  =  a + \hat w_1 & ; &
w_2  =  c+\hat w_2
\label{cosevarie}
\end{eqnarray}
The only information we need to know about $\hat w_1, \hat w_2$ is that
they depend linearly on $b, b^t$.
Anyhow for completess we give their explicit form:
\begin{eqnarray}
\hat w_{1; a b}
&=& \left. { d (\sqrt{ \bfone + x})_{a b}\over d x_{c d}} \right|_{x=qq^t}
(bq^t-qb^t)_{c d}\nonumber\\
\hat w_{2; s t}
&=& \left. { d (\sqrt{ \bfone + y})_{s t}\over d y_{p q}} \right|_{y=q^tq}
(b^tq-q^tb)_{p q}
\end{eqnarray}
>From these expressions  we can obtain the Killing vector field
\begin{equation}
\vec k_g =  (\delta q)_{a t} \frac{\partial} {\partial q _{a t}}
\label{assassino}
\end{equation}
\subsection{The momentum map.}
We are now in a position to compute the triholomorphic momentum map ${\cal P}^x_g$
associated with the generic element \ref{donpeppone} of the $so(4,m)$ Lie algebra.
Given the vector field \ref{assassino}, we are supposed to solve the first order
linear differential equation:
\begin{equation}
{\bf i}_{{\vec k}_g} K^x = -\nabla {\cal P} ^x
\end{equation}
$\nabla$ denoting the exterior derivative covariant with respect to
the $su(2)$--connection $\omega^x$ and ${\bf i}_{{\vec k}_g} K^x$ being the contraction
of the 2--form $K^x$ along the Killing vector field ${\vec k}_g$.
By direct verification the general solution is given by:
\begin{equation}
{\cal P}^x_g =\frac{1}{2} tr(\left(\matrix{J^x & 0\cr 0 & 0}\right) C_{g} )= tr(J^x P_{g})
\label{fragolina}
\end{equation}
where for any element  of the ${\bf so(4,m)}$ Lie algebra conjugated
with the adjoint action of the coset representative (\ref{parametrizzazione}), we have
introduced the following block decomposition and notation:
\begin{equation}
\forall { g}\, \in \, { so(4,m)}: \quad C_{ g}\,  \equiv \, \IL(q)^{-1} g
\IL(q) = \matrix{2P_{ g} &
 {\bf i}_{\vec k}E_{ g} \cr {\bf i}_{\vec k}E^t_{ g} & 2 Q_{ g}}
\label{fondamentale}
\end{equation}
Furthermore if we decompose $g=g^\Lambda \, {\bf T}_\Lambda$
the generic element $g$ along a basis  $\{ {\bf T}_\Lambda \}$ of
generators of the $so(4,m)$ Lie algebra, we can write:
\begin{equation}
{\cal P}^x_\Lambda =\frac{1}{2} tr(\left(\matrix{J^x & 0 \cr 0 & 0}\right) 
C_{{\bf T}_\Lambda})= tr(J^x P_{{\bf T}_\Lambda})
\label{fragolona}
\end{equation}
\subsection{Solution of the breaking problem in a generic point of the
quaternionic manifold}
At this stage we can attempt to find a solution for our problem i.e. introducing
a gauging that yields a partial supersymmetry breaking.
\par
Recalling the supersymmetry variations of the Fermi fields (\ref{ciliegina})
  we evaluate them at the origin of the special K\"ahler
using the Calabi-Vesentini coordinates (\ref{calvesesec}):
\begin{eqnarray}
S_{A B}|_{y=0} &=& -{1\over 4} i (\sigma_x \epsilon)_{A B} tr(J^x(P_0+i P_1))\nonumber\\
W_{ s| AB }|_{y=0} &=& -{1\over 4 (Im S)^{3\over 2}} i (\sigma_x\epsilon)_{A B}
tr(J^x(P_0+i P_1))\nonumber\\
W_{ \alpha | A B}|_{y=0} &=& 0\nonumber \\
N_A^\alpha |_{y=0}&\propto& ({\bf i}_{0}E^{a t} -i {\bf i}_{1}E^{a t}_1) (e_a)^A_B
\label{variations}
\end{eqnarray}
where in the last equation we have identified $\alpha\equiv a B$ as
explained after eq. (\ref{3.4}).
\par
In the next subsection we  explicitly compute a solution for the matrices $P_0,
P_1$ and ${\bf i}_{0}E^{a t}, {\bf i}_{1}E^{a t}$ at the origin of the
quaternionic manifold $\cQ_{m}$.
Starting from this result it can be shown that any point $q\neq 0$ can define
a vacuum of the theory in which SUSY is broken to $N=1$, provided a suitable
gauging is performed. Indeed we can find the general solution for any point $q$ by requiring
\begin{equation}
C_\Lambda(q)=C_\Lambda(q=0) ~~~~ \Lambda=0,1
\end{equation}
that is the group generators of the group we are gauging at a generic
point $q$  are given by
\begin{equation}
T_{\Lambda}(q)= \IL(q) T_\Lambda(0) \IL^{-1}(q)
\end{equation}
This result is very natural and just reflects the homogeneity of the
coset manifold,
i.e. all of its points are equivalent.
\subsection{Solution of the problem at the origin of $\cQ_{m}$.}
In what follows all the earlier defined quantities, related to
$\cQ_{m}$, will be computed near the origin $q=0$. The right-hand side
of equations (\ref{cosevarie},\ref{assassino}) is expanded in powers
of $q$ as it follows:
\begin{eqnarray} \delta q = b+aq-qc +O(q^2) & ; & \vec k_a \, =
\, (aq)_{a t} \frac{\partial} {\partial q _{a t}}\nonumber\\ \vec k_b \, =\, b_{a t} \frac{\partial}
{\partial q_{a t}} & ; & \vec k_c \, =\, -(qc)_{a t} \frac{\partial} {\partial q _{a t}}
\label{umamma}
\end{eqnarray}
The expressions for the vielbein, the connections and the quaternionic
structure, to the approximation order  we work, are:
\begin{eqnarray}
E \, =\,  dq-\frac{1}{2} q dq^t q &;&
\theta\, =\,  \frac{1}{2} dq q^t -\frac{1}{2} q dq^t \nonumber\\
K^x &=&  tr\left( dq^t ~J^x~ dq - dq q^t ~ J^x ~ dq q^t \right) \nonumber\\
\omega^x \, =\,  -\frac{1}{2} tr \left(dq q^t J^x \right) & ; &
{\cal U}^{A \alpha} \, =\,  \frac{1}{\sqrt{2}} dq^{a t} (e_a)^A_B
\label{partenza}
\end{eqnarray}
Finally the triholomorphic momentum maps corresponding to the a,b,c generators, have
the following form,  respectively :
\begin{equation}
{\cal P}^x_a \, =\,  \frac{1}{2} tr( J^x a + J^x q t^t a) \quad ; \quad
{\cal P}^x_b \, =\, -\frac{1}{2} tr (J^x q b^t) \quad ; \quad
{\cal P}^x_c \, =\, -\frac{1}{2} tr (J^x q c q^t)
\label{tripappa}
\end{equation}
Inserting eq.s (\ref{tripappa}) into equations (\ref{variations}) one finds the expressions
for the shift matrices  in the origin:
\begin{eqnarray}
S_{A B}|_{y=0} &=& -{1\over 4} i (\sigma_x\epsilon)_{A B} tr(J^x(a_0+i a_1))\nonumber\\
W_{ s| AB }|_{y=0} &=& -{1\over 4 (Im S)^{3\over 2}} i (\sigma_x\epsilon)_{A B}
tr(J^x(a_0+i a_1)) \nonumber\\
W_{ \alpha | A B}|_{y=0} &=& 0 \nonumber\\
N_A^\alpha |_{y=0} &\propto& (b_0^{a t} -i b^{a t}_1) (e_a)^A_B
\label{variatorig}
\end{eqnarray}
$a_{0,1}$ and $b_{0,1}$ being the a and b blocks of the matrices $P_{0,1}$,
respectively.
\par
It is now clear that in order to
break  supersymmetry we need $a\neq 0$ and $b \neq 0$ .
 From the first  two equations and from the requirement that the
gravitino mass--matrix $S_{AB}$ should have a zero
eigenvalue we get
\begin{equation}
\sum_x (a_0+i a_1)^2_x=0
\Rightarrow \vec a_0 \cdot \vec a_1= (\vec a_0)^2-(\vec a_1)^2=0
\end{equation}
We solve this constraint by setting ($ a_x =-{1\over 4} tr(J^x a)$ )
\begin{equation}
a_{0 x}= g_0 \delta_{x 1} ~~~~
a_{1 x}= g_1 \delta_{x 2}
\label{doncamillo}
\end{equation}
with $g_0=g_1$. These number are the gauge coupling constants of the
repeatedly mentioned gauge group $\IR^2$.
Note that due to the orthogonality of the antiself--dual t'Hooft matrices
${\bar J}^x$ to the self--dual ones $J^x$, the general solution is
not as in eq.~\ref{doncamillo}, but it involves additional arbitrary
combinations of  ${\bar J}^x$, namely:
\begin{equation}
a_0 \, =\,  g_0 J^1 + \bar a_{0 x} \bar {J}^x  \quad ; \quad
a_1 \, =\,  g_1 J^2 + \bar a_{1 x} \bar {J}^x
\end{equation}
For the conventions see \cite{fundpaper}.
To solve the last  of eq.s (\ref{variations}) we set
\begin{equation}
b_0 =\left(\matrix{ 0 \cr \vec \beta_0 \cr 0 \cr 0 } \right)
 ~~~~
b_1 =\left(\matrix{ 0 \cr 0 \cr \vec \beta_1 \cr 0 } \right)
\end{equation}
where $\vec \beta_0$ and $\vec \beta_1$ are $m$--vectors of the
in the fundamental representation of $SO(m)$.
\par
Now the essential questions are:
\begin{enumerate}
\item{\it  Can one find two commuting matrices belonging to
the ${\bf so(4,m)}$ Lie algebra and   satisfying the previous
constraints? These matrices are the generators of $\IR^2$.}
\item{\it If so, which is the maximal compact subalgebra ${\bf G}_{compact}\subset {\bf so(4,m)}$
commuting with them, namely the maximal compact subalgebra of the centralizer $Z(\IR^2)$?
${\bf G}_{compact}$ is the Lie algebra of the maximal compact gauge group that can survive
unbroken after the partial supersymmetry breaking. }
\end{enumerate}
\par
Let $a_{2},b_2,c_2$ denote the blocks of the commutator $[g_0,g_1]$.
To answer the first question we begin to seek a solution with $c=0$,
and it is easily checked that also their commutator have vanishing block $c_3$.
Let us now look at the condition $a_2=0$, namely
\begin{equation}
a_2= [a_0, a_1]+ b_0 b_1^t -b_1 b_0^t =
2 g_0 g_1 J^3 +2 \epsilon^{x y z} \bar a_0^x \bar a_1^y \bar J^z
+\frac{1}{2} \vec \beta_0 \cdot \vec \beta_1 (J^3 + \bar J^3) = 0
\end{equation}
This equation can be solved if we set
\begin{equation}
\bar a_0^x \, =\,  \gamma g_0 \delta_{x 1} \quad ; \quad
\bar a_1^x \, =\,  \gamma g_1 \delta_{x 2} \quad ; \quad
\vec \beta_0 \cdot \vec \beta_1 =-4 g_0 g_1
\label{eq1}
\end{equation}
and if $\gamma^2=1 \, \leftarrow \, \gamma=\pm 1$.
Now we are left to solve the equation $b_2=a_0 b_1 -a_1 b_0= 0$, which
implies
\begin{equation}
(\gamma-1)g_0\vec{b}_{1}=(\gamma-1)g_1\vec{b}_{0}
\label{eq2}
\end{equation}
This equation is automatically fulfilled if $\gamma=1$ for any choice of
the $\vec{b}_{i}$s, while it is not consistent with the last of
equations (\ref{eq1}) if $\gamma = - 1$.
  Thus the only admissible value for $\gamma$ is $1$.
\par
Choosing $\vec \beta_0= (\beta_0,0 \dots 0)$ we can immediately answer the
second question: for the normalizer of the $\IR^2$ algebra we have
 $Z_{\bf so(4,m)}\left (\IR^2 \right )= {\bf so(m-1)}$, so that of the $m$
hypermultiplets one is eaten by the superHiggs mechanism and the remaining
$m-1$ can be assigned to any linear representation of a compact gauge group
that can be as large as $SO(m-1)$.
\section{The $N=2 \to N=1$ breaking problem in
Alekseevski\v{i}'s formalism.}
There are two main motivations for the choice of Alekseevski\v{i}'s
formalism \cite{alex},\cite{cec}
while dealing with the partial breaking of $N=2$ supersymmetry:
\begin{itemize}
\item {it provides a description of the quaternionic manifold as a group
manifold on which the generators of the $\IR^{2}$ act as traslation
operators on two (say $t_{0},t_{1}$) of the $4m$ scalar fields,
 parametrizing ${\cal Q}_{m}$;}
\item {as it will be apparent in the sequel, the fields  $t_{0},t_{1}$ define flat
directions for the scalar potential. They
can be identified with the {\it hidden sector}
of the theory, for their coupling to the gauge fields $A_\mu^0$ and $A_\mu^1$ is
what determines the partial supersymmetry breaking.}
\end{itemize}
Thus Alekseevski\v{i}'s description of quaternionic manifolds provides a
conceptually very powerful tool to deal  with the SUSY
breaking problem, even in
the case in which ${\cal QM}$ is not a symmetric homogeneous manifold \cite{beria}.
In what follows I will refer to the Alekseevski\v{i}'s description of
 quaternionic manifolds introduced in Chapter 1, specializing it 
to the symmetric manifold ${\cal QM}_{m}$, and in terms of  its
 $4m$--dimensional quaternionic algebra $V_m$, an explicit realization
 of the gauge group generators will be found.
\subsection{The Quaternionic Algebra and Partial SUSY Breaking.}
Alekseevski\v{i}'s description of $V_m$ is given by \eqn{Vstru}, \eqn{f0tf0}
and \eqn{Vstru2}.
The K\"ahlerian algebra $W_m$ corresponding to $V_{m}$ through the C--map
has the general structure illustrated in \eqn{Wstru} in which the
 subspaces X and Y have dimension zero:
\begin{equation}
W_m\, = \, F_{1}\oplus F_{2}\oplus F_{3}\oplus Z\,\, ;\,\, Z\, =\, Z^+\oplus Z^-
\label{strutto}
\end{equation}
 The $m-4$--dimensional spaces $Z^\pm$ and its image
$\tilde{Z}^\pm$ through $J_{2}$ in $\tilde{U}$ are the only parts of the whole algebra whose
dimension depends on $m= \# ~of~ hypermultiplets$, and thus it is natural to choose
the fields parametrizing them in some representation of $\cG_{compact}$, for, as it was
shown earlier, $\cG_{compact}\subset SO(m-1)$. At  fixed $m$ the $Z$-sector
can provide enough scalar fields as to fill a representation of the compact
gauge group. On the other hand the fields of the {\it hidden sector}
are to be chosen in the orthogonal complement with respect to $Z$ and $\tilde{Z}$  and
to be singlets with respect to  $\cG_{compact}$. Indeed, by
definition the fields of the hidden sector interact only with $A_\mu^0$ and $A_\mu^1$ \\
 
Using Alekseevski\v{i}'s notation, an orthonormal basis for the U subspace of 
$V_m$ is provided
by $\{ h_{i},g_{i}\quad (i=0,1,2,3), z^{\pm}_{k} (k=1,...,m-4)\}$, while for
$\tilde{U}$ an orthonormal basis is given by
 $\{p_{i},q_{i}\quad (i=0,1,2,3), \tilde{z}^{\pm}_{k}\quad (k=1,...,m-4)\}$.
The elements $\{h_{i},g_{i},p_{i},q_{i}\quad (i=0,1,2,3)\}$ generate the 
maximally non--compact subalgebra $O(4,4)/O(4)\times O(4)$ of $V_m$ which is 
characteristic of the large class of quaternionic manifolds we are considering
and which include also the $E_{6(2)}/SU(2)\times U(6)$ manifold to be 
studied in more detail in next chapter.\par
All these generators have a simple
representation in terms of the canonical basis of the full isometry algebra ${\bf so(4,m)}$.
\par
The next step is to determine  within this formalism, the generators of the
gauge group. As far as the $\IR^{2}$ factor is concerned we demand it to act
by means of translations on the coordinates of the coset representative $\IL(q)$. To attain this
purpose
the generators of the translations $T_{0},T_{1}$ will be chosen within
an abelian ideal $\cA\subset V_{m}$ and the representative of ${\cal QM}_{m}$
 will be defined in the following way:
\begin{eqnarray}
\IL(t,b)\, =\, e^{T(t)}e^{G(b)} \quad (t) & = &t^{0},t^{1} \quad ;
\quad (b)=b^{1},...,b^{4m-2}\nonumber\\
T(t)\, =\,t^{\Lambda}T_{\Lambda}\quad \Lambda=0,1& ; &
G(t)\, =\, b^{a}G_{i}\quad a=1,...,4m-2\nonumber\\
V_{m}&=&T\oplus G
\label{margarina}
\end{eqnarray}
It is apparent from eq.(\ref{margarina}) that the left action of a transformation generated by $T_{\Lambda}$ amounts to a translation of the coordinates $t^{0},t^{1}$.
The latter  will define flat directions for the scalar potential.
As the scalar potential depends on the quaternionic
coordinates only through ${\cal P}^{x}_{\alpha}$ and the corresponding killing vector, to prove
the truth of the above statement
it suffices to show that the momentum-map ${\cal P}^{x}_{\alpha}(t,b)$ on ${\cal QM}_{m}$
does not depend on the variables $t^{\Lambda}$.
Indeed a general expression for ${\cal P}^{x}_{\alpha}$ associated to the generator
$T_{\alpha}$ of ${\bf so(4,m)}$ is given by equation (\ref{fragolina}):
\begin{equation}
{\cal P}^{x}_{\alpha}(t,b)=\frac{1}{2}tr(\IL^{-1}(t,b)T_{\alpha}\IL(t,b) {\cal J}^{x})=
\frac{1}{2}tr(e^{-G}e^{-T}T_{\alpha}e^{T}e^{G}{\cal J}^{x})
\label{masterfrutta}
\end{equation}
If $T_{\alpha}$ is a generator of the gauge group, then either it is in $T$
or it is in the compact subalgebra. In both cases it commutes with $T$
 allowing the exponentials of $T(t)$ to cancel against each other.
It is also straightforward to prove that the Killing vector components
on ${\cal QM}_{m}$ do not depend on $t^{\Lambda}$.
\par
A maximal abelian ideal $\cA$ in $V_{m}$ can be shown to have dimension $m+2$.
Choosing
\begin{equation}
{\cal A}=\left\{ e_{1}, g_{3}, p_{0}, p_{3},q_{1},q_{2},\tilde{z}_{+}^{k} \right \} \quad
\quad (k=1,...,m-4)
\end{equation}
one can show that possible candidates for the role of translation generators
are either $T_{\Lambda}$ or $T'_{\Lambda}$ defined below:
\begin{eqnarray}
T_{0}&=&e_{1}-g_{3}=E_{\epsilon_{l}+\epsilon_{3}}-E_{\epsilon_{1}-
\epsilon_{3}},
\quad T_{1}=p_{0}-p_{3}=E_{\epsilon_{1}-\epsilon_{3}}-
E_{\epsilon_{1}+\epsilon_{3}}\nonumber\\
T'_{0}&=&p_{0}-p_{3}=E_{\epsilon_{1}-\epsilon_{3}}-
E_{\epsilon_{1}+\epsilon_{3}}\, ,\,  T'_{1}=q_{2}-q_{1}=E_{\epsilon_{1}+
\epsilon_{4}}-E_{\epsilon_{1}-\epsilon_{4}}
\label{lacosa}
\end{eqnarray}
where $\epsilon_{k},\quad k=1,...,l=\mbox{rank}({\bf so(4,m)})$
is an orthonormal basis of
$\IR^{l}$ and $\epsilon_{i}\pm\epsilon_{j}$ are roots of $so(4,m)$. In the previous
section we found constraints  on the form of the $\IR^{2}$ generators in order for
partial SUSY breaking to occur on a vacuum defined at the origin of ${\cal QM}_{m}$.
 The matrices
$T_{\Lambda}$ fulfill such requirements. On the other hand
one can check that also $T'_{\Lambda}$ fit the purpose as well,
 even if they do not
 have the form predicted in the previous section.Indeed 
specializing the fermion shift matrices in \eqn{shifmatrici}
to a background defined by $y=0\, ;\, b=0$ one finds:
\begin{eqnarray}
W^{\bar{s}}_{AB}\, &\propto&\, (\sigma^x\epsilon^{-1})_{AB}({\cal P}^x_0+{\rm i}
{\cal P}^x_1)\nonumber \\
W^{\bar{\alpha}}_{AB}\, &\propto&\, (\sigma^x\epsilon^{-1})_{AB}
{\cal P}^x_{\bar{\alpha}}\nonumber\\
S_{AB}\, &\propto&\,(\sigma^x\epsilon^{-1})_{AB}({\cal P}^x_0+{\rm i}
{\cal P}^x_1)\nonumber \\
N^A_\alpha\, &\propto&\,(\epsilon e^a\epsilon^{-1})^A_{B}
(g_0{\cal U}^{a\vert t}_0-{\rm i}
g_1{\cal U}^{a\vert t}_1)\nonumber\\
\mbox{where: }\qquad\qquad\qquad&&\nonumber\\
 e^0 &=& Id_{2\times 2}\,\, ;\,\,e^x=-i\sigma^x\,\,\alpha=(t,B)=1,...,2m
\label{shifmatrici2}
\end{eqnarray}
Gauging for instance $T_{\Lambda}$ as generators of $\IR^2$, 
from \eqn{masterfrutta} it follows that
\begin{eqnarray} 
{\cal P}^x_0\, &=&\, -g_0\delta^x_3\,\, ;\,\,{\cal P}^x_1\,=\,g_1\delta^x_2
\,\, ;\,\,{\cal P}^x_{\bar{\alpha}}\,=\,0\nonumber\\
{\cal U}^{2\vert t}_0\, &=&\,2\,\, ;\,\,{\cal U}^{3\vert t}_1\, =\,-2\nonumber\\
\matrix{S_{AB}\cr W^{\bar{s}}_{AB}} \, &\propto&\, \left(\matrix{g_1 & g_0\cr g_0 & g_1}\right)\nonumber\\
W^{\bar{\alpha}}_{AB}\,&=&\,0
\label{shifmatrici3} 
\end{eqnarray}
Setting $g_0=g_1$ the shift matrices in \eqn{shifmatrici3} have the following 
killing spinor: $\eta^A= (1,-1)$. Computing the gaugino shift matrices
one finds that $ \eta^A N_A \propto (g_0-g_1)$ which vanishes for $g_0=g_1$.\par
Moreover, taking $T_{\Lambda}$ as generators of $\IR^{2}$, it follows from their matrix form that the largest compact subalgebra of ${\bf so(4,m)}$ suitable for generating $\cG_{compact}$ is ${\bf so(m-1)}$. 
\par
 As it was also pointed out in the previous section, any point
on ${\cal QM}_{m}$ described by $q=(t,b)$ can define a vacuum on which SUSY
is partially broken, provided that a suitable choice for the isometry generators to be gauged is
done, e.g.:
\begin{equation}
T_{\alpha}(t,b)=\IL(t,b)T_{\alpha}\IL(t,b)^{-1}
\label{demografia}
\end{equation}
To this extent all the points on the surface spanned by $t^0,t^1$ and containing the origin,
require the same kind of gauging (as it is apparent from
equation (\ref{demografia})), and this is another way of justifying the {\it flatness} of the
scalar potential along these two directions.
\par
The existence of a Killing spinor guarantees that on the corresponding constant scalar
field configuration
$N=2$ SUSY is broken to $N=1$, and the scalar potential vanishes.
Furthermore, as explained in \cite{kilspinold} and already recalled, the existence of
at least one Killing spinor implies
the stability of the background,namely that it is an extremum of
the scalar potential. We have explicitly verified that
the configuration $(y=0;b=0)$ are minima of the scalar potential by plotting
its projection on planes, corresponding to different choices
of pairs $(y,b)$ of coordinates which were let to vary while keeping all the others to zero.
The behaviour of all these curves shows a minimum in the origin, where the
potential vanishes, as expected. Two very typical representatives of the general
behaviour of such  plots are shown in figure 1 and 2.
\begin{figure}
\epsfxsize=5.truecm
\epsfysize=5.truecm
\centerline{\hbox{\epsffile{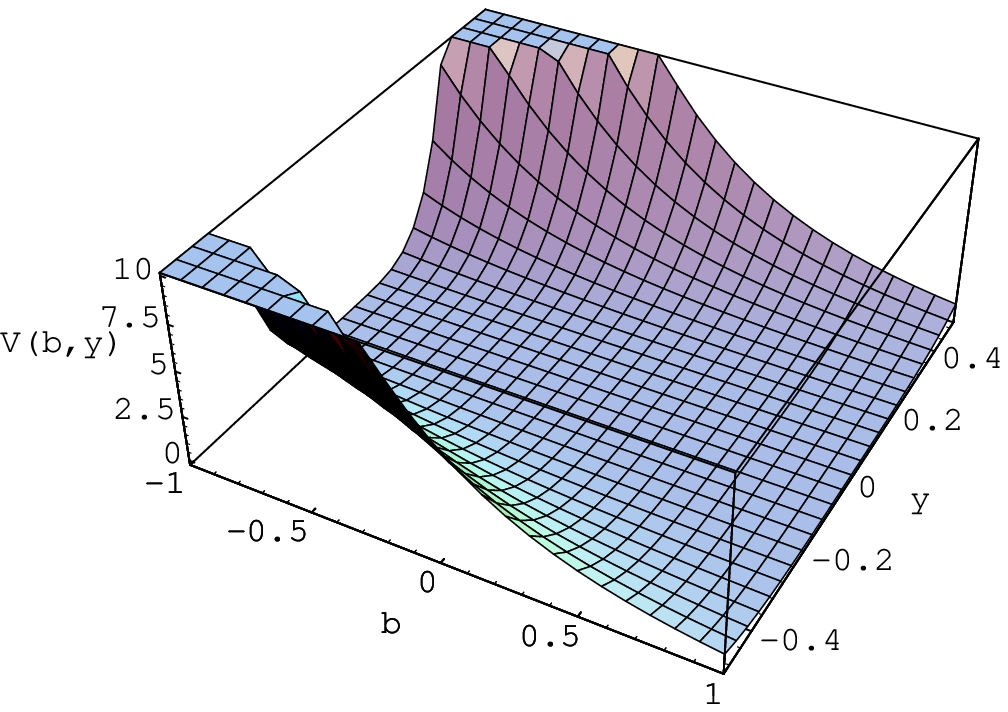}}}
\caption{ Scalar Potential Vs y generic and
b coefficient of $e_0$}
\epsfxsize=5.truecm
\epsfysize=5.truecm
\centerline{\hbox{\epsffile{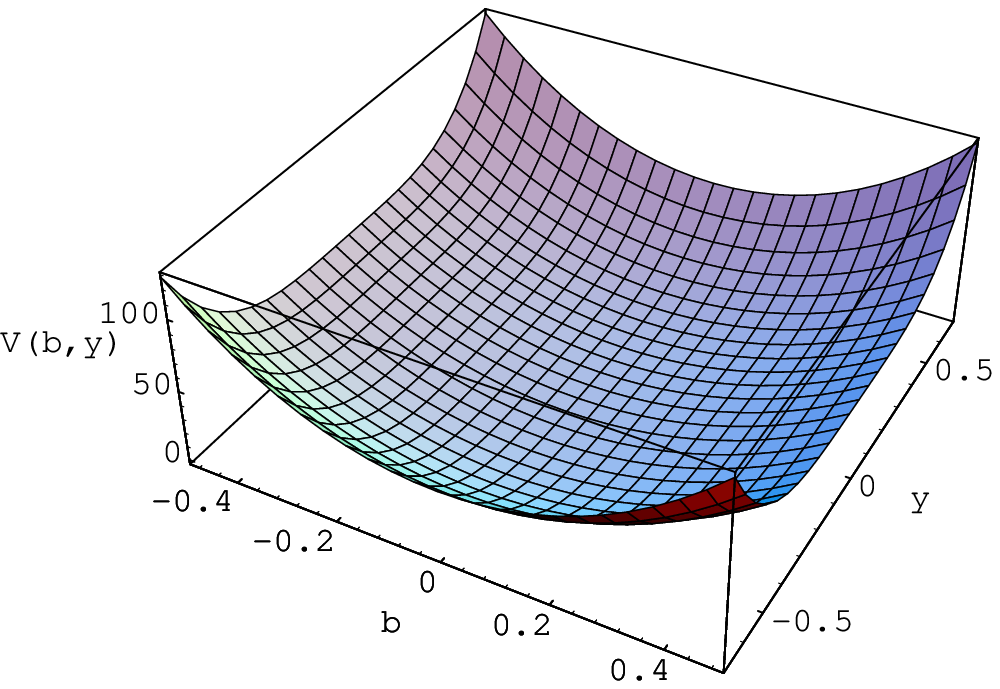}}}
\caption{ Scalar Potential Vs y generic and
b coefficient of $h_1$.}
\end{figure}
\chapter{BPS Black Holes in Supergravity}
In this last chapter I want to consider the application of
solvable Lie algebras to the derivation of the differential
equations that characterize BPS states as classical supergravity
solutions. The main reference for the present discussion is a recent
paper by L. Andrianopoli, R. D'Auria, S.Ferrara, P. Fr\'e and myself
\cite{noialtri3}.\par
Interest in the extremal black hole solutions of $D=4$ supergravity
theories has been quite vivid in the last couple of years
\cite{malda,gensugrabh} and it is just part of a more general interest in the
$p$--brane classical solutions of  supergravity theories in all dimensions
$4 \le D \le 11$ \cite{kstellec,duffrep}.
This interest streams from the interpretation of the classical solutions of
supergravity that preserve a fraction of the original supersymmetries
as the BPS non perturbative states necessary to complete the
perturbative string spectrum and make it invariant under the 
conjectured duality symmetries discussed in the first chapters.
This identification has become quite circumstantial with the advent of
$D$--branes \cite{polc} and the possibility raised by them of
a direct construction of the BPS states within the language of
perturbative string theory extended by the choice of Dirichlet
boundary conditions \cite{polc}.
\par
\subsection{BPS--saturated states in supergravity: extremal Black Holes}
From an abstract viewpoint BPS saturated states are characterized by the
fact that they preserve, in modern parlance, $1/2$ (or $1/4$, or $1/8$) of the original
supersymmetries. What this actually means is that there is a suitable
projection operator $\IP^2_{BPS} =\IP_{BPS}$ acting on the supersymmetry charge
$Q_{SUSY}$, such that:
\begin{equation}
 \left(\IP_{BPS} \,Q_{SUSY} \right) \, \vert \, \mbox{BPS state} \, >
 \,=  \, 0
 \label{bstato}
\end{equation}
Since the supersymmetry transformation rules of any supersymmetric
field theory are linear in the first derivatives of the fields
eq.\eqn{bstato} is actually a {\it system of first order differential
equations}. This system has to be combined with the second order
field equations of supergravity and the common solutions to both
system of equations is a classical BPS saturated state. That it is
actually an exact state of non--perturbative string theory follows
from supersymmetry representation theory. The classical BPS state is
by definition an element of a {\it short supermultiplet}
and, if supersymmetry is unbroken, it cannot be renormalized to
a {\it long supermultiplet}.
\par
Translating eq. \eqn{bstato} into an explicit first order
differential system requires knowledge of the supersymmetry
transformation rules of supergravity and it is at this level that
solvable Lie algebras can play an important role. In order to grasp the
significance of the above statement let us first rapidly review, as
an example, the algebraic definition of $D=4$, $N=2\nu$ BPS states and then the idea
of the solvable Lie algebra representation of the scalar sector.

The $D=4$ supersymmetry algebra with an even number
$N=2\nu$ of supersymmetry charges  can be written in the following
form:
\begin{eqnarray}
&\left\{ {\bar Q}_{aI \vert \alpha }\, , \,{\bar Q}_{bJ \vert \beta}
\right\}\, = \,  {\rm i} \left( C \, \gamma^\mu \right)_{\alpha \beta} \,
P_\mu \, \delta_{ab} \, \delta_{IJ} \, - \, C_{\alpha \beta} \,
\epsilon_{ab} \, \times \, \ZZ_{IJ}& \nonumber\\
&\left( a,b = 1,2 \qquad ; \qquad I,J=1,\dots, \nu \right)&
\label{susyeven}
\end{eqnarray}
where the SUSY charges ${\bar Q}_{aI}\equiv Q_{aI}^\dagger \gamma_0=
Q^T_{ai} \, C$ are Majorana spinors, $C$ is the charge conjugation
matrix, $P_\mu$ is the 4--momentum operator, $\epsilon_{ab}$ is the
two--dimensional Levi Civita symbol and the symmetric tensor
$\ZZ_{IJ}=\ZZ_{JI}$ is the central charge operator. It
can always be diagonalized $\ZZ_{IJ}=\delta_{IJ} \, Z_J$ and its $\nu$ eigenvalues
$Z_J$ are the central charges.
\par
The Bogomolny bound on the mass of a generalized monopole state:
\begin{equation}
M \, \ge \, \vert \, Z_I \vert \qquad \forall Z_I \, , \,
I=1,\dots,\nu
\label{bogobound}
\end{equation}
is an elementary consequence of the supersymmetry algebra and of the
identification between {\it central charges} and {\it topological
charges}. To see this it is convenient to introduce the following
reduced supercharges:
\begin{equation}
{\bar S}^{\pm}_{aI \vert \alpha }=\frac{1}{2} \,
\left( {\bar Q}_{aI} \gamma _0 \pm \mbox{i} \, \epsilon_{ab} \,  {\bar Q}_{bI}\,
\right)_\alpha
\label{redchar}
\end{equation}
They can be regarded as the result of applying
a projection operator to the supersymmetry
charges:
\begin{eqnarray}
{\bar S}^{\pm}_{aI} &=& {\bar Q}_{bI} \, \IP^\pm_{ba} \nonumber\\
 \IP^\pm_{ba}&=&\frac{1}{2}\, \left({\bf 1}\delta_{ba} \pm \mbox{i} \epsilon_{ba}
 \gamma_0 \right)
 \label{projop}
\end{eqnarray}
Combining eq.\eqn{susyeven} with the definition \eqn{redchar} and
choosing the rest frame where the four momentum is $P_\mu$ =$(M,0,0,0)$, we
obtain the algebra:
\begin{equation}
\left\{ {\bar S}^{\pm}_{aI}  \, , \, {\bar S}^{\pm}_{bJ} \right\} =
\pm \epsilon_{ac}\, C \, \IP^\pm_{cb} \, \left( M \mp Z_I \right)\,
\delta_{IJ}
\label{salgeb}
\end{equation}
By positivity of the operator $\left\{ {S}^{\pm}_{aI}  \, , \, {\bar S}^{\pm}_{bJ} \right\} $
it follows that on a generic state the Bogomolny bound \eqn{bogobound} is
fulfilled. Furthermore it also follows that the states which saturate
the bounds:
\begin{equation}
\left( M\pm Z_I \right) \, \vert \mbox{BPS state,} i\rangle = 0
\label{bpstate1}
\end{equation}
are those which are annihilated by the corresponding reduced supercharges:
\begin{equation}
{\bar S}^{\pm}_{aI}   \, \vert \mbox{BPS state,} i\rangle = 0
\label{susinvbps}
\end{equation}
On one hand eq.\eqn{susinvbps} defines {\sl short multiplet
representations} of the original algebra \eqn{susyeven} in the
following sense: one constructs a linear representation of \eqn{susyeven}
where all states are identically
annihilated by the operators ${\bar S}^{\pm}_{aI}$ for $I=1,\dots,n_{max}$.
If $n_{max}=1$ we have the minimum shortening, if $n_{max}=\nu$ we
have the maximum shortening. On the other hand eq.\eqn{susinvbps}
can be translated into a first order differential equation on the
bosonic fields of supergravity.
Indeed, let us consider
a configuration where all the fermionic fields are zero.
Setting the fermionic SUSY rules appropriate to such a background equal
to zero we find the following Killing spinor equation:
\begin{equation}
0=\delta \mbox{fermions} = \mbox{SUSY rule} \left( \mbox{bosons},\epsilon_{AI} \right)
\label{fermboserule}
\end{equation}
where the SUSY parameter satisfies the following conditions:
\begin{equation}
\begin{array}{rclcl}
\xi^\mu \, \gamma_\mu \,\epsilon_{aI} &=& \mbox{\rm i}\, \varepsilon_{ab}
\,  \epsilon^{bI}   & ; &   I=1,\dots,n_{max}\\
\epsilon_{aI} &=& 0  &;&   I > n_{max} \\
\end{array}
\end{equation}
Here $\xi^\mu$ is a time-like Killing vector for the space-time metric and
$ \epsilon _{aI}, \epsilon^{aI}$ denote the two chiral projections of
a single Majorana spinor:
\begin{equation}
\gamma _5 \, \epsilon _{aI} \, = \, \epsilon _{aI} \quad ; \quad
\gamma _5 \, \epsilon ^{aI} \, = - \epsilon ^{aI}
\label{chiralpro}
\end{equation}
Eq.\eqn{fermboserule} has two features which we want to stress
as main motivations for the developments presented in later sections:
\begin{enumerate}
\item{It requires an efficient parametrization of the scalar field
sector}
\item{It breaks the original $SU(2\nu)$ automorphism of the
supersymmetry algebra to the subgroup $SU(2)\times SU(2\nu-2)\times U(1)$}
\end{enumerate}
The first feature is the reason why the use of the solvable Lie
algebra $Solv$ associated with $U/SU(2\nu)\times H^\prime$ is of great help in this problem.
The second feature is the
reason why the solvable Lie algebra $Solv$ has to be decomposed in
a way appropriate to the decomposition of the isotropy group $H=SU(2\nu)\times H^\prime$
with respect to the subgroup $SU(2)\times SU(2\nu-2)\times U(1) \times H^\prime$.\par
Before continuing our analysis  it is worth giving, in 
this introductory section,
 a sketchy overview of some aspects of extremal Black Hole
solutions in supergravity. Black Holes are classical solutions of 
Einstein--Maxwell equations whose space--time structure is
asimptotically flat and has a singularity hidden by an {\it event
horizon}.
As it is well known, supergravity, being invariant under {\it local} super--Poincar\'e 
transformations, includes General Relativity, i.e. describes
gravitation
coupled to other fields in a supersymmetric framework. In particular, among
its classical solutions there are Black Holes.\par
We are interested in BPS saturated states among the charged, stationary,
asymptotically flat, spherically symmetric solutions. 
They are described by a particular kind
of Black Holes which have a {\it solitonic} interpretation. Indeed their 
configurations turn out to interpolate between two vacuum states of
the theory (as one expects from a soliton): the trivial Minkowski flat metric at spatial infinity and
the Bertotti--Robinson conformally--flat metric near the horizon.
When no scalar fields are coupled to gravity, this solution reduces to
an extreme Reissner--Nordstr\"om configuration described by the
following metric:
\begin{equation}
d s^2\, =\, \left(1-\frac{|q|}{r}\right)^2dt^2\, -\,
\left(1-\frac{|q|}{r}\right)^{-2}dr^2 -r^2d\Omega^2
\label{RNsol}
\end{equation}
being $r$ the radial distance from the singularity $r=0$, 
$d\Omega^2$ the solid angle element and $q$ the electric charge of the
Black Hole which, in the chosen units, equals its ADM mass: $M=\vert q\vert$.
This solution has a unique horizon defined as the sphere on which
$g_{tt}=0$, i.e. whose radius is $r_o=|q|$.  Changing the radial
variable $r\rightarrow \rho=r-r_o$, equation \eqn{RNsol} is rewritten
in the following way:
\begin{equation}
d s^2\, =\, \left(1+\frac{|q|}{\rho}\right)^{-2}dt^2\, -\,
\left(1+\frac{|q|}{\rho}\right)^{2}\left(d\rho^2 +\rho^2d\Omega^2\right)
\label{RNsol2}
\end{equation}
This solution has an apparent flat limit for $\rho\rightarrow \infty$
and near the horizon $\rho\approx 0$ eq. \eqn{RNsol2} reduces to:
\begin{equation}
d s^2\, =\, \frac{\rho^2}{M_{BR}^2}dt^2\, -\,
\frac{M_{BR}^2}{\rho^2}\left(d\rho^2 +\rho^2d\Omega^2\right)
\label{BRsolut}
\end{equation}
which is the Bertotti--Robinson metric ($M_{BR}$ is the
Bertotti--Robinson mass which equals, for this particular solution,
 the ADM mass i.e. $M_{BR}=|q|$).\par
When gravity is coupled with scalar fields $\phi$ the BPS saturated solution
is more complicate as we will see. The boundary conditions  on the
sphere at infinity are defined by a flat Minkowsky metric and a
constant value of the scalars $\phi (S^2_\infty)\equiv \phi_0$
corresponding to a point on the moduli space of the theory. The BPS
condition \eqn{bpstate1} on the ADM mass of the solution may be
rewritten in the following way:
\begin{equation}
M_{ADM}\, =\, \vert Z_I(p,q,\phi_0)\vert\,\,\,\, I=1,...,n_{max}
\label{BRSADM}
\end{equation}
This condition is then translated in a system of first order
differential equations derived from \eqn{fermboserule} in the bosonic
fields. It has been shown two years ago, \cite{FKS}, \cite{FK} that
the scalar solution to this system plus the equations of motion,
describes a path in the moduli space along which the scalars are driven
to a fixed point $\phi\rightarrow \phi_{fix}$ corresponding to the horizon of the Black Hole. In
other words, near the horizon, the scalars ``loose memory'' of their
initial values on $S^2_\infty$. The values of the scalars at the fixed
point depend, as we are going to see later, only on the ratios of the
electric and magnetic charges $q,p$ of the Black Hole 
measured on $S^2_\infty$ and are
defined by minimizing  $Z_I(p,q,\phi)$ with respect to the
scalars $\phi$. In the proximity of the horizon, the metric is again
the Bertotti--Robinson one, but this time $M_{BR}$ turns out to be
equal to the value of the central charge 
$M_{BR}=|Z_{min}(p,q,\phi_{fix}(p,q))|/4\pi$ at the horizon, i.e. its
minimum with respect to the scalar fields.\par
Bekenstein--Hawking formula relates the entropy of a Black Hole to 
the area of its horizon. Applying it to our solution, one obtains:
\begin{equation}
S\,=\, \frac{A}{4\pi}\,=\,\frac{1}{4\pi}\int_{horizon}d\theta
d\phi\sqrt{g_{\theta\theta}g_{\phi\phi}}\,=\, M_{BR}^2\,=\,
\frac{|Z_{min}(p,q)|^2}{4\pi}
\label{bekhaw}
\end{equation}
where we used the Bertotti--Robinson metric to compute the area of the
horizon. Since $Z_{min}$ and therefore the entropy does not depend on
the values of the scalars at infinity $\phi_0$, they can be expressed in
terms of moduli--independent topological quantities $I$ depending only
on the U--duality group and on the representation of the electric and
magnetic charges and defined on the whole moduli
space \cite{ADAFentropy}. 
As far as the $N=8$, $D=4$ theory is concerned, 
we know that the U--duality group is
$E_{7(7)}$ which has a quartic invariant $I_4$. Since $I_4$ depends
only on the charge vector of the Black Hole, in order for it to be
non--zero (i.e. in order for the Black Hole to have a non--vanishing entropy),
it can be shown that the solution should have at least four charges.
This is the kind of solutions that we are going to study in what follows.
\subsection{N=2 decomposition in the $N=8$ theory}
In the present chapter I shall
 concentrate on the maximally extended four--dimensional
theory, namely $N=8$ supergravity.
In Chapter 2 we studied in detail some important properties of the scalar
 manifold $E_{7(7)}/SU(8)$ of this theory, by analysing its generating
 solvable Lie algebra denoted by $Solv_7$.
According to the
previous discussion the Killing spinor equation for $N=8$ BPS states
requires that $Solv_7$ should be decomposed  according to the decomposition
of the isotropy subgroup: $SU(8) \longrightarrow SU(2)\times U(6)$. We
show in later sections that the corresponding decomposition of the solvable
Lie algebra is the following one:
\begin{equation}
Solv_7   =  Solv_3 \, \oplus \, Solv_4
\label{7in3p4}
\end{equation}
\begin{equation}
\begin{array}{rclrcl}
Solv_3 & \equiv & Solv \left( SO^\star(12)/U(6) \right) & Solv_4 &
\equiv & Solv \left( E_{6(2)}/SU(2)\times SU(6) \right) \\
\mbox{rank }\, Solv_3 & = & 3 & \mbox{rank }\,Solv_4 &
= & 4 \\
\mbox{dim }\, Solv_3 & = & 30 & \mbox{dim }\,Solv_4 &
= & 40 \\
\end{array}
\label{3and4defi}
\end{equation}
In the present chapter I shall slightly change the notation introduced
in Chapter 2, and denote by $Solv_3$ and $Solv_4$ non
maximally--non--compact solvable algebras.
The rank three  Lie algebra $Solv_3$ defined above describes the
thirty dimensional scalar sector of $N=6$ supergravity, while the rank four
solvable Lie algebra $Solv_4$ contains the remaining fourty scalars
belonging to $N=6$ spin $3/2$ multiplets. It should be noted
that, individually, both manifolds $ \exp \left[ Solv_3 \right]$ and
$ \exp \left[ Solv_4 \right]$ have also an $N=2$ interpretation since we have:
\begin{eqnarray}
\exp \left[ Solv_3 \right] & =& \mbox{homogeneous special K\"ahler}
\nonumber \\
\exp \left[ Solv_4 \right] & =& \mbox{homogeneous quaternionic}
\label{pincpal}
\end{eqnarray}
so that the first manifold can describe the interaction of
$15$ vector multiplets, while the second can describe the interaction
of $10$ hypermultiplets. Indeed if we decompose the $N=8$ graviton
multiplet in $N=2$ representations we find:
\begin{equation}
\mbox{N=8} \, \mbox{\bf spin 2}  \,\stackrel{N=2}{\longrightarrow}\,
 \mbox{\bf spin 2} + 6 \times \mbox{\bf spin 3/2} + 15 \times \mbox{\bf vect. mult.}
 +
 10 \times \mbox{\bf hypermult.}
 \label{n8n2decompo}
\end{equation}
Although at the level of linearized representations of supersymmetry
we can just delete the $6$ spin $3/2$ multiplets and obtain a
perfectly viable $N=2$ field content, at the full interaction level
this truncation is not consistent. Indeed, in order to get a consistent
$N=2$ truncation the complete scalar manifold must be the {\sl direct product}
of {\sl a special K\"ahler} manifold with {\sl a quaternionic manifold}. This is
not true in our case since putting together $ \exp \left[ Solv_3 \right]$
with $ \exp \left[ Solv_4 \right]$ we reobtain the $N=8$ scalar
manifold $E_{7(7)}/SU(8)$ which is neither a direct product nor
K\"ahlerian, nor quaternionic. The blame for this can be put on the
decomposition \eqn{7in3p4} which is a direct
sum of vector spaces but not a direct sum of Lie algebras: in other
words we have
\begin{equation}
\left[ Solv_3 \, , \, Solv_4 \right] \, \ne \, 0
\label{nocommut}
\end{equation}
The problem of deriving consistent $N=2$ truncations is most
efficiently addressed in the language of Alekseveeskian solvable
algebras \cite{alex}. $Solv_3$ is a K\"ahler solvable Lie algebra,
while $Solv_4$ is a quaternionic solvable Lie algebra. We must
determine a K\"ahler subalgebra ${\cal K}\,  \subset\, Solv_3$ and a
quaternionic subalgebra ${\cal Q}\, \subset\, Solv_4$ in such a way
that:
\begin{equation}
\left[ {\cal K} \, , \, {\cal Q} \right] \, = \, 0
\label{ycommut}
\end{equation}
Then the truncation to the vector multiplets described by ${\cal K}$ and
the hypermultiplets described by ${\cal Q}$ is consistent at the
interaction level. An obvious solution is to take no vector
multiplets ( ${\cal K}=0 $) and all hypermultiplets ( ${\cal Q}=Solv_4 $) or
viceversa  ( ${\cal K}=Solv_3 $), ( ${\cal Q}=0 $). Less obvious is
what happens if we introduce just one hypermultiplet,
corresponding to the minimal one--dimensional quaternionic algebra.
In later sections we show  that in that case the maximal number of
admitted vector multiplets is $9$. The corresponding K\"ahler
subalgebra is of rank 3 and it is given by:
\begin{equation}
 Solv_3 \, \supset \, {\cal K}_3 \, \equiv \, Solv \left(
 SU(3,3)/SU(3)\times U(3) \right)
 \label{su33ka}
\end{equation}
Note that, as we will discuss in the following, the 18 scalars
parametrizing the manifold $ SU(3,3)/SU(3)\times U(3) $ are all the scalars in
the NS-NS sector of $SO^*(12)$.
A thoroughful discussion of the N=2 truncation problem and of its
solution in terms of solvable Lie algebra decompositions
is discussed in section \ref{secsu33}. At the level
of the present introductory section we want to stress the relation
of the decomposition \eqn{7in3p4} with the
Killing spinor equation for BPS black-holes.

Indeed, as just pointed out, the decomposition \eqn{7in3p4} is implied by
the $SU(2) \times U(6)$ covariance of the Killing spinor equation.
As we show in section (3.2) this equation splits into
various components corresponding to different $SU(2) \times U(6)$
irreducible representations. Introducing the decomposition \eqn{7in3p4}
we will find that the $40$ scalars belonging to $Solv_4$ are constants
independent of the radial variable $r$. Only the $30$ scalars in the
K\"ahler algebra $Solv_3$ can have a radial dependence. In fact their
radial dependence is governed by a first order differential equation
that can be extracted from a suitable component of the Killing spinor
equation. In this way we see that the same solvable Lie algebra
decompositions occurring in the problem of N=2 truncations of N=8
supergravity occur also in the problem of constructing N=8 BPS black holes.
\vskip .5cm
\par
We present now our plan for the next sections.
\par
In section \ref{n8sugra} we discuss the structure of the scalar
sector in $N=8$ supergravity and its supersymmetry transformation
rules. As just stated, our goal is to develop methods for
the analysis of BPS states as classical solutions of supergravity
theories in all dimensions and for all values of $N$. Many results
exist in the literature for the $N=2$ case in four dimensions \cite{feka,n2kal,n2lust},
where the number of complex scalar fields involved just equals the number of
differential equations one obtains from the Killing spinor condition.
Our choice to focus on the $N=8$, $D=4$ case is motivated by the
different group--theoretical structure of the Killing spinor equation
in this case and by the fact that it is the maximally extended
supersymmetric theory.
\par
In section \ref{bhansazzo} we introduce the black--hole ansatz and we
show how using roots and weights of the $E_{7(7)}$ Lie algebra we can
rewrite in a very intrinsic way the Killing spinor equation. We
analyse its components corresponding to irreducible representations
of the isotropy subalgebra $U(1)\times SU(2) \times SU(6)$ and we show
the main result, namely that the $40$ scalars in the $Solv_4$
subalgebra are constants.
\par
In section \ref{cartadila} we exemplify our method by explicitly
solving the simplified model where the only non--zero fields are
the dilatons in the Cartan subalgebra. In this way we retrieve the
known $a$--model solutions of N=8 supergravity.
\par
The following two  sections 5 and \ref{solvodecompo} are concerned with the method and the results of our
computer aided calculations on the embedding of the subalgebras
$ U(1)\times SU(2)\times SU(6) \subset SU(8)$ in $E_{7(7)}$ and with the structure of the
 solvable Lie algebra decomposition  already introduced in eq.\eqn{7in3p4}.
 In particular we study the problem of consistent N=2 truncations using
 Alekseevski formalism.
 These two sections, being rather technical, can be skipped in a
 first reading by the non interested reader.
 We note however  that many of the results there obtained  are used for
 the discussion of the subsequent section. In particular, these results are
 preliminary for the allied
 project of gauging the maximal gaugeable abelian ideal ${\cal G}_{abel}$,
 which I have outlined in the previous chapters and which is still
work in progress. Note that such a gauging
 should on one hand produce spontaneous partial breaking of $N=8$
 supersymmetry and on the other hand be interpretable as due to the
 condensation of $N=8$ BPS black-holes.
 \par
 In the final section \ref{concludo} we address the question of the
 most general BPS black-hole. Using the little group of the charge
 vector in its normal form which, following \cite{lastserg}, is
 identified with $SO(4,4)$, we are able to conclude that the only
 relevant scalar fields are those associated with the solvable Lie
 subalgebra:
 \begin{equation}
Solv \left( \frac{SL(2,\IR)^3}{U(1)^3} \right) \, \subset \, Solv_3
\label{rilevanti}
\end{equation}
where the 6 scalars that parametrize the manifold $ \frac{SL(2,\IR)^3}{U(1)^3}$ are all in the
NS-NS sector.

Moreover, with an appropriate
 identification, we show how the calculation of the fixed scalars
 performed by the authors of \cite{STUkallosh} in the N=2 STU model  amounts to
 a solution of the same problem also in the N=8 theory.
 The most general solution can be actually generated by U-duality
 rotations of $E_{7(7)}$.

Therefore, the final result of our whole analysis, summarized in
the conclusions is that
up to U-duality transformations the most general $N=8$ black-hole
is actually an $N=2$ black-hole corresponding however to a very
specific choice of the special K\"ahler manifold, namely $ \frac{SO^*(12)}{U(6)}$
as in eq.\eqn{3and4defi}, \eqn{pincpal}.

\section{N=8 Supergravity and its scalar manifold $E_{7(7)}/SU(8)$ }
\label{n8sugra}
The bosonic Lagrangian of $N=8$ supergravity contains,
besides the metric $28$ vector fields and $70$ scalar fields
spanning the $E_{7(7)}/SU(8)$ coset manifold. This lagrangian falls
into the general type of lagrangians admitting
electric--magnetic duality rotations  considered in
\cite{fundpaper},\cite{pietrolectures},\cite{cdfvp}.
For the case where all the scalar fields of the coset manifold have been
switched on the Lagrangian, according to the normalizations of
\cite{fundpaper} has the form:
\begin{equation}
{\cal L} = \sqrt{-g} \, \left( 2\, R[g] + \frac{1}{4}\, \mbox{Im}\,
{\cal N}_{\Lambda\Sigma} \, {\cal F}^{\Lambda\vert \mu \nu}\,{\cal
F}^{\Sigma}_{\mu \nu}  + \frac{1}{4}
\mbox{Re}\, {\cal N}_{\Lambda\Sigma} \, {\cal F}^{\Lambda}_{\mu \nu} \, {\cal
F}^{\Sigma}_{\rho \sigma}\, \epsilon^{\mu \nu\rho\sigma}
+ \frac{\alpha^2}{2}\, g_{ij}(\phi) \, \partial_\mu \phi^i \, \partial^\mu \phi^j \right)
\label{lagrared}
\end{equation}
where the indeces $\Lambda,\Sigma$ enumerate the $28$ vector fields,
$g_{ij}$ is the $E_{7(7)}$ invariant metric on the scalar coset
manifold, $\alpha$ is a real number fixed by supersymmetry
and the period matrix ${\cal N}_{\Lambda\Sigma}$ has the
following general expression holding true for all symplectically embedded
coset manifolds \cite{gz}:
\begin{equation}
{\cal N}_{\Lambda\Sigma}=h \cdot f^{-1}
\label{gaiazuma}
\end{equation}
The complex $ 28 \times 28 $ matrices $ f,h$ are
defined by the $Usp(56)$ realization $\IL_{Usp} \left(\phi\right)$ of the
coset representative which is related to its $Sp(56,\IR)$ counterpart
$\IL_{Sp}(\phi)$ through a Cayley transformation, as dispayed in the
following formula \cite{amicimiei}:
\begin{eqnarray}
\IL_{Usp} \left(\phi\right) &=& \frac{1}{\sqrt{2}}\left(\matrix{ f+ {\rm i}h & \bar f+ {\rm i}\bar h \cr f- {\rm i}h
& \bar f - {\rm i}\bar h  \cr } \right) \nonumber\\
& \equiv & {\cal C} \, \IL_{Sp}\left(\phi\right) {\cal C}^{-1} \nonumber\\
\IL_{Sp}(\phi) & \equiv & \exp \left[ \phi^i \, T_i \right ] \, = \,
\left(\matrix{ A(\phi) & B(\phi) \cr C(\phi) & D(\phi) \cr } \right) \nonumber\\
{\cal C}& \equiv & \frac{1}{\sqrt{2}} \, \left(\matrix{ \bfone & {\rm i} \, \bfone \cr
\bfone & - \, {\rm i} \, \bfone \cr } \right)
\label{cayleytra}
\end{eqnarray}
In eq. \eqn{cayleytra} we have implicitly utilized the solvable
Lie algebra parametrization of the coset, by assuming that the
matrices $T_i$ ($i=1,\dots ,70$) constitute some basis of the solvable Lie
algebra $Solv_7=Solv\left(E_{7(7)}/SU(8)\right)$.
\par
Obviously, in order to make eq.\eqn{cayleytra} explicit one has to choose a
basis for the ${\bf 56}$ representation of $E_{7(7)}$. In the sequel,
according to our convenience, we utilize
two different bases for a such a representation.
\begin{enumerate}
\item {{\it The Dynkin basis}. In this case,
hereafter referred to as $SpD(56)$, the basis vectors of the real
symplectic representation  are eigenstates
of the Cartan generators with eigenvalue one of the $56$ weight vectors
($\pm {\vec \Lambda } =\{ \Lambda_1 , \dots , \Lambda_7 \}$ pertaining
to the representation:
\begin{eqnarray}
\label{dynkbas}
(W=1,\dots\, 56 )& :&  \vert \, {W} \, \rangle \, = \, \cases
{ \vert {\vec \Lambda} \rangle ~~~~\, : \quad H_i \vert \, {\vec \Lambda} \, \rangle
~~~~ = ~~~\Lambda_i \,
\vert \, {\vec \Lambda} \, \rangle   ~~~\quad (\Lambda= 1,\dots \, 28)
\cr
\vert \, -{\vec \Lambda} \rangle \, :
\quad H_i \vert \, -{\vec \Lambda} \, \rangle = \, -\Lambda_i \,
\vert \, -{\vec \Lambda} \, \rangle   \quad (\Lambda= 1,\dots \, 28)
\cr}\nonumber\\
\vert \, V  \, \rangle & = & f^{\Lambda} \,   \vert {\vec \Lambda} \rangle
\, \,  \oplus \,\,  g_{\Lambda} \,   \vert {- \vec \Lambda} \rangle
\nonumber\\
\mbox{or in matrix notation} && \nonumber\\
\nonumber\\
{\vec V}_{SpD} & = & \left(\matrix{ f^\Lambda \cr g_\Sigma \cr }\right )
\end{eqnarray}
}
\item{{\it The Young basis}. In this case, hereafter referred to as $UspY(56)$,
the basis vectors of the complex pseudounitary
representation correspond to the natural basis of the
${\bf 28}$ + ${\bar {\bf 28}}$ antisymmetric representation
of the maximal compact subgroup $SU(8)$. In other
words, in this  realization of the fundamental $E_{7(7)}$
representation a generic vector is of the following form:
\begin{eqnarray}
\label{youngbas}
\vert {V} \rangle &=& u^{AB} \,
\mbox{$\begin{array}{|c|}
\hline
\stackrel{ }{A}\cr
\hline
\stackrel{}{B}\cr
\hline
\end{array}
$} \,\,    \oplus \, \,   {v}_{AB} \,  \mbox{$\begin{array}{|c|}
\hline
\stackrel{ }{\bar A}\cr
\hline
\stackrel{ }
{\bar B}\cr
\hline
\end{array} $} \quad ; \quad (A,B=1,\dots,8) \nonumber\\
&&\null\nonumber\\
 \mbox{ or in matrix notation}&& \nonumber\\
&&\null\nonumber\\
{\vec V}_{UspY} &=&  \left( \matrix { u^{AB} \cr  v_{AB} \cr }\right)
\end{eqnarray}
}
\end{enumerate}
Although their definitions are respectively  given in terms of the real and the
complex case, via a Cayley transformation  each of the two basis has both a real
symplectic and a complex pseudounitary  realization. Hence
we will  actually deal with four bases:
\begin{enumerate}
\item {The $SpD(56)$--basis}
\item {The $Usp_D(56)$--basis}
\item {The $SpY(56)$--basis}
\item {The $UspY(56)$--basis}
\end{enumerate}
Each of them has distinctive advantages depending on the aspect of
the theory one addresses. In particular the $UspY(56)$ basis is
that originally utilized by de Wit and Nicolai in their construction
of gauged $N=8$ supergravity \cite{dwni}. In the considerations
of the present analysis the $SpD(56)$ basis will often offer the best picture
since it is that where the structure of the solvable Lie algebra is
represented in the simplest way. In order to use the best features
of each basis we just need to have full control on the
matrix that shifts fron one to the other. We name such matrix
${\cal S}$ and we write:
\begin{eqnarray}
\label{bfSmat}
\left( \matrix { u^{AB} \cr  v_{AB} \cr }\right) & = & {\cal S} \,
\left( \matrix { f^\Lambda \cr  g_\Sigma \cr }\right) \nonumber\\
&& \null\nonumber\\
 \mbox{where} && \nonumber\\
&& \null\nonumber\\
{\cal S}&=& \left( \matrix{ {\bf S} & {\bf 0} \cr {\bf 0} & {\bf S^\star}
\cr } \right) \, {\cal C} = \frac{1}{\sqrt{2}}\left( \matrix{ {\bf S} & {\rm i}\,{\bf S} \cr
{\bf S^\star} & -{\rm i}\,{\bf S^\star} \cr } \right)\nonumber\\
&& \null \nonumber\\
\mbox{the $ 28 \times 28 $ matrix ${\bf S}$ being unitary} && \nonumber\\
&& \null \nonumber\\
 {\bf S}^\dagger {\bf S} &=& \bfone
\end{eqnarray}
The explicit form of the $U(28)$ matrix ${\bf S}$ is given in
section 5.4. The weights of the $E_{7(7)}$ ${\bf 56}$ representation are
listed in table \ref{e7weight} of appendix E.
\par
\subsection{Supersymmetry transformation rules and central charges}
In order to obtain the $N=8$ BPS saturated Black Holes we cannot
confine ourselves to the bosonic lagrangian, but we also need the
the explicit expression for the supersymmetry transformation rules of the
fermions. Since the $N=8$ theory has no matter multiplets the
fermions are just  the ${\bf 8}$ spin $3/2$ gravitinos and the
${\bf 56}$ spin $1/2$ dilatinos. The two numbers ${\bf 8}$ and ${\bf
56}$ have been written boldfaced since they also single out the
dimensions of the two irreducible $SU(8)$ representations to which
the two kind of fermions are respectively assigned, namely the fundamental and the
three times antisymmetric:
\begin{equation}
 \psi_{\mu\vert A} \, \leftrightarrow \,
 \mbox{$ \begin{array}{|c|}
\hline
\stackrel{ }{A}\cr
\hline
\end{array}$} \, \equiv \,  {\bf 8}\quad ; \quad \chi _{ABC}  \, \leftrightarrow \,
\mbox{$ \begin{array}{|c|}
\hline
\stackrel{ }{A}\cr
\hline
\stackrel{}{B}\cr
\hline
\stackrel{}{C}\cr
\hline
\end{array}
$}   \, \equiv \,  {\bf 56}
\end{equation}
Following the conventions and formalism of
 \cite{amicimiei} and \cite{castdauriafre}
the relevant supersymmetry transformation rules can be written as follows:
\begin{eqnarray}
\delta \psi _{A\mu }&=&\nabla_\mu \epsilon_A - \frac{k}{4} \, c\,
T^{-}_{AB\vert \rho\sigma} \,\gamma^{\rho\sigma} \, \gamma_\mu  \,
\epsilon^B  + \cdots \nonumber\\
\delta \chi_{ABC} &=& a P_{ABCD\vert i } \, \partial_\mu \phi^i
\, \gamma^\mu  \, \epsilon^D + b \,
T^{-}_{[AB \vert \rho\sigma} \,\gamma^{\rho\sigma}
\epsilon_{C ]}  + \cdots
\label{trasforma}
\end{eqnarray}
where $a,b,c$ are numerical coefficients fixed
by superspace  Bianchi identities while, by definition,
$T^{-}_{AB\vert \mu\nu}$ is the antiselfdual part of the
graviphoton field strength
and  $P_{ABCD\vert i }$ is the vielbein of the scalar coset manifold,
completely antisymmetric in $ABCD$ and satisfying the pseudoreality
condition:
\begin{equation}
P_{ABCD}=\frac{1}{4!}\epsilon_{ABCDEFGH}\bar P^{EFGH}.
\end{equation}
What we need is the explicit expression of these objects in terms
of coset representatives. For the vielbein $ P_{ABCD\vert i }$ this
is easily done. Using the $UspY(56)$ basis the left invariant 1--form
has the following form:
\begin{equation}
\IL(\phi)^{-1} \,d \IL(\phi) \, = \, \left( \matrix{
\delta^{[A }_{[C} \, Q^{\phantom{D} B]}_{D]}
& P^{ABEF} \cr P_{CDGH} &   \delta^{[E}_{[G} \, Q^{ F]}_{\phantom{F}H]}\cr }\right)
\label{uspYconnec}
\end{equation}
where the $1$--form $Q^{\phantom{D} B}_{D}$ in the ${\bf 63}$ adjoint
representation of $SU(8)$ is the connection while the $1$--form
$P_{CDGH}$ in the ${\bf 70}$ four times antisymmetric representation
of $SU(8)$ is the vielbein of the coset manifold $E_{7(7)}/SU(8)$.
Later we need to express the same objects in different basis but
their definition is clear from eq.\eqn{uspYconnec}. A little more
care is needed to deal with the graviphoton field strenghts.
To this effect we begin by introducing
the multiplet of electric and magnetic field strengths according
to the definitions given in the first chapter:
\begin{equation}
{\vec V}_{\mu\nu} \equiv \left(\matrix {
F^{\Lambda}_{\mu\nu} \cr G_{\Sigma\vert\mu\nu} \cr}\right)
\label{symvecft}
\end{equation}
where
\begin{eqnarray}
G_{\Sigma\vert\mu\nu} &=& -\mbox{Im}{\cal N}_{\Lambda\Sigma} \,
{\widetilde F}^{\Sigma}_{\mu\nu} -\mbox{Re}{\cal N}_{\Lambda\Sigma} \,
{ F}^{\Sigma}_{\mu\nu}\nonumber\\
{\widetilde F}^{\Sigma}_{\mu\nu}&=&\frac{1}{2}\, \epsilon_{\mu
\nu\rho\sigma} \, F^{\Sigma\vert \rho\sigma}
\end{eqnarray}
The $56$--component field strenght vector ${\vec V}_{\mu\nu}$
transforms in the real symplectic representation of the U--duality
group $E_{7(7)}$. We can also write  a column vector containing
the $ 28 $ components of the graviphoton field strenghts and their
complex conjugate:
\begin{equation}
{\vec T}_{\mu\nu} \equiv \left(\matrix{
T^{\phantom{\mu \nu}\vert AB}_{\mu \nu}  \cr
T_{\mu \nu \vert AB} \cr }\right) \quad T^{\vert AB}_{\mu \nu} =
\left(T_{\mu \nu \vert AB}\right)^\star
\label{gravphotvec}
\end{equation}
in which the upper and lower components  transform in the canonical {\it Young
basis} of $SU(8)$ for the ${\bar {\bf 28}}$ and  ${\bf 28}$
representation respectively.\par
The relation between the graviphoton field strength vectors and the
electric magnetic field strenght vectors involves the coset
representative in the $SpD(56)$ representation and it is the following one:
\begin{equation}
{\vec T}_{\mu \nu} = - {\cal S} \, \IC \, \IL_{SpD}^{-1}(\phi) \, {\vec
V}_{\mu \nu}
\label{T=SCLV}
\end{equation}
The matrix
\begin{equation}
\IC =\left(\matrix { {\bf 0} & \bfone \cr -\bfone & {\bf 0} }\right)
\label{sympinv}
\end{equation}
is the symplectic invariant matrix. Eq.\eqn{T=SCLV} reveal that the
graviphotons transform under the $SU(8)$ compensators associated
with the $E_{7(7)}$ rotations. To show this let $g \in  E_{7(7)}$ be an
element of the U--duality group, $g(\phi)$ the action of $g$ on the
$70$ scalar fields and $\ID (g)$ be the $56$ matrix representing $g$
in the real Dynkin basis. Then by definition of coset representative
we can write:
\begin{equation}
  \ID (g) \, \IL_{SpD}(\phi) = \IL_{SpD}\left(g(\phi)\right) \,
  W_D(g,\phi)  \quad ; \quad  W_D(g,\phi) \, \in \, SU(8) \subset \,
  E_{7(7)}
  \label{compensa}
\end{equation}
where $W_D(g,\phi)$ is the $SU(8)$ compensator in the Dynkin basis.
If we regard the graviphoton composite field as a functional of the
scalars and vector field strenghts, from eq.\eqn{compensa} we
derive:
\begin{eqnarray}
T_{\mu \nu} \left( g(\phi), \ID(g) {\vec V} \right)& =& W^*_Y (g,\phi)
\,  T_{\mu \nu} \left(  \phi ,  {\vec V} \right) \nonumber\\
W_Y(g,\phi) & \equiv & {\cal S}^\star \, W_D(g,\phi) \, {\cal S}^T
\end{eqnarray}
where $W_Y(g,\phi)$ is the $SU(8)$ compensator in the Young basis.
\par
It is appropriate to express the upper and lower components of $\vec{T}$
in terms of the self--dual and antiself--dual parts of the graviphoton
field strenghts, since only the latters enter (\ref{trasforma}) and
therefore the equations for the BPS Black--Hole.\\
These components are defined as follows:
\begin{eqnarray}
T^{+\vert AB}_{ \mu \nu}&=& \frac{1}{2}\left(T^{ \vert AB}_{\mu \nu}+\frac{{\rm i}}{2}
\, \epsilon_{\mu\nu\rho\sigma}g^{\rho\lambda} g^{\sigma\pi}
\, T^{ \vert AB}_{\lambda \pi}\right) \nonumber\\
T^-_{ AB \vert \mu \nu}&=& \frac{1}{2}\left(T_{  AB \vert\mu \nu}-\frac{{\rm i}}{2}
\, \epsilon_{\mu\nu\rho\sigma}g^{\rho\lambda} g^{\sigma\pi}
\, T_{ AB \vert\lambda \pi} \right)
\label{selfdual}
\end{eqnarray}
\par
Indeed the following equalities hold true:
\begin{eqnarray}
 T_{\mu  \nu}^{\phantom{\mu  \nu}\vert AB}&=& T^{+ \vert AB}_{\mu  \nu} \nonumber\\
 T_{ \mu  \nu\vert AB}&=& T^{-}_{ \mu  \nu\vert AB}
\label{symprop}
\end{eqnarray}
In order to understand the above properties \cite{amicimiei}, let us first rewrite equation
(\ref{T=SCLV}) in components:
\begin{eqnarray}
 &&T^{\phantom{\mu  \nu} AB}_{\mu  \nu} =
 \bar{h}^{AB}_{\phantom{AB}\Sigma}F^{\Sigma}_{\mu  \nu}-
 \bar{f}^{ AB \vert\Lambda}G_{\Lambda\vert \mu  \nu}\nonumber \\
 &&T^{-}_{ \mu  \nu\vert AB} = h_{AB \vert \Sigma}F^{\Sigma}_{\mu  \nu}-
 f_{AB}^{\phantom{AB}\Lambda}G_{\Lambda\vert \mu  \nu}
 \label{T=SCLVcomp}
 \end{eqnarray}
having defined the matrices $ h,\, f,\,\bar{h},\,\bar{f}$ in
the following way:
\begin{equation}
 - {\cal S} \, \IC \, \IL_{SpD}^{-1}(\phi) =\left(\matrix{\bar{h} &
 -\bar{f}\cr h & -f }\right)
 \label{LM}
\end{equation}
Next step is to express the self--dual and antiself--dual components
of $G_\Sigma $ (defined in the same way as for $T_{\mu \nu}$ in
(\ref{selfdual}))in terms of the corresponding components of $F^\Sigma$
through the period matrix
\begin{eqnarray}
&& G^{+}_{\Sigma}={\cal N}_{\Sigma\Lambda}F^{+ \Lambda}\nonumber\\
&& G^{-}_{\Sigma}=\bar{{\cal N}}_{\Sigma\Lambda}F^{- \Lambda}
\label{Gpm=NFpm}
\end{eqnarray}
Projecting the two equations (\ref{T=SCLVcomp}) along its self--dual
and antiself--dual parts, and taking into account
(\ref{Gpm=NFpm}), one can deduce the following conditions:
\begin{eqnarray}
 &&T^{- \vert AB}_{\mu  \nu} =0 \nonumber \\
 &&T^{+}_{ \mu  \nu\vert AB} =0
\label{TpTm}
\end{eqnarray}
which imply in turn equations (\ref{symprop}). The symplectic vector
${\vec T}_{\mu \nu}$ of the graviphoton field strenghts
may therefore be rewritten in the following form:
\begin{equation}
{\vec T}_{\mu \nu} \equiv \left(\matrix{T^{+\vert AB}_{\mu \nu}  \cr
T^{-}_{\mu \nu \vert AB} \cr }\right)
\end{equation}
These preliminaries completed we are now ready to consider the
Killing spinor equation and its general implications.
\section{The Black Hole ansatz and the Killing spinor equation}
\label{bhansazzo}
As mentioned in the first section, 
the BPS saturated black holes we are interested in are classical
field configurations with rotational symmetry and time translation
invariance. As expected on general grounds we must allow for the
presence of both electric and magnetic charges. Hence we introduce
the following ansatz for the elementary bosonic fields of the theory
\subsection{The Black Hole ansatz}
We introduce isotropic coordinates:
\begin{eqnarray}
\left\{ x^\mu \right\} &=&\left\{ t , {\vec x} = x^a \right\}  \quad ;
\quad a=1,2,3 \nonumber\\
 r&=&\sqrt{ {\vec x}\, \cdot \, {\vec x} }
 \label{isocoord}
\end{eqnarray}
and we parametrize the metric, the vector fields and the scalar fields as follows:
\begin{eqnarray}
ds^2 &=& \exp \left [ 2U(r)\right ]\,  dt^2 - \exp \left [ -2U(r)\right ] \, d{\vec x}^2
\label{metansaz}\\
F^{-{\vec \Lambda}}_{\mu\nu}&=& \frac{1}{4\pi} \, t^{{\vec \Lambda}}(r) \,
E^{-}_{\mu\nu}\label{vecansaz}\\
\phi^i &=& \phi^i (r)
\label{ansaz}
\end{eqnarray}
where
\begin{equation}
 t^{{\vec \Lambda}}(r) \, \equiv \, 2 \pi \, \left( 2 \, p^{{\vec \Lambda}}+
 \mbox{i}\, q^{{\vec \Lambda}}(r) \right)
\label{tvecvero}
\end{equation}
is a $28$--component  complex vector whose real part is constant,
while the imaginary part is a radial fucntion to be determined. We
will see in a moment the physical interpretation of $p^{{\vec \Lambda}}$
and $q^{{\vec \Lambda}}(r)$.
$E^{-}_{\mu\nu}$ is the unique antiself--dual $2$--form in the background of
the chosen metric and  it reads
as follows \cite{pietrobh1}, \cite{pietrobh2} :
\begin{equation}
E^- \, = \, E^{-}_{\mu\nu} \, dx^\mu  \wedge dx^\nu =
\mbox{i} \frac{e^{2U(r)}}{r^3} \, dt \wedge {\vec x}\cdot
d{\vec x}  +  \frac{1}{2}   \frac{x^a}{r^3} \, dx^b   \wedge
dx^c   \epsilon_{abc}
\label{eaself}
\end{equation}
and it is normalized so that:
  $ \int_{S^2_\infty} \, E^-  =   2 \, \pi $.

Combining eq.\eqn{eaself} with eq.\eqn{vecansaz} and \eqn{tvecvero}
we conclude that
\begin{eqnarray}
 p^{{\vec \Lambda}} &=&\mbox{magnetic charge} \\
 q^{{\vec \Lambda}}(r=\infty) &=&\mbox{electric charge}
 \label{pqinterpret}
\end{eqnarray}
At the same time we can also identify:
\begin{equation}
 q^{{\vec \Lambda}}(r) \, = \, r^2 \, \frac{d C^{{\vec \Lambda}}(r)}{dr} \,
 \exp \left[C^{{\vec \Lambda}}(r) - 2 U \right]
 \label{identqc}
\end{equation}
where $\exp \left[C^{{\vec \Lambda}}(r)\right]$ is a function
parametrizing the electric potential:
\begin{equation}
A^{{\vec \Lambda}}_{elec} \, = \, dt \, \exp \left[C^{{\vec \Lambda}}(r)\right].
\end{equation}
\subsection{The Killing spinor equations}
We can now analyse the Killing spinor equation combining the results
  (\ref{symprop}), (\ref{Gpm=NFpm}) with our ansatz (\ref{ansaz}). This
allows us to rewrite (\ref{T=SCLVcomp}) in the following form:
\begin{eqnarray}
 &&T^{+\phantom{\mu  \nu} AB}_{\mu  \nu} = \frac{1}{4\pi}
 \left(\bar{h}^{AB}_{\phantom{AB}\Sigma}t^{\star\Sigma}-
 \bar{f}^{ AB \vert\Lambda}{\cal N}_{\Lambda\Sigma}t^{\star\Sigma}\right)E^{+}_{\mu \nu}
 \nonumber \\
 &&T^{-}_{ \mu  \nu\vert AB} = \frac{1}{4\pi}
 \left(h_{AB \vert \Sigma}t^{\Sigma}-
 f_{AB}^{\phantom{AB}\Lambda}\bar{{\cal N}}_{\Lambda\Sigma}t^{\Sigma}\right)
 E^{-}_{\mu \nu}
  \label{ThfE}
\end{eqnarray}
where
$ E^+_{\mu\nu}  =\left( E^-_{\mu\nu} \right)^\star $.

Then we can use the general result (obtained in \cite{pietrobh1}, \cite{pietrobh2}):
\begin{eqnarray}
&&E^-_{\mu\nu} \, \gamma^{\mu\nu}  = 2 \,\mbox{i} \frac{e^{2U(r)}}{r^3}  \,
\gamma_a x^a \, \gamma_0 \, \frac{1}{2}\left[ {\bf 1}+\gamma_5 \right] \nonumber \\
&&E^+_{\mu\nu} \, \gamma^{\mu\nu}  =-2 \,\mbox{i} \frac{e^{2U(r)}}{r^3}  \,
\gamma_a x^a \, \gamma_0 \, \frac{1}{2}\left[ {\bf 1}-\gamma_5 \right] \nonumber\\
\label{econtr}
\end{eqnarray}
and contracting both sides of (\ref{ThfE}) with $\gamma^{\mu\nu}$
one finally gets:
\begin{eqnarray}
&&T^{+\phantom{\mu  \nu} AB}_{\mu  \nu}\gamma^{\mu\nu}  = -\frac{{\rm i}}{2\pi}
\frac{e^{2U(r)}}{r^3}  \,
\gamma_a x^a \, \gamma_0 \, \left(\bar{h}^{AB}_{\phantom{AB}\Sigma}t^{\star\Sigma}-
 \bar{f}^{ AB \vert\Lambda}{\cal N}_{\Lambda\Sigma}t^{\star\Sigma}\right)
 \frac{1}{2}\left[ {\bf 1}-\gamma_5 \right]  \nonumber \\
 &&T^{-}_{ \mu  \nu\vert AB} \gamma^{\mu\nu} =\frac{{\rm i}}{2\pi} \frac{e^{2U(r)}}{r^3}
 \, \gamma_a x^a \, \gamma_0 \, \left(h_{AB \vert \Sigma}t^{\Sigma}-
 f_{AB}^{\phantom{AB}\Lambda}\bar{{\cal N}}_{\Lambda\Sigma}t^{\Sigma}\right)
 \frac{1}{2}\left[ {\bf 1}+\gamma_5 \right]
 \label{Tpmgamma}
\end{eqnarray}
At this point we specialize the supersymmetry parameter to be of the
form analogue to the form utilized in \cite{pietrobh1,pietrobh2}:
\begin{equation}
\epsilon_A = e^{f(r)} \xi_A
\label{kilspinor1}
\end{equation}
It is useful
to split the $SU(8)$ index $A=1,\dots ,8$ into an $SU(6)$ index
$X=1,\dots ,6$ and an $SU(2)$ index $a=7,8$. Since we look
 for BPS states belonging to {\sl just once shortened multiplets}
 ({\it i.e.} with  $N=2$ residual supersymmetry)
we require that $ \xi^X=0,\, X=1,\dots ,6$ and furthermore that:
\begin{equation}
\gamma_0\, \xi_a  = -\mbox{i} \, \epsilon_{ab} \, \xi^b
\label{kilspinor2}
\end{equation}
The vanishing of the gravitino transformation rule implies conditions on both functions
$U(r)$ and $f(r)$. The equation for the latter is uninteresting since
it simply fixes the form of the Killing spinor parameter.
The equation for $U$ instead is relevant since it yields the form
of the black hole metric.  It can be written in the following form:
\begin{equation}
\frac{dU}{dr}= - {k}\frac{e^U}{r^2}\left(h_\Sigma t^\Sigma - f^\Lambda \bar\cN_{\Lambda\Sigma}t^\Sigma \right)
\label{gratra}
\end{equation}

Furthermore the $56$ differential equations from the dilatino sector can
be written in the form:
\begin{eqnarray}
&& a\, P_{ABCa\vert i} \, \frac{d\phi^i}{dr} =\frac{b}{2\pi} \frac{e^{U(r)}}{r^2}
\left(h_{\Sigma}t^{\Sigma}-
 f^{\Lambda}\bar{{\cal N}}_{\Lambda\Sigma}t^{\Sigma}\right)_{[AB}
 \delta_{C]}^{b}\epsilon_{ba}\nonumber\\
&&a\, P^{ABCa\vert i} \, \frac{d\phi^i}{dr} =\frac{b}{2\pi} \frac{e^{U(r)}}{r^2}
\left(\bar{h}_{\Sigma}t^{\star\Sigma}-
 \bar{f}^{\Lambda}{\cal N}_{\Lambda\Sigma}t^{\star\Sigma}\right)^{[AB}
 \delta^{C]}_{b}\epsilon^{ba}
\label{BPS56eq1}
\end{eqnarray}

Suppose now that the triplet of indices $(A,\, B,\, C)$ is of the
 type   $(X,\, Y,\,Z)$.
 This corresponds to projecting eq. \eqn{BPS56eq1}  into the
 representation $({\bf 1},{\bf 2}, {\bf 20})$ of  $U(1) \times SU(2)\times SU(6) \subset SU(8)$.
 In this case however the right hand side vanishes identically:
 \begin{equation}
  a\, P_{XYZ a\vert i} \, \frac{d\phi^i}{dr} =\frac{b}{2\pi} \frac{e^{U(r)}}{r^2}
\left(h_{\Sigma}t^{\Sigma}-
 f^{\Lambda}\bar{{\cal N}}_{\Lambda\Sigma}t^{\Sigma}\right)_{[XY}
 \delta_{Z]}^{b}\epsilon_{ba} \equiv 0
 \label{pxyz}
 \end{equation}
 so that we find that the corresponding 40 scalar fields are actually
 constant.

In the case where the triplet of indices $(A,\, B,\, C)$ is $(X,\, Y,\, a )$
the equations may be put in the following matrix form:
\begin{eqnarray}
\left(\matrix{ P^{XY\vert i}\cr P_{XY\vert i}}\right)\,
\frac{d\phi^i}{dr}=\frac{b}{ 3 \, a \,\pi} \frac{e^{U(r)}}{r^2}
\left(\matrix{\bar{h} & -\bar{f}\cr h & -f }\right)_{\vert XY}\,\left(
\matrix{{\bf Re}(\, t\, )\cr {\bf Re}(\, \bar{{\cal N}}t \,)}\right)
\label{pxy}
\end{eqnarray}
The above equations are obtained by projecting the terms on the left and on the right
side of eq. \eqn{BPS56eq1} (transforming respectively in the ${\bf 70}$ and in
the ${\bf 56}$ of $SU(8)$) on the common representation ${\bf
(1, 1,15) \oplus \bar{(1,1,15)}}$ of the subgroup
$U(1) \times SU(2)\times SU(6) \subset SU(8)$).

Finally, when the  triplet  $(A,\, B,\, C)$ takes the values $(X,\, b,\, c )$,
the l.h.s. of eq. \eqn{BPS56eq1}  vanishes   and we are left with the
equation:
\begin{equation}
 0=\frac{e^{U(r)}}{r^2}
\left(h_{\Sigma}t^{\Sigma}-
 f^{\Lambda}\bar{{\cal N}}_{\Lambda\Sigma}t^{\Sigma}\right)_{[Xb}
 \epsilon_{c]a}
 \label{xbc}
\end{equation}
This corresponds to the projection of eq. \eqn{BPS56eq1}  into the
representation $({\bf 1},{\bf 2},{\bf 6}) + \bar{({\bf 1},{\bf 2},{\bf 6})} \subset \bf{28} +\bar{{ \bf{28}}}$.

Let us consider the basis vectors $ \vert \vec\Lambda > \, , \, \vert - \vec\Lambda > \in {\bf 56}= {\bf 28 + \bar 28} $
defined in ref. \eqn{dynkbas} and let us introduce the eigenmatrices $\IK^{\vec\lambda}$ of the $SU(8)$ Cartan generators
${\cal H}_i$ diagonalized on the subspace of non-compact $E_{7(7)}$ generators $\IK $ defined by the
Lie algebra Cartan decomposition $E_{7(7)}=\IK \oplus \IH$ ($\vec \lambda $
being  the weights of the ${\bf 70}$ of $SU(8)$).
It is convenient to use  a real basis for both representations ${\bf 56}$ and ${\bf 70}$, namely:
\begin{eqnarray}
  \vert \vec\Lambda_x > &=& \frac {\vert \vec\Lambda >  + \vert - \vec\Lambda >}{2} \nonumber\\
   \vert \vec\Lambda_y > &=& \frac {\vert \vec\Lambda >  - \vert - \vec\Lambda >}{2{\rm i}} \nonumber\\
   \IK^{\vec \lambda}_x &=& \frac {\IK^{\vec \lambda}  + \IK^{-\vec \lambda}}{2} \nonumber\\
   \IK^{\vec \lambda}_y &=& \frac {\IK^{\vec \lambda}  - \IK^{-\vec \lambda}}{2 {\rm i}}
   \label{realbas}
\end{eqnarray}
such that they satisfy the following relations:
\begin{equation}
\begin{array}{lcrcll}
\mbox{projectors on irrep ${\bf 70}$} &:&
 \left[{\cal H}_i\, ,\, \IK^{\vec \lambda}_{x}\right]\,
 &=& \null & \left({\vec \lambda},
{\vec a}_i\right)\IK^{\vec \lambda}_{y}  \\
\null & \null &
\left[{\cal H}_i\, ,\, \IK^{\vec \lambda}_{y}\right]\, &=&  - & \left({\vec \lambda},
{\vec a}_i\right)\IK^{\vec \lambda}_{x}  \\
\mbox{projectors on irrep ${\bf 28}$ } &:&
{\cal H}_i\vert \,{\vec \Lambda}_{x} \, \rangle\, &=& \null & \left({\vec \Lambda},
{\vec a}_i\right)\vert \,{\vec \Lambda}_{y} \, \rangle  \\
\null & \null &
{\cal H}_i\vert \,{\vec \Lambda}_{y} \, \rangle\, &=&  - & \left({\vec \Lambda},
{\vec a}_i\right)\vert \,{\vec \Lambda}_{x} \, \rangle \\
\end{array}
\label{procione}
\end{equation}

 As it will be explained in section \ref{su8ine7}, the  seven Cartan
 generators ${\cal H}_i$
are given by appropriate linear combinations of the $E_{7(7)}$ step
operators (see eq.s \eqn{su8cartan}).
\par
Using the definitions given above, it is possible
to rewrite equations \eqn{gratra} and \eqn{BPS56eq1} in an
algebraically intrinsic way, which, as we will see, will prove to be useful following.
This purpose can be achieved by expressing the l.h.s. and r.h.s. of
eqs. \eqn{gratra} and \eqn{BPS56eq1}
in terms of suitable projections on the real bases $\IK^{\vec\lambda}_{x,y}$
and $\vert \vec \Lambda_{x,y} >$ respectively.
In particular eqn.  \eqn{pxyz},  \eqn{xbc} become respectively:
 \begin{equation}
\begin{array}{rclcrcl}
\mbox{Tr}\left ( \IK^{{\vec \lambda}}_x \, \IL^{-1} d\IL \right )  & = &  0
& ; & {\vec \lambda} & \in & {\bf (1,2,20)}\subset {\bf 70}\\
0 &=&\langle \, {\vec \Lambda}_x={\vec \lambda}_D \vert \, \IC \, \IL^{-1} (\phi) \,
\vert \, {\vec t}, \phi \, \rangle &; & \quad {\vec \Lambda} & \in &
{\bf (1,2,6)} \subset {\bf 28}  \nonumber\\
0 &=&\langle \, {\vec \Lambda}_y={\vec \lambda}_D \vert \, \IC \, \IL^{-1} (\phi) \,
\vert \, {\vec t}, \phi \, \rangle &; & \quad {\vec \Lambda} & \in &
{\bf (1,2,6)} \subset {\bf 28}+ \bar{\bf 28}
\end{array}
\label{22026eq}
\end{equation}
where we have set:
\begin{equation}
\vert \, {\vec t}, \phi \, \rangle = \left (
\matrix{{\bf Re}(\, t\, )\cr {\bf Re}(\, \bar{{\cal N}}t \,)}\right)
\end{equation}
The real vectors $ \vert \vec\Lambda_x >$ and $ \vert \vec\Lambda_y >$ in eq. \eqn{22026eq} are related
by eq. \eqn{realbas} to $ \vert \vec\Lambda >$ and $ -\vert \vec\Lambda >$,  which are now restricted to
the subrepresentations ${\bf (1,2,6)} $ and $\bar{{\bf (1,2,6)}}$
respectively.
The pair of eq.s \eqn{22026eq} can now be  read as very clear statements. The
first in the pair tells us that $40$ out of the $70$ scalar fields in
the theory must be constants in the radial variable. Comparison with
the results of section \ref{solvodecompo} shows that the fourty
constant fields are those belonging to the solvable subalgebra
$Solv_4 \subset Sol_7$ defined in eq.\eqn{3and4defi}.
\par
Hence those scalars that in an N=2 truncation belong to hypermultiplets are
constant in any BPS black hole solution. \par
The second equation in the pair \eqn{22026eq} can  be read as
a statement on the available charges. Indeed since it must be zero
everywhere, the right hand side of this equation can be evaluated at
infinity where the vector $\vert {\vec t} ,\phi \rangle$ becomes the
$56$-component vector of electric and magnetic charges defined as:
\begin{equation}
\left(\matrix { g^{\vec {\Lambda}} \cr e_{\vec {\Sigma}}\cr } \right)
= \left( \matrix { \int_{S^2_\infty} \, F^{\vec {\Lambda}} \cr
 \int_{S^2_\infty} \, G_{\vec {\Sigma}} \cr }\right)
 \label{gedefi}
\end{equation}
As we will see in the following, the explicit evaluation of  \eqn{pappaequa}
implies that $24$ combinations of  the charges are zero.
This, together  with the fact that the last of eqn. \eqn{22026eq}
yields the vanishing of $24$ more independent combinations, implies that there are only
$8$ surviving charges.
\par
On the other hand, eq. \eqn{pxy} can be rewritten in the following form:
\begin{equation}
\mbox{Tr}\left ( \IK^{{\vec \lambda}} \, \IL^{-1} d\IL \right ) =
\frac{b}{ 3 \, a \,\pi} \frac{e^{U(r)}}{r^2} \, \langle \,
{\vec \Lambda}={\vec \lambda}  \vert_D \, \IC \, \IL^{-1} (\phi) \,
\vert \, {\vec t}, \phi \, \rangle
\label{pappaequa}
\end{equation}
In equation \eqn{pappaequa}  both the weights ${\vec \lambda}$ and ${\vec \Lambda}$ defining the projections in the
l.h.s. and r.h.s. belong to the common representation $ {\bf (1, 1,15) \oplus \bar{(1,1,15)}}$ (
$\lambda \, \in \, {\bf (1, 1,15) \oplus \bar{(1,1,15)}} \,
\subset \, {\bf 70} $,
$\left({\vec \Lambda}\, ,\, -{\vec \Lambda}\right)
\, \in \, {\bf (1, 1,15) \oplus \bar{(1,1,15)}} \,
\subset \, {\bf 28+\bar{28}} $).
Finally, in this intrinsic formalism eq. \eqn{gratra}  takes the
form:
\begin{equation}
\frac{dU}{dr}\,=\,2k \frac{e^U}{r^2} \langle {\vec \Lambda}\vert_D\, \IC \, \IL^{-1} (\phi) \,
\vert \, {\vec t}, \phi \, \rangle \quad; \quad {\vec \Lambda}\in
{\bf (1,1,1)}\subset {\bf 28}
\label{Ueq}
\end{equation}
Eq. (\ref{Ueq}) implies that the projection on $\vert \,{\vec \Lambda}_{x} \,
\rangle$ of the right-hand side equals $\frac{dU}{dr}$ and thus gives
the differential equation for $U$, while the projection on
$\vert \,{\vec \Lambda}_{y} \, \rangle$ equals zero, which means in turn that
the central charge must be real.
\par

In the next section, restricting our attention to a simplified case
where the only non-zero scalar fields are taken in the Cartan subalgebra,
we show how the above implications of the Killing spinor equation can
be made explicit.
\section{A simplified model: BPS black-holes reduced to the Cartan subalgebra}
\label{cartadila}
\par
Just as an illustrative exercise
in the present section we consider the explicit BPS black--hole
solutions where the only scalar fields excited out of zero are those
in the Cartan subalgebra of $E_{7(7)}$.
Having set to zero all the fields except these seven, we will see
that
the Killing spinor equation implies that $52$ of the $56$ electric and
magnetic charges are also zero. So by restricting our attention to
the Cartan subalgebra we retrieve a BPS black hole solution that
depends only on $4$ charges. Furthermore in this solution $4$ of
the $7$ Cartan fields are actually set to constants and only the
remaining $3$ have a non trivial radial dependence. Which is which
is related to the basic solvable Lie algebra decomposition \eqn{7in3p4}:
the $4$ Cartan fields in $Solv_4$ are constants, while the $3$ Cartan fields in
$Solv_3$ are radial dependent. This type of solution reproduces the
so called $a$-model black-holes studied in the literature, but it is
not the most general. However, as we shall argue in the last section, the toy
model presented here misses full generality by little. Indeed
the general solution that depends on $8$, rather than $4$ charges,
involves, besides the $3$ non trivial Cartan fields other $3$
nilpotent fields which correspond to axions of the compactified string
theory.
\par
Let us then begin examining this simplified case.
For its study it is
particularly useful to utilize the Dynkin basis since there the
Cartan generators have a diagonal action on the ${\bf 56}$ representation
and correspondingly on the vector fields. Namely in the Dynkin basis
the matrix ${\cal N}_{\Lambda\Sigma}$ is purely imaginary and
diagonal once the coset representative is restricted to the Cartan
subalgebra.  Indeed if we name $h^i(x)$ ($i=1,\dots\, 7$ ) the
Cartan subalgebra scalar fields, we can write:
\begin{eqnarray}
\label{LSpD}
\IL_{SpD}\left( h \right)& \equiv &\exp \left[ {\vec h} \, \cdot \,
{\vec H} \right ] \, = \, \left(\matrix{ {\bf A}(h) &  {\bf 0} \cr
{\bf 0} & {\bf D}(h) \cr }\right) \nonumber\\
&& \null \nonumber\\
&&\mbox{where}\nonumber\\
&& \null \nonumber\\
 {\bf A}(\phi)^{{\vec \Lambda}}_{\phantom{{\vec \Lambda}}{\vec \Sigma}}&=&
 \delta^{{\vec \Lambda}}_{\phantom{{\vec \Lambda}}{\vec \Sigma}} \,
 \exp\left[ {\vec \Lambda} \, \cdot \, {\vec h} \right]\nonumber\\
 {\bf D}(\phi)_{{\vec \Lambda}}^{\phantom{{\vec \Lambda}}{\vec \Sigma}}&=&
 \delta_{{\vec \Lambda}}^{\phantom{{\vec \Lambda}}{\vec \Sigma}} \,
 \exp\left[-\, {\vec \Lambda} \, \cdot \, {\vec h} \right]\nonumber\\
\end{eqnarray}
so that combining equation \eqn{gaiazuma} with \eqn{cayleytra} and
\eqn{LSpD} we obtain:
\begin{equation}
{\cal N}_{{\vec \Lambda}{\vec \Sigma}} = {\rm i} \,
\left( {\bf A}^{-1} \, {\bf D} \right)_{
{\vec \Lambda}{\vec \Sigma}}
= {\rm i} \, \delta_{{\vec \Lambda}{\vec \Sigma}} \,\exp\left[-\,2 \,
{\vec \Lambda} \, \cdot \, {\vec h} \right]
\label{diagenne}
\end{equation}
Hence in the Dynkin basis the lagrangian \eqn{lagrared} reduced to the
Cartan sector takes the following form:
\begin{equation}
{\cal L} = \sqrt{-g} \, \left( 2\,   R[g] -\frac{1}{4}
\sum_{{\vec \Lambda} \in \Pi^+}
\, \exp\left[-\,2 \,
{\vec \Lambda} \, \cdot \, {\vec h} \right] \, {\cal F}^{{\vec \Lambda}\vert \mu \nu}\,{\cal
F}^{{\vec \Lambda}}_{\mu \nu}
+ \frac{\alpha ^2}{2}\, \sum_{i=1}^{7} \, \partial_\mu h^i \, \partial^\mu h^i \right)
\label{lagranew}
\end{equation}
where by $\Pi^+$ we have denoted the set of positive weights for the
fundamental representation of the U--duality group $E_{7(7)}$ and
$\alpha$ is a real number fixed by supersymmetry already introduced in
eq.\eqn{lagrared}.

\par
Let us examine in detail the constraints imposed by (\ref{pappaequa}),(\ref{22026eq})
and (\ref{Ueq}) on the $28$ complex vectors $t^\Lambda$.  Note that these vectors are
naturally split  in two subsets:
\begin{itemize}
\item{The set: $t_z\equiv \{t^{17},t^{18},t^{23},t^{24}\}$ that
parametrizes $8$ real charges which, through 4 suitable linearly independent combinations,
transform
in the representations ${\bf (1,1,1)+\bar{(1,1,1)}}$
$\oplus{\bf (1,1,15)+\bar{(1,1,15)} }$   and  contribute to the 4 central
charges of the theory.}
\item{the remaining 24 complex vectors $t_\ell$.  Suitable linear combinations of these
vectors transform in the representations ${\bf (1,1,15)+\bar{(1,1,15)}}$ ${\bf  \oplus(1,2,6)
+\bar{(1,2,6)}}$ and are orthogonal to the set $t_z$
}
\end{itemize}

Referring now to the present simplified model we can analyze the
consequences of the projections
 (\ref{Ueq}) and (\ref{pappaequa}).
 They are $16$ complex conditions which
split into:
\begin{enumerate}
\item {a set of 4 equations whose coefficients depend only
on $t_z$ and give rise to 4 real conditions on
the real and immaginary parts of $t_z$ and 4 real differential equations on the
Cartan fields $h_{1,2,7}$ belonging to vector multiplets, namely to
the solvable Lie subalgebra $Solv_3$ (see section \ref{solvodecompo})
}
\item {a set of 12 equations which contribute,
together with the 12 conditions coming from the second
of eq.s (\ref{22026eq}), to set the 24 $t_\ell$ to zero}
\end{enumerate}
In obtaining the above results, we used the fact that all the Cartan fields in $Solv_4$,
(namely ($h_{3,4,5,6}$) which fall into
hypermultiplets when the theory is $N=2$ truncated) are constants, and thus can
be set to zero (modulo duality rotations) as was discussed in the previous section. \\
\\

After the projections have been taken into account we are left with
a reduced set of non vanishing fields that includes only four vectors and
three scalars, namely:
\begin{eqnarray}
\mbox{vector fields}&=&\cases{F^{\Lambda^{17}}_{\mu \nu} \equiv {\cal F}^{17}_{\mu \nu}\cr
F^{\Lambda^{18}}_{\mu \nu} \equiv {\cal F}^{18}_{\mu \nu}\cr
F^{\Lambda^{23}}_{\mu \nu} \equiv {\cal F}^{23}_{\mu \nu}\cr
F^{\Lambda^{24}}_{\mu \nu} \equiv {\cal F}^{24}_{\mu \nu}\cr}\nonumber\\
\mbox{scalar fields}&=&\cases{h_1   \cr
h_2   \cr
h_7   \cr }
\end{eqnarray}
In terms of these fields, using the scalar products displayed in table \ref{scalaprod},
the lagrangian has the following explicit expression:
\begin{eqnarray}
{\cal L}&=& \sqrt{-g} \, \Biggl \{ 2\, R[g] \, + \,
\frac{\alpha ^2}{2}\, \left[ \left(\partial_\mu h_1\right)^2 +
\left(\partial_\mu h_2\right)^2 +
\left(\partial_\mu h_3\right)^2 \right]\nonumber\\
&& -\, \frac{1}{4} \,
\exp \left[ 2\,\sqrt{\frac{2}{3}}\, h_1 \, -\,\frac{2}{\sqrt{3}}\, h_7 \right] \,
\left({\cal F}_{\mu \nu}^{17} \right)^2 \,- \frac{1}{4} \,
\exp\left[ 2\,\sqrt{\frac{2}{3}}\, h_2 \, -\,\frac{2}{\sqrt{3}}\, h_7 \right] \,
\left({\cal F}_{\mu \nu}^{18} \right)^2 \nonumber\\
&& -\, \frac{1}{4} \,
\exp\left[ 2\,\sqrt{\frac{2}{3}}\, h_1 \, +\,\frac{2}{\sqrt{3}}\, h_7 \right] \,
\left({\cal F}_{\mu \nu}^{23} \right)^2 \,- \frac{1}{4} \,
\exp\left[ 2\,\sqrt{\frac{2}{3}}\, h_2 \, +\,\frac{2}{\sqrt{3}}\, h_7 \right] \,
\left({\cal F}_{\mu \nu}^{24} \right)^2  \Biggr \}
\label{reduclagra}
\end{eqnarray}
\par
Introducing an index $\alpha$ that takes the four values $\alpha =17,18,23,24 $ for
the four field strenghts, and moreover four undetermined radial functions to
be fixed by the field equations:
 \begin{equation}
    q^\alpha(r)\, \equiv \, C_\alpha ^\prime e^{C_\alpha-2U} \, r^2
    \label{pseudocharge}
\end{equation}
and four real constants $p^\alpha$,
the ansatz for the vector fields can be parametrized as follows:
\begin{eqnarray}
 \cF^{\alpha }_{el}  &=& - \frac{q^\alpha(r) \, e^{2U(r)} }{r^3} \,
dt \wedge \vec x \cdot d\vec x \, \equiv \, - {C_\alpha ^\prime e^{C_\alpha}   \over r}
dt \wedge \vec x \cdot d\vec x
\nonumber\\
 \cF^{\alpha }_{mag}  &=& {p^\alpha   \over 2r^3}x^a dx^b \wedge d x^c \epsilon_{abc}
\nonumber\\
   \cF^{\alpha } &=&    \cF^{\alpha }_{el} +  \cF^{\alpha }_{mag} \nonumber\\
 \cF^{-\alpha }_{\mu\nu} &=& \frac{1}{4\pi} \, t^{\alpha } E^{-}_{\mu\nu}  \nonumber\\
   t^{\alpha }  &=& 2\pi\left(2p^\alpha  + {\rm i} q^\alpha (r) \right)\equiv
 2\pi (2p^\alpha  + {\rm i}  C^\prime_\alpha  e^{C_\alpha -2U}r^2)
 \end{eqnarray}
\par
The physical interpretation  of the above data is the following:
\begin{equation}
p^\alpha = \mbox{mag. charges} \quad ; \quad q^\alpha(\infty) =
\mbox{elec. charges}
\label{physinter}
\end{equation}
From the effective lagrangian of the reduced system we derive
the following set of Maxwell-Einstein-dilaton field equations, where in addition
to the index $ \alpha $ enumerating the vector fields an index $i$ taking the three values
$i=1,2,7$ for the corresponding three scalar fields has also been introduced:
\begin{eqnarray}
 \mbox{Einstein eq.} &:&
  -2 R_{MN} =T_{MN} = {1\over 2}\alpha^2 \sum_{i} \, \partial_M h_i
\partial_N h_i + S_{MN} \nonumber\\
\Biggl (  S_{MN} & \equiv &-{1 \over 2} \sum _{\alpha }  e^{-2\vec \Lambda_\alpha  \cdot h}
\left[ \cF^\alpha _{M\cdot}  \cF^\alpha _{N\cdot}  -{1 \over 4}
\cF^\alpha _{\cdot\cdot}  \cF^\alpha _{\cdot\cdot}   \eta_{MN}\right]  \Biggr ) \nonumber\\
  \mbox{Maxwell eq.}&:&
   \partial_\mu  \left( \sqrt{-g}   \exp \left[ -2\vec \Lambda_\alpha \cdot h
   \right ] \,
 F^{\alpha \vert \mu \nu} \right) =  0 \nonumber\\
 \mbox{Dilaton eq.s } &:&
    \frac{\alpha^2}{\sqrt{-g}}\, \partial_\mu  \left(\sqrt{-g}\,  g^{\mu \nu} \,
    \partial_\nu h_i(r) \right)
     +\frac{1}{2}\sum_{\alpha} \, \Lambda^\alpha_i \, \exp[-2 \,\Lambda^\alpha \cdot h] \,
      {\cal F}_{\cdot\cdot}^{\alpha} {\cal F}_{\cdot\cdot}^{\alpha} =0  \nonumber\\
 \label{campequa}
 \end{eqnarray}
 In eq.\eqn{campequa} we have denoted by dots the contraction of indices. Furthermore
 we have used the capital latin letters $M,N$ for the flat Lorentz indeces obtained through
 multiplication by the inverse or direct vielbein according to the case. For instance:
 \begin{equation}
\partial_M \equiv V_M^\mu \, \partial_\mu = \cases{\partial_0 = e^{-U}
\, \frac{\partial}{\partial t} \cr
\partial_I = e^{U}
\, \frac{\partial}{\partial x^i} \quad (I=1,2,3)\cr }
\end{equation}
Finally the components of the four relevant weights, restricted to
the three relevant scalar fields are:
\begin{eqnarray}
 \vec\Lambda_{17} &=& \left(-\sqrt{\frac{2}{3} } , 0  , \sqrt{\frac{1}{3} } \right)\nonumber\\
 \vec\Lambda_{18} &=& \left(   0, -\sqrt{\frac{2}{3} } , \sqrt{\frac{1}{3} } \right) \nonumber\\
 \vec\Lambda_{23} &=& \left(-\sqrt{\frac{2}{3} } , 0  , -\sqrt{\frac{1}{3} } \right)\nonumber\\
 \vec\Lambda_{24} &=& \left(   0, -\sqrt{\frac{2}{3} } , -\sqrt{\frac{1}{3} } \right) \nonumber\\
\end{eqnarray}
The flat indexed stress--energy tensor $T_{MN}$ can be evaluated by
direct calculation and we easily obtain:
\begin{eqnarray}
T_{00} &=& S_{00} = -{1\over 4}\sum_{\alpha} \exp\left[{-2\vec \Lambda_\alpha \cdot \vec h}
+4 U \right ] \, \frac{1}{r^4} \,
\left [ \left(q^\alpha(r)\right)^2  +  \left(p^\alpha \right)^2 \right]
\nonumber\\
   T_{\ell m} &=& \left(\delta_{\ell m} -2 {x_\ell x_m \over r^2}
\right)S_{00} + {\alpha^2\over 2} {x_\ell x_m \over r^2}{\partial \vec h
\over \partial r }\cdot {\partial \vec h \over \partial r }
\label{stresstens}
\end{eqnarray}
The next part of the calculation involves the evaluation of
the flat--indexed Ricci tensor for the metric in eq.\eqn{metansaz}.
From the definitions:
\begin{eqnarray}
0 &=& d V^M - \omega^{MN} \, \wedge \, V^N \, \eta_{NR}  \nonumber\\
R^{MN} & = & d\omega^{MN} - \omega^{MR} \, \wedge \, \omega^{SN} \,
\eta_{RS} \equiv R^{MN}_{RS} \,
V^R \, \wedge V^S \nonumber\\
V^R &=& \cases{V^0 = dt e^U\cr
V^I= dx^i e^{-U} \cr }
\label{curvdef}
\end{eqnarray}
we obtain the spin connection:
\begin{equation}
\omega^{0I} = -\frac{x^i}{r} \,  dx^i\, U^\prime\, \exp\left[ 2U \right]
\quad ; \quad \omega^{IJ} =  2 \frac{x^{[i}\,dx^{j]} }{r} \,  U^\prime\,
\label{conspin}
\end{equation}
and the Ricci tensor:
\begin{eqnarray}
R_{00} & = & -\frac{1}{2}\exp[2U] \, \left( U^{\prime\prime} +\frac{2}{r}
U^\prime \right)\nonumber\\
R_{ij} &=& \frac{x^i \, x^j}{r^2} \, \exp[2U] \, \left(U^\prime \right)^2
+ \delta_{ij} R_{00}
\label{rictens}
\end{eqnarray}
Correspondingly the field equations reduce to a set of first order
differential equations for the eight unknown functions:
\begin{equation}
U(r) \quad ; \quad h_i(r) \quad ; \quad q^{\alpha}(r)
\label{unknowns}
\end{equation}
From {\sl Einstein equations} in \eqn{campequa} we get the two
ordinary differential equations:
\begin{eqnarray}
U^{\prime\prime} +\frac{2}{r} U^\prime  &=& S_{00}\, \exp \left[\, -2U\,\right]
\nonumber\\
\left( U^\prime \right)^2 &=& \left(S_{00}  - \frac{\alpha^2}{4} \sum_{i}
\left(h^\prime_i\right)^2\right)\, \exp\left[\, -2U\,\right]
\label{paireinst}
\end{eqnarray}
 from which we can eliminate the contribution of the vector fields
 and obtain an equation involving only the scalar fields and the metric:
 \begin{equation}
 U^{\prime\prime} +\frac{2}{r} U^\prime - \left( U^\prime \right)^2 - \frac{1}{4} \sum_{i}
\left(h^\prime_i\right)^2 = 0
\label{ceccus}
\end{equation}
From {\sl the dilaton equations} in \eqn{campequa} we get the three
ordinary differential equations:
\begin{equation}
h^{\prime\prime}_i +\frac{2}{r} h^\prime_i  =
\frac{1}{\alpha^2} \sum_{\alpha} \, \Lambda^\alpha_i \, \exp \left[
-2\Lambda^\alpha \cdot h +2 U\right] \, \left[ \left(q^\alpha \right)^2 -
\left(p^\alpha \right)^2 \right]\frac{1}{r^4}
\label{dilequa}
\end{equation}
Finally {\sl form the Maxwell equations} in \eqn{campequa} we obtain:
\begin{equation}
0 =  \frac{d}{dr} \, \left (
\exp\left[-2\Lambda_\alpha \cdot h \right] \, q^\alpha (r)
\right )
\label{maxwelequa}
\end{equation}
\subsection{The first order equations from the projection ${\bf (1,1,15)}\oplus
{\bf (\bar 1,\bar 1,\bar {15})}$ }
If we reconsider the general form of eq.s \eqn{pappaequa}, \eqn{Ueq},
we find that out of these 32 equations $24$ are identically satisfied when the fields are restricted to
be non--zero only in the chosen sector.
The remaining $8$ non trivial equations take the following form:
\begin{eqnarray}
\label{nontriv15}
0& =& {{c\,{\frac{\bf e^U}{r^2}}\,
     \left( {e^{{{{\sqrt{2}}\,{\bf h_2} - {\bf h_7}}\over
             {{\sqrt{3}}}}}}\,{\bf p^{18}} +
       {e^{{{{\sqrt{2}}\,{\bf h_1} + {\bf h_7}}\over
             {{\sqrt{3}}}}}}\,{\bf p^{23}} -
       {e^{{{{\sqrt{2}}\,{\bf h_1} - {\bf h_7}}\over
             {{\sqrt{3}}}}}}\,{\bf q^{17}} +
       {e^{{{{\sqrt{2}}\,{\bf h_2} + {\bf h_7}}\over
             {{\sqrt{3}}}}}}\,{\bf q^{24}} \right) }  }\nonumber\\
0&=&{{c\,{\frac{\bf e^U}{r^2}}\, \left( {e^{{{{\sqrt{2}}\,{\bf h_1} -
              {\bf h_7}}\over {{\sqrt{3}}}}}
          }\,{\bf p^{17}} +
       {e^{{{{\sqrt{2}}\,{\bf h_2} + {\bf h_7}}\over {{\sqrt{3}}}}}
          }\,{\bf p^{24}} -
       {e^{{{{\sqrt{2}}\,{\bf h_2} -
              {\bf h_7}}\over {{\sqrt{3}}}}}
          }\,{\bf q^{18}} +
       {e^{{{{\sqrt{2}}\,{\bf h_1} +
              {\bf h_7}}\over {{\sqrt{3}}}}}
          }\,{\bf q^{23}} \right) }  } \nonumber \\
  0&=&{{c\,{\frac{\bf e^U}{r^2}}\,
     \left( {e^{{{{\sqrt{2}}\,{\bf h_2} -
              {\bf h_7}}\over {{\sqrt{3}}}}}
          }\,{\bf p^{18}} -
       {e^{{{{\sqrt{2}}\,{\bf h_1} +
              {\bf h_7}}\over {{\sqrt{3}}}}}
          }\,{\bf p^{23}} -
       {e^{{{{\sqrt{2}}\,{\bf h_1} -
              {\bf h_7}}\over {{\sqrt{3}}}}}
          }\,{\bf q^{17}} -
       {e^{{{{\sqrt{2}}\,{\bf h_2} +
              {\bf h_7}}\over {{\sqrt{3}}}}}
          }\,{\bf q^{24}} \right) }  } \nonumber \\
 0&=&{{c\,{\frac{\bf e^U}{r^2}}\,
     \left( {e^{{{{\sqrt{2}}\,{\bf h_1} -
              {\bf h_7}}\over {{\sqrt{3}}}}}
          }\,{\bf p^{17}} -
       {e^{{{{\sqrt{2}}\,{\bf h_2} +
              {\bf h_7}}\over {{\sqrt{3}}}}}
          }\,{\bf p^{24}} -
       {e^{{{{\sqrt{2}}\,{\bf h_2} -
              {\bf h_7}}\over {{\sqrt{3}}}}}
          }\,{\bf q^{18}} -
       {e^{{{{\sqrt{2}}\,{\bf h_1} +
              {\bf h_7}}\over {{\sqrt{3}}}}}
          }\,{\bf q^{23}} \right) }  } \nonumber \\
{\bf h^\prime_7}&=&
 {{ \frac{c}{2} \,{\frac{\bf e^U}{r^2}}\,\left( {e^{{{{\sqrt{2}}\,{\bf h_2} -
              {\bf h_7}}\over {{\sqrt{3}}}}}}\,{\bf p^{18}} -
       {e^{{{{\sqrt{2}}\,{\bf h_1} +
              {\bf h_7}}\over {{\sqrt{3}}}}}
          }\,{\bf p^{23}} +
       {e^{{{{\sqrt{2}}\,{\bf h_1} - {\bf h_7}}\over {{\sqrt{3}}}}}}\,
       {\bf q^{17}} + {e^{{{{\sqrt{2}}\,{\bf h_2} +
              {\bf h_7}}\over {{\sqrt{3}}}}}
          }\,{\bf q^{24}} \right) }  } \nonumber \\
\left( {\bf h^\prime_1} - {\bf h^\prime_2} \right) &=&
 {{ \frac{c}{{\sqrt{2}}} \,{\frac{\bf e^U}{r^2}}\,\left( {e^{{{{\sqrt{2}}\,{\bf h_1} -
              {\bf h_7}}\over {{\sqrt{3}}}}}
          }\,{\bf p^{17}} +
       {e^{{{{\sqrt{2}}\,{\bf h_2} + {\bf h_7}}\over {{\sqrt{3}}}}}
          }\,{\bf p^{24}} +
       {e^{{{{\sqrt{2}}\,{\bf h_2} -
              {\bf h_7}}\over {{\sqrt{3}}}}}
          }\,{\bf q^{18}} -
       {e^{{{{\sqrt{2}}\,{\bf h_1} +
              {\bf h_7}}\over {{\sqrt{3}}}}}
          }\,{\bf q^{23}} \right) }  } \nonumber \\
\left( {\bf h^\prime_1} + {\bf h^\prime_2} \right) &=&
 {{ \frac{c}{{\sqrt{2}}} \,{\frac{\bf e^U}{r^2}}\,\left( {e^{{{{\sqrt{2}}\,{\bf h_2} -
              {\bf h_7}}\over {{\sqrt{3}}}}}
          }\,{\bf p^{18}} + {e^{{{{\sqrt{2}}\,{\bf h_1} +
              {\bf h_7}}\over {{\sqrt{3}}}}}
          }\,{\bf p^{23}} +
       {e^{{{{\sqrt{2}}\,{\bf h_1} -
              {\bf h_7}}\over {{\sqrt{3}}}}}
          }\,{\bf q^{17}} -
       {e^{{{{\sqrt{2}}\,{\bf h_2} +
              {\bf h_7}}\over {{\sqrt{3}}}}}
          }\,{\bf q^{24}} \right) }  } \nonumber \\
\frac{dU}{dr}  &=&  - \frac{k}{4\sqrt{2}}{{{\frac{\bf e^U}{r^2}}\,\left( {e^{{{{\sqrt{2}}\,{\bf h_1} -
                       {\bf h_7}}\over {{\sqrt{3}}}}}}\,{\bf p^{17}} -
       {e^{{{{\sqrt{2}}\,{\bf h_2} + {\bf h_7}}\over
             {{\sqrt{3}}}}}}\,{\bf p^{24}} +
       {e^{{{{\sqrt{2}}\,{\bf h_2} - {\bf h_7}}\over
             {{\sqrt{3}}}}}}\,{\bf q^{18}} +
       {e^{{{{\sqrt{2}}\,{\bf h_1} + {\bf h_7}}\over
             {{\sqrt{3}}}}}}\,{\bf q^{23}} \right) }  }  \nonumber \\
\end{eqnarray}
where we have defined the coefficient
\begin{equation}
c \equiv \frac{b}{ 18 \times \sqrt{2} \, \pi }
\end{equation}
$b/a, k$ being the relative coefficient between the left and right hand
side of equations \eqn{gratra}, \eqn{BPS56eq1} respectively,
which are completely fixed by the supersymmetry
transformation rules of the $N=8$ theory \eqn{trasforma}.

From the homogeneous equations of the first order system we get:
\begin{eqnarray}
q_{17}(r)\,\exp\left[-{\vec \Lambda}_{17}\cdot {\vec h}\right]\, &=&\,
p_{18}\,\exp\left[-{\vec \Lambda}_{18}\cdot {\vec h}\right]\nonumber\\
q_{18}(r)\, \exp\left[-{\vec \Lambda}_{18}\cdot {\vec h}\right]\, &=&\,
p_{17}\, \exp\left[-{\vec \Lambda}_{17}\cdot {\vec h}\right]\nonumber\\
q_{23}(r)\, \exp\left[-{\vec \Lambda}_{23}\cdot {\vec h}\right]\, &=&\,
-p_{24}\, \exp\left[-{\vec \Lambda}_{24}\cdot {\vec h}\right]\nonumber\\
q_{24}(r)\, \exp\left[-{\vec \Lambda}_{24}\cdot {\vec h}\right]\, &=&\,
-p_{23}\, \exp\left[-{\vec \Lambda}_{23}\cdot {\vec h}\right]
\label{qpeq}
\end{eqnarray}
Then, from the Maxwell equations we get:
\begin{equation}
q_{\alpha}(r)\, =\, A_{\alpha} \, \exp\left[2{\vec \Lambda}_{\alpha}\cdot {\vec h}\right]
\end{equation}
where $A_\alpha$ are integration constants.
By substituting these into the inhomogeneous first order equations one obtains:
\begin{eqnarray}
q_{17}\, &=&\, q_{24}\, =\, p_{18}\, =\, p_{23}\, =\, 0\nonumber\\
h_{1}^{\prime}\, &=&\, -h_{2}^{\prime}\nonumber\\
h_{7}^{\prime}\, &=&\, 0
\end{eqnarray}
Introducing the field:
\begin{equation}
\phi =\sqrt{\frac{2}{3}}h_1-\frac{1}{\sqrt{3}}h_7
\end{equation}
so that:
\begin{eqnarray}
h_1\, &=&\, \sqrt{\frac{3}{2}}\left(\phi +\frac{1}{\sqrt{3}}h_7\right)\nonumber\\
h_2\, &=&\, -\sqrt{\frac{3}{2}}\left(\phi +\frac{1}{\sqrt{3}}h_7-\log{B}\right)
\nonumber\\
\phi^{\prime}\, &=&\, \sqrt{\frac{2}{3}}h_1^{\prime}\, =\, -\sqrt{\frac{2}{3}}h_2^{\prime}
\label{unosolo}
\end{eqnarray}
where $B$ is an arbitrary constant, the only independent first order equations become:
\begin{equation}
\phi^{\prime}\, =\,\frac{{c}}{\sqrt{3}}\frac{{\bf e^U}}{r^2}
\left(p_{17} e^{\phi}\, +\, p_{24} B e^{-\phi}\right)
\label{eqphi}
\end{equation}
\begin{equation}
U^\prime =\,- \frac{k}{2 \sqrt{2}}\frac{{\bf e^U}}{r^2}
\left(p_{17} e^{\phi}\, -\, p_{24} B e^{-\phi}\right)
\label{equ}
\end{equation}
and correspondingly the second order scalar field equations become:
\begin{eqnarray}
\phi^{\prime \prime}+\frac{2}{r}\phi^{\prime}\, &=&\, \frac{1}{3 \alpha^2 \pi^2}
\left(p_{17}^2 e^{2\phi}-p_{24}^2 B^2 e^{-2\phi}\right)\\
h^{\prime \prime}_7+\frac{2}{r}h^{\prime}_7\, &=&\, 0\\
\label{scaleveronesi}
\end{eqnarray}

The system of first and second order differential equations given by eqn.
\eqn{eqphi}, \eqn{equ},  the Einstein
equations  \eqn{paireinst} and the scalar fields equations
\eqn{scaleveronesi} can now be solved and gives:
\begin{eqnarray}
\phi &=& - \frac{1}{2} {\mbox {log}}\left(1+ \frac{b}{r}\right) +
\frac{1}{2} {\mbox {log}}\left(1+ \frac{d}{r}\right)
\label{phi}  \\
U &=& - \frac{1}{2} {\mbox {log}}\left(1+ \frac{b}{r}\right) -
\frac{1}{2} {\mbox {log}}\left(1+ \frac{d}{r}\right)
+  {\mbox {log}}A
\label{u}
\end{eqnarray}
 with:
 \begin{equation}
b= - \frac{1}{\pi \sqrt{2}}p_{17} \quad ; \quad d= - \frac{1}{\pi \sqrt{2}}Bp_{24}
\end{equation}
fixing at the same time the coefficients (which could be
alternatively fixed with supersymmetry techniques):

\begin{eqnarray}
\alpha^2\, &=&\, \frac{4}{3}\nonumber\\
c\, &=&\, -\frac{\sqrt{3}}{2}\nonumber\\
k \, &=& \, \sqrt{2}
\label{susi}
\end{eqnarray}
\par
This concludes our discussion of the simplified model.
\vskip 5mm

We can now identify the simplified $N=8$ model (reduced to the Cartan
subalgebra) that we have studied with a class of black holes well
studied in the literature. These are the black--hole generating
solutions of the heterotic string compactified on a six torus.
As described in \cite{extrortin}, these heterotic black--holes can be
found as solutions of the following truncated action:
\begin{eqnarray}
S^{het} &=& \int \, d^4 x \, \sqrt{-g} \, \, \Biggl \{
2 \, R + 2 \bigl [ (\partial \phi)^2 +(\partial\sigma)^2 +(\partial\rho)^2
\bigr ] \nonumber \\
&& -\frac{1}{4} \, e^{-2\phi} \, \Bigl [ e^{-2(\sigma+\rho) } (F_1)^2
 + e^{-2(\sigma-\rho) } (F_2)^2 \nonumber \\
&& + e^{ 2(\sigma+\rho) } (F_1)^2 + e^{2(\sigma-\rho) } (F_2)^2 \Bigr ]  \Biggr \}
\label{ortinact}
\end{eqnarray}
and were studied in \cite{12ref},\cite{13ref},\cite{14ref}. The
truncated action \eqn{ortinact} is nothing else but our truncated
action \eqn{reduclagra}. The translation vocabulary is given by:
\begin{equation}
\begin{array}{rclcrcl}
h_1 &=& \frac{\sqrt{3}}{2}\, \left( \sigma - \phi\right) & ; &
F_{17} &=& F_{4} \\
h_2 &=& \frac{\sqrt{3}}{2}\, \left( -\sigma - \phi\right) & ; &
F_{18} &=& F_{1} \\
h_7 &=&  \sqrt{3} \, \rho & ; &
F_{23} &=& F_{3} \\
\null &\null &  \null  & \null &
F_{24} &=& F_{2} \\
\end{array}
\label{zingarelli}
\end{equation}
As discussed in \cite{extrortin} the extreme multi black--hole solutions to the
truncated action \eqn{ortinact} depend on four harmonic functions
$H_i({\vec x})$
and for a single black hole solution the four harmonic functions are
simply:
\begin{equation}
H_i = 1+ \frac{\vert k_i \vert }{r}
\end{equation}
which introduces four electromagnetic charges. These are the four
surviving charges $p_{17},p_{23},q_{18},q_{24}$ that we have found in
our BPS saturated solution. It was observed in
\cite{extrortin} that among the general extremal solutions of this
model only a subclass are BPS saturated states, but in the way we
have derived them, namely through the Killing spinor equation, we
have automatically selected the BPS saturated ones.
\par
To make contact with the discussion in \cite{extrortin} let us
introduce the following four harmonic functions:
\begin{equation}
\begin{array}{rclcrcl}
H_{17}(r) &=& 1+\frac{\vert g^{17} \vert }{r} &;& H_{24}(r) &=&
1+\frac{\vert g^{24} \vert }{r} \\
H_{18}(r) &=& 1+\frac{\vert e_{18} \vert }{r} &;& H_{23}(r) &=&
1+\frac{\vert e_{23} \vert }{r} \\
\end{array}
\end{equation}
where $g^{17},g^{24},e_{18},e_{23}$ are four real parameters.
Translating the extremal general solution of the lagrangian \eqn{ortinact}
(see eq.s (30) of \cite{extrortin} )
into our notations we can write it as follows:
\begin{equation}
\begin{array}{rcrl}
h_1(r) &=& -\frac{\sqrt{3}}{4} & \log \,
\left[H_{17}/H_{23}\right]
\\
h_2(r) &=& -\frac{\sqrt{3}}{4} & \log \, \left[ H_{24}/ H_{18}\right] \\
h_7(r) &=& -\frac{\sqrt{3}}{4} & \log \, \left[ H_{18} \, H_{24}/ H_{17} \, H_{23} \right]
\\
U(r) &=& -\frac{1}{4} & \log \left[ H_{17} \, H_{18} \, H_{23} \,
H_{24} \right] \\
q^{18}(r) &=& -e_{18} & H_{18}^{-2}\\
q^{23}(r) &=& -e_{23} & H_{23}^{-2} \\
p^{24}    &=& g^{24} &\null \\
p^{17}    &=& g^{17} &\null \\
\end{array}
\label{idecariche}
\end{equation}
and we see that indeed $e_{18} = -q^{18}(\infty)$, $e_{23}= -q^{23}(\infty)$
are the electric charges, while $g^{24}=p^{24}$, $g^{17}=g^{17}$ are
the magnetic charges for the general extremal black--hole solution.
\par
Comparing now eq.\eqn{zingarelli} with our previous result
\eqn{unosolo} we see that having enforced the Killing spinor
equation, namely the BPS condition, we have the restrictions:
\begin{equation}
 h_1+h_2 = \mbox{const} \quad ; \quad h_7 = \mbox{const}
\end{equation}
which yield:
\begin{equation}
H_{17}^{2} = H_{18}^{2} \quad ; \quad H_{23}^{2} = H_{24}^{2}
\end{equation}
and hence
\begin{equation}
  e_{18} = g^{17} \quad ; \quad e_{23} = g^{24}
\end{equation}
Hence the BPS condition imposes that the electric charges are
pairwise equal to the magnetic charges.

\section{Solvable Lie algebra representation}
\subsection{${\bf  U(1)\times SU(2) \times SU(6) \subset SU(8) \subset E_{7(7)}}$}
\label{su8ine7}
In order to make formulae \eqn{pappaequa} \eqn{22026eq},\eqn{Ueq} explicit and
in order to derive the solvable Lie algebra decompositions we are
interested in  a preliminary work  based  on standard Lie
algebra techniques.
\par
The  ingredients that we have already tacitly used in the previous sections
and that are needed for a thoroughful discussion of the solavable Lie
algebra splitting in \eqn{7in3p4} are:
\begin{enumerate}
\item{The explicit listing of all the positive roots of the $E_{7(7)}$
Lie algebra}
\item{The explicit listing of all the weight vectors of the fundamental
${\bf 56}$ representation of $E_{7(7)}$}
\item{The explicit construction of the $56 \times 56$ matrices
realizing the 133 generators of $E_{7(7)}$ real Lie algebra in the fundamental
representation}
\item{The canonical Weyl--Cartan decomposition of the $SU(8)$
maximally compact subalgebra of $E_{7(7)}$. This involves the
construction of a Cartan subalgebra of $A_7$ type made out of
$E_{7(7)}$ step operators and the construction of all $A_7$
step operators also in terms of suitable combinations of
$E_{7(7)}$ step operators.}
\item{The determination of the embedding ${\bf SU(2)\times U(6)\subset
SU(8)}\subset E_{7(7)}$.}
\item{The decomposition of the maximal non compact subspace
$\IK\subset E_{7(7)}$
with respect to ${\bf U(1)} \times {\bf SU(2)\times SU(6)}$: $${\bf 70}\rightarrow {\bf
(1,1,15)}\oplus {\bf\bar{(1,1,15)}}\oplus {\bf (1,2,20)}$$}
\item{Using the ${\bf 56}$ representation of $E_{7(7)}$ in the {\bf Usp}-basis,
the construction of the subalgebra ${\bf SO^{*}(12)}$ }
\end{enumerate}
The work-plan described in the above points has been completed by
means of a computer programme written in  MATHEMATICA \cite{Wolfram}.
In the present section we just outline the logic of our calculations
and we describe the results that are summarized in various tables
in appendix E. In particular we explain the method to
generate the matrices of the ${\bf 56}$ representation whose explicit form is
the basic tool of our calculations.
\par
\subsection{Roots and Weights and the fundamental representation of $E_{7(7)}$}
\label{pesanti}
Let us begin with the construction of the fundamental representation
of the U--duality group. \par
Let us adopt the decomposition of the $63$--dimensional
positive part $\Phi^+(E_7)$ of the $E_7$ root space described in \eqn{decompo}.
The filtration \eqn{decompo} provides indeed a convenient way to
enumerate the $63$ positive roots which in Appendix D
are paired in one--to--one way with the massless bosonic fields of
compactified string theory (for instance the TypeIIA theory).
We name the roots as follows:
\begin{equation}
 {\vec \alpha }_{i,j} \, \in \, \ID_i \quad ; \quad \cases { i=1,
 \dots , 6 \cr
 j=1,\dots , \mbox{dim} \, \ID_i \cr }
 \label{filtroname}
\end{equation}
Each positive root can be decomposed along a basis of simple roots
$\alpha_\ell$ (i=1,\dots, 7):
\begin{equation}
{\vec \alpha }_{i,j} = n_{i,j}^\ell \, \alpha_\ell  \, \quad  n_{i,j}^\ell \in \ZZ
\end{equation}
It turns out that as simple roots we can choose:
\begin{equation}
\begin{array}{rclcrclcrcl}
\alpha_1 & = & {\vec \alpha }_{6,2} & ; & \alpha_2 & = & {\vec \alpha }_{5,2}
& ; & \alpha_3 & = & {\vec \alpha }_{4,2} \\
\alpha_4 & = & {\vec \alpha }_{3,2} & ; & \alpha_5 & = & {\vec \alpha }_{2,2}
& ; & \alpha_6 & = & {\vec \alpha }_{2,1} \\
\alpha_7 & = & {\vec \alpha }_{1,1} & \null & \null & \null & \null
& \null & \null & \null & \null \\
\end{array}
\label{simplerut}
\end{equation}
Having fixed this basis, each root is intrinsically identified by its
Dynkin labels, namely by its integer valued components in the basis
\eqn{simplerut}. The listing of all positive roots is given in
table \ref{dideals} were we give their name \eqn{filtroname} according to the abelian
ideal filtration, their Dynkin labels and the correspondence with massless
fileds in a TypeIIA toroidal compactification.
\par
Having identified the roots, the next step for the construction of
real fundamental representation $SpD(56)$  of our
U--duality Lie algebra $E_{7(7)}$ is the knowledge of
the corresponding weight vectors ${\vec W}$.\par
A particularly relevant property of the maximally non--compact
real sections of a simple complex Lie algebra is that all
its irreducible representations are real.  $E_{7(7)}$ is the
maximally non compact real section of the complex Lie algebra $E_7$, hence
all its irreducible representations $\Gamma$  are real.
This implies that if an element of the  weight lattice ${\vec W} \, \in \, \Lambda_w$ is
a weight of a given irreducible representation
${\vec W}\in \Gamma$ then also its negative is a weight of the
same representation: $-{\vec W}\in \Gamma$. Indeed changing sign to
the weights corresponds to complex conjugation.
\par
According to standard Lie algebra lore
every irreducible representation of a simple Lie algebra $\IG$ is identified by a unique
{\it highest} weight ${\vec W}_{max}$. Furthermore all weights can be expressed as
integral non--negative linear  combinations of the {\it simple}
weights ${\vec W}_{\ell}\,(\ell=1,...,r=\mbox{rank}(\IG)) $, whose components
are named the Dynkin labels of the weight.
The simple weights ${\vec W}_{i}$ of $\IG$ are the generators of the
dual lattice to the root lattice and are defined by the condition:
\begin{equation}
\frac{2 ({\vec W}_{i}\, ,\, {\vec \alpha}_{j})}{({\vec \alpha}_{j}\, ,\,
{\vec \alpha}_{j})}=\delta_{ij}
\end{equation}
In the simply laced $E_{7(7)}$ case, the previous equation simplifies as follows
\begin{equation}
({\vec W}_{i}\, ,\, {\vec \alpha}_{j})=\delta_{ij}
\label{simw}
\end{equation}
where ${\vec \alpha}_{j}$ are the the simple roots.
Since we are no more interested in the geometrical description of
dimensional reduction on tori, as we were in Chapter 2, for the sake
of our present analysis, it is more
convenient to use a
labeling for the simple roots in the Dynkin diagram of
$E_{7(7)}$ (see Figure \ref{stande7}) 
different from that represented in Figure \ref{standar}.
With this convention and using eq.\eqn{simplerut} and table \ref{dideals},
 from eq.\eqn{simw} we can easily
obtain the explicit expression of the simple weights.
\iffigs
\begin{figure}
\caption{$E_7$ Dynkin diagram}
\label{stande7}
\epsfxsize = 10cm
\epsffile{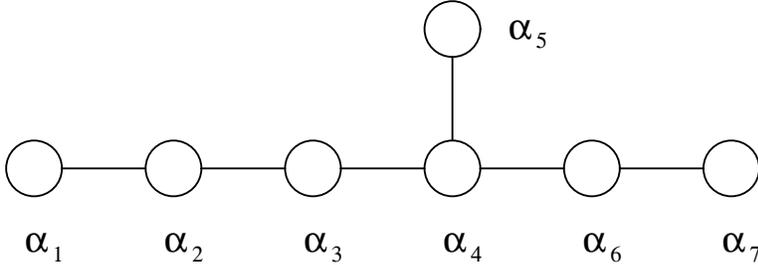}
\vskip -0.1cm
\unitlength=1mm
\end{figure}
\fi
The Dynkin labels of the highest weight of an irreducible
representation $\Gamma$ gives the Dynkin labels of the
representation. Therefore the representation is usually denoted by
$\Gamma[n_1,...,n_{r}]$. All the weights ${\vec W}$ belonging
to the representation $\Gamma$ can be described by $r$ integer non--negative numbers
$q^\ell$ defined by the following equation:
\begin{equation}
{\vec W}_{max}-{\vec W}=\sum_{\ell=1}^{r}q^\ell{\vec \alpha}_{\ell}
\label{qi}
\end{equation}
where $\alpha_\ell$ are the simple roots.
According to this standard formalism the fundamental real representation $SpD(56)$
of $E_{7(7)}$ is $\Gamma[1,0,0,0,0,0,0]$
and the expression of its weights in terms of $q^\ell$ is given in table
\ref{e7weight}, the highest weight being ${\vec W}^{(51)}$.
\par
We can now explain the specific ordering of the weights we have
adopted.
\par
First of all we have separated the $56$ weights in two
groups of $28$ elements so that the first group:
\begin{equation}
{\vec \Lambda}^{(n)}={\vec W}^{(n)} \quad n=1,...,28
\label{elecweight}
\end{equation}
are the weights for  the irreducible {\bf 28} dimensional representation of the
{\sl electric} subgroup {\bf $SL(8,\IR) \subset E_{7(7)}$}.
The remaining group of $28$ weight vectors are the weights for the
transposed representation of the same group that we name ${\bf \bar{28}}$.
\par
Secondly the $28$ weights ${\vec \Lambda}$
have been arranged according to the decomposition with respect to the
T--{\it duality} subalgebra $SO(6,6)\subset E_7(7)$:  the first $16$
correspond to R--R vectors and are the weights of the spinor
representation of $SO(6,6)$ while the last $12$ are associated with N--S fields
and correspond to the weights of the vector representation of $SO(6,6)$.
\par
Eq.\eqn{elecweight} makes explicit the adopted labeling for the
electric gauge fields $A_\mu^{{\vec \Lambda}  }$ and their field
strenghts $F_{\mu\nu}^{{\vec \Lambda}}$ adopted throughout the
previous sections.
\par
Equipped with the weight vectors we can now proceed to the explicit
construction of the ${\bf SpD(56)}$ representation
of $E_{7(7)}$. In our construction the basis vectors
are the $56$ weights, according to the enumeration of table \ref{e7weight}.
What we need are the $56\times 56$ matrices associated with the  $7$
Cartan generators $H_{{\vec \alpha}_i}$ ($i=1,\dots , 7$) and with
the $126$ step operators $E^{\vec \alpha}$ that are defined by:
\begin{eqnarray}
\left[ SpD_{56}\left( H_{{\vec \alpha}_i} \right)\right]_{nm}&  \equiv &  \langle
{\vec W}^{(n)} \vert \,  H_{ {\vec \alpha}_i } \,\vert {\vec W}^{(m)}
\rangle \nonumber\\
\left[ SpD_{56}\left( E^{ {\vec \alpha} } \right)\right]_{nm}&  \equiv &  \langle
{\vec W}^{(n)} \vert \,  E^{ {\vec \alpha} }  \,\vert {\vec W}^{(m)}
\rangle
\label{sp56defmat}
\end{eqnarray}
Let us begin with the Cartan generators. As a basis of the
Cartan subalgebra we use the generators $H_{\vec \alpha_i}$ defined by the
commutators:
\begin{equation}
\left[ E^{{\vec \alpha }_i}, E^{-{\vec \alpha }_i} \right] \, \equiv
\, H_{\vec \alpha_i}
\label{cartbadefi}
\end{equation}
In the $SpD(56)$ representation the corresponding matrices are
diagonal and of the form:
\begin{equation}
\langle {\vec W}^{(p)} \vert\, H_{\vec \alpha_{i}}\, \vert
{\vec W}^{(q)}\rangle\, =\,\left({\vec W}^{(p)},{\vec \alpha_{i}}\right)
\delta_{p\, q}\quad ; \quad ( p,q\, =\, 1,...,56)
\label{cartane7}
\end{equation}
The scalar products
\begin{equation}
\left({\vec \Lambda}^{(n)} \, \cdot \, {\vec h},
-{\vec \Lambda}^{(m)} \, \cdot \, {\vec h}\right)\, =\, \left({\vec W}^{(p)}
 \, \cdot \,
{\vec h}\right) \quad ; \quad(n,m=1,...,28\, ;\, p=1,...,56)
\end{equation}
appearing in the definition
\ref{LSpD} of the coset representative restricted to the Cartan fields, are
therefore to be understood in the following way:
\begin{equation}
{\vec W}^{(p)} \, \cdot \, {\vec h}\, =\, \sum_{i=1}^{7}\left({\vec W}^{(p)},
{\vec \alpha_{i}}\right)h^i
\end{equation}
The explicit form of these scalar products is given in table
\ref{scalaprod}  \par
Next we construct the matrices associated with the step operators. Here the first
observation is that it suffices to consider the positive roots. Indeed because
of the reality of the representation, the matrix associated with the
negative of a root is just the transposed of that associated with the
root itself:
\begin{equation}
E^{-\alpha} = \left[ E^\alpha \right]^T \, \leftrightarrow \, \langle
{\vec W}^{(n)} \vert \,  E^{- {\vec \alpha} }  \,\vert {\vec W}^{(m)} \rangle \, = \,
\langle {\vec W}^{(m)} \vert \,  E^{ {\vec \alpha} }  \,\vert {\vec W}^{(n)} \rangle
\label{transpopro}
\end{equation}
The method we have followed to obtain the matrices for all the
positive roots is that of constructing first the $56\times 56$
matrices for the step operators $E^{\vec \alpha_{\ell}}\, (\ell=1,...,7)$
associated with the simple roots and then generating all the others
through their commutators. The construction rules for the $SpD(56)$
representation of  the six operators $E^{\alpha_{\ell}} \, (\ell\neq 5)$
are:
\begin{equation}
\ell \, \neq \, 5 \quad \,
\Biggl \{ \matrix {\langle {\vec W}^{(n)} \vert\, E^{\vec \alpha_{\ell}}\, \vert
{\vec W}^{(m)}\rangle & = & \delta_{{\vec W}^{(n)},
{\vec W}^{(m)}+{\vec \alpha}_\ell} &;& n,m=1,\dots, 28 \cr
\langle {\vec W}^{(n+28)} \vert\, E^{\vec \alpha_{\ell}}\, \vert
{\vec W}^{(m+28)}\rangle & = & -\delta_{{\vec W}^{(n+28)},
{\vec W}^{(m+28)}+{\vec \alpha}_\ell} & ; & n,m=1,\dots, 28 \cr }
\label{repineq5}
\end{equation}
The six simple roots ${\vec \alpha_{\ell}} $ with $ \ell \neq 5$
belong also to the the Dynkin diagram of the electric subgroup {\bf SL(8,\IR)}
(see fig.\ref{compsu8}).  Thus their shift
operators have a block diagonal action on the {\bf 28} and ${\bf \bar{28}}$
subspaces of the $SpD(56)$ representation that are irreducible
under the electirc subgroup. Indeed from eq.\eqn{repineq5} we conclude
that:
\begin{equation}
\ell \, \neq \, 5 \quad \,SpD_{56}\left(E^{{\vec \alpha}_\ell} \right)=
\left(\matrix { A[{{\vec \alpha}_\ell}]
& {\bf 0} \cr {\bf 0} & - A^T[{{\vec \alpha}_\ell}] \cr} \right)
\label{spdno5}
\end{equation}
the $28 \times 28$ block  $A[{{\vec \alpha}_\ell}]$ being defined
by the first line of eq.\eqn{repineq5}.\par
On the contrary the operator $E^{\vec \alpha_{5}}$, corresponding to the only
root of the $E_7$ Dynkin diagram that is not also part of the $A_7$
diagram is represented by a matrix whose non--vanishing $28\times 28$ blocks
are off--diagonal. We have
\begin{equation}
SpD_{56}\left(E^{{\vec \alpha}_5} \right)=\left(\matrix { {\bf 0} & B[{{\vec \alpha}_5}]
\cr  C[{{\vec \alpha}_5}] & {\bf 0} \cr} \right)
\label{spdyes5}
\end{equation}
where both $B[{{\vec \alpha}_5}]=B^T[{{\vec \alpha}_5}]$ and
$C[{{\vec \alpha}_5}]=C^T[{{\vec \alpha}_5}]$ are symmetric $28
\times 28$ matrices. More explicitly the  matrix
 $SpD_{56}\left(E^{{\vec \alpha}_5} \right)$
is given by:
\begin{eqnarray}
&& \langle {\vec W}^{(n)} \vert\, E^{\vec \alpha_{5}}\, \vert
{\vec W}^{(m+28)}\rangle \, =\,  \langle {\vec W}^{(m)} \vert\,
E^{\vec \alpha_{5}}\, \vert {\vec W}^{(n+28)}\rangle \nonumber \\
&& \langle {\vec W}^{(n+28)} \vert\, E^{\vec \alpha_{5}}\, \vert
{\vec W}^{(m)}\rangle \, =\,  \langle {\vec W}^{(m+28)} \vert\,
E^{\vec \alpha_{5}}\, \vert {\vec W}^{(n)}\rangle
\label{sim5}
\end{eqnarray}
with
\begin{equation}
\begin{array}{rcrcrcrcrcr}
  \langle {\vec W}^{(7)} \vert\, E^{\vec \alpha_{5}}\, \vert
{\vec W}^{(44)}\rangle & =& -1 & \null &
  \langle {\vec W}^{(8)} \vert\, E^{\vec \alpha_{5}}\, \vert
{\vec W}^{(42)}\rangle & = & 1  & \null &
   \langle {\vec W}^{(9)} \vert\, E^{\vec \alpha_{5}}\, \vert
{\vec W}^{(43)}\rangle & = & -1  \\
   \langle {\vec W}^{(14)} \vert\, E^{\vec \alpha_{5}}\, \vert
{\vec W}^{(36)}\rangle & = & 1 & \null &
   \langle {\vec W}^{(15)} \vert\, E^{\vec \alpha_{5}}\, \vert
{\vec W}^{(37)}\rangle & = & -1 & \null &
 \langle {\vec W}^{(16)} \vert\, E^{\vec \alpha_{5}}\, \vert
{\vec W}^{(35)}\rangle & = & -1  \\
   \langle {\vec W}^{(29)} \vert\, E^{\vec \alpha_{5}}\, \vert
{\vec W}^{(6)}\rangle  & = & -1 & \null &
   \langle {\vec W}^{(34)} \vert\, E^{\vec \alpha_{5}}\, \vert
{\vec W}^{(1)}\rangle  & = & -1 & \null &
  \langle {\vec W}^{(49)} \vert\, E^{\vec \alpha_{5}}\, \vert
{\vec W}^{(28)}\rangle & = & 1 \\
  \langle {\vec W}^{(50)} \vert\, E^{\vec \alpha_{5}}\, \vert
{\vec W}^{(27)}\rangle & = & -1 & \null &
   \langle {\vec W}^{(55)} \vert\, E^{\vec \alpha_{5}}\, \vert
{\vec W}^{(22)}\rangle & = &  -1 & \null &
   \langle {\vec W}^{(56)} \vert\, E^{\vec \alpha_{5}}\, \vert
{\vec W}^{(21)}\rangle & =& 1 \\
\end{array}
\label{sim5bis}
\end{equation}
In this way we have completed the construction of the $E^{{\vec \alpha}_\ell}$
operators associated with simple roots. For the matrices associated
with higher roots we just proceed iteratively in the following way.
As usual we organize the roots by height :
\begin{equation}
{\vec \alpha}=n^\ell \, {\vec \alpha}_\ell \quad \rightarrow
\quad \mbox{ht}\,{\vec \alpha} \, = \, \sum_{\ell=1}^{7} n^\ell
\label{altezza}
\end{equation}
and for the roots $\alpha_i + \alpha_j$ of height $\mbox{ht}=2$ we
set:
\begin{equation}
SpD_{56} \left( E^{ \alpha _i + \alpha _j} \right) \equiv \left[
SpD_{56}\left(E^{\alpha _i} \right) \, , \,
SpD_{56}\left(E^{\alpha _i} \right) \right] \quad ; \quad i<j
\label{alto2}
\end{equation}
Next for the roots of $\mbox{ht}=3$ that can be written as $\alpha_i
+ \beta $ where $\alpha_i$ is simple and $\mbox{ht}\, \beta\, =\, 2$
we write:
\begin{equation}
SpD_{56} \left( E^{ \alpha _i + \beta} \right) \equiv \left[
SpD_{56}\left(E^{\alpha _i} \right) \, , \,
SpD_{56}\left(E^{\beta} \right) \right]
\label{alto3}
\end{equation}
Obtained the matrices for the roots of $\mbox{ht}=3$ one proceeds in
a similar way for those of the next height and so on up to exhaustion
of all the $63$ positive roots.
\par
This concludes our description of the algorithm by means of which our
computer programme constructed all the $70$ matrices spanning the
solvable Lie algebra $Solv_7$ in the $SpD(56)$ representation. Taking
into account property \eqn{transpopro} once the representation of the
solvable Lie algebra is given, also the remaining $63$ operators
corresponding to negative roots are also given.
\par
The next point in our programme is the organization of the
maximal compact subalgebra $SU(8)$ in a canonical Cartan Weyl basis.
This is instrumental for a decomposition of the full algebra and of
the solvable Lie algebra in particular into irreducible
representations of the subgroup $U(1)\times SU(2) \times SU(6)$.
\subsection{Cartan Weyl decomposition of the maximal compact subalgebra $SU(8)$}
The Lie algebra $\IG$ of $E_{7(7)}$ is written, according to the Cartan decomposition,
in the form:
\begin{equation}
\IG=\IH\oplus \IK
\label{chicco}
\end{equation}
where $ \IH$ denotes its maximal {\sl compact} subalgebra (i.e. the Lie algebra of
${\bf SU(8)}$) and $\IK$ its maximal {\sl non-compact} subspace.
Starting from the knowledge of the $E_{7(7)}$ generators in the symplectic
$SpD(56)$ representation, for brevity now denoted  $H_{\alpha_{i}}\quad E^{\alpha}$,
the generators in $\IH$ and in $\IK$ are obtained from the following identifications:
\begin{eqnarray}
\IH&=&\{ E^{\alpha}-E^{-\alpha}\}=\{ E^{\alpha}-(E^{\alpha})^T\}\\
\IK&=&\{H_{\alpha_{i}};\quad E^{\alpha}+E^{-\alpha}\}=\{
H_{\alpha_{i}};\quad E^{\alpha}+(E^{\alpha})^T \}
\label{compnocomp}
\end{eqnarray}
In eq. \eqn{compnocomp} what is actually meant is that both $\IH$ and $\IK$
are the vector spaces generated by the linear combinations with
{\sl real} coefficients of the specified generators.
\par
In order to find out the generators belonging to
the subalgebra ${\bf U(1) \times SU(2)\times SU(6)}$  within $\IH$
the generators of $\IH$ have to be rearranged according to the canonical form of
the ${\bf SU(8)}$ algebra. This was achieved by first fixing seven commuting
matrices in $\IH$ to be the Cartan generators of ${\bf SU(8)}$ and then diagonalizing
with a computer programme their adjoint action over $\IH$. Their eigenmatrices were identified
with the shift operators of ${\bf SU(8)}$. In the sequel we will use
the following notation: $a$ will denote a generic root of $\IH$ of the form
$a=\pm\epsilon_i \pm \epsilon_j$, $E^a$ the corresponding shift operator,
${\cal H}_{a_i}$ the Cartan generator associated with the simple root $a_i$ and
$B^{\alpha_{i,j}}$ the compact combination $E^{\alpha_{i,j}}-E^{-\alpha_{i,j}}$ where
$ \alpha_{i,j} $ is the $j^{th}$ positive root in the $i^{th}$ abelian
ideal $\ID_i \quad i=1,...,6 $ of $ E_{7(7)} $, according to the enumeration
of table \ref{dideals}. \par
A basis ${\cal H}_i$ of Cartan operators  was chosen as follows:
\begin{eqnarray}
{\cal H}_1 &=&E^{{\vec \alpha}_{2,1} } - E^{-{\vec \alpha}_{2,1}}\nonumber \\
{\cal H}_2 &=&E^{{\vec \alpha}_{2,2} } - E^{-{\vec \alpha}_{2,2}}\nonumber \\
{\cal H}_3 &=&E^{{\vec \alpha}_{4,1} } - E^{-{\vec \alpha}_{4,1}}\nonumber \\
{\cal H}_4 &=&E^{{\vec \alpha}_{4,2} } - E^{-{\vec \alpha}_{4,2}}\nonumber \\
{\cal H}_5 &=&E^{{\vec \alpha}_{6,1} } - E^{-{\vec \alpha}_{6,1}}\nonumber \\
{\cal H}_6 &=&E^{{\vec \alpha}_{6,2} } - E^{-{\vec \alpha}_{6,2}}\nonumber \\
{\cal H}_7 &=&E^{{\vec \alpha}_{6,11} } - E^{-{\vec \alpha}_{6,11}}
\label{su8cartan}
\end{eqnarray}
The reason for this choice is that the seven roots:
\begin{equation}
\begin{array}{ccccc}
 \{1,2,2,2,1,1,0\}& \leftrightarrow & {\vec \alpha}_{6,1} & = &
 \epsilon _1 + \epsilon _2  \\
 \{1,0,0,0,0,0,0\}& \leftrightarrow & {\vec \alpha}_{6,2} & = &
 \epsilon _1 - \epsilon _2  \\
\{0,0,1,2,1,1,0\}& \leftrightarrow & {\vec \alpha}_{4,1} & = &
 \epsilon _3 + \epsilon _4  \\
\{0,0,1,0,0,0,0\}& \leftrightarrow & {\vec \alpha}_{4,1} & = &
 \epsilon _3 - \epsilon _4  \\
 \{0,0,0,0,1,0,0\}& \leftrightarrow & {\vec \alpha}_{2,1} & = &
 \epsilon _5 + \epsilon _6  \\
\{0,0,0,1,0,0,0\}& \leftrightarrow & {\vec \alpha}_{2,2} & = &
 \epsilon _5 -\epsilon _6  \\
\{1,2,3,4,2,3,2\}& \leftrightarrow & {\vec \alpha}_{6,11} & = &
 \sqrt{2} \, \epsilon_7 \\
\end{array}
\label{spiegcart}
\end{equation}
are all orthogonal among themselves as it is evident by the last
column of eq.\eqn{spiegcart} where $\epsilon_i$ denote the unit
vectors in a Euclidean $7$--dimensional space.
\par
The roots ${\vec a}$ were obtained by arranging into a vector the seven
eigenvalues associated with each ${\cal H}_i$ for a fixed
eigenmatrix $E^{\vec a}$. Following a very well known procedure,
the {\sl positive} roots were computed
as those vectors ${\vec a}$ that a positive projection along an arbitrarely fixed direction
(not parallel to any of them) and among them the simple roots ${\vec a}_i$ were
identified with the undecomposable ones \cite{gilmore}. Finally the Cartan
generator corresponding to a generic root ${\vec a}$ was worked out using
the following expression:
\begin{equation}
{\cal H}_{a}=a^j {\cal H}_{j}
\end{equation}
Once the generators of the ${\bf SU(8)}$ algebra were written in the
canonical form, the ${\bf U(1)\times}$ \hfill \break ${\bf SU(2)\times U(6)}$
subalgebra could be
easily extracted. Choosing ${\vec a}_1$ as the root of ${\bf SU(2)}$ and
${\vec a}_i \quad i=3,...,7 $ as the simple roots of ${\bf SU(6) }$, the ${\bf U(1)}$
generator was found to be the following combination of Cartan generators:
\begin{equation}
{\cal H}_{U(1)}= -3{\cal H}_{a_1}-6{\cal H}_{a_2}-5{\cal H}_{a_3}-4{\cal H}_{a_4}
-3{\cal H}_{a_5}-2{\cal H}_{a_6}-{\cal H}_{a_7}
\end{equation}
A suitable combination of the shift operators $E^{\vec a}$ allowed to define
the proper real compact form of the generators of ${\bf SU(8)}$,
denoted by $X^{a},\quad Y^{a}$. By definition, these latter fulfill the following
commutation rules:
\begin{eqnarray}
\left [{\cal H}_{i},X^{a}\right]&=&a^{i} Y^{a} \\
\left[{\cal H}_{i},Y^{a}\right]&=&-a^{i} X^{a} \\
\left[X^{a},Y^{a}\right]&=&a^{i} {\cal H}_{i}
\end{eqnarray}
In tables \ref{su8rutte}, \ref{su8rutte2}, \ref{su8rutte3}
the explicit expressions of the generators
$X^a$ and $Y^a$ are dislpayed as linear combinations of the step
operators $B^{i,j}$. These latter are labeled
according to the labeling of the $E_{7(7)}$ roots  as given in   table \ref{dideals}
where they are classified by the abelian ideal  filtration.
The labeling of the $SU(8)$ positive roots is the standard one
according to their height.
Calling ${\vec a}_1,\dots ,{\vec a}_7$ the simple roots, the full
set of the $28$ positive roots is the following one:
\begin{equation}
{\vec a}_{i,i+i,\dots,j-1,j} \, = \,{\vec a}_{i} +
 \,{\vec a}_{i+1} + \dots +{\vec a}_{j-1}+{\vec a}_{j}
 \quad ; \quad \forall 1\le i <j\le 7
 \label{allroots}
\end{equation}
We stress that the non--compact $A_7$ subalgebra $SL(8,\IR)$ of $E_{7(7)}$ is
regularly embedded, so that it shares the same Cartan subalgebra
and its roots are vectors in the same $7$--dimensional space as
the roots of $E_{7(7)}$. On the other hand the compact
$A_7$ subalgebra of $SU(8)$ is irregularly embedded  and its
Cartan subalgebra has actually intersection zero with the Cartan
subalgebra of $E_{7(7)}$. Hence the $SU(8)$ roots are vectors in
a $7$--dimensional totally different from the space where the
$E_{7(7)}$ roots live. Infact the Cartan generators of $SU(8)$
have been written as linear combinations of the step operators
$E_{7(7)}$. The difference is emphasized in fig.\ref{compsu8}
\iffigs
\begin{figure}
\caption{$A_7$ subalgebras of $E_{7(7)}$}
\label{compsu8}
\epsfxsize = 10cm
\epsffile{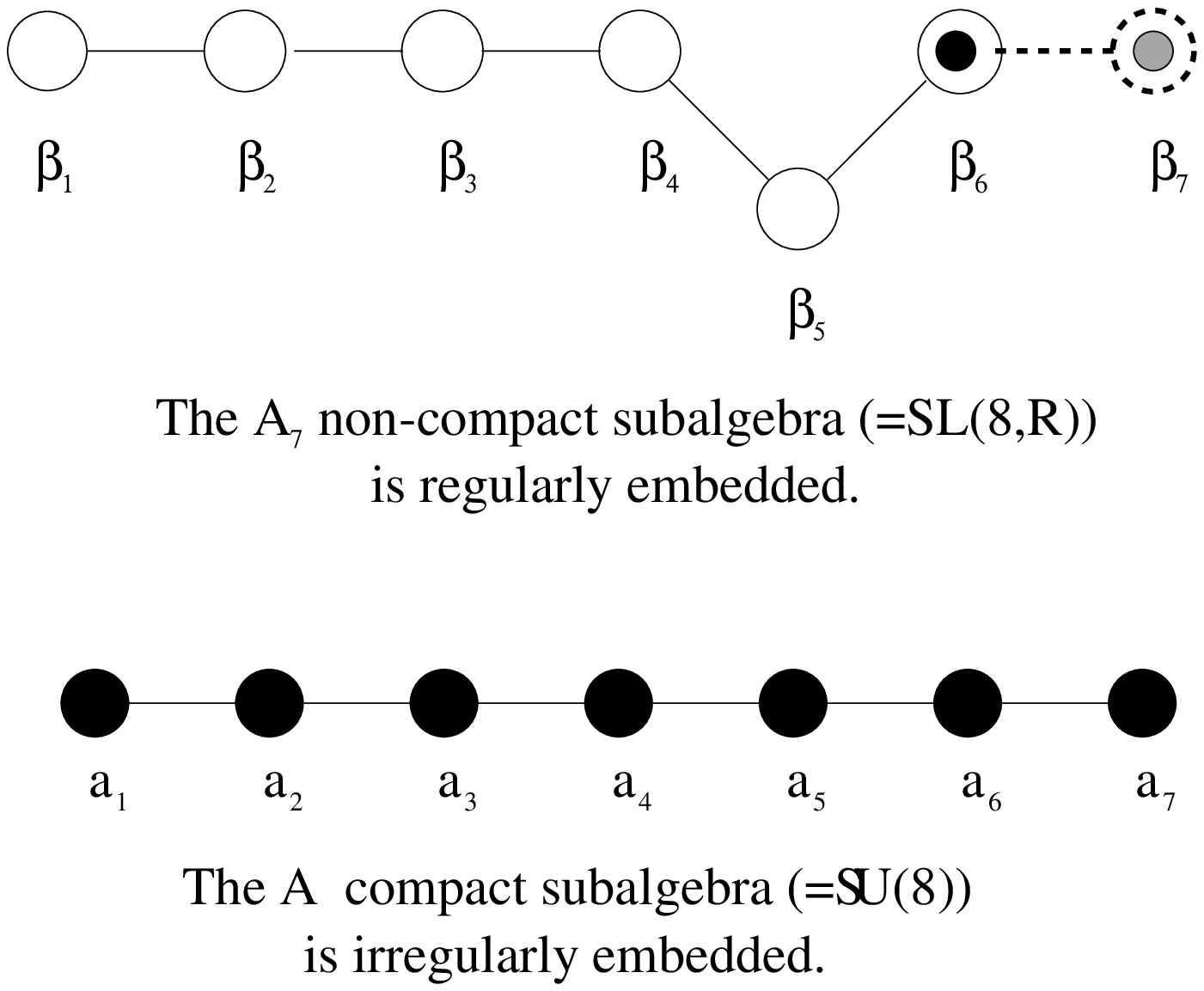}
\vskip -0.1cm
\unitlength=1mm
\end{figure}
\fi
\subsection{The ${\bf UspY(56)}$ basis for fundamental representation of
$E_{7(7)}$}
As outlined in the preceeding subsection, the generators of ${\bf U(1) \times SU(2)\times SU(6)}\subset SU(8)$
were found
in terms of the matrices $B^{\alpha_{ij}}= E^{\alpha_{ij}}-E^{-\alpha_{ij}} $
belonging to the real symplectic representation ${ \bf 56}$ of $E_{7(7)}$ ($SpD(56)$).
By the simultaneous diagonalization of ${\cal H}_{U(1)}$ and
the Casimir operator of ${\bf SU(2)}$.
it was then possible to  decompose
the  $SpD(56)$ with respect to ${\bf U(1)\times SU(2)\times SU(6)}$ ( i.e.
${ \bf 56}\rightarrow [(1,1,1)\oplus (1,1,15)\oplus (1,2,6)]\oplus
\bar{[...]}$). The eigenvector basis of this
decomposition provided the unitary symplectic representation ${ \bf UspY(56)}$
in which the first diagonal block has the standard form for the Young basis:
\begin{eqnarray}
T^{AB}_{CD}&=&\frac{1}{2} \delta^{[A}_{[C}q^{\phantom{D}B]}_{D]}\\
q^{\phantom{D}B}_{D}& \in &{\bf SU(8)} \quad A,...,D=1,...,8
\label{Yourepre}
\end{eqnarray}
the $8 \times 8$ matrix $q^{\phantom{D}B}_{D}$ being the fundamental
octet representation of the corresponding $SU(8)$
generator.
\par
Such a procedure amounts to the determination of the matrix ${\bf S}$
introduced in eq.\eqn{bfSmat}.
The explicit form of ${\bf S}$ is given below:
\begin{eqnarray}
& {\bf S}= & \nonumber
\end{eqnarray}
{\tiny
\begin{eqnarray}
&\frac{1}{\sqrt{2}}\left(\matrix{ 0 & 0 & 0 & 0 & 0 & 0 & 0 & 0 & 0 & 0 & 0 & 0 & 0 & 0 & 0 & 0 &
  {i\over {{\sqrt{2}}}} & {1\over {{\sqrt{2}}}} & 0 & 0 & 0 & 0 &
  {1\over {{\sqrt{2}}}} & {{-i}\over {{\sqrt{2}}}} & 0 & 0 & 0 & 0 \cr 0 & 0
   & 0 & 0 & 0 & 0 & 0 & 0 & 0 & 0 & 0 & 0 & 0 & 0 & 0 & 0 & -i & 0 & 0 & 0 &
  0 & 0 & 0 & -i & 0 & 0 & 0 & 0 \cr 0 & 0 & 0 & 0 & 0 & 0 & 0 & 0 & 0 & 0 & 0
   & 0 & 0 & 0 & 0 & 0 & -{1\over {{\sqrt{3}}}} & 0 & 0 & 0 & 0 & 0 &
  {{-2\,i}\over {{\sqrt{3}}}} & {1\over {{\sqrt{3}}}} & 0 & 0 & 0 & 0 \cr 0 &
  0 & 0 & 0 & 0 & 0 & 0 & 0 & 0 & 0 & 0 & 0 & 0 & 0 & 0 & 0 &
  -{1\over {{\sqrt{6}}}} & -i\,{\sqrt{{3\over 2}}} & 0 & 0 & 0 & 0 &
  {i\over {{\sqrt{6}}}} & {1\over {{\sqrt{6}}}} & 0 & 0 & 0 & 0 \cr 1 & 0 & 0
   & 0 & 0 & 0 & -1 & 0 & 0 & 0 & 0 & 0 & 0 & 0 & 0 & 0 & 0 & 0 & 0 & 0 & 0 &
  0 & 0 & 0 & 0 & 0 & 0 & 0 \cr 0 & 1 & i & 0 & 0 & 0 & 0 & 0 & 0 & 0 & 0 & 0
   & 0 & 0 & 0 & 0 & 0 & 0 & 0 & 0 & 0 & 0 & 0 & 0 & 0 & 0 & 0 & 0 \cr 0 & 0
   & 0 & 1 & 0 & 0 & 0 & 0 & 0 & 0 & 0 & 0 & 1 & 0 & 0 & 0 & 0 & 0 & 0 & 0 & 0
   & 0 & 0 & 0 & 0 & 0 & 0 & 0 \cr 0 & 0 & 0 & 0 & 1 & 0 & 0 & 0 & 0 & 0 & 0
   & -1 & 0 & 0 & 0 & 0 & 0 & 0 & 0 & 0 & 0 & 0 & 0 & 0 & 0 & 0 & 0 & 0 \cr 0
   & 0 & 0 & 0 & 0 & 1 & 0 & 0 & 0 & 0 & 0 & 0 & 0 & 0 & 0 & 1 & 0 & 0 & 0 & 0
   & 0 & 0 & 0 & 0 & 0 & 0 & 0 & 0 \cr 0 & 0 & 0 & 0 & 0 & 0 & 0 & 1 & 0 & 0
   & 0 & 0 & 0 & 0 & i & 0 & 0 & 0 & 0 & 0 & 0 & 0 & 0 & 0 & 0 & 0 & 0 & 0
   \cr 0 & 0 & 0 & 0 & 0 & 0 & 0 & 0 & 1 & 0 & 0 & 0 & 0 & i & 0 & 0 & 0 & 0
   & 0 & 0 & 0 & 0 & 0 & 0 & 0 & 0 & 0 & 0 \cr 0 & 0 & 0 & 0 & 0 & 0 & 0 & 0
   & 0 & 1 & i & 0 & 0 & 0 & 0 & 0 & 0 & 0 & 0 & 0 & 0 & 0 & 0 & 0 & 0 & 0 & 0
   & 0 \cr 0 & 0 & 0 & 0 & 0 & 0 & 0 & 0 & 0 & 0 & 0 & 0 & 0 & 0 & 0 & 0 & 0
   & 0 & 1 & 0 & 0 & 0 & 0 & 0 & i & 0 & 0 & 0 \cr 0 & 0 & 0 & 0 & 0 & 0 & 0
   & 0 & 0 & 0 & 0 & 0 & 0 & 0 & 0 & 0 & 0 & 0 & 0 & 1 & 0 & 0 & 0 & 0 & 0 & i
   & 0 & 0 \cr 0 & 0 & 0 & 0 & 0 & 0 & 0 & 0 & 0 & 0 & 0 & 0 & 0 & 0 & 0 & 0
   & 0 & 0 & 0 & 0 & 1 & 0 & 0 & 0 & 0 & 0 & i & 0 \cr 0 & 0 & 0 & 0 & 0 & 0
   & 0 & 0 & 0 & 0 & 0 & 0 & 0 & 0 & 0 & 0 & 0 & 0 & 0 & 0 & 0 & 1 & 0 & 0 & 0
   & 0 & 0 & i \cr 1 & 0 & 0 & 0 & 0 & 0 & 1 & 0 & 0 & 0 & 0 & 0 & 0 & 0 & 0
   & 0 & 0 & 0 & 0 & 0 & 0 & 0 & 0 & 0 & 0 & 0 & 0 & 0 \cr 0 & 1 & -i & 0 & 0
   & 0 & 0 & 0 & 0 & 0 & 0 & 0 & 0 & 0 & 0 & 0 & 0 & 0 & 0 & 0 & 0 & 0 & 0 & 0
   & 0 & 0 & 0 & 0 \cr 0 & 0 & 0 & 1 & 0 & 0 & 0 & 0 & 0 & 0 & 0 & 0 & -1 & 0
   & 0 & 0 & 0 & 0 & 0 & 0 & 0 & 0 & 0 & 0 & 0 & 0 & 0 & 0 \cr 0 & 0 & 0 & 0
   & 1 & 0 & 0 & 0 & 0 & 0 & 0 & 1 & 0 & 0 & 0 & 0 & 0 & 0 & 0 & 0 & 0 & 0 & 0
   & 0 & 0 & 0 & 0 & 0 \cr 0 & 0 & 0 & 0 & 0 & 1 & 0 & 0 & 0 & 0 & 0 & 0 & 0
   & 0 & 0 & -1 & 0 & 0 & 0 & 0 & 0 & 0 & 0 & 0 & 0 & 0 & 0 & 0 \cr 0 & 0 & 0
   & 0 & 0 & 0 & 0 & 1 & 0 & 0 & 0 & 0 & 0 & 0 & -i & 0 & 0 & 0 & 0 & 0 & 0 &
  0 & 0 & 0 & 0 & 0 & 0 & 0 \cr 0 & 0 & 0 & 0 & 0 & 0 & 0 & 0 & 1 & 0 & 0 & 0
   & 0 & -i & 0 & 0 & 0 & 0 & 0 & 0 & 0 & 0 & 0 & 0 & 0 & 0 & 0 & 0 \cr 0 & 0
   & 0 & 0 & 0 & 0 & 0 & 0 & 0 & 1 & -i & 0 & 0 & 0 & 0 & 0 & 0 & 0 & 0 & 0 &
  0 & 0 & 0 & 0 & 0 & 0 & 0 & 0 \cr 0 & 0 & 0 & 0 & 0 & 0 & 0 & 0 & 0 & 0 & 0
   & 0 & 0 & 0 & 0 & 0 & 0 & 0 & 1 & 0 & 0 & 0 & 0 & 0 & -i & 0 & 0 & 0 \cr 0
   & 0 & 0 & 0 & 0 & 0 & 0 & 0 & 0 & 0 & 0 & 0 & 0 & 0 & 0 & 0 & 0 & 0 & 0 & 1
   & 0 & 0 & 0 & 0 & 0 & -i & 0 & 0 \cr 0 & 0 & 0 & 0 & 0 & 0 & 0 & 0 & 0 & 0
   & 0 & 0 & 0 & 0 & 0 & 0 & 0 & 0 & 0 & 0 & 1 & 0 & 0 & 0 & 0 & 0 & -i & 0
   \cr 0 & 0 & 0 & 0 & 0 & 0 & 0 & 0 & 0 & 0 & 0 & 0 & 0 & 0 & 0 & 0 & 0 & 0
   & 0 & 0 & 0 & 1 & 0 & 0 & 0 & 0 & 0 & -i \cr  }\right ) &
\end{eqnarray}
}
\subsection{Weights of the compact subalgebra $SU(8)$ }
Having gained control over the embedding of the subgroup ${\bf U(1)\times SU(2)
\times SU(6)}$, let us now come back to the fundamental representation
of $E_{7(7)}$ and consider the further decomposition of its
${\bf 28}$ and ${\bf \bar{28}}$ components, irreducible with respect to
${\bf SU(8)}$, when we reduce this latter to its subgroup  ${\bf U(1)\times SU(2)
\times SU(6)}$.
In the unitary symplectic basis (either $UspD(56)$ or $UspY(56)$)
the general form of an $E_{7(7)}$ Lie algebra matrix is
 \begin{equation}
{\cal S} = \left(\matrix{ T & V \cr V^* & T^*} \right)
\label{genformal}
\end{equation}
where $T$ and $V$ are $28 \times 28$ matrices respectively
antihermitean and symmetric:
\begin{equation}
T= -T^\dagger \quad ; \quad V=V^T
\label{vtpro}
\end{equation}
The subgalgebra $SU(8)$ is represented by matrices where $V=0$.
Hence the subspaces corresponding to the first and second blocks of
$28$ rows are ${\bf 28}$ and ${\bf \bar{28}}$ irreducible
representations, respectively. Under the subgroup
${\bf U(1)\times SU(2) \times SU(6)}$ each blocks decomposes as
follows:
\begin{eqnarray}
\null & \null & \null \nonumber\\
{\bf 28} & \rightarrow & {\bf (1,1,1)\oplus (1,1,15)\oplus (1,2,6)}\nonumber \\
\null & \null & \null \nonumber\\
 {\bf \bar{28}}& \rightarrow & {\bf \bar{(1,1,1)}\oplus \bar{(1,1,15)}\oplus
\bar{(1,2,6)}} \nonumber\\
\null & \null & \null
\label{28decomp}
\end{eqnarray}
This decomposition corresponds to a rearrangement of the
${\vec \Lambda}^{(n)}={\vec W}^{(n)}$ (and therefore $-{\vec \Lambda}^{(n)}=
{\vec W}^{(n+28)}$) in
a new sequence of weights ${\vec \Lambda}^{\prime( n)}$
($-{\vec \Lambda}^{\prime( n)}$), defined in the following way:
\begin{eqnarray}
{\vec \Lambda}^{\prime( n^{\prime})}\, &=&\, {\vec \Lambda}^{(n)}\nonumber \\
n=1,...,28\leftrightarrow n^{\prime}&=&\cases{[7] \, \leftarrow \,  {\bf (1,1,1)}\cr
[5,20,26,16,13,28,1,22,4,19,25,21,6,27,12]  \, \leftarrow \,  {\bf (1,1,15)}\cr
 [9,14,2,17,23,3,24,18,8,15,10,11] \, \leftarrow \,  {\bf (1,2,6)}\cr }
\label{lambdaprime}
\end{eqnarray}
Denoting by ${\vec \Lambda}_i\,(i=1,...,7)$ the simple weights of
 ({\bf SU(8)}), defined by an equation analogous to \ref{simw},
it turns out that:
${\bf 28}=\Gamma[0,0,0,0,0,1,0]$ and ${\bf \bar{28}}=\Gamma[0,1,0,0,0,0,0]$.
Using the labeling \eqn{qi}, the weights ${\vec \Lambda}^{\prime(n)}$ for the {\bf 28}
representation of $SU(8)$ and the weights $-{\vec \Lambda}^{\prime(n)}$ for the
${\bf \bar{28}}$ representation, ordered according to
 the decomposition \ref{28decomp}, have the  form listed in table
 \ref{28weights} and \ref{28bweights}.
\par
In fig.\ref{su2u6} we show the structure of the $SU(8)$ Lie algebra
 elements in the ${\bf UspY(56)}$
 basis for the fundamental representation of $E_{7(7)}$.
\iffigs
\begin{figure}
\caption{$SU(2)\otimes U(6)$  and $SU(8)$  generators}
\label{su2u6}
\epsfxsize = 10cm
\epsffile{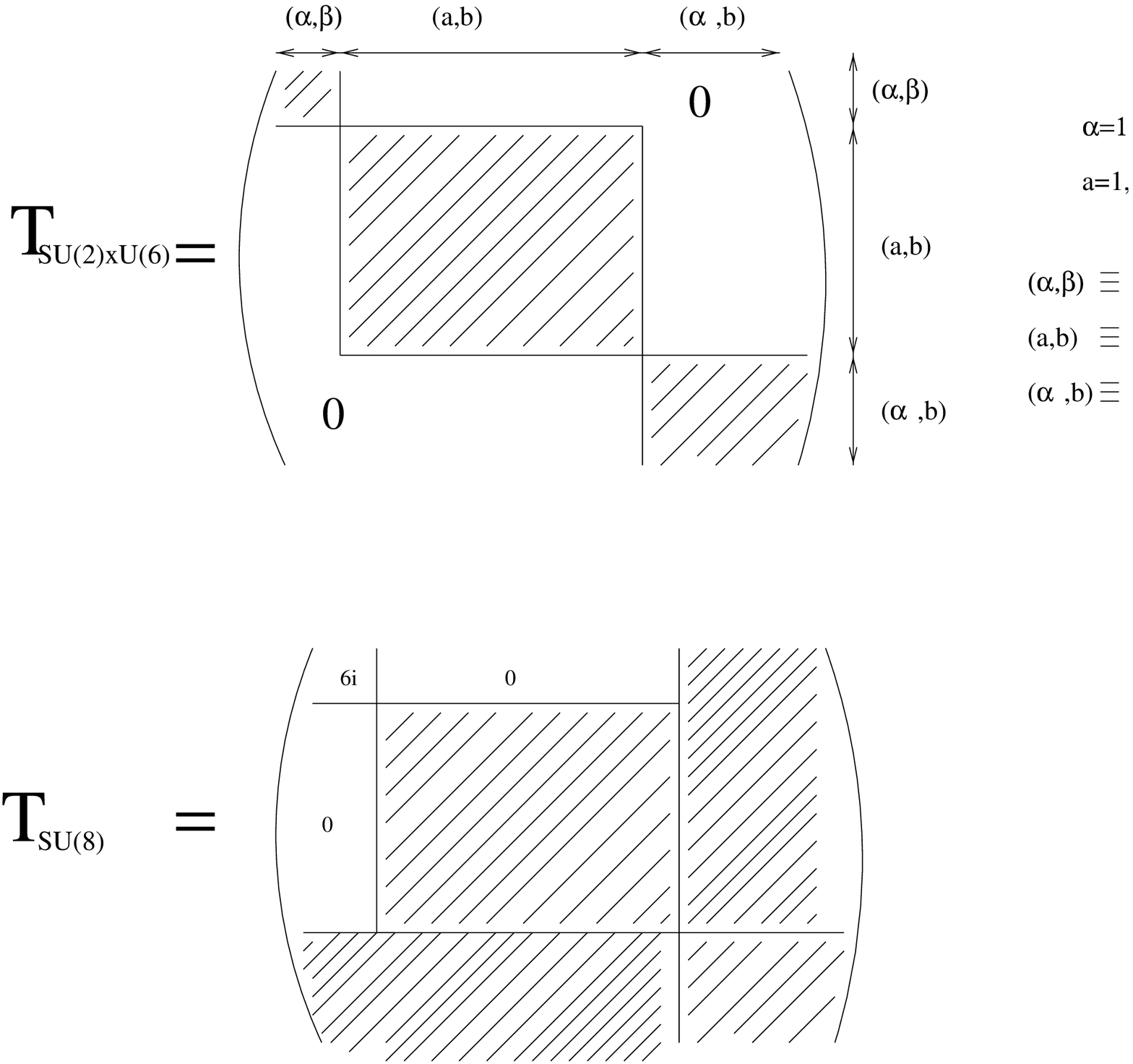}
\vskip -0.1cm
\unitlength=1mm
\end{figure}
\fi
\section{Solvable Lie algebra decompositions}
\label{solvodecompo}
In the present section, the construction of the $ SO^{\star}(12)$
and $ E_{6(2)}$ subalgebras of $E_{7(7)}$ will be discussed in detail.
The starting point of this analysis is eq. (\ref{7in3p4}).
This equality does not  uniquely define the embedding of  $ SO^{\star}(12)$
and $ E_{6(2)}$ into $E_{7(7)}$.
This embedding is  determined
by the requirement that the effective $N=2$ theories corresponding to a truncation of
the $N=8$ scalar manifold to either ${\cal M}_3\sim \exp Solv_3$
or ${\cal M}_4 \sim \exp Solv_4$, be obtained from $D=10 $ type IIA theory through
a compactification on a suitable Calabi--Yau manifold. This condition amounts
to imposing that the fields parametrizing $ Solv_3$ and $Solv_4$ should split
into R--R and N--S  in the following way:
\begin{eqnarray}
Solv_3\, &:& \quad {\bf 30}\rightarrow {\bf 18}\, (N-S) \, +\, {\bf 12} \,
(R-R)\nonumber \\
Solv_4\, &:& \quad {\bf 40}\rightarrow {\bf 20}\, (N-S) \, +\, {\bf 20} \,
(R-R)
\label{rrnssplit}
\end{eqnarray}
The above equations should correspond, according to a procedure defined in
(\cite{noialtri2}) and (\cite{witten}), to the decomposition of $ Solv_3$ and
$Solv_4$ with respect to the solvable algebra of the ST--duality group $O(6,6)
\,\otimes\,  SL(2,\IR)$, which is parametrized
by the whole set of N--S fields in  $D=4$ maximally extended supergravity.
The decomposition is the following one
\footnote{ For notational brevity in this section we use
$Solv \, G \equiv \, Solv \, G/H$, $H$ being the maximal compact subgroup
of $G$}:
\begin{eqnarray}
Solv\left( O(6,6)\,\otimes\, SL(2,\IR)\right)\,&=&\, Solv\left(SU(3,3)_1\right)
\, \oplus\, Solv\left(SU(3,3)_2\right)\, \oplus\,  Solv\left(SL(2,\IR)\right)
\nonumber \\
Solv_3\, &=&\, Solv\left(SU(3,3)_1\right)\, \oplus \, {\cal W}_{12}\nonumber \\
Solv_4\, &=&\,  Solv\left(SL(2,\IR)\right)\, \oplus \, Solv\left(SU(3,3)_2\right)\,
\oplus \, {\cal W}_{20}
\label{nsrrdecom}
\end{eqnarray}
where ${\cal W}_{12}$ and ${\cal W}_{20}$ consist of nilpotent generators
in $ Solv_3$ and
 $Solv_4$ respectively, describing R--R fields in the {\bf 12} and {\bf 20}
irreducible representations of $SU(3,3)$.
\subsection{Structure of $SO^\star(12)$ and $E_{6(2)}$ subalgebras of $E_{7(7)}
$ and some consistent $N=2$ truncations.}
\label{secsu33}
The subalgebras $SO^\star(12)$ and $E_{6(2)}$ in
$E_{7(7)}$ were explicitly constructed starting from their maximal compact subalgebras,
namely $U(6)$ and $ SU(2)\otimes SU(6)\subset \IH$, respectively.
The construction of the algebra $U(1)\otimes SU(2)\otimes SU(6)\subset SU(8)$
was discussed in section $5.2$. As mentioned in earlier sections, by diagonalizing the adjoint
action of $U(1)$ on the $70$--dimensional vector space
$\IK$, (see e.\eqn{chicco}) we could decompose it into
irreducible representations of $U(1)\otimes SU(2)\otimes SU(6)\subset SU(8)$,
namely:
\begin{equation}
\IK\, =\, \IK_{(1,1,15)}\, \oplus\, \IK_{{\bar{(1,1,15)}}}\, \oplus\,
\IK_{(1,2,20)}
\label{ncdec}
\end{equation}
The algebras $SO^\star(12)$ and $E_{6(2)}$ were then constructed as follows:
\begin{eqnarray}
SO^\star(12)\, &=&\, \IK_{(1,1,15)}\, \oplus\, \IK_{{\bar {(1,1,15)}}}\,\oplus
\, U(1)\, \oplus\, SU(6)\nonumber\\
E_{6(2)}\, &=&\,\IK_{(1,2,20)}\, \oplus\, SU(2)\, \oplus\, SU(6)
\label{so12const}
\end{eqnarray}
Unfortunately this construction does not define an embedding $Solv_3\, , \,
Solv_4 \hookrightarrow Solv_7$  fulfilling the requirements (\ref{nsrrdecom}).
However this is not a serious problem. Indeed it suffices to write a
new conjugate solvable Lie algebra $Solv_{7}^{\prime} \, =\,
U^{-1}\, Solv_7 \, U\\ \,\,\, \left( U\in SU(8)/U(1)\otimes SU(2)\otimes SU(6)
\right)$ (recall that $Solv_7$ is not stable with respect to the action of  $SU(8)$)
such that the new embedding $Solv_3\, , \, Solv_4 \hookrightarrow
Solv_{7}^{\prime}
$ fulfills (\ref{nsrrdecom}). We could easily determine
such a matrix $U$. The unitary transformation $U$ depends of course
on the particular embedding $U(1)\otimes SU(2)\otimes SU(6) \hookrightarrow
SU(8)$ chosen to define $SO^\star(12)$ and $E_{6(2)}$. Therefore, in order
to achieve an interpretation of the generators of $Solv_3\, , \, Solv_4$
in terms of $N=8$ fields, the positive roots defining
the two solvable algebras must be viewed as roots of $Solv_7^\prime$ whose
Dynkin diagram consists of the following new simple roots:
\begin{eqnarray}
\tilde{\alpha}_1\, &=&\, {\vec \alpha}_5\quad\tilde{\alpha}_2\, =\,
 {\vec \alpha}_{3,6}\quad \tilde{\alpha}_3\, =\,{\vec \alpha}_{3}\nonumber \\
\tilde{\alpha}_4\, &=&\,{\vec \alpha}_2\quad\tilde{\alpha}_5\, =\,
 {\vec \alpha}_{1}\quad \tilde{\alpha}_6\, =\,{\vec \alpha}_{4,1}\nonumber \\
\tilde{\alpha}_7\, &=&\,-{\vec \alpha}_{6,27}
\label{newdynk}
\end{eqnarray}
Since $Solv_3$ and $Solv_4$  respectively define a special K\"ahler and
a quaternionic manifold, it is useful to describe them  in the Alekseevski's
formalism \cite{alex}.
The algebraic structure of $Solv_3$ and $Solv_4$ can be described in the following way:
\begin{eqnarray}
Solv_3\, :\quad\quad\quad\quad\quad\quad\quad\quad & &\nonumber\\
Solv_3\, &=&\, F_1\, \oplus\, F_2\, \oplus\, F_3\, \oplus\, {\bf X}\, \oplus\,
{\bf Y}\, \oplus\, {\bf Z}\nonumber\\
F_i\, &=&\, \{{\rm h}_i\, ,\,{\rm g}_i\}\quad i=1,2,3\nonumber\\
{\bf X}\, &=&\, {\bf X}^+\, \oplus\, {\bf X}^-\, =\, {\bf X}_{NS}\, \oplus\,
{\bf X}_{RR}\nonumber\\
{\bf Y}\, &=&\, {\bf Y}^+\, \oplus\, {\bf Y}^-\, =\, {\bf Y}_{NS}\, \oplus\,
{\bf Y}_{RR}\nonumber\\
{\bf Z}\, &=&\, {\bf Z}^+\, \oplus\, {\bf Z}^-\, =\, {\bf Z}_{NS}\, \oplus\,
{\bf Z}_{RR}\nonumber\\
Solv\left(SU(3,3)_1\right)\, &=&\, F_1\, \oplus\, F_2\, \oplus\, F_3\, \oplus\,
{\bf X}_{NS}\, \oplus\,
{\bf Y}_{NS}\, \oplus\, {\bf Z}_{NS}\nonumber\\
Solv\left(SL(2,\IR)^3\right)\, &=&\, F_1\, \oplus\, F_2\, \oplus\, F_3\nonumber
\\
{\cal W}_{12}\, &=&\, {\bf X}_{RR}\, \oplus\,{\bf Y}_{RR}\, \oplus\,{\bf Z}_{RR}
\nonumber\\
dim\left( F_i\right)\, &=&\, 2\, ;\quad dim\left( {\bf X}_{NS/RR}\right)\, =\,
dim\left( {\bf X}^{\pm}\right)\, =\,4\nonumber \\
 dim\left( {\bf Y}_{NS/RR}\right)\, &=&\,dim\left( {\bf Y}^{\pm}\right)\, =\,
dim\left( {\bf Z}_{NS/RR}\right)\, =\,dim\left( {\bf Z}^{\pm}\right)\,
=\,4\nonumber\\
\left[{\rm h}_i\, ,\,{\rm g}_i\right]\, &=&\, {\rm g}_i\quad i=1,2,3\nonumber\\
\left[F_i\, ,\,F_j\right]\, &=&\, 0 \quad i\neq j\nonumber\\
\left[{\rm h}_3\, ,\,{\bf Y}^{\pm}\right]\, &=&\,\pm\frac{1}{2}{\bf Y}^{\pm}
\nonumber\\
\left[{\rm h}_3\, ,\,{\bf X}^{\pm}\right]\, &=&\,\pm\frac{1}{2}{\bf X}^{\pm}
\nonumber\\
\left[{\rm h}_2\, ,\,{\bf Z}^{\pm}\right]\, &=&\,\pm\frac{1}{2}{\bf Z}^{\pm}
\nonumber\\
\left[{\rm g}_3\, ,\,{\bf Y}^{+}\right]\, &=&\,\left[{\rm g}_2\, ,\,{\bf Z}^{+}
\right]\, =\,\left[{\rm g}_3\, ,\,{\bf X}^{+}\right]\, =\, 0\nonumber\\
 \left[{\rm g}_3\, ,\,{\bf Y}^{-}\right]\, &=&\,{\bf Y}^+\, ;\,\,
\left[{\rm g}_2\, ,\,{\bf Z}^{-}\right]\, =\,{\bf Z}^+\, ;\,\,
\left[{\rm g}_3\, ,\,{\bf X}^{-}\right]\, =\,{\bf X}^+\nonumber\\
\left[F_1\, ,\,{\bf X}\right]\, &=&\,\left[F_2\, ,\,{\bf Y}\right]\, =\,
\left[F_3\, ,\,{\bf Z}\right]\, =\, 0\nonumber\\
\left[{\bf X}^-\, ,\,{\bf Z}^{-}\right]\, &=&\, {\bf Y}^-
\label{Alekal}
\end{eqnarray}
\begin{eqnarray}
Solv_4\, :\quad\quad\quad\quad\quad\quad\quad\quad\quad\quad\quad\quad\quad
 & &\nonumber\\
Solv_4 & =& F_0\, \oplus  \, F_1^{\prime}\,\oplus \, F_2^{\prime}
\,\oplus \, F_2^{\prime}\,\nonumber\\
 & &\oplus \, {\bf X}_{NS}^{\prime}\,\oplus \,
{\bf Y}_{NS}^{\prime}\,\oplus \, {\bf Z}_{NS}^{\prime}\,\oplus \,
{\cal W}_{20}\nonumber\\
Solv\left(SL(2,\IR)\right)\,\oplus \,Solv\left(SU(3,3)_2\right)\, &=&\,
\left[F_0\right]\,\oplus\, \Biggl [ F_1^{\prime}\,\oplus \, F_2^{\prime}
\,\oplus \, F_2^{\prime} \nonumber \\
 & & \,\oplus \, {\bf X}_{NS}^{\prime}\,\oplus \,\Biggr ]\nonumber\\
F_0\, &=&\, \{{\rm h}_0\, ,\,{\rm g}_0\}\quad \left[{\rm h}_0\, ,\,
{\rm g}_0\right]\, =\, {\rm g}_0\nonumber\\
F_i^\prime\, &=&\, \{{\rm h}_i^\prime\, ,\,{\rm g}_i^\prime\}
\quad i=1,2,3\nonumber\\
\left[F_0\, , \, Solv\left(SU(3,3)_2\right)\right]\,&=& \, 0\, ;\quad
\left[{\rm h}_0\, ,\,{\cal W}_{20}\right]\, =\, \frac{1}{2}{\cal W}_{20}\nonumber\\
\left[{\rm g}_0\, ,\,{\cal W}_{20}\right]\, &=&\, \left[{\rm g}_0\, , \,
 Solv\left(SU(3,3)_2\right)\right]\,=\, 0\nonumber\\
\left[Solv\left(SL(2,\IR)\right)\,\oplus \,Solv\left(SU(3,3)_2\right)\, ,\,
{\cal W}_{20}\right]\, &=&\, {\cal W}_{20}
\label{Alekquat}
\end{eqnarray}
The operators ${\rm h}_i\quad i=1,2,3$ are the Cartan generators of
$SO^\star(12)$ and ${\rm g}_i$ the corresponding axions which together
with
${\rm h}_i$ complete the solvable algebra $Solv\left(SL(2,\IR)^3\right)$
For reasons that will be apparent in the next sections we name:
\begin{equation}
 Solv\left(SL(2,\IR)^3\right) = \mbox{the STU algebra}
\end{equation}
In order to achieve a characterization of all the $Solv\left(SO^\star(12)\right)$
 generators
in terms of fields, the next step is to write down the explicit expression of
the $Solv\left(SO^\star(12)\right)$ generators in terms of  roots of
$Solv_7^\prime$, whose field interpretation can be read directly from Table
\ref{dideals}. We have:
\begin{eqnarray}
{\rm h}_1\, &=& \, \frac{1}{2} H_{{\vec \alpha}_{6,1}}\quad {\rm g}_1\, = \,
E^{{\vec \alpha}_{6,1}}\nonumber\\
{\rm h}_2\, &=& \, \frac{1}{2} H_{{\vec \alpha}_{4,1}}\quad {\rm g}_2\, = \,
E^{{\vec \alpha}_{4,1}}\nonumber\\
{\rm h}_3\, &=& \, \frac{1}{2} H_{{\vec \alpha}_{2,2}}\quad {\rm g}_3\, = \,
E^{{\vec \alpha}_{2,2}}\nonumber\\
{\bf X}_{NS}^+\, &=&\, \left( \begin{array}{c}
E^{{\vec \alpha}_{4,3}}+E^{{\vec \alpha}_{3,4}}\\
E^{{\vec \alpha}_{3,1}}-E^{{\vec \alpha}_{4,6}}\end{array}\right)\,\,\,
{\bf X}_{NS}^-\, =\, \left( \begin{array}{c}
E^{{\vec \alpha}_{4,5}}+E^{{\vec \alpha}_{3,2}}\\
E^{{\vec \alpha}_{3,3}}-E^{{\vec \alpha}_{4,4}}\end{array}\right)\nonumber\\
{\bf X}_{RR}^+\, &=&\, \left( \begin{array}{c}
E^{{\vec \alpha}_{6,21}}+E^{-{\vec \alpha}_{6,17}}\\
E^{{\vec \alpha}_{5,16}}-E^{-{\vec \alpha}_{5,12}}\end{array}\right)\,\,\,
{\bf X}_{RR}^-\, =\, \left( \begin{array}{c}
E^{{\vec \alpha}_{6,20}}+E^{-{\vec \alpha}_{6,16}}\\
E^{{\vec \alpha}_{5,15}}-E^{-{\vec \alpha}_{5,11}}\end{array}\right)\nonumber\\
{\bf Y}_{NS}^+\, &=&\, \left( \begin{array}{c}
E^{{\vec \alpha}_{6,10}}+E^{{\vec \alpha}_{5,5}}\\
E^{{\vec \alpha}_{5,8}}-E^{{\vec \alpha}_{6,7}}\end{array}\right)\,\,\,
{\bf Y}_{NS}^-\, =\, \left( \begin{array}{c}
E^{{\vec \alpha}_{6,8}}+E^{{\vec \alpha}_{5,7}}\\
E^{{\vec \alpha}_{5,6}}-E^{{\vec \alpha}_{6,9}}\end{array}\right)\nonumber\\
{\bf Y}_{RR}^+\, &=&\, \left( \begin{array}{c}
E^{{\vec \alpha}_{6,24}}+E^{-{\vec \alpha}_{3,6}}\\
E^{{\vec \alpha}_{6,26}}-E^{-{\vec \alpha}_{4,9}}\end{array}\right)\,\,\,
{\bf Y}_{RR}^-\, =\, \left( \begin{array}{c}
E^{{\vec \alpha}_{6,23}}+E^{-{\vec \alpha}_{3,5}}\\
E^{{\vec \alpha}_{6,25}}-E^{-{\vec \alpha}_{4,8}}\end{array}\right)\nonumber\\
{\bf Z}_{NS}^+\, &=&\, \left( \begin{array}{c}
E^{{\vec \alpha}_{6,5}}+E^{{\vec \alpha}_{5,1}}\\
E^{{\vec \alpha}_{5,3}}-E^{{\vec \alpha}_{6,3}}\end{array}\right)\,\,\,
{\bf Z}_{NS}^-\, =\, \left( \begin{array}{c}
E^{{\vec \alpha}_{6,4}}-E^{{\vec \alpha}_{5,4}}\\
E^{{\vec \alpha}_{5,2}}+E^{{\vec \alpha}_{6,6}}\end{array}\right)\nonumber\\
{\bf Z}_{RR}^+\, &=&\, \left( \begin{array}{c}
E^{{\vec \alpha}_{6,12}}-E^{-{\vec \alpha}_{2,3}}\\
E^{{\vec \alpha}_{6,27}}+E^{-{\vec \alpha}_{1,1}}\end{array}\right)\,\,\,
{\bf Z}_{RR}^-\, =\, \left( \begin{array}{c}
E^{{\vec \alpha}_{6,22}}+E^{-{\vec \alpha}_{4,10}}\\
E^{{\vec \alpha}_{6,13}}+E^{-{\vec \alpha}_{4,7}}\end{array}\right)
\label{so12struc}
\end{eqnarray}
One can finally check  that the axions associated with the STU--algebra, i.e.
with the generators ${\rm g}_i$ are $B_{5,6}\, ,\, B_{7,8}\, ,\, g_{9,10}$.
Futhermore it is worthwhile noticing that the bidimensional subalgebra
$F_0$ of $Solv_4$ is the solvable algebra of the S--duality group $SL(2,\IR)$
of the $N=8$ theory, and therefore is parametrized by the following fields:
\begin{eqnarray}
\phi \,(dilaton)\,&\leftrightarrow& {\rm h}_0\nonumber\\
B_{\mu \nu} \, \,\,\, &\leftrightarrow &{\rm g}_0
\label{Skey}
\end{eqnarray}
Some consistent  $N=2 $ truncations of the $N=8$ theory can be described
in terms of their scalar content in the following way:
\begin{eqnarray}
{\cal M}_{N=8}&\sim& Solv_7^\prime\,\rightarrow \, {\cal M}_{N=2}\equiv
{\cal M}_{vec}\, \otimes\, {\cal M}_{quat}\nonumber\\
{\cal M}_{vec}\, &\sim&\, Solv_3 \,\,\,\,\,\,\,\,\,\,\,\,\,\,\,\,\,\,\,\,\,\,\,
\,\,\,\,\quad {\cal M}_{quat}\, \sim\, \bfone
\nonumber\\
{\cal M}_{vec}\, &\sim&\, Solv\left(SU(3,3)_1\right) \quad {\cal M}_{quat}\, \sim\,
Solv\left(SU(2,1)\right)\nonumber\\
{\cal M}_{vec}\, &\sim&\, Solv\left(Sl(2,\IR)^3\right) \quad {\cal M}_{quat}\, \sim\,
Solv\left(SO(4,6)\right)\nonumber\\
{\cal M}_{vec}\, &\sim&\, \bfone \,\,\,\,\,\,\,\,\,\,\,\,\,\,\,\,\,\quad\qquad
\,\,\,\,\quad {\cal M}_{quat}\, \sim\, Solv\left(
E_{6(2)}\right)
\end{eqnarray}
\section{The general solution}
\label{concludo}
In this treatment we have considered two separate but closely related issues:
\begin{enumerate}
\item{The $N=2$ decomposition of the $N=8$ solvable Lie algebra
$Solv_7 \equiv Solv \left(E_{7(7)}/SU(8)\right)$}
\item{The system of first and second order equations characterizing
$BPS$ black--holes in the $N=8$ theory}
\end{enumerate}
With respect to issue (1) our treatment has been exhaustive and
we have shown how the decomposition \eqn{7in3p4},\eqn{3and4defi}
corresponds to the splitting of the $N=8$ scalar fields into vector
multiplet scalars and hypermultiplet scalars. We have also shown how
the alekseevskian analysis of the decomposed solvable Lie algebra
$Solv_7$ is the key to determine the consistent $N=2$ truncations of
the $N=8$ theory at the interaction level. In addition the algebraic results
on the embedding of the $U(1)\times SU(2) \times SU(6)$ Lie algebra
into $E_{7(7)}$ and the solvable counterparts of this embedding are
instrumental for the completion of the programme already outlined in
the previous chapters, namely the gauging of the
maximal gaugeable abelian ideal  ${\cal G}_{abel} \subset Solv_7$ which turns out to be of
dimension $7$. \par
With respect to issue (2) we made a general group--theoretical
analysis of the Killing vector equations and we proved that the
hypermultiplet scalars corresponding to the solvable Lie subalgebra
$Solv_4 \subset Solv_7$ are constant in the most general solution.
Next we analysed a simplified model where the only non--zero fields
are those in the Cartan subalgebra $H \, \subset \, Solv_7$ and we
showed how the algebraically decomposed Killing spinor equations work
in an explicit way. In particular by means of this construction we
retrieved the $N=8$ embedding of the $a$--model black-hole solutions known
in the literature \cite{gensugrabh}. It remains to be seen how general
the presented solutions are, modulo U--duality rotations. That they are not
fully general is evident from the fact that by restricting the non--zero
fields to be in the Cartan subalgebra we obtain constraints on the
electric and magnetic charges such that the solution is parametrized
by only four charges: two electric $(q_{18},q_{23})$ and two magnetic
$p_{17},p_{24}$.  We are therefore lead to consider the question
\par
{ \it How many more scalar fields besides those associated with the
Cartan subalgebra have to be set non zero in order to generate the
most general solution modulo U--duality rotations?}
\par
An answer can be given in terms of solvable Lie algebra once again.
The argument is the following.
\par
Let
\begin{equation}
{\vec Q} \equiv \left( \matrix { g^{\vec {\Lambda}} \cr e_{\vec {\Sigma}}\cr } \right)
\label{chavecto}
\end{equation}
be the vector of electric and magnetic charges (see eq.\eqn{gedefi})
that transforms in the ${\bf 56}$ dimensional real representation
of the U duality group $E_{7(7)}$.
Through the Cayley matrix we can convert it to the ${\bf Usp(56)}$ basis namely to:
\begin{equation}
\left(\matrix { t^{{\vec {\Lambda}}_1}=
g^{{\vec {\Lambda}}_1}+ {\rm i} \, e_{{\vec {\Lambda}}_1}\, \cr
{\bar t}_{{\vec {\Lambda}}_1}=
g^{{\vec {\Lambda}}_1}- {\rm i} \, e_{{\vec {\Lambda}}_1}\,\cr
} \right)
\label{uspqvec}
\end{equation}
Acting on ${\vec Q}$ by means of suitable $Solv \left(E_{7(7)}\right)$
transformations, we can reduce it to the following {\it normal} form:
\begin{equation}
{\vec Q}\rightarrow {\vec Q}^N \equiv \left(\matrix { t^0_{(1,1,1)}\cr t^1_{(1,1,15)}
\cr t^2_{(1,1,15)}\cr t^3_{(1,1,15)}\cr 0\cr
\dots \cr 0 \cr  \cr {\bar t}^0_{{\bar (1,1,1)}}\cr {\bar t}^1_{{\bar (1,1,15)}
} \cr
{\bar t}^2_{{\bar (1,1,15)}}\cr {\bar t}^3_{{\bar (1,1,15)}}\cr 0 \cr
\dots \cr 0 \cr  } \right)
\label{qnormalf}
\end{equation}
Consequently also the central charge ${\vec Z}\equiv \left( Z^{AB}\, , \,
Z_{CD}\right)$, which depends on ${\vec Q}$ through the coset representative
in a symplectic--invariant way, will be brought to the {\it normal} form
\begin{equation}
{\vec Z}\rightarrow {\vec Z}^N \equiv \left(\matrix { z^0_{(1,1,1)}\cr z^1_{(1,1,15)}
\cr z^2_{(1,1,15)}\cr z^3_{(1,1,15)}\cr 0\cr
\dots \cr 0 \cr  \cr {\bar z}^0_{{\bar (1,1,1)}}\cr {\bar z}^1_{{\bar (1,1,15)}
} \cr
{\bar z}^2_{{\bar (1,1,15)}}\cr {\bar z}^3_{{\bar (1,1,15)}}\cr 0 \cr
\dots \cr 0 \cr  } \right)
\label{znormalf}
\end{equation}
through a suitable $SU(8)$ transformation. It was shown in
\cite{lastserg}, \cite{savoy} that ${\vec Q}^N$ is invariant with respect to
the action of an $O(4,4)$ subgroup of $ E_{7(7)}$ and its {\it normalizer}
is an $SL(2,\IR)^3 \subset E_{7(7)}$ commuting with it.
 Indeed it turns out that
the eight real parameters in ${\vec Q}^N$ are singlets with respect to
$O(4,4)$ and in a ${\bf (2,2,2)}$ irreducible representation of
$SL(2,\IR)^3$  as it is shown in the following
decomposition of the ${\bf 56}$ with respect to $O(4,4)\otimes SL(2,\IR)^3$:
\begin{equation}
{\bf 56}\rightarrow {\bf (8_v,2,1,1)\, \oplus \, (8_s,1,2,1)\,\oplus \,
(8_{s^\prime},1,1,2)}\, \oplus\, {\bf (1,2,2,2)}
\label{56normaldec}
\end{equation}
The corresponding subgroup of $SU(8)$ leaving ${\vec Z}^N$ invariant
is therefore $SU(2)^4$ which is the maximal compact subgroup of $O(4,4)$.

Note that $SL(2,\IR)^3$ contains a $U(1)^3$ which is in $SU(8)$ and
which can be further used to classify the general normal frame
black-holes by five real parameters, namely four complex numbers with
the same phase.
This corresponds to write the 56 dimensional generic vector in terms
of the five normal frame parameters plus 51 ``angles'' which
parametrize the 51 dimensional compact space $\frac{SU(8)}{SU(2)^4}$, where $ SU(2)^4 $
is the maximal compact subgroup of the stability group
$O(4,4)$ \cite{cvet}.

Consider now the scalar ``geodesic'' potential (see eq.\eqn{T=SCLV}):
\begin{eqnarray}
V(\phi) & \equiv & {\bar Z}^{AB}(\phi) \, Z_{AB}(\phi) \nonumber\\
&=& {\vec Q}^T \, \left[ \IL^{-1}\left(\phi\right) \right]^T \,
\IL^{-1}\left(\phi\right) \, {\vec Q}
\end{eqnarray}
whose minimization determines the fixed values of the scalar fields
at the horizon of the black--hole.
Because of its invariance properties  the scalar potential $V(\phi)$ depends on
  ${\vec Z}$ and therefore on ${\vec Q}$ only  through their normal forms. Since the
 fixed scalars at the horizon of the Black--Hole are obtained minimizing
$V(\phi)$, it can be inferred that the most general solution of this kind will depend
(modulo duality transformations) only on those scalar fields associated with the
{\it normalizer} of the normal form ${\vec Q}^N$. Indeed the dependence of $V(\phi)$ on
a scalar field is achieved by acting on ${\vec Q}$ in the expression
of $V(\phi)$ by means of the transformations in $Solv_7$ associated with that field.
Since at any point of the scalar manifold $V(\phi)$ can be made to depend only on
${\vec Q}^N$, its minimum will be defined only by those scalars that correspond
to trasformations acting on the non--vanishing components of the normal form
({\it normalizer} of ${\vec Q}^N$). Indeed
 all the other isometries were used to rotate  ${\vec Q}$ to the normal form ${\vec Q}^N$.
Among those scalars which are not determined by the fixed point conditions
there are the {\it flat direction fields} namely
those on which the scalar potential does not  depend  at all:
\begin{equation}
  \mbox{flat direction field } \, q_f \quad \leftrightarrow \quad
  \frac{\partial}{\partial q_f} \, V(\phi) =0
\end{equation}
Some of these fields parametrize $Solv\left(O(4,4)\right)$  since
they are associated with isometries leaving ${\vec Q}^N$ invariant, and the remaining ones
are  obtained from the latter by means of duality transformations.
In order to identify the scalars which are {\it flat} directions of $V(\phi)$, let
us consider the way in which  $Solv\left(O(4,4)\right)$ is embedded into $Solv_7$,
referring to the description of $Solv_4$ given in eqs. (\ref{Alekquat}):
\begin{eqnarray}
Solv\left(O(4,4)\right)&\subset& Solv_4\nonumber\\
Solv\left(O(4,4)\right)\, &=&\,  F_0\,\oplus \, F_1^{\prime}\,\oplus \,
F_2^{\prime}\,\oplus \, F_3^{\prime}\, \oplus \, {\cal W}_{8}
\label{o44}
\end{eqnarray}
where the R--R part ${\cal W}_{8}$ of $Solv\left(O(4,4)\right)$ is the quaternionic image
of $ F_0\,\oplus \, \\F_1^{\prime}\,\oplus \, F_2^{\prime}\,\oplus \, F_3^{\prime}$
in ${\cal W}_{20}$. Therefore $Solv\left(O(4,4)\right)$
is parametrized by the $4$ {\it hypermultiplets} containing the Cartan fields of
$Solv\left(E_{6(2)}\right)$. One finds that the other flat directions are all
the remaining parameters of $Solv_4$, that is all the hyperscalars.
\par
Alternatively we can observe that since the hypermultiplet
scalars are flat directions of the potential, then we can use the solvable Lie algebra
$ Solv_4$    to set them to zero at the horizon.
Since we know from the Killing spinor equations that these
$40$ scalars  are  constants
it follows that we can safely set them to zero and forget
about their existence (modulo U--duality transformations).
Hence the non zero scalars required for a general
solution have to be looked for among the vector multiplet scalars
that is in the solvable Lie algebra $Solv_3$. In other words
the most general $N=8$ black--hole (up to U--duality rotations) is
given by the most general $N=2$ black--hole based on the $15$--dimensional
special K\"ahler manifold:
\begin{equation}
{\cal SK }_{15}  \, \equiv \, \exp \left[ Solv_3 \right] \, =
\frac{SO^\star(12)}{U(1) \times SU(6)}
\label{mgeneral}
\end{equation}
Having determined the little group of the normal form enables us
to decide which among the above $30$ scalars have to be
kept alive in order to generate the most general BPS black--hole
solution (modulo U--duality).
\par
We argue as follows. The {\it normalizer} of the normal form
is contained in the largest subgroup
of $E_{7(7)}$ commuting with $O(4,4)$.
Indeed, a necessary condition for a group $G^N$ to be the {\it normalizer}
of ${\vec Q}^N$ is to commute with the {\it little group} $G^L=O(4,4)$ of
${\vec Q}^N$:
\begin{eqnarray}
{\vec Q}^{\prime N}\, &=&\, G^N\cdot {\vec Q}^N\quad {\vec Q}^N\, =\,
 G^L\cdot {\vec Q}^{N}\nonumber\\
{\vec Q}^{\prime N}\, &=&\, G^L\cdot {\vec Q}^{\prime N}\Rightarrow
\left[G^N\, ,\, G^L\right]\, =\, 0
\label{gngl}
\end{eqnarray}
As previously mentioned, it was proven that $G^N\, =\, SL(2,\IR)^{3} \subset SO^{\star}(12)$ whose solvable
algebra is defined by the last
 of eqs. (\ref{Alekal}). Moreover $G^N$ coincides with the largest subgroup of $Solv_7$
 commuting with $G^L$.\\ The duality transformations associated with
 the $SL(2,\IR)^{3}$ isometries act only on the eight non vanishing components
of ${\vec Q}^N$ and therefore belong to ${\bf Sp(8)}$.\par
{\it In conclusion the most general $N=8$ black--hole solution is  described
by the 6 scalars
parametrizing  $Solv\left(SL(2,\IR)^{3}\right)$, which are the only ones involved in the
fixed point conditions at the horizon.}
\par
 Another way of seeing this is to
notice that all the other $64$ scalars are either the $16$ parameters of
$Solv\left(O(4,4)\right)$
which are flat directions of $V\left(\phi\right)$, or coefficients of the
$48=56-8$ transformations needed to rotate ${\vec Q}$ into ${\vec Q}^N$ that is to set $48$
components of ${\vec Q}$ to zero as shown in eq. (\ref{qnormalf}).
\par
Let us then reduce our attention to
the Cartan vector multiplet sector, namely to the 6 vectors
corresponding to the solvable Lie algebra $Solv \left ( SL(2,\IR)
\right )$.
\subsection{$SL(2,\IR)^3$ and the fixed scalars at the horizon}
\label{fixascala}
So far we have elaborated 
 group--theoretical apparatus  for studying generic BPS Black Holes 
and in section \ref{cartadila} we have worked out the simplified
example where the only non--zero fields are in the Cartan subalgebra.
From the  viewpoint of string toroidal compactifications this means that we have
just introduced the dilaton and the $6$ radii $R_i$ of the torus $T^6$.
An item that so far was clearly missing  are the $3$ commuting axions $B_{5,6}$,
$B_{7,8}$ and $g_{9,10}$. As already pointed out, by looking at table \ref{dideals}
we realize
that they correspond to the roots $ \alpha_{6,1},\alpha_{4,1},
\alpha_{2,2}$. So, as it is evident from eq.\eqn{so12struc} the nilpotent
generators associated with these fields are the $g_1, g_2, g_3$
partners of the Cartan generators $h_1,h_2,h_3$ completing the three
2--dimensional {\it key algebras} $F_1, F_2, F_3$ in the
Alekseveeskian decomposition of the K\"ahler algebra
$Sol_3=Solv \left( SO^\star(12) \right)$
(see eq.\eqn{Alekal}):
\begin{equation}
Solv_3=F_1 \oplus F_2 \oplus F_3 \oplus {\bf X} \oplus {\bf Y} \oplus
{\bf Z}
\label{keyalg}
\end{equation}
This triplet of key algebras is nothing else but the Solvable Lie
algebra of $ \left[ SL(2,\IR)/U(1) \right]^3$ defined above as the
normalizer of the little group of the normal form ${\vec Q}^N$:
\begin{equation}
F_1 \oplus F_2 \oplus F_3 = Solv \left( SL(2,\IR)\otimes SL(2,\IR)
\otimes SL(2,\IR) \right)
\label{guarunpo}
\end{equation}
The above considerations have reduced the quest for the most
general $N=8$ black--hole to the solution of the model containing only the
$6$ scalar fields associated with the triplet of key algebras \eqn{guarunpo}.
This model is nothing else but the model of $STU$ N=2 black-holes
studied in \cite{STUkallosh}. Hence we can utilize the results of
that paper and insert them in the general set up we have derived.
In particular we can utilize the determination of the fixed values of the
scalars at the horizon in terms of the charges given in
\cite{STUkallosh}. To make a complete connection between the results
of that paper and our framework we just need to derive the relation
between the fields of the solvable Lie algebra parametrization and
the standard $S,T,U$ complex fields utilized as coordinates of the
special K\"ahler manifold:
\begin{equation}
 {\cal ST}[2,2]\,  \equiv  \, \frac{SU(1,1)}{U(1)} \, \otimes \,
 \frac{SO(2,2)}{SO(2) \times SO(2)}
 \label{st22}
\end{equation}
To this effect we consider the embedding
of the Lie algebra $SL(2,\IR)_1 \times SL(2,\IR)_2 \times SL(2,\IR)_3$
into $Sp(8,\IR)$   such  that the fundamental
${\bf 8}$--dimensional representation of $Sp(8,\IR)$ is irreducible
under the three subgroups and is
\begin{equation}
   {\bf 8}= \left({\bf 2},{\bf 2},{\bf 2} \right)
   \label{222}
\end{equation}
The motivation of this embedding is that the $SO(4,4)$ singlets
in the decomposition \eqn{56normaldec} transform under $SL(2,\IR)^3$
as the representation mentioned in eq.\eqn{222}. Therefore the
requested embedding corresponds to the action of the {\it key
algebras} $ F_1 \oplus F_2 \oplus F_3$ on the non vanishing
components of the charge vector in its normal form.
We obtain the desired result from the standard embedding of
$SL(2,\IR) \, \times \, SO(2,n)$ in $Sp \left( 2\times (2+n) ,\IR \right)$:
\begin{eqnarray}
{\bf A} \, \in \,SO(2,n) & \hookrightarrow &
\left( \matrix { {\bf A} & {\bf 0} \cr {\bf 0} & -{\bf A}^T \cr } \right) \nonumber\\
\left( \matrix { a & b \cr c & d \cr } \right)\,  \in \, SL(2,\IR)
& \hookrightarrow & \left( \matrix { a \, \bfone & b\, \eta  \cr c \, \eta  & d \, \bfone \cr }
\right)
\label{standaimbed}
\end{eqnarray}
used to derive the Calabi Vesentini parametrization of the Special
K\"ahler manifold:
\begin{equation}
{\cal ST}[2,n] \, \equiv \,  \frac{SU(1,1)}{U(1)}\times
\frac{SO(2,n)}{SO(2)\times SO(n)}
\end{equation}
It suffices to set $n=2$ and to use the accidental isomorphism:
\begin{equation}
SO(2,2) \, \sim \, SL(2,\IR) \, \times  \, SL(2,\IR)
\label{accisoph}
\end{equation}
Correspondingly we can write an explicit realization of the
$SL(2,\IR)^3$ Lie algebra:
\begin{equation}
\begin{array}{rcrl}
\left[ L^{(i)}_0 \,  , \, L^{(i)}_\pm \right ] & = &  \pm \,  L^{(i)}_\pm  & i=1,2,3
 \\
\left[ L^{(i)}_+ \,  , \, L^{(i)}_- \right ] & = &  2\,  L^{(i)}_0  & i=1,2,3
\nonumber\\
 \left[ L^{(i)}_A \,  , \, L^{(j)}_B \right ] &=& 0 & i \neq j   \\
 \end{array}
 \label{treLie}
\end{equation}
by means of $8 \times  8$ symplectic matrices satisfying:
\begin{equation}
\left[ L^{(i)}_A \right]^T \,  \IC  + \IC \, L^{(i)}_A = 0
\end{equation}
where
\begin{equation}
  \IC = \left( \matrix{ {\bf 0}_{4 \times 4} & \bfone_{4 \times 4}
  \cr  - \bfone_{4 \times 4} &  {\bf 0}_{4 \times 4} } \right)
\label{symp8}
\end{equation}

Given this structure of the algebra, we can easily construct the
coset representatives by writing:
\begin{eqnarray}
\IL^{(i)}\left(h_i,a_i \right) & \equiv & \exp[ 2 \, h_i \, L_0^{(i)} ] \,
\exp[ a_i \, e^{-h_i} \, L_+^{(i)}] \nonumber \\
 & = & \left( Cosh[h_i] \bfone + Sinh [h_i] \, L_0^{(i)} \right)\, \left(\bfone +
 a_i \, e^{-h_i} \, L_+^{(i)}\right) \nonumber\\
\end{eqnarray}
which follows from the identitities:
\begin{eqnarray}
 L_0^{(i)}\, L_0^{(i)} &=& \frac{1}{4} \, \bfone \\
 L_+^{(i)}\, L_+^{(i)} &=& {\bf 0}\\
\end{eqnarray}
The explicit form of the matrices $L_a^{(i)}$ and $\IL^{(i)}$ is given in appendix D.

We are now ready to construct the central charges and their modulus
square whose minimization with respect to the fields yields the
values of the fixed scalars.
\par
Let us introduce the charge vector:
\begin{equation}
{\vec Q} = \left(\matrix{g^1 \cr g^2 \cr g^3 \cr g^4 \cr e_1 \cr e_2 \cr e_3
\cr e_4 }\right)
\end{equation}
Following our general formulae we can write the central charge vector
as follows:
\begin{equation}
{\vec Z} \, = \, {\cal S}\, \IC \, \prod_{i=1}^{3} \,
\IL^{(i)}(-h_1,-a_i) \, {\vec Q}
\label{centvec}
\end{equation}
where ${\cal S}$ is some unitary matrix and $\IC$ is the symplectic metric.
\par
At this point it is immediate to write down the potential,
whose minimization with respect to the scalar fields
yields the values of the fixed scalars at the horizon.
\par
We have:
\begin{equation}
V({\vec Q}, h, a) \equiv {\vec Z}^\dagger \, {\vec Z} \, = \,
{\vec Q} \, \prod_{i=1}^{3} \, M^{(i)} \left(h_i,a_i\right) \, {\vec
Q}
\label{potential}
\end{equation}
where:
\begin{equation}
 M^{(i)} \left(h_i,a_i \right) \, \equiv \,
 \left[ \IL^{(i)}\left(-h_i,-a_i\right) \right ]^T \, \IL^{(i)}\left(-h_i,-a_i \right)
 \label{mll}
\end{equation}
Rather then working out the derivatives of this potential and
equating them to zero, we can just use the results of paper
\cite{STUkallosh}. It suffices to write the correspondence between our
solvable Lie algebra fields and the $3$ complex scalar fields
$S,T,U$ used in the $N=2$ standard parametrization of the theory.
This correspondence is:
\begin{eqnarray}
T &=&\, a_1 \, + \mbox{\rm i} \, \exp[2 h_1]
\nonumber\\
U & =& \,a_2\,+ \mbox{\rm i} \, \exp[2 h_2]
\nonumber\\
S &=& \,a_3\, + \mbox{\rm i} \, \exp[2 h_3]
\label{pippobaudo}
\end{eqnarray}
and it is established with the following argument. The symplectic
section of special geometry $X^\Lambda$ is defined, in terms of the
$SO(2,2)$ coset representative $L^\Lambda_{\phantom{\Lambda}\Sigma}(\phi)$,
by the formula ( see eq.(C.1) of \cite{fundpaper}):
\begin{equation}
\frac{1}{\sqrt{X^\Lambda \, X^\Sigma}} \, X^\Lambda = \frac{1}{\sqrt{2}}
\, \left( L^\Lambda_{\phantom{\Lambda}1} + \mbox{\rm i}\,L^\Lambda_{\phantom{\Lambda}2}
\right )
\label{C1equa}
\end{equation}
Using for $L^\Lambda_{\phantom{\Lambda}\Sigma}(\phi)$ the upper $4
\times 4$ block of the product $\IL^{(1)}\left(h_1,a_1\right) \,
\IL^{(2)}\left(h_2,a_2\right) $ and using for the symplectic section $X^\Lambda$
that given in eq.(58) of \cite{STUkallosh} we obtain the first two lines
of eq.\eqn{pippobaudo}. The last line of the same equation is
obtained by identifying  the $SU(1,1)$ matrix:
\begin{equation}
{\cal C} \, \left(\matrix{ e^{h_3} & a_3 \cr 0 & e^{-h3} \cr }\right)  \, {\cal C}^{-1}
\end{equation}
where ${\cal C}$ is the $2$--dimensional Cayley matrix with the
matrix $M(S)$ defined in eq.(3.30) of \cite{fundpaper}.
\par
Given this identification of the fields, the fixed values at the horizon
are given by eq.(37) of \cite{STUkallosh}.
\par
We can therefore conclude that we have determined the fixed values of
the scalar fields at the horizon in a general $N=8$ BPS saturated
black--hole.

\vfill
\appendix
\chapter{ Scalars $\leftrightarrow$ generators correspondence}

By referring to the toroidal dimensional reduction of type IIA
superstring and Tables \ref{tabu1}, \ref{tabu2},
it is straightforward to establish a correspondence
between the scalar fields of either Neveu--Schwarz or Ramond--Ramond type
emerging at each step of the sequential compactification and the positive
roots of $\Phi^{+}(E_7)$, distributed into
the  $\ID^{+}_{r+1}$ subspaces.
This gives rise to the following list of solvable algebra generators.
The roots of $SO(6,6)$ are associated with N--S fields,
the spinor weights of $SO(6,6)$ are associated with
R--R fields.
\par
\vskip 0.2cm
The abelian ideal in $D=8$   $  E_3 \supset{\cal A}_3 \equiv \ID^{+}_{2}$  is given
by the roots:
{\flushleft{$\ID^{+}_2 = $}}
\begin{eqnarray}
B_{9,10} ~\rightarrow ~  D_{2} (1) &=&  \{ 0,0,0,0,1,1,0\} \nonumber \\
g_{9,10} ~\rightarrow ~  D_{2} (2) &=& \{ 0,0,0,0,1,-1,0\} \nonumber \\
A_9 ~\rightarrow ~  D_{2} (3) &=&\{ -{1\over 2},-{1\over 2},-{1\over 2},
-{1\over 2},{1\over 2},{1\over 2},
   {1\over {{\sqrt{2}}}} \}
\label{d2root}
\end{eqnarray}
The abelian ideal in $D=7$   $  E_4\supset {\cal A}_4\equiv \ID^{+}_{3}$  is given
by the roots:
{\flushleft{$\ID^{+}_3 = $}}
\begin{eqnarray}
B_{8,9} ~\rightarrow ~  D_{3} (1) &=&  \{ 0,0,0,1,1,0,0\} \nonumber \\
g_{8,9} ~\rightarrow ~  D_{3} (2) &=&\{ 0,0,0,1,-1,0,0\} \nonumber\\
B_{8,10} ~\rightarrow ~  D_{3} (3) &=& \{ 0,0,0,1,0,1,0\}\nonumber \\
g_{8,10}~\rightarrow ~  D_{3} (4) &=&   \{ 0,0,0,1,0,-1,0\} \nonumber \\
A_8 ~\rightarrow ~  D_{3} (5) &=& \{ -{1\over 2},-{1\over 2},-{1\over 2},{1\over 2},
   {1\over 2},-{1\over 2},{1\over {{\sqrt{2}}}}\} \nonumber \\
A_{8,9,10} ~\rightarrow ~  D_{3} (6) &=&  \{ -{1\over 2},-{1\over 2},-{1\over 2},{1\over 2},
-{1\over 2},{1\over 2},
   {1\over {{\sqrt{2}}}}\}
\label{d3root}
\end{eqnarray}
The abelian ideal in $D=6$   $ E_5\supset  {\cal A}_5 \equiv \ID^{+}_{4}$  is given
by the roots:
{\flushleft{$\ID^{+}_4 = $} }
\begin{eqnarray}
B_{7,8} ~\rightarrow ~  D_{4} (1) &=&  \{ 0,0,1,1,0,0,0\}\nonumber \\
g_{7,8} ~\rightarrow ~  D_{4} (2) &=& \{ 0,0,1,-1,0,0,0\}\nonumber \\
B_{7,9} ~\rightarrow ~  D_{4} (3) &=&\{ 0,0,1,0,1,0,0\}\nonumber \\
g_{7,9} ~\rightarrow ~  D_{4} (4) &=&  \{ 0,0,1,0,-1,0,0\}\nonumber \\
B_{7,10} ~\rightarrow ~  D_{4} (5) &=& \{ 0,0,1,0,0,1,0\}\nonumber \\
g_{7,10} ~\rightarrow ~  D_{4} (6) &=& \{ 0,0,1,0,0,-1,0\}\nonumber \\
A_{7,9,10}~\rightarrow ~  D_{4} (7) &=&  \{ -{1\over 2},-{1\over 2},{1\over 2},
{1\over 2},-{1\over 2},-{1\over 2},
   {1\over {{\sqrt{2}}}}\}\nonumber \\
A_{7,8,10}~\rightarrow ~  D_{4} (8) &=&\{ -{1\over 2},-{1\over 2},{1\over 2},-{1\over 2},
   {1\over 2},-{1\over 2},{1\over {{\sqrt{2}}}}\}\nonumber \\
 A_{7,8,9}~\rightarrow ~   D_{4}(9)&=& \{ -{1\over 2},-{1\over 2},
 {1\over 2},-{1\over 2},-{1\over 2},{1\over 2},
   {1\over {{\sqrt{2}}}}\}\nonumber \\
A_{7} ~\rightarrow ~   D_{4}(10)&=&\{ -{1\over 2},-{1\over 2},{1\over 2},{1\over 2},
   {1\over 2},{1\over 2},{1\over {{\sqrt{2}}}} \}
\label{d4root}
\end{eqnarray}
The abelian ideal in $D=5$   $ E_6 \supset  {\cal A}_6\equiv \ID^{+}_{5}$  is given
by the roots:
{\flushleft{$\ID^{+}_5 = $}}
\begin{eqnarray}
B_{6,7}~\rightarrow ~   D_{5}(1)&=& \{ 0,1,1,0,0,0,0\} \nonumber \\
g_{6,7}~\rightarrow ~   D_{5}(2)&=& \{ 0,1,-1,0,0,0,0\} \nonumber \\
B_{6,8}~\rightarrow ~   D_{5}(3)&=& \{ 0,1,0,1,0,0,0\} \nonumber \\
g_{6,8}~\rightarrow ~   D_{5}(4)&=&  \{ 0,1,0,-1,0,0,0\} \nonumber \\
B_{6,9}~\rightarrow ~   D_{5}(5)&=& \{ 0,1,0,0,1,0,0\} \nonumber \\
g_{6,9}~\rightarrow ~   D_{5}(6)&=&\{ 0,1,0,0,-1,0,0\} \nonumber \\
B_{6,10}~\rightarrow ~   D_{5}(7)&=&  \{ 0,1,0,0,0,1,0\} \nonumber \\
g_{6,10}~\rightarrow ~   D_{5}(8)&=&\{ 0,1,0,0,0,-1,0\} \nonumber \\
A_{6,8,9}~\rightarrow ~   D_{5}(9)&=&  \{ -{1\over 2},{1\over 2},{1\over 2},-{1\over 2},
-{1\over 2},-{1\over 2},
   {1\over {{\sqrt{2}}}}\} \nonumber \\
A_{6,7,9}~\rightarrow ~   D_{5}(10)&=& \{ -{1\over 2},{1\over 2},-{1\over 2},{1\over 2},
   -{1\over 2},-{1\over 2},{1\over {{\sqrt{2}}}}\} \nonumber \\
A_{6,7,8}~\rightarrow ~   D_{5}(11)&=&  \{ -{1\over 2},{1\over 2},-{1\over 2},-{1\over 2},
{1\over 2},
-{1\over 2},
   {1\over {{\sqrt{2}}}}\} \nonumber \\
A_{\mu\nu\rho}~\rightarrow ~   D_{5}(12)&=& \{ -{1\over 2},{1\over 2},-{1\over 2},-{1\over 2},
   -{1\over 2},{1\over 2},{1\over {{\sqrt{2}}}}\} \nonumber \\
A_{6,7,10}~\rightarrow ~   D_{5}(13)&=& \{ -{1\over 2},{1\over 2},-{1\over 2},{1\over 2},
{1\over 2},{1\over 2},
   {1\over {{\sqrt{2}}}}\} \nonumber \\
A_{6,8,10}~\rightarrow ~   D_{5}(14)&=& \{ -{1\over 2},{1\over 2},{1\over 2},-{1\over 2},
   {1\over 2},{1\over 2},{1\over {{\sqrt{2}}}}\} \nonumber \\
A_{6,9,10}~\rightarrow ~   D_{5}(15)&=&  \{ -{1\over 2},{1\over 2},{1\over 2},{1\over 2},
-{1\over 2},{1\over 2},
   {1\over {{\sqrt{2}}}}\} \nonumber \\
A_6 ~\rightarrow ~   D_{5}(16)&=&\{ -{1\over 2},{1\over 2},{1\over 2},{1\over 2},
   {1\over 2},-{1\over 2},{1\over {{\sqrt{2}}}} \}
\label{d5root}
\end{eqnarray}
The abelian ideal in $D=4$   $  E_7\supset {\cal A}_7 \equiv \ID^{+}_{6}$  is given
by the roots:
{\flushleft{$\ID^{+}_6 = $}}
\begin{eqnarray}
B_{5,6}~\rightarrow ~   D_{6}(1)&=& \{ 1,1,0,0,0,0,0\} \nonumber \\
g_{5,6}~\rightarrow ~   D_{6}(2)&=& \{ 1,-1,0,0,0,0,0\} \nonumber \\
B_{5,7}~\rightarrow ~   D_{6}(3)&=& \{ 1,0,1,0,0,0,0\} \nonumber \\
g_{5,7}~\rightarrow ~   D_{6}(4)&=&  \{ 1,0,-1,0,0,0,0\} \nonumber \\
B_{5,8}~\rightarrow ~   D_{6}(5)&=& \{ 1,0,0,1,0,0,0\} \nonumber \\
g_{5,8}~\rightarrow ~   D_{6}(6)&=& \{ 1,0,0,-1,0,0,0\} \nonumber \\
B_{5,9}~\rightarrow ~   D_{6}(7)&=&  \{ 1,0,0,0,1,0,0\} \nonumber \\
g_{5,9}~\rightarrow ~   D_{6}(8)&=& \{ 1,0,0,0,-1,0,0\} \nonumber \\
B_{5,10}~\rightarrow ~   D_{6}(9)&=& \{ 1,0,0,0,0,0,1\} \nonumber \\
g_{5,10}~\rightarrow ~   D_{6}(10)&=& \{ 1,0,0,0,0,0,-1\} \nonumber \\
B_{\mu\nu}~\rightarrow ~   D_{6}(11)&=& \{ 0,0,0,0,0,0,{1\over {{\sqrt{2}}}}\} \nonumber \\
A_5 ~\rightarrow ~   D_{6}(12)&=& \{ {1\over 2},{1\over 2},{1\over 2},{1\over 2},{1\over 2},
{1\over 2},
   {1\over {{\sqrt{2}}}}\} \nonumber \\
A_{\mu\nu 6}~\rightarrow ~   D_{6}(13)&=& \{ {1\over 2},{1\over 2},-{1\over 2},-{1\over 2},
   -{1\over 2},-{1\over 2},{1\over {{\sqrt{2}}}}\} \nonumber \\
A_{\mu\nu 7} ~\rightarrow ~   D_{6}(14)&=&  \{ {1\over 2},-{1\over 2},{1\over 2},-{1\over 2},
-{1\over 2},-{1\over 2},
   {1\over {{\sqrt{2}}}}\} \nonumber \\
A_{\mu\nu 8} ~\rightarrow ~   D_{6}(15)&=& \{ {1\over 2},-{1\over 2},-{1\over 2},{1\over 2},
   -{1\over 2},-{1\over 2},{1\over {{\sqrt{2}}}}\} \nonumber \\
A_{\mu\nu 9} ~\rightarrow ~   D_{6}(16)&=&  \{ {1\over 2},-{1\over 2},-{1\over 2},-{1\over 2},
{1\over 2},-{1\over 2},
   {1\over {{\sqrt{2}}}}\} \nonumber \\
A_{\mu\nu 10} ~\rightarrow ~   D_{6}(17)&=& \{ {1\over 2},-{1\over 2},-{1\over 2},-{1\over 2},
   -{1\over 2},{1\over 2},{1\over {{\sqrt{2}}}}\} \nonumber \\
 A_{5,6,7} ~\rightarrow ~   D_{6}(18)&=& \{ {1\over 2},-{1\over 2},-{1\over 2},{1\over 2},
 {1\over 2},{1\over 2},
   {1\over {{\sqrt{2}}}}\} \nonumber \\
 A_{5,6,8} ~\rightarrow ~   D_{6}(19)&=&\{ {1\over 2},-{1\over 2},{1\over 2},-{1\over 2},
   {1\over 2},{1\over 2},{1\over {{\sqrt{2}}}}\} \nonumber \\
  A_{5,6,9} ~\rightarrow ~   D_{6}(20)&=& \{ {1\over 2},-{1\over 2},{1\over 2},{1\over 2},
  -{1\over 2},{1\over 2},
   {1\over {{\sqrt{2}}}}\} \nonumber \\
 A_{5,6,10} ~\rightarrow ~   D_{6}(21)&=&  \{ {1\over 2},-{1\over 2},{1\over 2},{1\over 2},
   {1\over 2},-{1\over 2},{1\over {{\sqrt{2}}}}\} \nonumber \\
  A_{5,7,8} ~\rightarrow ~   D_{6}(22)&=&  \{ {1\over 2},{1\over 2},-{1\over 2},-{1\over 2},
  {1\over 2},{1\over 2},
   {1\over {{\sqrt{2}}}}\} \nonumber \\
 A_{5,7,9} ~\rightarrow ~   D_{6}(23)&=&  \{ {1\over 2},{1\over 2},-{1\over 2},{1\over 2},
   -{1\over 2},{1\over 2},{1\over {{\sqrt{2}}}}\} \nonumber \\
 A_{5,7,10} ~\rightarrow ~   D_{6}(24)&=&   \{ {1\over 2},{1\over 2},-{1\over 2},{1\over 2},
 {1\over 2},-{1\over 2},
   {1\over {{\sqrt{2}}}}\} \nonumber \\
 A_{5,8,9} ~\rightarrow ~   D_{6}(25)&=& \{ {1\over 2},{1\over 2},{1\over 2},-{1\over 2},
   -{1\over 2},{1\over 2},{1\over {{\sqrt{2}}}}\} \nonumber \\
  A_{5,8,10} ~\rightarrow ~   D_{6}(26)&=&  \{ {1\over 2},{1\over 2},{1\over 2},-{1\over 2},{1\over 2},-{1\over 2},
   {1\over {{\sqrt{2}}}}\} \nonumber \\
  A_{5,9,10} ~\rightarrow ~   D_{6}(27)&=& \{ {1\over 2},{1\over 2},{1\over 2},{1\over 2},
   -{1\over 2},-{1\over 2},{1\over {{\sqrt{2}}}} \}
\label{d6root}
\end{eqnarray}
Finally, in $D=9$ we have the only root of the $E_2$ root space:
\begin{equation}
A_{10} ~\rightarrow ~ \Phi^+(E_2) \, = \,
\{ -{1\over 2},-{1\over 2},-{1\over 2},-{1\over 2},-{1\over 2},-{1\over 2},
  {1\over {{\sqrt{2}}}}\}
  \label{e2root}
\end{equation}
\chapter{Gaugeable isometries in ${\cal A}$.}
In this appendix we list the matrices describing the action of the
abelian ideals on the space of vector fields. For all cases $D \ge
5$ the numbering of rows and columns of the matrix corresponds to the
listing of generators $D_r(i)$ given in the previous appendix. In the
case $D=4$ we need more care. The vector fields are associated with
a subset of $28$ weights of the $56$ fundamental weights of $E_7$.
For these weights we have chosen a conventional numbering that 
will be defined in the last chapter. Using this numbering
the following matrix describes the action of the 10 dimensional  subspace of $\ID^+_6$ made of
``electric'' generators (that is the intersection of the abelian ideal $\cA_7$ with
the ``electric'' subgroup $SL(8,\IR)$ of the U--duality group) on the 28 dimensional
column vector of the
 ``electric'' field strengths.
It is a linear combination $\sum s_i N_i$, where $N_i$ are the  ten nilpotent  generators
and $s_i$ the corresponding parameters of the solvable Lie algebra.
The maximal number of vector fields which correspond to gauging translational isometries
 is  found by looking at the maximal number of vectors which are annihilated by the maximal subset of abelian $N_i$
generators.
It turns out that in the present four dimensional case this number is 7.
{ \tiny
\begin{flushleft}
\begin{equation}
  \label{rapsei}
\left ( \matrix{ 0 & s1 & s2 & s3 & s4 & 0 & 0 & 0 & 0 & 0 & 0 & 0 & 0 & 0 & 0 & 0 & 0 & 0
   & 0 & 0 & 0 & 0 & 0 & 0 & 0 & 0 & 0 & 0 \cr 0 & 0 & 0 & 0 & 0 & 0 & 0 & 0
   & 0 & 0 & 0 & 0 & 0 & 0 & 0 & 0 & 0 & 0 & 0 & 0 & 0 & 0 & 0 & 0 & 0 & 0 & 0
   & 0 \cr 0 & 0 & 0 & 0 & 0 & 0 & 0 & 0 & 0 & 0 & 0 & 0 & 0 & 0 & 0 & 0 & 0
   & 0 & 0 & 0 & 0 & 0 & 0 & 0 & 0 & 0 & 0 & 0 \cr 0 & 0 & 0 & 0 & 0 & 0 & 0
   & 0 & 0 & 0 & 0 & 0 & 0 & 0 & 0 & 0 & 0 & 0 & 0 & 0 & 0 & 0 & 0 & 0 & 0 & 0
   & 0 & 0 \cr 0 & 0 & 0 & 0 & 0 & 0 & 0 & 0 & 0 & 0 & 0 & 0 & 0 & 0 & 0 & 0
   & 0 & 0 & 0 & 0 & 0 & 0 & 0 & 0 & 0 & 0 & 0 & 0 \cr 0 & 0 & 0 & 0 & 0 & 0
   & 0 & 0 & 0 & 0 & 0 & 0 & 0 & 0 & 0 & 0 & 0 & 0 & 0 & 0 & 0 & 0 & 0 & 0 & 0
   & 0 & 0 & 0 \cr 0 & 0 & 0 & 0 & 0 & 0 & 0 & s1 & s2 & s3 & s4 & 0 & s7 & s8 & s9
   & s10 & 0 & 0 & 0 & 0 & 0 & 0 & s5 & 0 & 0 & 0 & 0 & s6 \cr 0 & 0 & 0 & 0 & 0
   & 0 & 0 & 0 & 0 & 0 & 0 & 0 & 0 & 0 & 0 & 0 & s8 & s9 & s10 & 0 & 0 & 0 & 0 &
  5 & 0 & 0 & 0 & 0 \cr 0 & 0 & 0 & 0 & 0 & 0 & 0 & 0 & 0 & 0 & 0 & 0 & 0 & 0
   & 0 & 0 & s7 & 0 & 0 & s9 & s10 & 0 & 0 & 0 & s5 & 0 & 0 & 0 \cr 0 & 0 & 0 & 0
   & 0 & 0 & 0 & 0 & 0 & 0 & 0 & 0 & 0 & 0 & 0 & 0 & 0 & s7 & 0 & s8 & 0 & s10 &
  0 & 0 & 0 & s5 & 0 & 0 \cr 0 & 0 & 0 & 0 & 0 & 0 & 0 & 0 & 0 & 0 & 0 & 0 & 0
   & 0 & 0 & 0 & 0 & 0 & s7 & 0 & s8 & s9 & 0 & 0 & 0 & 0 & s5 & 0 \cr 0 & 0 & 0
   & 0 & 0 & 0 & 0 & 0 & 0 & 0 & 0 & 0 & 0 & 0 & 0 & 0 & 0 & 0 & 0 & 0 & 0 & 0
   & 0 & s7 & s8 & s9 & s10 & 0 \cr 0 & s6 & 0 & 0 & 0 & 0 & 0 & 0 & 0 & 0 & 0 & 0
   & 0 & 0 & 0 & 0 & s2 & s3 & s4 & 0 & 0 & 0 & 0 & 0 & 0 & 0 & 0 & 0 \cr 0 & 0
   & s6 & 0 & 0 & 0 & 0 & 0 & 0 & 0 & 0 & 0 & 0 & 0 & 0 & 0 & s1 & 0 & 0 & s3 & s4
   & 0 & 0 & 0 & 0 & 0 & 0 & 0 \cr 0 & 0 & 0 & s6 & 0 & 0 & 0 & 0 & 0 & 0 & 0
   & 0 & 0 & 0 & 0 & 0 & 0 & s1 & 0 & s2 & 0 & s4 & 0 & 0 & 0 & 0 & 0 & 0 \cr 0
   & 0 & 0 & 0 & s6 & 0 & 0 & 0 & 0 & 0 & 0 & 0 & 0 & 0 & 0 & 0 & 0 & 0 & s1 & 0
   & s2 & s3 & 0 & 0 & 0 & 0 & 0 & 0 \cr 0 & 0 & 0 & 0 & 0 & 0 & 0 & 0 & 0 & 0
   & 0 & 0 & 0 & 0 & 0 & 0 & 0 & 0 & 0 & 0 & 0 & 0 & 0 & 0 & 0 & 0 & 0 & 0
   \cr 0 & 0 & 0 & 0 & 0 & 0 & 0 & 0 & 0 & 0 & 0 & 0 & 0 & 0 & 0 & 0 & 0 & 0
   & 0 & 0 & 0 & 0 & 0 & 0 & 0 & 0 & 0 & 0 \cr 0 & 0 & 0 & 0 & 0 & 0 & 0 & 0
   & 0 & 0 & 0 & 0 & 0 & 0 & 0 & 0 & 0 & 0 & 0 & 0 & 0 & 0 & 0 & 0 & 0 & 0 & 0
   & 0 \cr 0 & 0 & 0 & 0 & 0 & 0 & 0 & 0 & 0 & 0 & 0 & 0 & 0 & 0 & 0 & 0 & 0
   & 0 & 0 & 0 & 0 & 0 & 0 & 0 & 0 & 0 & 0 & 0 \cr 0 & 0 & 0 & 0 & 0 & 0 & 0
   & 0 & 0 & 0 & 0 & 0 & 0 & 0 & 0 & 0 & 0 & 0 & 0 & 0 & 0 & 0 & 0 & 0 & 0 & 0
   & 0 & 0 \cr 0 & 0 & 0 & 0 & 0 & 0 & 0 & 0 & 0 & 0 & 0 & 0 & 0 & 0 & 0 & 0
   & 0 & 0 & 0 & 0 & 0 & 0 & 0 & 0 & 0 & 0 & 0 & 0 \cr 0 & 0 & 0 & 0 & 0 & s6
   & 0 & 0 & 0 & 0 & 0 & 0 & 0 & 0 & 0 & 0 & 0 & 0 & 0 & 0 & 0 & 0 & 0 & s1 & s2
   & s3 & s4 & 0 \cr 0 & 0 & 0 & 0 & 0 & 0 & 0 & 0 & 0 & 0 & 0 & 0 & 0 & 0 & 0
   & 0 & 0 & 0 & 0 & 0 & 0 & 0 & 0 & 0 & 0 & 0 & 0 & 0 \cr 0 & 0 & 0 & 0 & 0
   & 0 & 0 & 0 & 0 & 0 & 0 & 0 & 0 & 0 & 0 & 0 & 0 & 0 & 0 & 0 & 0 & 0 & 0 & 0
   & 0 & 0 & 0 & 0 \cr 0 & 0 & 0 & 0 & 0 & 0 & 0 & 0 & 0 & 0 & 0 & 0 & 0 & 0
   & 0 & 0 & 0 & 0 & 0 & 0 & 0 & 0 & 0 & 0 & 0 & 0 & 0 & 0 \cr 0 & 0 & 0 & 0
   & 0 & 0 & 0 & 0 & 0 & 0 & 0 & 0 & 0 & 0 & 0 & 0 & 0 & 0 & 0 & 0 & 0 & 0 & 0
   & 0 & 0 & 0 & 0 & 0 \cr 0 & s7 & s8 & s9 & s10 & s5 & 0 & 0 & 0 & 0 & 0 & 0 & 0
   & 0 & 0 & 0 & 0 & 0 & 0 & 0 & 0 & 0 & 0 & 0 & 0 & 0 & 0 & 0 \cr  }\right )
\end{equation}
\end{flushleft}
}
In $D=5$ the matrix $\sum s_i N_i$ is 27 dimensional and using
the numbering of eq.\eqn{d6root} is given by:
{\scriptsize
\begin{equation}
\label{rap5}
\pmatrix{ 0 & 0 & s2 & s1 & s4 & s3 & s6 & s5 & 0 & 0 & 0 & 0 & 0 & 0 & 0 & 0 & 0 & 0
   & 0 & 0 & 0 & 0 & 0 & 0 & 0 & 0 & 0 \cr 0 & 0 & 0 & 0 & 0 & 0 & 0 & 0 & 0
   & 0 & 0 & 0 & 0 & 0 & 0 & 0 & 0 & 0 & 0 & 0 & 0 & 0 & 0 & 0 & 0 & 0 & 0
   \cr 0 & s1 & 0 & 0 & 0 & 0 & 0 & 0 & 0 & 0 & 0 & 0 & 0 & 0 & 0 & 0 & 0 & 0
   & 0 & 0 & 0 & 0 & 0 & 0 & 0 & 0 & 0 \cr 0 & s2 & 0 & 0 & 0 & 0 & 0 & 0 & 0
   & 0 & 0 & 0 & 0 & 0 & 0 & 0 & 0 & 0 & 0 & 0 & 0 & 0 & 0 & 0 & 0 & 0 & 0
   \cr 0 & s3 & 0 & 0 & 0 & 0 & 0 & 0 & 0 & 0 & 0 & 0 & 0 & 0 & 0 & 0 & 0 & 0
   & 0 & 0 & 0 & 0 & 0 & 0 & 0 & 0 & 0 \cr 0 & s4 & 0 & 0 & 0 & 0 & 0 & 0 & 0
   & 0 & 0 & 0 & 0 & 0 & 0 & 0 & 0 & 0 & 0 & 0 & 0 & 0 & 0 & 0 & 0 & 0 & 0
   \cr 0 & s5 & 0 & 0 & 0 & 0 & 0 & 0 & 0 & 0 & 0 & 0 & 0 & 0 & 0 & 0 & 0 & 0
   & 0 & 0 & 0 & 0 & 0 & 0 & 0 & 0 & 0 \cr 0 & s6 & 0 & 0 & 0 & 0 & 0 & 0 & 0
   & 0 & 0 & 0 & 0 & 0 & 0 & 0 & 0 & 0 & 0 & 0 & 0 & 0 & 0 & 0 & 0 & 0 & 0
   \cr 0 & 0 & 0 & 0 & 0 & 0 & 0 & 0 & 0 & 0 & 0 & 0 & 0 & 0 & 0 & 0 & 0 & 0
   & 0 & 0 & 0 & 0 & 0 & 0 & 0 & 0 & 0 \cr 0 & 0 & 0 & 0 & 0 & 0 & 0 & 0 & 0
   & 0 & 0 & 0 & 0 & 0 & 0 & 0 & 0 & 0 & 0 & 0 & 0 & 0 & 0 & 0 & 0 & 0 & 0
   \cr 0 & 0 & 0 & 0 & 0 & 0 & 0 & 0 & 0 & 0 & 0 & 0 & 0 & 0 & 0 & 0 & 0 & 0
   & 0 & 0 & 0 & 0 & 0 & 0 & 0 & 0 & 0 \cr 0 & 0 & s13 & 0 & s14 & 0 & s15 & 0 &
  0 & 0 & 0 & 0 & 0 & 0 & 0 & 0 & 0 & s1 & s3 & s5 & s7 & 0 & 0 & 0 & 0 & 0 & 0
   \cr 0 & 0 & 0 & s9 & 0 & s10 & 0 & s11 & 0 & 0 & 0 & 0 & 0 & s2 & s4 & s6 & s8 & 0
   & 0 & 0 & 0 & 0 & 0 & 0 & 0 & 0 & 0 \cr 0 & s9 & 0 & 0 & 0 & 0 & 0 & 0 & 0
   & 0 & 0 & 0 & 0 & 0 & 0 & 0 & 0 & 0 & 0 & 0 & 0 & 0 & 0 & 0 & 0 & 0 & 0
   \cr 0 & s10 & 0 & 0 & 0 & 0 & 0 & 0 & 0 & 0 & 0 & 0 & 0 & 0 & 0 & 0 & 0 & 0
   & 0 & 0 & 0 & 0 & 0 & 0 & 0 & 0 & 0 \cr 0 & s11 & 0 & 0 & 0 & 0 & 0 & 0 & 0
   & 0 & 0 & 0 & 0 & 0 & 0 & 0 & 0 & 0 & 0 & 0 & 0 & 0 & 0 & 0 & 0 & 0 & 0
   \cr 0 & s12 & 0 & 0 & 0 & 0 & 0 & 0 & 0 & 0 & 0 & 0 & 0 & 0 & 0 & 0 & 0 & 0
   & 0 & 0 & 0 & 0 & 0 & 0 & 0 & 0 & 0 \cr 0 & s13 & 0 & 0 & 0 & 0 & 0 & 0 & 0
   & 0 & 0 & 0 & 0 & 0 & 0 & 0 & 0 & 0 & 0 & 0 & 0 & 0 & 0 & 0 & 0 & 0 & 0
   \cr 0 & s14 & 0 & 0 & 0 & 0 & 0 & 0 & 0 & 0 & 0 & 0 & 0 & 0 & 0 & 0 & 0 & 0
   & 0 & 0 & 0 & 0 & 0 & 0 & 0 & 0 & 0 \cr 0 & s15 & 0 & 0 & 0 & 0 & 0 & 0 & 0
   & 0 & 0 & 0 & 0 & 0 & 0 & 0 & 0 & 0 & 0 & 0 & 0 & 0 & 0 & 0 & 0 & 0 & 0
   \cr 0 & s16 & 0 & 0 & 0 & 0 & 0 & 0 & 0 & 0 & 0 & 0 & 0 & 0 & 0 & 0 & 0 & 0
   & 0 & 0 & 0 & 0 & 0 & 0 & 0 & 0 & 0 \cr 0 & 0 & 0 & s14 & 0 & s13 & s12 & 0 &
  0 & 0 & 0 & 0 & 0 & 0 & 0 & s7 & s5 & s4 & s2 & 0 & 0 & 0 & 0 & 0 & 0 & 0 & 0
   \cr 0 & 0 & 0 & s15 & s12 & 0 & 0 & s13 & 0 & 0 & 0 & 0 & 0 & 0 & s7 & 0 & s3 &
  6 & 0 & s2 & 0 & 0 & 0 & 0 & 0 & 0 & 0 \cr 0 & 0 & 0 & s16 & s11 & 0 & s10 & 0
   & 0 & 0 & 0 & 0 & 0 & 0 & s5 & s3 & 0 & s8 & 0 & 0 & s2 & 0 & 0 & 0 & 0 & 0 & 0
   \cr 0 & 0 & s12 & 0 & 0 & s15 & 0 & s14 & 0 & 0 & 0 & 0 & 0 & s7 & 0 & 0 & s1 &
  0 & s6 & s4 & 0 & 0 & 0 & 0 & 0 & 0 & 0 \cr 0 & 0 & s11 & 0 & 0 & s16 & s9 & 0 &
  0 & 0 & 0 & 0 & 0 & s5 & 0 & s1 & 0 & 0 & s8 & 0 & s4 & 0 & 0 & 0 & 0 & 0 & 0
   \cr 0 & 0 & s10 & 0 & s9 & 0 & 0 & s16 & 0 & 0 & 0 & 0 & 0 & s3 & s1 & 0 & 0 & 0
   & 0 & s8 & s6 & 0 & 0 & 0 & 0 & 0 & 0 \cr  }
\end{equation}}
The maximal number of gaugeable translational isometries is 12.

We list in the following, with the same notations as before, the analogous matrices in $D=6,7,8,9$,
which have dimensions 16, 10, 6 and 3 respectively.We number the rows
and columns according to eq.s\eqn{d5root},\eqn{d4root},\eqn{d3root}
and \eqn{e2root}. (In the last case, corresponding to D=9 there are
two additional vector fields besides the one corresponding to the
$E_2$ root.
\par
In each case the number of gaugeable translational isometries turns out to be 6,4,3,1 respectively.

$D=6$:
{\scriptsize
\begin{equation}
\label{rap4}
\pmatrix{ 0 & 0 & s2 & s1 & s4 & s3 & s6 & s5 & 0 & 0 & 0 & 0 & 0 & 0 & 0 & 0 \cr 0
   & 0 & 0 & 0 & 0 & 0 & 0 & 0 & 0 & 0 & 0 & 0 & 0 & 0 & 0 & 0 \cr 0 & s1 & 0
   & 0 & 0 & 0 & 0 & 0 & 0 & 0 & 0 & 0 & 0 & 0 & 0 & 0 \cr 0 & s2 & 0 & 0 & 0
   & 0 & 0 & 0 & 0 & 0 & 0 & 0 & 0 & 0 & 0 & 0 \cr 0 & s3 & 0 & 0 & 0 & 0 & 0
   & 0 & 0 & 0 & 0 & 0 & 0 & 0 & 0 & 0 \cr 0 & s4 & 0 & 0 & 0 & 0 & 0 & 0 & 0
   & 0 & 0 & 0 & 0 & 0 & 0 & 0 \cr 0 & s5 & 0 & 0 & 0 & 0 & 0 & 0 & 0 & 0 & 0
   & 0 & 0 & 0 & 0 & 0 \cr 0 & s6 & 0 & 0 & 0 & 0 & 0 & 0 & 0 & 0 & 0 & 0 & 0
   & 0 & 0 & 0 \cr 0 & 0 & 0 & s7 & 0 & s8 & 0 & s9 & 0 & s2 & s4 & s6 & 0 & 0 & 0
   & 0 \cr 0 & s7 & 0 & 0 & 0 & 0 & 0 & 0 & 0 & 0 & 0 & 0 & 0 & 0 & 0 & 0 \cr 0
   & s8 & 0 & 0 & 0 & 0 & 0 & 0 & 0 & 0 & 0 & 0 & 0 & 0 & 0 & 0 \cr 0 & s9 & 0
   & 0 & 0 & 0 & 0 & 0 & 0 & 0 & 0 & 0 & 0 & 0 & 0 & 0 \cr 0 & s10 & 0 & 0 & 0
   & 0 & 0 & 0 & 0 & 0 & 0 & 0 & 0 & 0 & 0 & 0 \cr 0 & 0 & 0 & s10 & s9 & 0 & s8
   & 0 & 0 & 0 & s5 & s3 & s2 & 0 & 0 & 0 \cr 0 & 0 & s9 & 0 & 0 & s10 & s7 & 0 & 0
   & s5 & 0 & s1 & s4 & 0 & 0 & 0 \cr 0 & 0 & s8 & 0 & s7 & 0 & 0 & s10 & 0 & s3 & s1
   & 0 & s6 & 0 & 0 & 0 \cr  }
 \end{equation}}
$D=7$:
\begin{equation}
\label{rap3}
\pmatrix{ 0 & 0 & s2 & s1 & s4 & s3 & 0 & 0 & 0 & 0 \cr 0 & 0 & 0 & 0 & 0 & 0 & 0
   & 0 & 0 & 0 \cr 0 & s1 & 0 & 0 & 0 & 0 & 0 & 0 & 0 & 0 \cr 0 & s2 & 0 & 0 & 0
   & 0 & 0 & 0 & 0 & 0 \cr 0 & s3 & 0 & 0 & 0 & 0 & 0 & 0 & 0 & 0 \cr 0 & s4 & 0
   & 0 & 0 & 0 & 0 & 0 & 0 & 0 \cr 0 & 0 & 0 & s5 & 0 & s6 & 0 & s2 & s4 & 0 \cr 0
   & s5 & 0 & 0 & 0 & 0 & 0 & 0 & 0 & 0 \cr 0 & s6 & 0 & 0 & 0 & 0 & 0 & 0 & 0
   & 0 \cr 0 & 0 & s6 & 0 & s5 & 0 & 0 & s3 & s1 & 0 \cr  }
\end{equation}

$D=8$:
\begin{equation}
\label{rap2}
\pmatrix{ 0 & 0 & s2 & s1 & 0 & 0 \cr 0 & 0 & 0 & 0 & 0 & 0 \cr 0 & s1 & 0 & 0 & 0
   & 0 \cr 0 & s2 & 0 & 0 & 0 & 0 \cr 0 & 0 & 0 & s3 & 0 & s2 \cr 0 & s3 & 0 & 0
   & 0 & 0 \cr  }
\end{equation}

$D=9$:
\begin{equation}
\label{rap1}
\pmatrix{ 0 & 0 & s1 \cr 0 & 0 & 0 \cr 0 & 0 & 0 \cr   }
\end{equation}
\chapter{Alekseevskii structure of $SO(4,m)/SO(4)\times SO(m)$ manifold}
In the present Appendix I will represent the generators of 
the quaternionic algebra $V_m$ generating the manifold
\begin{equation}
{\cal QM}_m\, =\, \frac{SO(4,m)}{SO(4)\otimes SO(m)}
\label{qmm}
\end{equation} 
in terms of the ${\bf so(4,m)}$ generators in the 
{\it canonical representation}: $H_{\alpha}\, ,\, E_{\pm\beta}$,
(H being the Cartan generators, $E_\alpha$ the {\it shift} operators and 
$\alpha$ ; $\beta$ positive roots). \par
Let us consider first the case {\it m odd}, i.e. $m=2k-1$.\par
The algebra ${\bf so(4,m)}$ has rank $k+1$ and is described by the Dynkin
diagram $B_{k+1}$. With respect to an orthonormal basis
$(\epsilon_i)$ of $\IR^{k+1}$, its {\it simple roots} have the following 
expression:
\begin{equation}
\alpha_1=\epsilon_1-\epsilon_2\, , \,\alpha_2=\epsilon_2-\epsilon_3\, ,\,
......,\alpha_k=\epsilon_k-\epsilon_{k+1}\, , \,\alpha_{k+1}=\epsilon_{k+1}
\label{srso4m}
\end{equation}
The space $\Phi^+$ of positive roots has the following content:
\begin{equation}
\Phi^+\, =\,\cases{\epsilon_i\pm \epsilon_j & \mbox{long roots}\cr \epsilon_i & \mbox{short roots}}\,\,i,j=1,...,k+1
\label{psrot}
\end{equation} 
Choosing as {\it non--compact}  Cartan generators (i.e. belonging to
 ${\cal C}_K$) those corresponding to the roots $\epsilon_1\pm\epsilon_2\,
;\,\epsilon_3\pm\epsilon_4$, the positive roots expressed only in terms
of $\epsilon_i=5,...,k+1$ will not enter the set $\Delta^+$ introduced in 
the Chapter 1, or equivalently they are ruled out from the
general definition \ref{iwa} of $V_m$. Since ${\cal C}_K\neq {\cal C}$
$V_m$ is an other example of non--maximally non compact solvable Lie algebra.
\par
It is worth recalling the general structure of quaternionic algebras 
defined by \eqn{Vstru}, \eqn{f0tf0}, eqs{Vstru2}:
\begin{eqnarray}
V_m\, &=&\, U\oplus \tilde{U}\nonumber\\
U\, &=&\, F_0 \oplus W_m\nonumber\\
\tilde{U}\, &=&\, \tilde{F}_0 \oplus \tilde{W}_m\nonumber\\
J^1\cdot U\, &=&\,U\,\,\,J^2\cdot U\, =\, \tilde{U}
\end{eqnarray}
The K\"ahlerian algebra $W_m$ has the form given in \eqn{strutto}. Following 
the conventions adopted in Chapter 3, let us denote by $\{p_{i},q_{i}\quad (i=0,1,2,3), \tilde{z}^{\pm}_{k}\quad (k=1,...,2k-5)\}$ and $\{ h_{i},g_{i}\quad (i=0,1,2,3), z^{\pm}_{k} (k=1,...,2k-5)\}$ orthonormal bases 
of $\tilde{U}$ and $U$ respectively. The subset $\{h_{i},g_{i}\}$ generates
the {\it key algebras}  $F_i\,\, i=0,1,2,3$ of $U$ while $\{p_{i},q_{i}\}$
generate the image of $F_i$ through $J^2$ in $\tilde{U}$, namely
$\tilde{F}_0\oplus\tilde{F}_1\oplus\tilde{F}_2\oplus\tilde{F}_3$. Finally
$z^{\pm}_{k}$ are the generators of $Z^+\oplus Z^-$ and $\tilde{z}^{\pm}_{k}$
generate $\tilde{Z}^+\oplus \tilde{Z}^-= J^2\cdot (Z^+\oplus Z^-)$.
The explicit form of the above generators is:
\begin{eqnarray}
\mbox{{\bf U:}}\qquad\qquad\qquad &&\nonumber\\
h_0 \,&=&\,\frac{1}{2}H_{\epsilon_1+\epsilon_2}\,\, ; \,\,\, g_0\, =\, 
E_{\epsilon_1+\epsilon_2}\nonumber\\
h_1 \,&=&\,\frac{1}{2}H_{\epsilon_3+\epsilon_4}\,\, ; \,\,\, g_1\, =\, 
E_{\epsilon_3+\epsilon_4}\nonumber\\
h_2 \,&=&\,\frac{1}{2}H_{\epsilon_3-\epsilon_4}\,\, ; \,\,\, g_2\, =\, 
E_{\epsilon_3-\epsilon_4}\nonumber\\
h_3 \,&=&\,\frac{1}{2}H_{\epsilon_1-\epsilon_2}\,\, ; \,\,\, g_3\, =\, 
E_{\epsilon_1-\epsilon_2}\nonumber\\
\matrix{ z^+_{i\vert (1,2)} \cr z^-_{i\vert (1,2)}}\,&=&\,\matrix{
E_{\epsilon_3+\epsilon_{i+4}}\pm E_{\epsilon_3-\epsilon_{i+4}}\cr  
E_{\epsilon_4+\epsilon_{i+4}}\mp 
E_{\epsilon_4-\epsilon_{i+4}}} \,\,(i=1,...,k-3)\nonumber\\
z^+_{2k-5}\,&=&\,E_{\epsilon_3}\,\,;\,\,
z^-_{2k-5}\,=\,E_{\epsilon_4}\nonumber\\
\mbox{{\bf $\tilde{U}$:}}\qquad\qquad\qquad &&\nonumber\\
p_0 \,&=&\,E_{\epsilon_1+\epsilon_3}\,\, ; \,\,\, q_0\, =\, 
E_{\epsilon_2-\epsilon_3}\nonumber\\
p_1 \,&=&\,E_{\epsilon_2+\epsilon_4}\,\, ; \,\,\, q_1\, =\, 
E_{\epsilon_1-\epsilon_4}\nonumber\\
p_2 \,&=&\,E_{\epsilon_2-\epsilon_4}\,\, ; \,\,\, q_2\, =\, 
E_{\epsilon_1+\epsilon_4}\nonumber\\
p_3 \,&=&\,E_{\epsilon_1-\epsilon_3}\,\, ; \,\,\, q_3\, =\, 
E_{\epsilon_2+\epsilon_3}\nonumber\\
\matrix{ \tilde{z}^+_{i\vert (1,2)} \cr \tilde{z}^-_{i\vert (1,2)}}\,&=&\,\matrix{
E_{\epsilon_1+\epsilon_{i+4}}\pm E_{\epsilon_1-\epsilon_{i+4}}\cr  
E_{\epsilon_2+\epsilon_{i+4}}\mp 
E_{\epsilon_2-\epsilon_{i+4}}} \,\,(i=1,...,k-3)\nonumber\\
\tilde{z}^+_{2k-5}\,&=&\,E_{\epsilon_1}\,\,;\,\,
\tilde{z}^-_{2k-5}\,=\,E_{\epsilon_2}
\label{exprep}
\end{eqnarray}
In the case {\it m even}, namely $m=2(k-1)$, the algebra ${\bf so(4,m)}$
is described by the Dynkin diagram $D_{k+1}$. The positive--root 
space consists now of all the {\it long} roots in \eqn{psrot} and therefore 
the form of the $V_m$ generators is the same as given in \eqn{exprep}
except for the fact that there are $4$ generators less, namely:
$z^{\pm}_{2k-5}$ and $\tilde{z}^\pm_{2k-5}$.

\chapter{$Sp(8,\IR)$ representation of the STU--isometries.}
The explicit expression for the generators of  $SL(2,\IR)^3$ of section 7.1 is:
\begin{eqnarray}
L_{0 }^{(1)}&=&
\left(
\matrix{ 0 & 0 & 0 & {1\over 2} & 0 & 0 & 0 & 0
   \cr 0 & 0 & -{1\over 2} & 0 & 0 & 0 & 0 & 0
   \cr 0 & -{1\over 2} & 0 & 0 & 0 & 0 & 0 & 0
   \cr {1\over 2} & 0 & 0 & 0 & 0 & 0 & 0 & 0
   \cr 0 & 0 & 0 & 0 & 0 & 0 & 0 & -{1\over 2}
   \cr 0 & 0 & 0 & 0 & 0 & 0 & {1\over 2} & 0
   \cr 0 & 0 & 0 & 0 & 0 & {1\over 2} & 0 & 0
   \cr 0 & 0 & 0 & 0 & -{1\over 2} & 0 & 0 & 0
   \cr  } \right) \nonumber\\ \null & \null & \null \nonumber\\
L_{+ }^{(1)}&=& \left(
\matrix{ 0 & -{1\over 2} & -{1\over 2} & 0 & 0 &
  0 & 0 & 0 \cr {1\over 2} & 0 & 0 & -{1\over 2}
   & 0 & 0 & 0 & 0 \cr -{1\over 2} & 0 & 0 &
  {1\over 2} & 0 & 0 & 0 & 0 \cr 0 & -{1\over 2}
   & -{1\over 2} & 0 & 0 & 0 & 0 & 0 \cr 0 & 0 &
  0 & 0 & 0 & -{1\over 2} & {1\over 2} & 0 \cr 0
   & 0 & 0 & 0 & {1\over 2} & 0 & 0 & {1\over 2}
   \cr 0 & 0 & 0 & 0 & {1\over 2} & 0 & 0 &
  {1\over 2} \cr 0 & 0 & 0 & 0 & 0 & {1\over 2}
   & -{1\over 2} & 0 \cr  } \right)\nonumber\\ \null & \null & \null \nonumber\\
L_{- }^{(1)}&=& \left(
\matrix{ 0 & {1\over 2} & -{1\over 2} & 0 & 0 & 0
   & 0 & 0 \cr -{1\over 2} & 0 & 0 & -{1\over 2}
   & 0 & 0 & 0 & 0 \cr -{1\over 2} & 0 & 0 &
  -{1\over 2} & 0 & 0 & 0 & 0 \cr 0 & -{1\over 2}
   & {1\over 2} & 0 & 0 & 0 & 0 & 0 \cr 0 & 0 & 0
   & 0 & 0 & {1\over 2} & {1\over 2} & 0 \cr 0 &
  0 & 0 & 0 & -{1\over 2} & 0 & 0 & {1\over 2}
   \cr 0 & 0 & 0 & 0 & {1\over 2} & 0 & 0 &
  -{1\over 2} \cr 0 & 0 & 0 & 0 & 0 & {1\over 2}
   & {1\over 2} & 0 \cr  }\right)\nonumber\\ \null & \null & \null \nonumber\\
L_{0 }^{(2)}&=& \left(
 \matrix{ 0 & 0 & 0 & -{1\over 2} & 0 & 0 & 0 & 0
    \cr 0 & 0 & -{1\over 2} & 0 & 0 & 0 & 0 & 0
    \cr 0 & -{1\over 2} & 0 & 0 & 0 & 0 & 0 & 0
    \cr -{1\over 2} & 0 & 0 & 0 & 0 & 0 & 0 & 0
    \cr 0 & 0 & 0 & 0 & 0 & 0 & 0 & {1\over 2}
\   \cr 0 & 0 & 0 & 0 & 0 & 0 & {1\over 2} & 0
    \cr 0 & 0 & 0 & 0 & 0 & {1\over 2} & 0 & 0
    \cr 0 & 0 & 0 & 0 & {1\over 2} & 0 & 0 & 0
\   \cr  }  \right)\nonumber\\ \null & \null & \null \nonumber\\
L_{+ }^{(2)}&=& \left(
\matrix{ 0 & -{1\over 2} & -{1\over 2} & 0 & 0 &
  0 & 0 & 0 \cr {1\over 2} & 0 & 0 & {1\over 2}
   & 0 & 0 & 0 & 0 \cr -{1\over 2} & 0 & 0 &
  -{1\over 2} & 0 & 0 & 0 & 0 \cr 0 & {1\over 2}
   & {1\over 2} & 0 & 0 & 0 & 0 & 0 \cr 0 & 0 & 0
   & 0 & 0 & -{1\over 2} & {1\over 2} & 0 \cr 0
   & 0 & 0 & 0 & {1\over 2} & 0 & 0 & -{1\over 2}
   \cr 0 & 0 & 0 & 0 & {1\over 2} & 0 & 0 &
  -{1\over 2} \cr 0 & 0 & 0 & 0 & 0 & -{1\over 2}
   & {1\over 2} & 0 \cr  } \right) \nonumber\\ \null & \null & \null \nonumber\\
L_{- }^{(2)}&=& \left(
\matrix{ 0 & {1\over 2} & -{1\over 2} & 0 & 0 & 0
   & 0 & 0 \cr -{1\over 2} & 0 & 0 & {1\over 2}
   & 0 & 0 & 0 & 0 \cr -{1\over 2} & 0 & 0 &
  {1\over 2} & 0 & 0 & 0 & 0 \cr 0 & {1\over 2}
   & -{1\over 2} & 0 & 0 & 0 & 0 & 0 \cr 0 & 0 &
  0 & 0 & 0 & {1\over 2} & {1\over 2} & 0 \cr 0
   & 0 & 0 & 0 & -{1\over 2} & 0 & 0 &
  -{1\over 2} \cr 0 & 0 & 0 & 0 & {1\over 2} & 0
   & 0 & {1\over 2} \cr 0 & 0 & 0 & 0 & 0 &
  -{1\over 2} & -{1\over 2} & 0 \cr  } \right)\nonumber\\ \null & \null & \null \nonumber\\
L_{0 }^{(3)}&=& \left(
\matrix{ {1\over 2} & 0 & 0 & 0 & 0 & 0 & 0 & 0
   \cr 0 & {1\over 2} & 0 & 0 & 0 & 0 & 0 & 0
   \cr 0 & 0 & {1\over 2} & 0 & 0 & 0 & 0 & 0
   \cr 0 & 0 & 0 & {1\over 2} & 0 & 0 & 0 & 0
   \cr 0 & 0 & 0 & 0 & -{1\over 2} & 0 & 0 & 0
   \cr 0 & 0 & 0 & 0 & 0 & -{1\over 2} & 0 & 0
   \cr 0 & 0 & 0 & 0 & 0 & 0 & -{1\over 2} & 0
   \cr 0 & 0 & 0 & 0 & 0 & 0 & 0 & -{1\over 2}
   \cr  } \right)\nonumber\\ \null & \null & \null \nonumber\\
L_{+ }^{(3)}&=& \left(
\matrix{ 0 & 0 & 0 & 0 & 1 & 0 & 0 & 0 \cr 0 & 0
   & 0 & 0 & 0 & 1 & 0 & 0 \cr 0 & 0 & 0 & 0 & 0
   & 0 & -1 & 0 \cr 0 & 0 & 0 & 0 & 0 & 0 & 0 &
  -1 \cr 0 & 0 & 0 & 0 & 0 & 0 & 0 & 0 \cr 0 & 0
   & 0 & 0 & 0 & 0 & 0 & 0 \cr 0 & 0 & 0 & 0 & 0
   & 0 & 0 & 0 \cr 0 & 0 & 0 & 0 & 0 & 0 & 0 & 0
   \cr  } \right)\nonumber\\ \null & \null & \null \nonumber\\
L_{- }^{(3)}&=& \left(
\matrix{ 0 & 0 & 0 & 0 & 0 & 0 & 0 & 0 \cr 0 & 0
   & 0 & 0 & 0 & 0 & 0 & 0 \cr 0 & 0 & 0 & 0 & 0
   & 0 & 0 & 0 \cr 0 & 0 & 0 & 0 & 0 & 0 & 0 & 0
   \cr 1 & 0 & 0 & 0 & 0 & 0 & 0 & 0 \cr 0 & 1 &
  0 & 0 & 0 & 0 & 0 & 0 \cr 0 & 0 & -1 & 0 & 0 &
  0 & 0 & 0 \cr 0 & 0 & 0 & -1 & 0 & 0 & 0 & 0
   \cr  } \right) \nonumber\\ \null & \null & \null \nonumber\\
\end{eqnarray}

Furthermore, the explicit expression for the coset representatives
of  $\frac{SL(2,\IR)^3}{U(1)^3}$ in the same section is:
 \begin{eqnarray}
&\IL^{(1)}\left(h_1,a_1\right)=& \nonumber\\
&\null& \nonumber\\
& \left(
\matrix{ \cosh h_1 & -a_1 & -a_1 & \sinh h_1 &
  0 & 0 & 0 & 0 \cr a_1 & \cosh h_1 & -\sinh h_1 &
  -a_1 & 0 & 0 & 0 & 0 \cr -a_1 & -\sinh h_1 &
  \cosh h_1 & a_1 & 0 & 0 & 0 & 0 \cr \sinh h_1 &
  -a_1 & -a_1 & \cosh h_1 & 0 & 0 & 0 & 0 \cr 0 & 0 & 0 &
  0 & \cosh h_1 & -a_1 & a_1 & -\sinh h_1 \cr 0
   & 0 & 0 & 0 & a_1 & \cosh h_1 & \sinh h_1 &
  a_1 \cr 0 & 0 & 0 & 0 & a_1 & \sinh h_1 &
  \cosh h_1 & a_1 \cr 0 & 0 & 0 & 0 & -\sinh h_1 &
  a_1 & -a_1 & \cosh h_1 \cr  }  \right)&\nonumber\\
  &\null & \nonumber\\
 &\IL^{(2)}\left(h_2,a_2\right)=& \nonumber\\
 & \left(
 \matrix{ \cosh   h_2 & - a_2 & - a_2 & -\sinh  h_2
    & 0 & 0 & 0 & 0 \cr  a_2 & \cosh   h_2 & -\sinh  h_2
    &  a_2 & 0 & 0 & 0 & 0 \cr - a_2 & -\sinh  h_2 &
   \cosh   h_2 & - a_2 & 0 & 0 & 0 & 0 \cr -\sinh  h_2 &
    a_2 &  a_2 & \cosh   h_2 & 0 & 0 & 0 & 0 \cr 0 & 0 & 0 & 0
    & \cosh   h_2 & - a_2 &  a_2 & \sinh  h_2 \cr 0
    & 0 & 0 & 0 &  a_2 & \cosh   h_2 & \sinh  h_2 &
   - a_2 \cr 0 & 0 & 0 & 0 &  a_2 & \sinh  h_2 &
   \cosh   h_2 & - a_2 \cr 0 & 0 & 0 & 0 & \sinh  h_2 &
   - a_2 &  a_2 & \cosh   h_2 \cr  } \right)& \nonumber\\
   & \null & \nonumber \\
  &\IL^{(3)}\left(h_3,a_3\right)=& \nonumber\\
  & \left(
  \matrix{ 1\,{e^{h_3}} & 0 & 0 & 0 & 2\,a_3 & 0 & 0 & 0 \cr 0 &
    1\,{e^{h_3}} & 0 & 0 & 0 &  \,a_3 & 0 & 0 \cr 0 & 0 &
    1\,{e^{h_3}} & 0 & 0 & 0 & - \,a_3 & 0 \cr 0 & 0 & 0 &
    1\,{e^{h_3}} & 0 & 0 & 0 & - \,a_3 \cr 0 & 0 & 0 & 0 &
    {1\over {{e^{h_3}}}} & 0 & 0 & 0 \cr 0 & 0 & 0 & 0 & 0 &
    {1\over {{e^{h_3}}}} & 0 & 0 \cr 0 & 0 & 0 & 0 & 0 & 0 &
    {1\over {{e^{h_3}}}} & 0 \cr 0 & 0 & 0 & 0 & 0 & 0 & 0 &
    {1\over {{e^{h_3}}}} \cr  }\right)& \nonumber\\
\end{eqnarray}


\par
\vfill
\eject


\chapter{Tables on $E_{7(7)}$ and $SU(8)$.}
In this appendix we give several tables concerning various results
obtained by computer-aided computations about roots and weights  and
their relations to the physical fields of the solvable Lie algebra of
$E_{7(7)}/SU(8)$.

\begin{table}[ht]
\caption{{\bf
The abelian ideals $ \ID^{+}_{r}$ and the roots of $E_{7(7)}$:}}
\label{dideals}
\begin{center}
\begin{tabular}{||lcl|c|lcl||}
\hline
\hline
  Type IIA & Root & Dynkin & \null  &Type IIA & Root & Dynkin \\
  field    & name & labels & \null  &field    & name & labels \\
\hline
\null & \null & \null & $\ID^+_1$ & \null & \null & \null \\
$ A_{10}$ &  ${\vec \alpha}_{1,1}$ & $\{ 0,0,0,0,0,0,1\}$ & \null &
\null & \null & \null \\
\hline
\null & \null & \null & $\ID^+_2$ & \null & \null & \null \\
$B_{9,10}$ & ${\vec \alpha}_{2,1}$ &$\{ 0,0,0,0,0,1,0\}$ & \null &
$g_{9,10}$ & ${\vec \alpha}_{2,2}$& $\{ 0,0,0,0,1,0,0\}$ \\
$A_9$ & ${\vec \alpha}_{2,3}$ & $\{ 0,0,0,0,0,1,1\}$ & \null &
\null & \null & \null \\
\hline
\null & \null & \null & $\ID^+_3$ & \null & \null & \null \\
$B_{8,9}$ & ${\vec \alpha}_{3,1}$ & $\{ 0,0,0,1,1,1,0\}$ & \null &
$g_{8,9}$ & ${\vec \alpha}_{3,2}$ & $\{ 0,0,0,1,0,0,0\}$ \\
$B_{8,10} $ & ${\vec \alpha}_{3,3} $ & $\{ 0,0,0,1,0,1,0\} $ & \null &
$g_{8,10}$ & ${\vec \alpha}_{3,4} $ & $\{ 0,0,0,1,1,0,0\} $    \\
$A_8 $ & ${\vec \alpha}_{3,5} $ & $\{ 0,0,0,1,1,1,1\} $    & \null &
$A_{8,9,10} $ & ${\vec \alpha}_{3,6} $ & $\{ 0,0,0,1,0,1,1\} $ \\
\hline
\null & \null & \null & $\ID^+_4$ & \null & \null & \null \\
$B_{7,8} $ & ${\vec \alpha}_{4,1} $ & $ \{ 0,0,1,2,1,1,0\}  $ & \null &
$g_{7,8} $ & ${\vec \alpha}_{4,2} $ & $ \{ 0,0,1,0,0,0,0\}  $ \\
$B_{7,9} $ & ${\vec \alpha}_{4,3} $ & $ \{ 0,0,1,1,1,1,0\}  $ & \null &
$g_{7,9} $ & ${\vec \alpha}_{4,4} $ & $ \{ 0,0,1,1,0,0,0\}  $ \\
$B_{7,10} $ & ${\vec \alpha}_{4,5} $ & $ \{ 0,0,1,1,0,1,0\}  $ & \null &
$g_{7,10} $ & ${\vec \alpha}_{4,6} $ & $ \{ 0,0,1,1,1,0,0\}  $ \\
$A_{7,9,10}$ & ${\vec \alpha}_{4,7} $ & $ \{ 0,0,1,2,1,1,1\}  $ & \null &
$A_{7,8,10}$ & ${\vec \alpha}_{4,8} $ & $ \{ 0,0,1,1,1,1,1\}  $ \\
 $A_{7,8,9}$ & ${\vec \alpha}_{4,9} $ & $ \{ 0,0,1,1,0,1,1\}  $ & \null &
$A_{7} $ & ${\vec \alpha}_{4,10} $ & $ \{ 0,0,1,2,1,2,1\} $ \\
\hline
\null & \null & \null & $\ID^+_5$ & \null & \null & \null \\
$ B_{6,7}$ & ${\vec \alpha}_{5,1} $ & $ \{ 0,1,2,2,1,1,0\}  $ & \null &
$ g_{6,7}$ & ${\vec \alpha}_{5,2} $ & $ \{ 0,1,0,0,0,0,0\}  $ \\
$ B_{6,8}$ & ${\vec \alpha}_{5,3} $ & $ \{ 0,1,1,2,1,1,0\}  $ & \null &
$ g_{6,8}$ & ${\vec \alpha}_{5,4} $ & $ \{ 0,1,1,0,0,0,0\}  $ \\
$ B_{6,9}$ & ${\vec \alpha}_{5,5} $ & $ \{ 0,1,1,1,1,1,0\}  $ & \null &
$ g_{6,9}$ & ${\vec \alpha}_{5,6} $ & $ \{ 0,1,1,1,0,0,0\}  $ \\
$ B_{6,10}$ & ${\vec \alpha}_{5,7} $ & $ \{ 0,1,1,1,0,1,0\}  $ & \null &
$ g_{6,10}$ & ${\vec \alpha}_{5,8} $ & $ \{ 0,1,1,1,1,0,0\}  $ \\
$ A_{6,8,9}$ & ${\vec \alpha}_{5,9} $ & $ \{ 0,1,2,2,1,1,1\}  $ & \null &
$ A_{6,7,9}$ & ${\vec \alpha}_{5,10} $ & $ \{ 0,1,1,2,1,1,1\}  $ \\
$ A_{6,7,8}$ & ${\vec \alpha}_{5,11} $ & $ \{ 0,1,1,1,1,1,1\}  $ & \null &
$ A_{\mu\nu\rho}$ & ${\vec \alpha}_{5,12} $ & $ \{ 0,1,1,1,0,1,1\}  $ \\
$ A_{6,7,10}$ & ${\vec \alpha}_{5,13} $ & $ \{ 0,1,1,2,1,2,1\}  $ & \null &
$ A_{6,8,10}$ & ${\vec \alpha}_{5,14} $ & $ \{ 0,1,2,2,1,2,1\}  $ \\
$ A_{6,9,10}$ & ${\vec \alpha}_{5,15} $ & $ \{ 0,1,2,3,1,2,1\}  $ & \null &
$ A_6 $ & ${\vec \alpha}_{5,16} $ & $ \{ 0,1,2,3,2,2,1\}$ \\
\hline
\null & \null & \null & $\ID^+_6$ & \null & \null & \null \\
$ B_{5,6}$ & ${\vec \alpha}_{6,1} $ & $ \{ 1,2,2,2,1,1,0\}  $ & \null &
$ g_{5,6}$ & ${\vec \alpha}_{6,2} $ & $ \{ 1,0,0,0,0,0,0\}  $ \\
$ B_{5,7}$ & ${\vec \alpha}_{6,3} $ & $ \{ 1,1,2,2,1,1,0\}  $ & \null &
$ g_{5,7}$ & ${\vec \alpha}_{6,4} $ & $ \{ 1,1,0,0,0,0,0\}  $ \\
$ B_{5,8}$ & ${\vec \alpha}_{6,5} $ & $ \{ 1,1,1,2,1,1,0\}  $ & \null &
$ g_{5,8}$ & ${\vec \alpha}_{6,6} $ & $ \{ 1,1,1,0,0,0,0\}  $ \\
$ B_{5,9}$ & ${\vec \alpha}_{6,7} $ & $ \{ 1,1,1,1,1,1,0\}  $ & \null &
$ g_{5,9}$ & ${\vec \alpha}_{6,8} $ & $ \{ 1,1,1,1,0,0,0\}  $ \\
$ B_{5,10}$ & ${\vec \alpha}_{6,9} $ & $ \{ 1,1,1,1,0,1,0\}  $ & \null &
$ g_{5,10}$ & ${\vec \alpha}_{6,10} $ & $ \{ 1,1,1,1,1,0,0\}  $ \\
$ B_{\mu\nu}$ & ${\vec \alpha}_{6,11} $ & $ \{ 1,2,3,4,2,3,2\}  $ & \null &
$ A_5 $ & ${\vec \alpha}_{6,12} $ & $ \{ 1,2,3,4,2,3,1\}  $ \\
$ A_{\mu\nu 6}$ & ${\vec \alpha}_{6,13} $ & $ \{ 1,2,2,2,1,1,1\}  $ & \null &
$ A_{\mu\nu 7} $ & ${\vec \alpha}_{6,14} $ & $ \{ 1,1,2,2,1,1,1\}  $ \\
$ A_{\mu\nu 8} $ & ${\vec \alpha}_{6,15} $ & $ \{ 1,1,1,2,1,1,1\}  $ & \null &
$ A_{\mu\nu 9} $ & ${\vec \alpha}_{6,16} $ & $ \{ 1,1,1,1,1,1,1\}  $ \\
$ A_{\mu\nu 10} $ & ${\vec \alpha}_{6,17} $ & $ \{ 1,1,1,1,0,1,1\}  $ & \null &
$ A_{5,6,7} $ & $ {\vec \alpha}_{6,18} $ & $ \{ 1,1,1,2,1,2,1\}  $ \\
$ A_{5,6,8} $ & ${\vec \alpha}_{6,19} $ & $ \{ 1,1,2,2,1,2,1\}  $ & \null &
$ A_{5,6,9} $ & ${\vec \alpha}_{6,20} $ & $ \{ 1,1,2,3,1,2,1\}  $ \\
$ A_{5,6,10} $ & ${\vec \alpha}_{6,21} $ & $ \{ 1,1,2,3,2,2,1\}  $ & \null &
$ A_{5,7,8} $ & ${\vec \alpha}_{6,22} $ & $ \{ 1,2,2,2,1,2,1\}  $ \\
$ A_{5,7,9} $ & ${\vec \alpha}_{6,23} $ & $ \{ 1,2,2,3,1,2,1\}  $ & \null &
$ A_{5,7,10} $ & ${\vec \alpha}_{6,24} $ & $ \{ 1,2,2,3,2,2,1\}  $ \\
$ A_{5,8,9} $ & ${\vec \alpha}_{6,25} $ & $ \{ 1,2,3,3,1,2,1\}  $ & \null &
$ A_{5,8,10} $ & ${\vec \alpha}_{6,26} $ & $ \{ 1,2,3,3,2,2,1\}  $ \\
$ A_{5,9,10} $ & ${\vec \alpha}_{6,27} $ & $ \{ 1,2,3,4,2,2,1\}  $ & \null &
\null & \null & \null \\
\hline
\hline
\end{tabular}
\end{center}
\end{table}
\begin{table}[ht]\caption{{\bf
Weights of the ${\bf 56}$ representation of $E_{7(7)} $:}}
\label{e7weight}
\begin{center}
\begin{tabular}{||cl|c|cl||}
\hline
\hline
    Weight & $q^\ell$  & \null & Weight & $q^\ell$ \\
      name & vector  & \null & name & vector \\
\hline
\null & \null & \null & \null & \null \\
$ {\vec W}^{(1)} \, =\, $ &$  \{ 2,3,4,5,3,3,1\} $  & \null &
$ {\vec W}^{(2)} \, =\, $ &$  \{ 2,2,2,2,1,1,1\} $   \\
$ {\vec W}^{(3)} \, =\, $ &$  \{ 1,2,2,2,1,1,1\} $  & \null &
$ {\vec W}^{(4)} \, =\, $ &$  \{ 1,1,2,2,1,1,1\} $    \\
$ {\vec W}^{(5)} \, =\, $ &$  \{ 1,1,1,2,1,1,1\} $  & \null &
$ {\vec W}^{(6)} \, =\, $ &$  \{ 1,1,1,1,1,1,1\} $    \\
$ {\vec W}^{(7)} \, =\, $ &$  \{ 2,3,3,3,1,2,1\} $  & \null &
$ {\vec W}^{(8)} \, =\, $ &$  \{ 2,2,3,3,1,2,1\} $    \\
$ {\vec W}^{(9)} \, =\, $ &$  \{ 2,2,2,3,1,2,1\} $  & \null &
$ {\vec W}^{(10)} \, =\, $ &$  \{ 2,2,2,2,1,2,1\} $   \\
$ {\vec W}^{(11)} \, =\, $ &$  \{ 1,2,2,2,1,2,1\} $  & \null &
$ {\vec W}^{(12)} \, =\, $ &$  \{ 1,1,2,2,1,2,1\} $    \\
$ {\vec W}^{(13)} \, =\, $ &$  \{ 1,1,1,2,1,2,1\} $  & \null &
$ {\vec W}^{(14)} \, =\, $ &$  \{ 1,2,2,3,1,2,1\} $    \\
$ {\vec W}^{(15)} \, =\, $ &$  \{ 1,2,3,3,1,2,1\} $  & \null &
$ {\vec W}^{(16)} \, =\, $ &$  \{ 1,1,2,3,1,2,1\} $    \\
$ {\vec W}^{(17)} \, =\, $ &$  \{ 2,2,2,2,1,1,0\} $  & \null &
$ {\vec W}^{(18)} \, =\, $ &$  \{ 1,2,2,2,1,1,0\} $    \\
$ {\vec W}^{(19)} \, =\, $ &$  \{ 1,1,2,2,1,1,0\} $  & \null &
$ {\vec W}^{(20)} \, =\, $ &$  \{ 1,1,1,2,1,1,0\} $    \\
$ {\vec W}^{(21)} \, =\, $ &$  \{ 1,1,1,1,1,1,0\} $  & \null &
$ {\vec W}^{(22)} \, =\, $ &$  \{ 1,1,1,1,1,0,0\} $    \\
$ {\vec W}^{(23)} \, =\, $ &$  \{ 3,4,5,6,3,4,2\} $  & \null &
$ {\vec W}^{(24)} \, =\, $ &$  \{ 2,4,5,6,3,4,2\} $    \\
$ {\vec W}^{(25)} \, =\, $ &$  \{ 2,3,5,6,3,4,2\} $  & \null &
$ {\vec W}^{(26)} \, =\, $ &$  \{ 2,3,4,6,3,4,2\} $    \\
$ {\vec W}^{(27)} \, =\, $ &$  \{ 2,3,4,5,3,4,2\} $  & \null &
$ {\vec W}^{(28)} \, =\, $ &$  \{ 2,3,4,5,3,3,2\} $    \\
$ {\vec W}^{(29)} \, =\, $ &$  \{ 1,1,1,1,0,1,1\} $  & \null &
$ {\vec W}^{(30)} \, =\, $ &$  \{ 1,2,3,4,2,3,1\} $    \\
$ {\vec W}^{(31)} \, =\, $ &$  \{ 2,2,3,4,2,3,1\} $  & \null &
$ {\vec W}^{(32)} \, =\, $ &$  \{ 2,3,3,4,2,3,1\} $    \\
$ {\vec W}^{(33)} \, =\, $ &$  \{ 2,3,4,4,2,3,1\} $  & \null &
$ {\vec W}^{(34)} \, =\, $ &$  \{ 2,3,4,5,2,3,1\} $    \\
$ {\vec W}^{(35)} \, =\, $ &$  \{ 1,1,2,3,2,2,1\} $  & \null &
$ {\vec W}^{(36)} \, =\, $ &$  \{ 1,2,2,3,2,2,1\} $   \\
$ {\vec W}^{(37)} \, =\, $ &$  \{ 1,2,3,3,2,2,1\} $  & \null &
$ {\vec W}^{(38)} \, =\, $ &$  \{ 1,2,3,4,2,2,1\} $    \\
$ {\vec W}^{(39)} \, =\, $ &$  \{ 2,2,3,4,2,2,1\} $  & \null &
$ {\vec W}^{(40)} \, =\, $ &$  \{ 2,3,3,4,2,2,1\} $   \\
$ {\vec W}^{(41)} \, =\, $ &$  \{ 2,3,4,4,2,2,1\} $  & \null &
$ {\vec W}^{(42)} \, =\, $ &$  \{ 2,2,3,3,2,2,1\} $    \\
$ {\vec W}^{(43)} \, =\, $ &$  \{ 2,2,2,3,2,2,1\} $  & \null &
$ {\vec W}^{(44)} \, =\, $ &$  \{ 2,3,3,3,2,2,1\} $    \\
$ {\vec W}^{(45)} \, =\, $ &$  \{ 1,2,3,4,2,3,2\} $  & \null &
$ {\vec W}^{(46)} \, =\, $ &$  \{ 2,2,3,4,2,3,2\} $    \\
$ {\vec W}^{(47)} \, =\, $ &$  \{ 2,3,3,4,2,3,2\} $  & \null &
$ {\vec W}^{(48)} \, =\, $ &$  \{ 2,3,4,4,2,3,2\} $    \\
$ {\vec W}^{(49)} \, =\, $ &$  \{ 2,3,4,5,2,3,2\} $  & \null &
$ {\vec W}^{(50)} \, =\, $ &$  \{ 2,3,4,5,2,4,2\} $    \\
$ {\vec W}^{(51)} \, =\, $ &$  \{ 0,0,0,0,0,0,0\} $  & \null &
$ {\vec W}^{(52)} \, =\, $ &$  \{ 1,0,0,0,0,0,0\} $    \\
$ {\vec W}^{(53)} \, =\, $ &$  \{ 1,1,0,0,0,0,0\} $  & \null &
$ {\vec W}^{(54)} \, =\, $ &$  \{ 1,1,1,0,0,0,0\} $   \\
$ {\vec W}^{(55)} \, =\, $ &$  \{ 1,1,1,1,0,0,0\} $  & \null &
$ {\vec W}^{(56)} \, =\, $ &$  \{ 1,1,1,1,0,1,0\} $   \\
\hline
\end{tabular}
\end{center}
\end{table}
\begin{table}[ht]\caption{{\bf
Scalar products of weights and Cartan dilatons}:}
\label{scalaprod}
\begin{center}
\begin{tabular}{||rcl|rcl||}
\hline
\hline
\null & \null & \null & \null & \null & \null \\
$ {\vec \Lambda}^{(1)} \, \cdot \,   {\vec h} $ &=& $ {{-h_1  - h_2  - h_3  - h_4  - h_5  +
     h_6 }\over {{\sqrt{6}}}} $ &
$ {\vec \Lambda}^{(2)} \, \cdot \,   {\vec h} $ &=& $ {{-h_1  + h_2  + h_3  + h_4  + h_5  +
     h_6 }\over {{\sqrt{6}}}} $ \\
$ {\vec \Lambda}^{(3)} \, \cdot \,
  {\vec h} $ &=& $ {{h_1  - h_2  + h_3  + h_4  + h_5  + h_6 }\over {{\sqrt{6}}}} $ &
$ {\vec \Lambda}^{(4)} \, \cdot \,
  {\vec h} $ &=& $ {{h_1  + h_2  - h_3  + h_4  + h_5  + h_6 }\over {{\sqrt{6}}}} $ \\
$ {\vec \Lambda}^{(5)} \, \cdot \,
  {\vec h} $ &=& $ {{h_1  + h_2  + h_3  - h_4  + h_5  + h_6 }\over {{\sqrt{6}}}} $ &
$ {\vec \Lambda}^{(6)} \, \cdot \,
  {\vec h} $ &=& $ {{h_1  + h_2  + h_3  + h_4  - h_5  + h_6 }\over {{\sqrt{6}}}}  $\\
$ {\vec \Lambda}^{(7)} \, \cdot \,
  {\vec h} $ &=& $ {{-h_1  - h_2  + h_3  + h_4  + h_5  -
     h_6 }\over {{\sqrt{6}}}} $ &
$ {\vec \Lambda}^{(8)} \, \cdot \,   {\vec h} $ &=& $ {{-h_1  + h_2  - h_3  + h_4  + h_5  -
     h_6 }\over {{\sqrt{6}}}} $ \\
$ {\vec \Lambda}^{(9)} \, \cdot \,   {\vec h} $ &=& $ {{-h_1  + h_2  + h_3  - h_4  + h_5  -
     h_6 }\over {{\sqrt{6}}}} $ &
$ {\vec \Lambda}^{(10)} \, \cdot \,   {\vec h} $ &=& $ {{-h_1  + h_2  + h_3  + h_4  - h_5  -
     h_6 }\over {{\sqrt{6}}}} $ \\
$ {\vec \Lambda}^{(11)} \, \cdot \,   {\vec h} $ &=& $ {{h_1  - h_2  + h_3  + h_4  - h_5  -
     h_6 }\over {{\sqrt{6}}}} $ &
$ {\vec \Lambda}^{(12)} \, \cdot \,   {\vec h} $ &=& $ {{h_1  + h_2  - h_3  + h_4  - h_5  -
     h_6 }\over {{\sqrt{6}}}} $ \\
$ {\vec \Lambda}^{(13)} \, \cdot \,   {\vec h} $ &=& $ {{h_1  + h_2  + h_3  - h_4  - h_5  -
     h_6 }\over {{\sqrt{6}}}} $ &
$ {\vec \Lambda}^{(14)} \, \cdot \,   {\vec h} $ &=& $ {{h_1  - h_2  + h_3  - h_4  + h_5  -
     h_6 }\over {{\sqrt{6}}}} $ \\
$ {\vec \Lambda}^{(15)} \, \cdot \,   {\vec h} $ &=& $ {{h_1  - h_2  - h_3  + h_4  + h_5  -
     h_6 }\over {{\sqrt{6}}}} $ &
$ {\vec \Lambda}^{(16)} \, \cdot \,   {\vec h} $ &=& $ {{h_1  + h_2  - h_3  - h_4  + h_5  -
     h_6 }\over {{\sqrt{6}}}} $ \\
$ {\vec \Lambda}^{(17)} \, \cdot \,   {\vec h} $ &=& $ {{-\left( {\sqrt{2}}\,h_1  \right )  +
     h_7 }\over {{\sqrt{3}}}} $ &
$ {\vec \Lambda}^{(18)} \, \cdot \,   {\vec h} $ &=& $ {{-\left( {\sqrt{2}}\,h_2  \right )  +
     h_7 }\over {{\sqrt{3}}}} $ \\
$ {\vec \Lambda}^{(19)} \, \cdot \,   {\vec h} $ &=& $ {{-\left( {\sqrt{2}}\,h_3  \right )  +
     h_7 }\over {{\sqrt{3}}}} $ &
$ {\vec \Lambda}^{(20)} \, \cdot \,   {\vec h} $ &=& $ {{-\left( {\sqrt{2}}\,h_4  \right )  +
     h_7 }\over {{\sqrt{3}}}} $ \\
$ {\vec \Lambda}^{(21)} \, \cdot \,   {\vec h} $ &=& $ {{-\left( {\sqrt{2}}\,h_5  \right )  +
     h_7 }\over {{\sqrt{3}}}} $ &
$ {\vec \Lambda}^{(22)} \, \cdot \,   {\vec h} $ &=& $ {{{\sqrt{2}}\,h_6  +
h_7 }\over {{\sqrt{3}}}}$
  \\
$ {\vec \Lambda}^{(23)} \, \cdot \,   {\vec h} $ &=& $ -{{{\sqrt{2}}\,h_1  +
h_7 }\over {{\sqrt{3}}}} $
&
$ {\vec \Lambda}^{(24)} \, \cdot \,   {\vec h} $ &=& $ -{{{\sqrt{2}}\,h_2  + h_7
}\over {{\sqrt{3}}}} $
  \\
$ {\vec \Lambda}^{(25)} \, \cdot \,   {\vec h} $ &=& $ -{{{\sqrt{2}}\,h_3  + h_7 }
\over {{\sqrt{3}}}} $
&
$ {\vec \Lambda}^{(26)} \, \cdot \,   {\vec h} $ &=& $ -{{{\sqrt{2}}\,h_4  + h_7 }
\over {{\sqrt{3}}}} $
  \\
$ {\vec \Lambda}^{(27)} \, \cdot \,   {\vec h} $ &=& $ -{{{\sqrt{2}}\,h_5  + h_7 }
\over {{\sqrt{3}}}} $
&
$ {\vec \Lambda}^{(28)} \, \cdot \,   {\vec h} $ &=& $ {{{\sqrt{2}}\,h_6  - h_7 }
\over {{\sqrt{3}}}} $ \\
\null & \null & \null & \null & \null & \null \\
\hline
\hline
\end{tabular}
\end{center}
\end{table}
\begin{table}[ht]\caption{{\bf
The step operators of the $SU(8)$ subalgebra of $E_{7(7)}$:}}
\label{su8rutte}
\begin{center}
\begin{tabular}{||ccl|l||}
\hline
\hline
  $\#$ & Root & Root & SU(8) step operator in\\
       & name & vector & terms of $E_{7(7)}$ step oper. \\
\hline
1 & ${\vec a}_{1}$ & $ \left \{{-1, -1, 1, 1, 0, 0, 0} \right \} $
& $\cases { \null \cr
X^{a_{1}} =
2 \left ( B^{\alpha_{3,3} } + B^{\alpha_{3,4} }
- B^{\alpha_{4,3} } + B^{\alpha_{4,4} } \right )
\cr
Y^{a_{1}} =
2 \left ( B^{\alpha_{3,1} }
- B^{\alpha_{3,2} } + B^{\alpha_{4,5} } + B^{\alpha_{4,6} } \right )
\cr \null \cr}$ \\ \hline
 2& ${\vec a}_{2}$ & $
\left \{{0, 0, -1, -1, 1, 1, 0} \right \}
$
& $\cases { \null \cr
X^{a_{2}} =
2 \left ( B^{\alpha_{5,3} } +
B^{\alpha_{5,4} } - B^{\alpha_{6,3} } + B^{\alpha_{6,4} } \right )
\cr
Y^{a_{2}} =
2 \left ( B^{\alpha_{5,1} } - B^{\alpha_{5,2} }
+ B^{\alpha_{6,5} } + B^{\alpha_{6,6} } \right )
\cr \null \cr}$ \\ \hline
3& ${\vec a}_{3}$ & $
\left \{{1, 1, 1, 1, 0, 0, 0} \right \}
$
& $\cases { \null \cr
X^{a_{3}} =
2 \left ( B^{\alpha_{3,3} }
+ B^{\alpha_{3,4} } + B^{\alpha_{4,3} } - B^{\alpha_{4,4} } \right )
\cr
Y^{a_{3}} =
2 \left ( -B^{\alpha_{3,1} }
+ B^{\alpha_{3,2} } + B^{\alpha_{4,5} } + B^{\alpha_{4,6} } \right )
\cr \null \cr}$ \\ \hline
4& ${\vec a}_{4}$ & $
\left \{{-1, 0, -1, 0, -1, 0, -1} \right \}
$
& $\cases { \null \cr
X^{a_{4}} =
2 \left ( B^{\alpha_{2,3} }
+ B^{\alpha_{4,7} } - B^{\alpha_{6,12} } + B^{\alpha_{6,13} } \right )
\cr
Y^{a_{4}} =
2 \left ( B^{\alpha_{1,1} } + B^{\alpha_{4,10} } +
B^{\alpha_{6,22} } + B^{\alpha_{6,27} } \right )
\cr \null \cr}$ \\ \hline
5& ${\vec a}_{5}$ & $
\left \{{1, -1, 0, 0, 1, -1, 0} \right \}
$
& $\cases { \null \cr
X^{a_{5}} =
2 \left ( B^{\alpha_{5,5} } +
B^{\alpha_{5,6} } - B^{\alpha_{6,9} } + B^{\alpha_{6,10} } \right )
\cr
Y^{a_{5}} =
-2 \left ( B^{\alpha_{5,7} }
- B^{\alpha_{5,8} } + B^{\alpha_{6,7} } + B^{\alpha_{6,8} } \right )
\cr \null \cr}$ \\ \hline
6 & ${\vec a}_{6}$ & $
\left \{{0, 0, 1, -1, -1, 1, 0} \right \}
$
& $\cases { \null \cr
X^{a_{6}} =
2 \left ( -B^{\alpha_{5,1} } -
B^{\alpha_{5,2} } - B^{\alpha_{6,5} } + B^{\alpha_{6,6} } \right )
\cr
Y^{a_{6}} =
-2 \left ( -B^{\alpha_{5,3} }
+ B^{\alpha_{5,4} } + B^{\alpha_{6,3} } + B^{\alpha_{6,4} } \right )
\cr \null \cr}$ \\ \hline
7& ${\vec a}_{7}$ & $
\left \{{-1, 1, 0, 0, 1, -1, 0} \right \}
$
& $\cases { \null \cr
X^{a_{7}} =
2 \left ( B^{\alpha_{5,5} }
+ B^{\alpha_{5,6} } + B^{\alpha_{6,9} } - B^{\alpha_{6,10} } \right )
\cr
Y^{a_{7}} =
-2 \left ( -B^{\alpha_{5,7} }
+ B^{\alpha_{5,8} } + B^{\alpha_{6,7} } + B^{\alpha_{6,8} } \right )
\cr \null \cr}$ \\ \hline
8 & ${\vec a}_{12}$ & $
\left \{{-1, -1, 0, 0, 1, 1, 0} \right \}
$
& $\cases { \null \cr
X^{a_{12}} =
2 \left ( B^{\alpha_{5,7} } +
B^{\alpha_{5,8} } - B^{\alpha_{6,7} } + B^{\alpha_{6,8} } \right )
\cr
Y^{a_{12}} =
2 \left ( B^{\alpha_{5,5} } -
B^{\alpha_{5,6} } + B^{\alpha_{6,9} } + B^{\alpha_{6,10} } \right )
\cr \null \cr}$ \\ \hline
9 & ${\vec a}_{23}$ & $
\left \{{1, 1, 0, 0, 1, 1, 0} \right \}
$
& $\cases { \null \cr
X^{a_{23}} =
2 \left ( B^{\alpha_{5,7} } +
B^{\alpha_{5,8} } + B^{\alpha_{6,7} } - B^{\alpha_{6,8} } \right )
\cr
Y^{a_{23}} =
2 \left ( -B^{\alpha_{5,5} } +
B^{\alpha_{5,6} } + B^{\alpha_{6,9} } + B^{\alpha_{6,10} } \right )
\cr \null \cr}$ \\ \hline
10 & ${\vec a}_{34}$ & $
\left \{{0, 1, 0, 1, -1, 0, -1} \right \}
$
& $\cases { \null \cr
X^{a_{34}} =
2 \left ( B^{\alpha_{3,6} } +
B^{\alpha_{4,8} } - B^{\alpha_{6,24} } - B^{\alpha_{6,25} } \right )
\cr
Y^{a_{34}} =
2 \left ( -B^{\alpha_{3,5} } +
B^{\alpha_{4,9} } - B^{\alpha_{6,23} } + B^{\alpha_{6,26} } \right )
\cr \null \cr}$ \\ \hline
11 & ${\vec a}_{45}$ & $
\left \{{0, -1, -1, 0, 0, -1, -1} \right \}
$
& $\cases { \null \cr
X^{a_{45}} =
2 \left ( B^{\alpha_{5,11} }
+ B^{\alpha_{5,15} } - B^{\alpha_{6,17} } + B^{\alpha_{6,21} } \right )
\cr
Y^{a_{45}} =
-2 \left ( B^{\alpha_{5,12} } -
B^{\alpha_{5,16} } + B^{\alpha_{6,16} } + B^{\alpha_{6,20} } \right )
\cr \null \cr}$ \\ \hline
\hline
\end{tabular}
\end{center}
\end{table}
\par
\begin{table}[ht]\caption{{\bf
The step operators of the $SU(8)$...}{\sl continued $2^{nd}$}:}
\label{su8rutte2}
\begin{center}
\begin{tabular}{||ccl|l||}
\hline
\hline
  $\#$ & Root & Root & SU(8) step operator in\\
       & name & vector & terms of $E_{7(7)}$ step oper. \\
\hline
12 & ${\vec a}_{56}$ & $
\left \{ {1, -1, 1, -1, 0, 0, 0} \right \}
$
& $\cases { \null \cr
X^{a_{56}} =
2 \left ( B^{\alpha_{3,1} } +
B^{\alpha_{3,2} } - B^{\alpha_{4,5} } + B^{\alpha_{4,6} } \right )
\cr
Y^{a_{56}} =
-2 \left ( B^{\alpha_{3,3} } -
B^{\alpha_{3,4} } + B^{\alpha_{4,3} } + B^{\alpha_{4,4} } \right )
\cr \null \cr}$ \\ \hline
13 & ${\vec a}_{67}$ & $
\left \{{-1, 1, 1, -1, 0, 0, 0} \right \}
$
& $\cases { \null \cr
X^{a_{67}} =
2 \left ( B^{\alpha_{3,1} } +
B^{\alpha_{3,2} } + B^{\alpha_{4,5} } - B^{\alpha_{4,6} } \right )
\cr
Y^{a_{67}} =
-2 \left ( -B^{\alpha_{3,3} } + B^{\alpha_{3,4} }
+ B^{\alpha_{4,3} } + B^{\alpha_{4,4} } \right )
\cr \null \cr}$ \\ \hline
14 & ${\vec a}_{123}$ & $
\left \{{0, 0, 1, 1, 1, 1, 0} \right \}
$
& $\cases { \null \cr
X^{a_{123}} =
2 \left ( B^{\alpha_{5,3} } + B^{\alpha_{5,4} }
+ B^{\alpha_{6,3} } - B^{\alpha_{6,4} } \right )
\cr
Y^{a_{123}} =
2 \left ( -B^{\alpha_{5,1} } + B^{\alpha_{5,2} }
+ B^{\alpha_{6,5} } + B^{\alpha_{6,6} } \right )
\cr \null \cr}$ \\ \hline
15 & ${\vec a}_{234}$ & $
\left \{{0, 1, -1, 0, 0, 1, -1} \right \}
$
& $\cases { \null \cr
X^{a_{234}} =
2 \left ( B^{\alpha_{5,12} } + B^{\alpha_{5,16} }
+ B^{\alpha_{6,16} } - B^{\alpha_{6,20} } \right )
\cr
Y^{a_{234}} =
2 \left ( -B^{\alpha_{5,11} } + B^{\alpha_{5,15} }
+ B^{\alpha_{6,17} } + B^{\alpha_{6,21} } \right )
\cr \null \cr}$ \\ \hline
16 & ${\vec a}_{345}$ & $
\left \{{1, 0, 0, 1, 0, -1, -1} \right \}
$
& $\cases { \null \cr
X^{a_{345}} =
2 \left ( B^{\alpha_{5,9} } + B^{\alpha_{5,13} }
+ B^{\alpha_{6,15} } - B^{\alpha_{6,19} } \right )
\cr
Y^{a_{345}} =
-2 \left ( -B^{\alpha_{5,10} } + B^{\alpha_{5,14} }
+ B^{\alpha_{6,14} } + B^{\alpha_{6,18} } \right )
\cr \null \cr}$ \\ \hline
17 & ${\vec a}_{456}$ & $
\left \{{0, -1, 0, -1, -1, 0, -1} \right \}
$
& $\cases { \null \cr
X^{a_{456}} =
2 \left ( B^{\alpha_{3,6} } + B^{\alpha_{4,8} }
+ B^{\alpha_{6,24} } + B^{\alpha_{6,25} } \right )
\cr
Y^{a_{456}} =
2 \left ( B^{\alpha_{3,5} } - B^{\alpha_{4,9} }
- B^{\alpha_{6,23} } + B^{\alpha_{6,26} } \right )
\cr \null \cr}$ \\ \hline
18 & ${\vec a}_{567}$ & $
\left \{{0, 0, 1, -1, 1, -1, 0} \right \}
$
& $\cases { \null \cr
X^{a_{567}} =
2 \left ( B^{\alpha_{5,1} } + B^{\alpha_{5,2} }
- B^{\alpha_{6,5} } + B^{\alpha_{6,6} } \right )
\cr
Y^{a_{567}} =
-2 \left ( B^{\alpha_{5,3} } - B^{\alpha_{5,4} }
+ B^{\alpha_{6,3} } + B^{\alpha_{6,4} } \right )
\cr \null \cr}$ \\ \hline
\hline
\end{tabular}
\end{center}
\end{table}
\begin{table}[ht]\caption{{\bf
The step operators of the $SU(8)$...}{\sl continued $3^{rd}$}:}
\label{su8rutte3}
\begin{center}
\begin{tabular}{||ccl|l||}
\hline
\hline
  $\#$ & Root & Root & SU(8) step operator in\\
       & name & vector & terms of $E_{7(7)}$ step oper. \\
\hline
19 & ${\vec a}_{1234}$ & $
\left \{{-1, 0, 0, 1, 0, 1, -1} \right \}
$
& $\cases { \null \cr
X^{a_{1234}} =
2 \left ( B^{\alpha_{5,10} } +
B^{\alpha_{5,14} } + B^{\alpha_{6,14} } - B^{\alpha_{6,18} } \right )
\cr
Y^{a_{1234}} =
2 \left ( -B^{\alpha_{5,9} } +
B^{\alpha_{5,13} } + B^{\alpha_{6,15} } + B^{\alpha_{6,19} } \right )
\cr \null \cr}$ \\ \hline
20 & ${\vec a}_{2345}$ & $
\left \{{1, 0, -1, 0, 1, 0, -1} \right \}
$
& $\cases { \null \cr
X^{a_{2345}} =
2 \left ( -B^{\alpha_{1,1} } +
B^{\alpha_{4,10} } - B^{\alpha_{6,22} } + B^{\alpha_{6,27} } \right )
\cr
Y^{a_{2345}} =
-2 \left ( B^{\alpha_{2,3} } -
B^{\alpha_{4,7} } + B^{\alpha_{6,12} } + B^{\alpha_{6,13} } \right )
\cr \null \cr}$ \\ \hline
21 & ${\vec a}_{3456}$ & $
\left \{{1, 0, 1, 0, -1, 0, -1} \right \}
$
& $\cases { \null \cr
X^{a_{3456}} =
2 \left ( B^{\alpha_{2,3} } +
B^{\alpha_{4,7} } + B^{\alpha_{6,12} } - B^{\alpha_{6,13} } \right )
\cr
Y^{a_{3456}} =
2 \left ( -B^{\alpha_{1,1} } -
B^{\alpha_{4,10} } + B^{\alpha_{6,22} } + B^{\alpha_{6,27} } \right )
\cr \null \cr}$ \\ \hline
22 & ${\vec a}_{4567}$ & $
\left \{{-1, 0, 0, -1, 0, -1, -1} \right \}
$
& $\cases { \null \cr
X^{a_{4567}} =
2 \left ( B^{\alpha_{5,9} } +
B^{\alpha_{5,13} } - B^{\alpha_{6,15} } + B^{\alpha_{6,19} } \right )
\cr
Y^{a_{4567}} =
-2 \left ( B^{\alpha_{5,10} } -
B^{\alpha_{5,14} } + B^{\alpha_{6,14} } + B^{\alpha_{6,18} } \right )
\cr \null \cr}$ \\ \hline
23 & ${\vec a}_{12345}$ & $
\left \{{0, -1, 0, 1, 1, 0, -1} \right \}
$
& $\cases { \null \cr
X^{a_{12345}} =
2 \left ( B^{\alpha_{3,5} } +
B^{\alpha_{4,9} } + B^{\alpha_{6,23} } + B^{\alpha_{6,26} } \right )
\cr
Y^{a_{12345}} =
-2 \left ( B^{\alpha_{3,6} } -
B^{\alpha_{4,8} } - B^{\alpha_{6,24} } + B^{\alpha_{6,25} } \right )
\cr \null \cr}$ \\ \hline
24 & ${\vec a}_{23456}$ & $
\left \{{1, 0, 0, -1, 0, 1, -1} \right \}
$
& $\cases { \null \cr
X^{a_{23456}} =
2 \left ( B^{\alpha_{5,10} } +
B^{\alpha_{5,14} } - B^{\alpha_{6,14} } + B^{\alpha_{6,18} } \right )
\cr
Y^{a_{23456}} =
2 \left ( B^{\alpha_{5,9} } -
B^{\alpha_{5,13} } + B^{\alpha_{6,15} } + B^{\alpha_{6,19} } \right )
\cr \null \cr}$ \\ \hline
25 & ${\vec a}_{34567}$ & $
\left \{{0, 1, 1, 0, 0, -1, -1} \right \}
$
& $\cases { \null \cr
X^{a_{34567}} =
2 \left ( B^{\alpha_{5,11} } + B^{\alpha_{5,15} }
+ B^{\alpha_{6,17} } - B^{\alpha_{6,21} } \right )
\cr
Y^{a_{34567}} =
-2 \left ( -B^{\alpha_{5,12} } +
B^{\alpha_{5,16} } + B^{\alpha_{6,16} } + B^{\alpha_{6,20} } \right )
\cr \null \cr}$ \\ \hline
26 & ${\vec a}_{123456}$ & $
\left \{{0, -1, 1, 0, 0, 1, -1} \right \}
$
& $\cases { \null \cr
X^{a_{123456}} =
2 \left ( B^{\alpha_{5,12} } +
B^{\alpha_{5,16} } - B^{\alpha_{6,16} } + B^{\alpha_{6,20} } \right )
\cr
Y^{a_{123456}} =
2 \left ( B^{\alpha_{5,11} } -
B^{\alpha_{5,15} } + B^{\alpha_{6,17} } + B^{\alpha_{6,21} } \right )
\cr \null \cr}$ \\ \hline
27 & ${\vec a}_{234567}$ & $
\left \{{0, 1, 0, -1, 1, 0, -1} \right \}
$
& $\cases { \null \cr
X^{a_{234567}} =
2 \left ( -B^{\alpha_{3,5} } -
B^{\alpha_{4,9} } + B^{\alpha_{6,23} } + B^{\alpha_{6,26} } \right )
\cr
Y^{a_{234567}} =
2 \left ( -B^{\alpha_{3,6} } +
B^{\alpha_{4,8} } - B^{\alpha_{6,24} } + B^{\alpha_{6,25} } \right )
\cr \null \cr}$ \\ \hline
28 & ${\vec a}_{1234567}$ & $
\left \{{-1, 0, 1, 0, 1, 0, -1} \right \}
$
& $\cases { \null \cr
X^{a_{1234567}} =
2 \left ( -B^{\alpha_{1,1} } + B^{\alpha_{4,10} } +
B^{\alpha_{6,22} } - B^{\alpha_{6,27} } \right )
\cr
Y^{a_{1234567}} =
-2 \left ( -B^{\alpha_{2,3} } + B^{\alpha_{4,7} } +
B^{\alpha_{6,12} } + B^{\alpha_{6,13} } \right )
\cr \null \cr}$ \\ \hline
\hline
\end{tabular}
\end{center}
\end{table}
\begin{table}[ht]\caption{{\bf
Weights of the $28$ representation of $SU(8)$}:}
\label{28weights}
\begin{center}
\begin{tabular}{||cl|c|cl||}
\hline
\hline
   Weight & Weight & \null & Weight & Weight  \\
   name & vector & \null & name & vector  \\
\hline
\null & \null & {\bf (1,1,1)} & \null & \null \\
$ {\vec \Lambda}^{\prime(1)} \, = \, $ & $\{ 1,2,2,2,2,2,1\} $   & \null &
\null & \null \\
\hline
\null & \null & {\bf (1,1,15)} & \null & \null \\
$ {\vec \Lambda}^{\prime(2)} \, = \, $ & $\{ 0,0,0,1,1,2,1\} $   & \null &
$ {\vec \Lambda}^{\prime(3)} \, = \, $ & $\{ 0,0,0,1,1,1,1\} $    \\
$ {\vec \Lambda}^{\prime(4)} \, = \, $ & $\{ 0,0,0,1,1,1,0\} $   & \null &
$ {\vec \Lambda}^{\prime(5)} \, = \, $ & $\{ 0,0,1,2,2,2,1\} $     \\
$ {\vec \Lambda}^{\prime(6)} \, = \, $ & $\{ 0,0,0,1,2,2,1\} $   & \null &
$ {\vec \Lambda}^{\prime(7)} \, = \, $ & $\{ 0,0,0,0,0,1,0\} $    \\
$ {\vec \Lambda}^{\prime(8)} \, = \, $ & $\{ 0,0,0,0,0,0,0\} $   & \null &
$ {\vec \Lambda}^{\prime(9)} \, = \, $ & $\{ 0,0,0,0,0,1,1\} $     \\
$ {\vec \Lambda}^{\prime(10)} \, = \, $ & $\{ 0,0,1,1,1,2,1\} $   & \null &
$ {\vec \Lambda}^{\prime(11)} \, = \, $ & $\{ 0,0,1,1,1,1,1\} $     \\
$ {\vec \Lambda}^{\prime(12)} \, = \, $ & $\{ 0,0,1,1,1,1,0\} $   & \null &
$ {\vec \Lambda}^{\prime(13)} \, = \, $ & $\{ 0,0,0,0,1,1,1\} $     \\
$ {\vec \Lambda}^{\prime(14)} \, = \, $ & $\{ 0,0,0,0,1,2,1\} $   & \null &
$ {\vec \Lambda}^{\prime(15)} \, = \, $ & $\{ 0,0,0,0,1,1,0\} $     \\
$ {\vec \Lambda}^{\prime(16)} \, = \, $ & $\{ 0,0,1,1,2,2,1\} $   & \null &
\null & \null \\
\hline
\null & \null & {\bf (1,2,6)} & \null & \null \\
$ {\vec \Lambda}^{\prime(17)} \, = \, $ & $\{ 1,1,1,2,2,2,1\} $   & \null &
$ {\vec \Lambda}^{\prime(18)} \, = \, $ & $\{ 0,1,1,2,2,2,1\} $     \\
$ {\vec \Lambda}^{\prime(19)} \, = \, $ & $\{ 1,1,1,1,1,2,1\} $   & \null &
$ {\vec \Lambda}^{\prime(20)} \, = \, $ & $\{ 1,1,1,1,1,1,1\} $     \\
$ {\vec \Lambda}^{\prime(21)} \, = \, $ & $\{ 1,1,1,1,1,1,0\} $   & \null &
$ {\vec \Lambda}^{\prime(22)} \, = \, $ & $\{ 0,1,1,1,1,2,1\} $    \\
$ {\vec \Lambda}^{\prime(23)} \, = \, $ & $\{ 0,1,1,1,1,1,0\} $   & \null &
$ {\vec \Lambda}^{\prime(24)} \, = \, $ & $\{ 0,1,1,1,1,1,1\} $     \\
$ {\vec \Lambda}^{\prime(25)} \, = \, $ & $\{ 1,1,2,2,2,2,1\} $   & \null &
$ {\vec \Lambda}^{\prime(26)} \, = \, $ & $\{ 0,1,2,2,2,2,1\} $    \\
$ {\vec \Lambda}^{\prime(27)} \, = \, $ & $\{ 1,1,1,1,2,2,1\} $   & \null &
$ {\vec \Lambda}^{\prime(28)} \, = \, $ & $\{ 0,1,1,1,2,2,1\} $     \\
\hline
\end{tabular}
\end{center}
\end{table}
\begin{table}[ht]\caption{{\bf
Weights of the ${\bar 28}$ representation of $SU(8)$}:}
\label{28bweights}
\begin{center}
\begin{tabular}{||cl|c|cl||}
\hline
\hline
   Weight & Weight & \null & Weight & Weight  \\
   name & vector & \null & name & vector  \\
\hline
\null & \null & \null & \null & \null \\
\null & \null & ${\bf  {\overline {(1,1,1) }} }$  & \null & \null \\
$ {-\vec \Lambda}^{\prime(1)} \, = \, $ & $\{ 0,0,0,0,0,0,0\} $ & \null &
\null & \null \\
\hline
\null & \null & \null & \null & \null \\
\null & \null &  ${\bf  {\overline {(1,1,15)} } }$ & \null & \null \\
$ {-\vec \Lambda}^{\prime(2)} \, = \, $ & $\{ 1,2,2,1,1,0,0\} $ & \null &
$ {-\vec \Lambda}^{\prime(3)} \, = \, $ & $\{ 1,2,2,1,1,1,0\} $  \\
$ {-\vec \Lambda}^{\prime(4)} \, = \, $ & $\{ 1,2,2,1,1,1,1\} $ & \null &
$ {-\vec \Lambda}^{\prime(5)} \, = \, $ & $\{ 1,2,1,0,0,0,0\} $   \\
$ {-\vec \Lambda}^{\prime(6)} \, = \, $ & $\{ 1,2,2,1,0,0,0\} $ & \null &
$ {-\vec \Lambda}^{\prime(7)} \, = \, $ & $\{ 1,2,2,2,2,1,1\} $   \\
$ {-\vec \Lambda}^{\prime(8)} \, = \, $ & $\{ 1,2,2,2,2,2,1\} $ & \null &
$ {-\vec \Lambda}^{\prime(9)} \, = \, $ & $\{ 1,2,2,2,2,1,0\} $   \\
$ {-\vec \Lambda}^{\prime(10)} \, = \, $ & $\{ 1,2,1,1,1,0,0\} $ & \null &
$ {-\vec \Lambda}^{\prime(11)} \, = \, $ & $\{ 1,2,1,1,1,1,0\} $   \\
$ {-\vec \Lambda}^{\prime(12)} \, = \, $ & $\{ 1,2,1,1,1,1,1\} $ & \null &
$ {-\vec \Lambda}^{\prime(13)} \, = \, $ & $\{ 1,2,2,2,1,1,0\} $  \\
$ {-\vec \Lambda}^{\prime(14)} \, = \, $ & $\{ 1,2,2,2,1,0,0\} $ & \null &
$ {-\vec \Lambda}^{\prime(15)} \, = \, $ & $\{ 1,2,2,2,1,1,1\} $   \\
$ {-\vec \Lambda}^{\prime(16)} \, = \, $ & $\{ 1,2,1,1,0,0,0\} $ & \null &
\null & \null \\
\hline
\null & \null & \null & \null & \null \\
\null & \null & ${\bf  {\overline {(1,2,6)} } }$  & \null & \null \\
$ {-\vec \Lambda}^{\prime(17)} \, = \, $ & $\{ 0,1,1,0,0,0,0\} $ & \null &
$ {-\vec \Lambda}^{\prime(18)} \, = \, $ & $\{ 1,1,1,0,0,0,0\} $   \\
$ {-\vec \Lambda}^{\prime(19)} \, = \, $ & $\{ 0,1,1,1,1,0,0\} $ & \null &
$ {-\vec \Lambda}^{\prime(20)} \, = \, $ & $\{ 0,1,1,1,1,1,0\} $   \\
$ {-\vec \Lambda}^{\prime(21)} \, = \, $ & $\{ 0,1,1,1,1,1,1\} $ & \null &
$ {-\vec \Lambda}^{\prime(22)} \, = \, $ & $\{ 1,1,1,1,1,0,0\} $   \\
$ {-\vec \Lambda}^{\prime(23)} \, = \, $ & $\{ 1,1,1,1,1,1,1\} $ & \null &
$ {-\vec \Lambda}^{\prime(24)} \, = \, $ & $\{ 1,1,1,1,1,1,0\} $   \\
$ {-\vec \Lambda}^{\prime(25)} \, = \, $ & $\{ 0,1,0,0,0,0,0\} $ & \null &
$ {-\vec \Lambda}^{\prime(26)} \, = \, $ & $\{ 1,1,0,0,0,0,0\} $   \\
$ {-\vec \Lambda}^{\prime(27)} \, = \, $ & $\{ 0,1,1,1,0,0,0\} $ & \null &
$ {-\vec \Lambda}^{\prime(28)} \, = \, $ & $\{ 1,1,1,1,0,0,0\} $   \\
\hline
\hline
\end{tabular}
\end{center}
\end{table}

\end{document}